\documentclass[11pt,a4paper]{article}
\pdfoutput=1

\usepackage{jheppub}
\usepackage{epstopdf}
\usepackage[T1]{fontenc}
\usepackage{extarrows}

\usepackage{graphicx}
\usepackage{cancel}
\usepackage{amsmath}

\def\beq{\begin{equation}}

\def\eeq{\end{equation}}

\newcommand{\bea}{\begin{eqnarray}}

\newcommand{\eea}{\end{eqnarray}}

\parskip=\itemsep             

\newcommand{\beqar}[1]{\begin{eqnarray}\label{#1}}

\newcommand{\eeqar}{\end{eqnarray}}

\title{High Energy QCD at NLO: from light-cone wave function to JIMWLK evolution.}
\author[a,b]{Michael Lublinsky}
\author[a]{and Yair Mulian}
\affiliation[a]{Department of Physics, Ben-Gurion University of the Negev,
Beer-Sheva 84105, Israel}
\affiliation[b]{Physics Department, University of Connecticut, 2152 Hillside
Road, Storrs, CT 06269-3046, USA}

\emailAdd{lublinm@bgu.ac.il}
\emailAdd{yair25m@gmail.com}

\abstract{

Soft components of the light cone wave-function of a fast moving projectile hadron is computed in perturbation theory to the third order in QCD coupling constant. At this order, the Fock space of the soft modes consists of one-gluon,  two-gluon, and a quark-antiquark states. The hard component of the wave-function acts as a non-Abelian background field for the soft modes and is represented by a valence charge distribution that accounts for non-linear density effects in the projectile. When scattered off a dense target, the diagonal element of the $S$-matrix reveals the Hamiltonian of high energy evolution, the JIMWLK Hamiltonian. This way we  provide a new  direct derivation of the JIMWLK Hamiltonian at the Next-to-Leading Order.

}


\begin{document}
\maketitle

\flushbottom

\bibliographystyle{JHEP}

\section{Introduction and Summary}
The phenomenon  of perturbative saturation is believed to occur in high energy hadronic collisions
and is most pronounced in processes involving nuclei. The idea dates back to the seminal paper 
\cite{GLR}. It has been long known theoretically \cite{BFKL}  that at high energy gluonic density in a hadron  grows exponentially. It was suggested in \cite{GLR} that this  growth does not continue indefinitely, but instead the gluon phase space density saturates when it reaches a critical value of order $1/\alpha_s$.  

In the years since the basic idea has been proposed the theoretical understanding of saturation has progressed significantly \cite{dipole,MV}. We now have a first principles QCD formulation of high energy evolution, including the saturation effects (see e.g. \cite{kovlevin,cgc2}). The basic framework was derived in \cite{balitsky} -- the Balitsky's hierarchy, and in a series of  papers \cite{JIMWLK} -- the   Jalilian-Marian, Iancu, McLerran, Weigert, Leonidov, Kovner (JIMWLK) equation or Colour Glass Condensate (CGC). 

The JIMWLK Hamiltonian predicts energy evolution of a hadronic observable $\cal O$ via  functional equation of the form 
\begin{equation}
\frac{d\cal O}{dY} \,=\,-\,H_{JIMWLK}\,{\cal O}\,.
\end{equation}
Here the rapidity $Y\sim \ln \,(energy)$. 
The JIMWLK Hamiltonian is  applicable when one of the colliding particles is dilute (small parton number), such as  in forward production in pA collisions  at the LHC
or in  deep inelastic scattering (DIS).  While  in the dilute-dilute limit, the JIMWLK equation reduces to the linear BFKL equation \cite{BFKL} and its BKP extension \cite{BKP}, it is free from the famous infrared diffusion problem. 
Furthermore, the major success associated with the JIMWLK is in that it  seemingly
restores the $s$-channel unitarity of scattering amplitudes (see, however, \cite{meunitarity} where the $s$-channel unitarity of the JIMWLK was argued to be incomplete). 

Phenomenology based on complete LO JIMWLK has  emerged  recently \cite{KRW,KKRW,Langevin,lang} thanks to its reformulation in terms of stochastic
 Langevin equation \cite{Langevin}. 
Yet, for most phenomenological applications JIMWLK is commonly replaced by  the Balitsky-Kovchegov (BK) equation \cite{balitsky,Kovchegov,KOV2}, 
which is a mean field/large $N_c$ reduction of the JIMWLK. 
During the last decade, these developments have become a basis for phenomenological studies of saturation physics applied to 
high energy/low  $x$  collision data. 
 
HERA  has accumulated a vast amount of precise  data
 in the low $x$ region,  which is particularly useful  for  calibration of  gluon distributions used as an input for calculations of  less inclusive processes at RHIC and LHC.   
 Several generations of successful saturation-based fits to low $x$ HERA data are available, from the first work \cite{lublinsky}, to the latest state of the art AAMQS phenomenology \cite{AAMSQ}. These provide comprehensive good quality description of the low x data, including the latest available combined H1-ZEUS set \cite{HERA}.   
 However, HERA  was not  convincing  in experimentally  confirming the saturation phenomenon.

Even though the main features of CGC are qualitatively  understood,  quantitative description  is still at an unsatisfactory  level as most of the studies 
have been performed to the lowest order  (LO) in $\alpha_s$.  It is, however,  known that higher order corrections could be very significant and their 
inclusion crucial for qualitatively reliable phenomenology.  Calculational control enables  to discriminate between  different approaches, some of which 
describe experimental data but do not involve saturation,  is critically important and   is the main challenge.
Precision   mandates inclusion of next-to-leading order (NLO) corrections to evolution  and improvement of fixed order calculations for a variety of 
observables.  

The era of  saturation-based phenomenology with NLO precision just started.  Among recent progress in fixed order NLO calculations it is important to 
mention  the virtual photon impact factor computed  in \cite{NLOIF,NLOIF1,NLOIF2} and a series of papers on forward particle production in pA  collisions \cite{CXY,Ioffe} (see  e.g \cite{Stastorev} for a review).  
The full set of NLO corrections to BK equation was completed in \cite{BC}  following on the  earlier works  \cite{BalNLO,BalNLO1}.  When linearised,
the NLO BK was shown to coincide with the NLO BFKL computed over a decade earlier \cite{NLOBFKL}.
Running coupling effects are known to be phenomenologically important. While they do not account for all the NLO effects in the evolution,  
they do carry important information about  high order corrections.  Those were analysed in \cite{BalNLO,Weigertrun,Gardi}. Beyond the NLO accuracy,
the BFKL was considered in \cite{CK,Simon2}.

The NLO extension of the JIMWLK Hamiltonian is imperative for calculation of more general amplitudes, beyond the quark dipole of the BK equation, 
which determine important  experimental observables like single- and double-inclusive particle production.  Moreover, precision 
also mandates going beyond the large $N_c$ approximation of the BK equation, which is realised by the JIMWLK.
The NLO JIMWLK Hamiltonian was obtained  in \cite{nlojimwlk,nlojimwlk1,nlojimwlk2} 
building upon the calculations of \cite{BC,Grab}, and in  \cite{BClast} 
by a direct calculation extending \cite{BC}.  A conformal part of the Hamiltonian  was also deduced in \cite{Simon}. 
The NLO JIMWLK contains information both about  the NLO BKP \cite{NLOBKP, BalGrab} and on $\alpha_s$ corrections to the triple Pomeron vertex.

While  improved  equations are known,  the key conceptual question  remains to be
  stability of the $\alpha_s$ expansion. In the framework of the linear BFKL 
equation the NLO corrections are big  \cite{NLOBFKL} and even  may  lead to negative cross sections.   
Numerical calculations  \cite{Stasto} of  forward particle production in pA collisions done in the hybrid formalism of 
\cite{hybrid} and including  NLO
corrections \cite{CXY} displayed similar pathology.
The recent study of the NLO BK \cite{lappi} indicates that the problem is not cured by the non-linear effects.
The origin of the problem is understood reasonably well:  the LO resummation picks up enhanced logarithmic
contributions from the  region in the gluon emission phase space, where the underlying approximations  break down.
These spurious contributions are subtracted in higher orders, thus leading to large negative corrections.
To stabilise the expansion, one is normally forced to perform additional collinear resummations  
introducing  kinematical constraints \cite{Kutak,Motyka,Beuf,Iancucons,SabioVera} or more involved schemes compatible with the 
DGLAP equation \cite{dglap}. It is clear that a good understanding of a systematic approach to rectifying this problem is urgently needed for 
reliable phenomenological applications of the NLO BK-JIMWLK framework.

In this paper, we report  our contribution to the effort in establishing NLO accuracy in saturation physics. 
Our results are split into two parts. In the first part, the light cone wave function (LCWF) of  a fast moving hadron  is computed.
More precisely, in a typical Born-Oppenheimer approximation,  the Hilbert space of a fast moving projectile is split
into a ``valence'' and a ``soft'' sectors. The valence gluons have longitudinal momenta greater than some ''cutoff" implicitly defined 
by the collision energy.  In the LCWF, these  gluons are  characterised by correlators 
of the colour charge densities $\rho^a(x)$ ($x$ is a transverse coordinate).

Soft gluons with momenta  smaller than the cutoff  do not participate in scattering.    Increasing the energy of the hadron increases longitudinal momenta of the soft gluons and they emerge from below the cutoff. Having done so they contribute to physical observables, such as scattering amplitude, particle production {\it etc}.
This "soft" component of the LCWF at NLO schematically has the form 
\begin{eqnarray}\label{psi}
|\psi\rangle&=& (1\,-\,g_s^2\,\kappa_0\, \rho\rho\,-\,g_s^4(\delta_1\,\rho\rho\,+\,\delta_2\,\rho\rho\rho\,+\,\delta_3\,\rho\rho\rho\rho)\,|\,no\,soft\, gluons \rangle +
\nonumber \\
&+&(\,g_s \kappa_1\,\rho\,+\,g_s^3\epsilon_1\, \rho\,+\, g_s^3\,\epsilon_2\,\rho\rho\,+\, g_s^3\,\epsilon_3\,\rho\rho\rho)\,|\,one\, soft \,gluon\rangle
\nonumber \\
&+&g_s^2 (\epsilon_4\, \rho\,\,+\,\epsilon_5\, \rho\rho)\,|\,two\, soft \,gluons\rangle
+g_s^2\,\epsilon_6\,\rho\,|\,quark-antiquark\rangle\,.
\end{eqnarray}
The coefficient $\kappa_1$ is nothing else but the LO Weizsacker Williams gluon emission vertex. 
Below we compute all the remaining coefficients  in the third order perturbation theory in which the valence charges $\rho^a$ act as background fields. 
A major complexity related to this calculation  is that these charges  are non-commuting  operators on the valence Hilbert space 
forming a local $SU(N_c)$ algebra. Neither commute 
the matrix elements of perturbation operators averaged over the soft states.  As a result, we had to rederive the perturbation theory while carefully controlling over
the  ordering between the matrix elements.  Most of the coefficients in the first line in (\ref{psi}) are computed from the normalisation of the state. 
Yet, $\delta_2$ turns out to be a tricky one: it vanishes in perturbation theory.   In our calculation it emerges as a phase of the wavefunction, which could be 
determined from a constraint on applicability of the Born-Oppenheimer approximation. This phase is somewhat similar to the Berry phase, though its nature
is apparently rooted in the non-commutativity of $\rho$ rather than being sourced by any topology. 

While our calculation and presentation of the results below are fully self-contained, admittedly some of them are not new and could be 
located  in the 
literature. Particularly, the entire quark sector has been worked out in \cite{Weigertrun} in a formalism that is identical to ours.  The coefficients in the second line in (\ref{psi})
are related  to single inclusive gluon production and we believe that  it might be possible to relate $\epsilon_2$ with partial result of \cite{CKW}. The comparison is however not straightforward and we have not pursued it. We also believe that some of our results could be extracted from the LCWF of a dense projectile \cite{foam}, in which 
$\alpha_s$ corrections enhanced by the density $\rho$ were resummed. 

There are many advantages in having the complete LCWF at NLO. 
First,  it provides an alternative  method for derivation of the NLO JIMWLK Hamiltonian. 
Given the complexity of the latter, it is critical to have an independent cross check of the results before moving forward with any phenomenological applications.
This constitutes the second part of our paper. Second, the LCWF formalism is possibly a good starting point to reformulate the NLO JIMWLK as stochastic evolution
suitable for numeric simulations.  Finally, the formalism makes it possible to simultaneously address NLO corrections   to the Hamiltonian and to semi-inclusive observables, such as single inclusive production.

As was mentioned above, with increase of collision energy,  the soft modes emerge above the cutoff and contribute to scattering cross sections.
Thus they are naturally identified with the source of high energy evolution \cite{KLremarks}.  Let's define the projectile averaged $\hat S$-matrix 
\begin{equation}\label{Sigma}
\Sigma\,\equiv\, \langle \psi| \hat S \,-\,1|\psi\rangle
\end{equation}
The averaging is over the soft modes only leaving $\Sigma[\rho]$ as operator on the valence Hilbert space.  In fact,  $\Sigma$ is a rapidity evolution operator
\begin{equation}\label{opr}
\Sigma\,=\, e^{-\delta Y\,H_{JIMWLK}}\,-1\,\simeq \,-\,\delta Y\,H_{JIMWLK} \,+\,{1\over 2} \,\delta Y^2 \,H_{JIMWLK}^2\, \ldots
\end{equation}
Here $\delta Y$ is a rapidity interval associated with the longitudinal phase space of the soft modes.
The JIMWLK Hamiltonian is expandable in $\alpha_s$:
\begin{equation}
H_{JIMWLK} \,= \,H^{LO}_{JIMWLK}(\alpha_s) \,+\, H^{NLO}_{JIMWLK}(\alpha_s^2) \, +\ldots
\end{equation}
Computing (\ref{Sigma}) with the LCWF (\ref{psi}) provides a direct method to determine $H^{NLO}_{JIMWLK}$ defined as
\begin{equation}\label{defH}
H_{JIMWLK}\,\equiv\,-\, {d\Sigma\over d\,\delta Y}\,|_{{\delta Y=0}}
\end{equation}
This is, however, not how $H^{NLO}_{JIMWLK}$
was obtained in \cite{nlojimwlk,nlojimwlk1,nlojimwlk2}. 
There, we only assumed the  structure (\ref{psi}) without computing any of the coefficients. This structure of the LCWF 
helped us to identify the most general form of $H^{NLO}_{JIMWLK}$.  More constraints on the form of the Hamiltonian came from the symmetries of the theory.
As discussed in detail in \cite{KLreggeon}, the theory must have $SU_L(N)\times SU_R(N)$ symmetry, which in QCD terms is the gauge symmetry of $|in\rangle$ and 
$|out\rangle$ states and two discrete symmetries: the charge conjugation, and another $Z_2$ symmetry, which in \cite{KLreggeon} was identified with signature, and can be understood as the combination of charge conjugation and time reversal symmetry \cite{iancutri}.

The Hamiltonian in \cite{nlojimwlk} was parametrised by five kernels. They were initially fixed by demanding that the evolution equations for the
quark dipole and SU(3) baryon generated by this Hamiltonian match the ones known from the literature \cite{BC,Grab}. The Hamiltonian thus constructed 
was suitable for evolution of gauge invariant operators only. It was  later amended in \cite{nlojimwlk2} to also include  non-gauge invariant pieces. Interestingly, the terms
in the Hamiltonian that originate from the previously discussed $\delta_2$ were found in \cite{nlojimwlk1} thanks to a constraint on conformal invariance of 
the JIMWLK Hamiltonian in ${\cal N}=4$.  We quote these results for the Hamiltonian in Section \ref{sdfla}.  It will become a benchmark for comparison with our direct 
calculation.

Below we compute (\ref{Sigma}) with the LCWF (\ref{psi}).  At  $\alpha_s^2$ order, we find terms that are proportional to $\delta Y$ and terms
that are proportional to $\delta Y^2$. The former are identified with the NLO JIMWLK Hamiltonian (\ref{defH})
and we find a complete agreement with our 
previous results. The latter are naturally identified with $(H^{LO}_{JIMWLK})^2$, which is a non-trivial cross check on our calculation.

This paper is organised as follows. Section 2  provides an introductory overview of the wavefunctional formalism.
We start with presenting the QCD Hamiltonian in the light cone gauge, introduce field quantisation and then discuss the eikonal approximation
for the LC Hamiltonian  and scattering matrix.  To prepare the stage, we derive the LO JIMWLK Hamiltonian and also quote the previous result for 
the NLO JIMWLK.  The LCWF is computed in Section 3. The reader who is interested in the end result only may skip most of the section and proceed 
directly to the  final result summarised in Section \ref{finres}.  
Section 4 is devoted to the calculation of $\Sigma$ and extraction of the NLO Hamiltonian. 
Several Appendices contain complementary materials and details of the calculations.

\section{Basics of JIMWLK}\label{sectio2}

In this section, we review the main elements of high energy QCD in the light cone Hamiltonian formalism.  

\subsection{Light Cone QCD Hamiltonian}
Our starting point is  QCD Hamiltonian  in  light-cone gauge \cite{bl,rajj,ligh1,ligh2,ligh4}. In  light-cone coordinates,  four-vectors are $x^{\mu}=\left(x^{+},\, x^{-},\, \mathbf{x}\right)$, where $x^{+}\,\equiv\, x^{0}\,+\, x^{3}$ and $x^{-}\,\equiv\, x^{0}\,-\, x^{3}$ stand for longitudinal, while $\mathbf{x}=\left(x_{1},\, x_{2}\right)$ for transverse components. We adopt mostly negative metric \cite{bl}. The light-cone gauge:
\begin{equation}
A^{a+}=A^{a0}+A^{a3}=0.
\end{equation}
Derivation of the  Hamiltonian in the light cone gauge can be found in many reviews and original publications. For self-completeness of our presentation
the derivation  is briefly sketched in Appendix \ref{hamder}:
\begin{equation}\begin{split}\label{qcdhamii}
&H_{LC\; QCD}\\
&\quad =\int dx^{-}d^{2}\mathbf{x}\left(\frac{1}{2}\Pi^{a}(x^{-},\, \mathbf{x})\,\Pi^{a}(x^{-},\, \mathbf{x})\,+\,\frac{1}{4}F_{ij}^{a}(x^{-},\, \mathbf{x})\, F_{ij}^{a}(x^{-},\, \mathbf{x})\,+\, i\overline{\psi}\gamma^{+}D_{+}\psi\right),\\
  \end{split}\end{equation}
 where  the electric and magnetic pieces have the form:
 \begin{equation}\begin{split}
&\Pi^{a}(x^{-},\,\mathbf{x})\,\equiv\,-\,\frac{1}{\partial^{+}}(D_{i}^{ab}\partial^{+}A_{i}^{b}\,-\,2g\psi_{+}^{\dagger}t^{a}\psi_{+}),\\
&F_{ij}^{a}(x^{-},\,\mathbf{x})\,\equiv\,\partial_{i}A_{j}^{a}\,-\,\partial_{i}A_{j}^{a}\,-\, gf^{abc}A_{i}^{b}A_{j}^{c}.\\
\end{split}\end{equation}
$i\,\in\,(1,\,2)$ is a transverse component index; $\psi$ denotes  Dirac's 4-component quark spinor\footnote{The flavour index is suppressed in this section.}, while
$\psi_+$ is a corresponding 2-component spinor (see Appendix  \ref{hamder}). After substitution of $\Pi^{a}(x^{-},\, \mathbf{x})$ and $F_{ij}^{a}(x^{-},\, \mathbf{x})$ in $H_{LC\; QCD}$, the result can be written as  $H_{LC\; QCD}=H_{0}+H_{int}$, with free Hamiltonian $H_{0}$ given by:
\begin{equation}\begin{split}\label{hzero}
&H_{0}\,\equiv\,\int dx^{-}\, d^{2}\mathbf{x}\,\left(\frac{1}{2}(\partial_{i}A_{j}^{a})^{2}\,+\, i\psi_{+}^{\dagger}\frac{\partial_{i}\partial_{i}}{\partial^{+}}\psi_{+}\right).\\
  \end{split}\end{equation}
The interaction Hamiltonian $H_{int}$ reads
\begin{eqnarray}\label{hint}
H_{int}&&\equiv\,\int dx^{-}\, d^{2}\mathbf{x}\,\left(-gf^{abc}A_{i}^{b}A_{j}^{c}\partial_{i}A_{j}^{a}\,+\,\frac{g^{2}}{4}f^{abc}f^{ade}A_{i}^{b}A_{j}^{c}A_{i}^{d}A_{j}^{e}\,\right.\nonumber\\
&&-\, gf^{abc}(\partial_{i}A_{i}^{a})\frac{1}{\partial^{+}}(A_{j}^{b}\partial^{+}A_{j}^{c})\,+\,\frac{g^{2}}{2}f^{abc}f^{ade}\frac{1}{\partial^{+}}(A_{i}^{b}\partial^{+}A_{i}^{c})\frac{1}{\partial^{+}}(A_{j}^{d}\partial^{+}A_{j}^{e})\\
&&\,+\,2g^{2}f^{abc}\frac{1}{\partial^{+}}(A_{i}^{b}\partial^{+}A_{i}^{c})\frac{1}{\partial^{+}}(\psi_{+}^{\dagger}t^{a}\psi_{+})+\,2g^{2}\frac{1}{\partial^{+}}(\psi_{+}^{\dagger}t^{a}\psi_{+})\frac{1}{\partial^{+}}(\psi_{+}^{\dagger}t^{a}\psi_{+})\nonumber\\
&&\,-\,2g(\partial_{i}A_{i}^{a})\frac{1}{\partial^{+}}(\psi_{+}^{\dagger}t^{a}\psi_{+})\,-\, g\psi_{+}^{\dagger}t^{a}(\sigma_{i}\partial_{i})\frac{1}{\partial^{+}}(\sigma_{j}A_{j}^{a}\psi_{+})-\, g\psi_{+}^{\dagger}t^{a}\sigma_{i}A_{i}^{a}\frac{1}{\partial^{+}}(\sigma_{j}\partial_{j}\psi_{+})\nonumber\\
&&\left.\,-\, ig^{2}\psi_{+}^{\dagger}t^{a}t^{b}\sigma_{i}A_{i}^{a}\frac{1}{\partial^{+}}(\sigma_{j}A_{j}^{b}\psi_{+})\right).\nonumber\end{eqnarray}
$\sigma^{i}$ are  Pauli matrices, the matrices $t^{a}$ denote the $SU(N_{c})$ gauge group generators in  fundamental representation (the gauge group generators in  adjoint representation will be denoted by $T^{a}$), which obey the algebra $\left[t^{a},t^{b}\right]=if^{abc}t^{c}$, where $f^{abc}$ are 
structure constants of the gauge group.

\subsection{Field Quantisation}

Quantisation of the fields is performed in  usual manner introducing creation/annihilation operators and imposing commutation (anti-commutation) relations among them.
For the gauge fields:
\begin{equation}\label{glufield}
A_{i}^{a}(x)=\int_{0}^{\infty}\frac{dk^{+}}{2\pi}\int\frac{d^{2}\mathbf{k}}{(2\pi)^{2}}\frac{1}{\sqrt{2k^{+}}}\left(a_{i}^{a}(k^{+},\mathbf{k})e^{-ik\cdot x}+a_{i}^{a\dagger}(k^{+},\mathbf{k})e^{ik\cdot x}\right).
  \end{equation}
The creation and annihilation operators obey the bosonic algebra:
\begin{equation}
\left[a_{i}^{a}(k^{+},\mathbf{k}),\, a_{j}^{b\dagger}(p^{+},\mathbf{p})\right]=(2\pi)^{3}\delta^{ab}\delta_{ij}\delta(k^{+}-p^{+})\delta^{(2)}(\mathbf{k}-\mathbf{p}).
 \end{equation}
Transforming to coordinate space,
     \begin{equation}\begin{split}\label{tran}
a_{i}^{a}(k^{+},\, \mathbf{k})=\int_{\mathbf{z}}e^{-i\mathbf{k}\cdot \mathbf{z}}\, a_{i}^{a}(k^{+},\, \mathbf{z})\;,\qquad\qquad a_{i}^{a\dagger}(k^{+},\, \mathbf{k})=\int_{\mathbf{z}}e^{i\mathbf{k}\cdot \mathbf{z}}\, a_{i}^{a\dagger}(k^{+},\, \mathbf{z}),
 \end{split}\end{equation}
the commutation relation becomes:
 \begin{equation}\label{composw}
 \left[a_{i}^{a}(k^{+},\mathbf{x}),\, a_{j}^{b\dagger}(p^{+},\mathbf{y})\right]=2\pi\delta^{ab}\delta_{ij}\delta(k^{+}-p^{+})\delta^{(2)}(\mathbf{x}-\mathbf{y}).
  \end{equation}
For the quark fields:
\begin{equation}\label{quarfield}
\psi_{+}^{\alpha}(x)=\sum_{\lambda=\pm\frac{1}{2}}\chi_{\lambda}\int_{0}^{\infty}\frac{dk^{+}}{2\pi}\int\frac{d^{2}\mathbf{k}}{(2\pi)^{2}}\frac{1}{\sqrt{2}}\left(b_{\lambda}^{\alpha}(k^{+},\mathbf{k})e^{-ik\cdot x}+d_{\lambda}^{\alpha\dagger}(k^{+},\mathbf{k})e^{ik\cdot x}\right)
 \end{equation}
The polarisation vectors are:
\begin{equation}\chi_{+\frac{1}{2}}\,=\,\left(\begin{array}{c}
1\\
0
\end{array}\right),\qquad\quad\chi_{-\frac{1}{2}}\,=\,\left(\begin{array}{c}
0\\
1
\end{array}\right),
 \end{equation}
 \begin{equation}
 \chi_{\lambda_{1}}^{\dagger}\, I\,\chi_{\lambda_{2}}\,=\,\delta_{\lambda_{1}\lambda_{2}},\qquad\quad\chi_{\lambda_{1}}^{\dagger}\,\sigma^{3}\,\chi_{\lambda_{2}}\,=\,2\lambda_{1}\delta_{\lambda_{1}\lambda_{2}}.
 \end{equation}
The anti-commutation relations:
 \begin{equation}\begin{split}
\left\{ b_{\lambda_{1}}^{\alpha}(k^{+},\,\mathbf{k}),\, b_{\lambda_{2}}^{\beta\dagger}(p^{+},\,\mathbf{p})\right\} &=\,\left\{ d_{\lambda_{1}}^{\alpha}(k^{+},\,\mathbf{k}),\, d_{\lambda_{2}}^{\beta\dagger}(p^{+},\,\mathbf{p})\right\} \\
&=\,(2\pi)^{3}\,\delta_{\lambda_{1}\lambda_{2}}\,\delta^{\alpha\beta}\,\delta^{(2)}(\mathbf{k}-\mathbf{p})\,\delta(k^{+}-p^{+}).\\
 \end{split}\end{equation}
 Transforming the fields to coordinate space a la (\ref{tran}):
 \begin{equation}\begin{split}
\left\{ b_{\lambda_{1}}^{\alpha}(k^{+},\,\mathbf{x}),\, b_{\lambda_{2}}^{\beta\dagger}(p^{+},\,\mathbf{y})\right\} &=\,\left\{ d_{\lambda_{1}}^{\alpha}(k^{+},\,\mathbf{x}),\, d_{\lambda_{2}}^{\beta\dagger}(p^{+},\,\mathbf{y})\right\} \,\\
&=\,2\pi\,\delta_{\lambda_{1}\lambda_{2}}\,\delta^{\alpha\beta}\,\delta^{(2)}(\mathbf{x}-\mathbf{y})\,\delta(k^{+}-p^{+}).
 \end{split}\end{equation}
Inserting the field expansions (\ref{glufield}) and (\ref{quarfield}) into (\ref{hzero}), the free  Hamiltonian  becomes:
  \begin{equation}\begin{split}
 H_{0}=\int_{0}^{\infty}\frac{dk^{+}}{2\pi}&\int\frac{d^{2}\mathbf{k}}{(2\pi)^{2}}\frac{\mathbf{k}^{2}}{2k^{+}}\bigg(a_{i}^{a\dagger}(k^{+},\mathbf{k})\, a_{i}^{a}(k^{+},\mathbf{k})\\
&\left.+\sum_{\lambda}\,\left[b_{\lambda}^{\alpha\dagger}(k^{+},\, \mathbf{k})\, b_{\lambda}^{\alpha}(k^{+},\, \mathbf{k})\,-\, d_{\lambda}^{\alpha}(k^{+},\, \mathbf{k})\, d_{\lambda}^{\alpha\dagger}(k^{+},\, \mathbf{k})\right]\right),\\
  \end{split}\end{equation}
from which the dispersion relation for free quarks and gluons is $E_{k}=\frac{\mathbf{k}^{2}}{2k^{+}}$.
 The interacting part of the Hamiltonian will be quantised  in subsection \ref{eikhami}.

  \subsection{Light Cone Wave-Function of a Fast Hadron}\label{softvalence}
  
Consider a fast moving projectile hadron at some rapidity $Y_0$. Its LCWF $\left|\Psi\right\rangle _{\mathsf{Y_0}}$  is an eigenfunction of $H_{LC\,QCD}$.
It is convenient to split the modes in the LCWF in accord to their longitudinal momenta, introducing a longitudinal cutoff 
$\Lambda$, which is implicitly related to the energy of the target hadron (see  \cite{kovnerbasic} for review of the formalism). We further assume that the soft modes in the LCWF whose longitudinal momenta
$k^+$ is smaller than $\Lambda$  are not energetic enough to contribute  significantly to the scattering process (their contribution is power suppressed by
the collision energy). The hard modes with $k^+>\Lambda$ are also referred as valence modes. 

\begin{figure}
  \centering
  \includegraphics[scale=0.8]{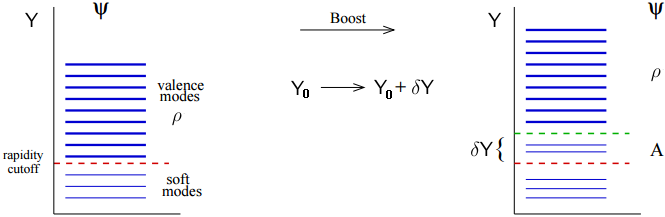}
  \caption{The LCWFs before and after boost. Here  $\rho$ stands for the colour charge density of the valence modes, and A for the soft gluon field.\label{rapidity_fig}}
\end{figure}

Total energy of the collision can be increased by boosting the projectile: $\mathsf{Y}\,=\,\mathsf{Y}_{0}\,+\, \delta\mathsf{Y}$, whereas $ \delta\mathsf{Y}$
is a boost parameter.  By boosting the projectile, all the longitudinal momenta of partons  in the LCWF get shifted up (Fig.\ref{rapidity_fig}):
   \begin{equation}
k^{+}\;\longrightarrow\; e^{\delta\mathsf{Y}}\, k^{+}.
 \end{equation}
As a result, some of the soft modes that lived below the cutoff $\Lambda$ get lifted above $\Lambda$ and start to contribute to the scattering process. 
After the boost, these modes occupy a window $\Lambda\,<\,k^+\,< \,\Lambda\, e^{\delta\mathsf{Y}}$. The valence modes get shifted above this window.

Ignoring the modes below $\Lambda$, the LCWF $\left|\Psi\right\rangle _{\mathsf{Y_0}}$ before the boost is dominated by the valence modes only, whose
distribution is approximately independent of their longitudinal momenta.
\begin{equation}\label{psisplit0}
\left|\Psi\right\rangle _{\mathsf{Y_0}}\,=\,\left|\upsilon\right\rangle .
 \end{equation}
Here $|v\rangle$ represents  a valence state with no soft gluons (vaccuum of soft gluons). We assume that this LCWF is known. The LCWF after the boost is 
 \begin{equation}\label{psisplit}
\left|\Psi\right\rangle _{\mathsf{Y}}\,=\,\left|\psi\right\rangle \,\otimes\,\left|\upsilon\right\rangle ,
 \end{equation}
where $\left|\psi\right\rangle$ denotes the soft part of the wave function living in the window $\Lambda\,<\,k^+\,< \,\Lambda\, e^{\delta\mathsf{Y}}$.
The factorisation  (\ref{psisplit}) is a typical Born-Oppenheimer approximation. 
Our goal below will be to find $\left|\psi\right\rangle$ by perturbatively diagonalising $H_{LC\,QCD}$ on the Hilbert space of the soft modes, 
assuming the valence sector is known and frozen.

 \subsection{Eigenstates of Free Hamiltonian }\label{states}

We denote by $\left|0\right\rangle$ the vacuum state of the free Hamiltonian $H_0$ on the soft Hilbert space.
The energy of this state will be set to zero, $E_{0}=0$. The vacuum is defined as:
     \begin{equation}
a_{i}^{a}(k^{+},\, \mathbf{k})\left|0\right\rangle =0,\qquad\qquad b_{i}^{\alpha}(k^{+},\, \mathbf{k})\left|0\right\rangle =0,\qquad\qquad d_{i}^{\alpha}(k^{+},\, \mathbf{k})\left|0\right\rangle =0,
  \end{equation}
for any $\Lambda<k^{+}<e^{\delta\mathsf{Y}}\Lambda$. 
The soft vacuum $\left|0\right\rangle$  should be understood as $\left|0\right\rangle\otimes|v\rangle$  and, because of its valence component, it 
differs from the QCD vacuum.  As explained above, our prime objective in the next section will be to develop a perturbation theory for this soft vacuum state. 
 Three  eigenstates of $H_0$  will emerge in the NLO calculation below:\\
$\bullet$ \textit{\textbf{One gluon state:}}
  \begin{equation}\label{gsta}
\left|g_{i}^{a}(k)\right\rangle \,\equiv\,\frac{a_{i}^{a\dagger}(k)}{(2\pi)^{3/2}}\left|0\right\rangle \qquad or\qquad\left|g_{i}^{a}(k^{+},\, \mathbf{z})\right\rangle \,\equiv\,\frac{a_{i}^{a\dagger}(k^{+},\, \mathbf{z})}{(2\pi)^{1/2}}\left|0\right\rangle ,
  \end{equation}\\
These is a normalised state with the normalisation   
$$\left\langle g_{j}^{b}(p)\left|g_{i}^{a}(k)\right.\right\rangle =\delta^{ab}\delta_{ij}\delta^{(2)}(\mathbf{k}-\mathbf{p})\delta(k^{+}-p^{+})\;$$ and $$\;\left\langle g_{j}^{b}(p^{+},\,\mathbf{z}^{\prime})\left|g_{i}^{a}(k^{+},\,\mathbf{z})\right.\right\rangle \,=\,\delta^{ab}\,\delta_{ij}\,\delta^{(2)}(\mathbf{z}-\mathbf{z}^{\prime})\,\delta(k^{+}-p^{+}).
$$
This state has the energy  $E_{g}(k)=\frac{\mathbf{k}^{2}}{2k^{+}}.$ \\
$\bullet$ \textit{\textbf{Two gluon state:}}
  \begin{equation}\begin{split}\label{ggst}
&\left|g_{i}^{a}(k)\, g_{j}^{b}(p)\right\rangle \,\equiv\,\frac{a_{i}^{a\dagger}(k)\, a_{j}^{b\dagger}(p)}{(2\pi)^{3}}\left|0\right\rangle \qquad or\\
&\left|g_{i}^{a}(k^{+},\,\mathbf{z})\, g_{j}^{b}(p^{+},\,\mathbf{z}^{\prime})\right\rangle \,\equiv\,\frac{a_{i}^{a\dagger}(k^{+},\,\mathbf{z})\, a_{j}^{b\dagger}(p^{+},\,\mathbf{z}^{\prime})}{2\pi}\left|0\right\rangle .\\
\end{split}\end{equation}\\
with the energy $E_{gg}(k,\, p)\,\equiv\,\frac{\mathbf{k}^{2}}{2k^{+}}+\frac{\mathbf{p}^{2}}{2p^{+}}$.\\
 $\bullet$ \textit{\textbf{Quark anti-quark state:}}
\begin{equation}\begin{split}\label{qqst}
&\left|\bar{q}_{\lambda_{2}}^{\beta}(p)\, q_{\lambda_{1}}^{\alpha}(k)\right\rangle \,\equiv\,\frac{b_{\lambda_{1}}^{\alpha\dagger}(k)\, d_{\lambda_{2}}^{\beta\dagger}(p)}{(2\pi)^{3}}\left|0\right\rangle \quad\; or\\
&\left|\bar{q}_{\lambda_{2}}^{\beta}(p^{+},\,\mathbf{z}^{\prime})\, q_{\lambda_{1}}^{\alpha}(k^{+},\,\mathbf{z})\right\rangle \,\equiv\,\frac{b_{\lambda_{1}}^{\alpha\dagger}(k^{+},\,\mathbf{z})\, d_{\lambda_{2}}^{\beta\dagger}(p^{+},\,\mathbf{z}^{\prime})}{2\pi}\left|0\right\rangle ,\\
\end{split}\end{equation}\\
with the energy $E_{q\bar{q}}(k,\, p)=\frac{\mathbf{k}^{2}}{2k^{+}}+\frac{\mathbf{p}^{2}}{2p^{+}}$.

 \subsection{Eikonal Approximation for  QCD Hamiltonian}\label{eikhami}
 
 In section \ref{softvalence}, we split the Hilbert space into soft and valence modes and introduced a Born-Oppenheimer approximation for the LCWF.  
 Accordingly, we have to split the Hamiltonian $H_{int}$ as defined by (\ref{hint}) into "soft" and "valence". First, we introduce field decomposition:
 $A_{i}^{a}(x)=\underline{A}_{i}^{a}(x)+\overline{A}_{i}^{a}(x)$ and $\psi_{+}^{\alpha}(x)=\underline{\psi}_{+}^{\alpha}(x)+\overline{\psi}_{+}^{\alpha}(x)$ where the underlined fields contain only modes with longitudinal momenta in the interval $k^{+}\in\left(\Lambda,\, e^{\delta\mathsf{Y}}\Lambda\right)$ (soft modes)
 while the fields with the bar  contain modes with $k^{+} >e^{\delta\mathsf{Y}}\Lambda$ (valence modes). Substitution of this decomposition into $H_{int}$ 
is done in  Appendix \ref{sec2}. 

Our next step is to introduce an eikonal approximation for  the interaction Hamiltonian. A typical longitudinal momentum of a valence mode is much larger than that
of a soft mode. Consequently, all the terms  in  $H_{int}$ that are suppressed by the ratio of soft to valence longitudinal momenta  can  be  systematically ignored.   
Keeping these terms would introduce non-eikonal power suppressed corrections to our calculation. 
Example of terms in the Hamiltonian which  are neglected 
$\frac{1}{\partial^{+}}(\underline{A}_{j}^{b}\partial^{+}\overline{A}_{j}^{c})$ or $\frac{1}{\partial^{+}}(\underline{\psi}_{+}^{\dagger}t^{a}\overline{\psi}_{+})$. 
The action of $\frac{1}{\partial^{+}}$ is equivalent to division by a very large valence momentum (i.e. $\frac{1}{p^{+}}\rightarrow0$ for any $p^{+}>e^{\delta\mathsf{Y}}\Lambda$). 

In Appendix \ref{sec2}, we first split the modes between soft and valence in $H_{int}$ and  then employ the eikonal approximation. The result is 
\begin{equation}
H_{int}=H_{g}+H_{gqq}+H_{ggg}+H_{gggg}+H_{ggqq}+H_{qq-inst}+H_{gg-inst}+H_{gq}+H_{V}.
  \end{equation}
The first eight contributions defined by (\ref{Hg}) $-$ (\ref{gq}), account for  soft-soft and soft-valence interactions, while  $H_{V}$ involves 
valence-valence interactions only.

Here $\rho^{a}(-\mathbf{p})$ denotes the total valence current 
  (charge density operator) which acts as a background for the soft  fields, and is defined by:
  \begin{equation}\begin{split}\label{rho}
\rho^{a}(-\mathbf{p})\,\equiv\,\rho_{g}^{a}(-\mathbf{p})+\rho_{q\bar{q}}^{a}(-\mathbf{p}),
 \end{split}\end{equation}
with
 \begin{equation}\begin{split}
\rho_{g}^{a}(-\mathbf{p})\,\equiv\,-if^{abc}\int_{e^{\delta\mathsf{Y}}\Lambda}^{\infty}\frac{dk^{+}}{2\pi}\int\frac{d^{2}\mathbf{k}}{(2\pi)^{2}}\, a_{j}^{\dagger b}(k^{+},\, \mathbf{k})\, a_{j}^{c}(k^{+},\, \mathbf{k}+\mathbf{p}),
  \end{split}\end{equation} 
and
    \begin{equation}
\rho_{q\bar{q}}^{a}(-\mathbf{p})\,\equiv\,t_{\alpha\beta}^{a}\int_{e^{\delta\mathsf{Y}}\Lambda}^{\infty}\frac{dk^{+}}{2\pi}\int\frac{d^{2}\mathbf{k}}{(2\pi)^{2}}\,\left(b_{\lambda}^{\alpha\dagger}(k^{+},\, \mathbf{k}-\mathbf{p})\, b_{\lambda}^{\beta}(k^{+},\, \mathbf{k})+d_{\lambda}^{\beta}(k^{+},\, \mathbf{k})\, d_{\lambda}^{\alpha\dagger}(k^{+},\, \mathbf{k}-\mathbf{p})\right). 
\end{equation} 
The valence current $\rho^{a}(\mathbf{p})$ satisfies the $SU(N_{c})$ algebra:
 \begin{equation}\begin{split}\label{rhoalge}
\left[\rho^{a}(\mathbf{x}),\,\rho^{b}(\mathbf{y})\right]=if^{abc}\rho^{c}(\mathbf{x})\delta^{(2)}(\mathbf{x}-\mathbf{y})\:;\quad\quad
\left[\rho^{a}(\mathbf{k})\,,\,\,\rho^{b}(\mathbf{p})\right]=if^{abc}\rho^{c}(\mathbf{k}+\mathbf{p})\:.
  \end{split}\end{equation} 
where  $\rho^{a}(\mathbf{x})$ is a Fourier transform of $\rho^{a}(\mathbf{p})$:
\begin{equation}\begin{split}\label{tran.rho}
&\rho^{a}(\mathbf{x})=\frac{1}{(2\pi)^{2}}\int d^{2}\mathbf{k}\, e^{i\mathbf{k}\cdot \mathbf{x}}\rho^{a}(-\mathbf{k})\;;\qquad\qquad\rho^{a}(-\mathbf{k})=\int d^{2}\mathbf{x}\, e^{-i\mathbf{k}\cdot \mathbf{x}}\rho^{a}(\mathbf{x})\:.
   \end{split}\end{equation}

As a preparation for the next section, it is useful to express (\ref{Hg}) $-$ (\ref{ggi}) in terms of the creation and annihilation operators. This is done by employing the expansions (\ref{glufield}) and (\ref{quarfield}). Only the terms, which would yield a non-trivial contribution when inserted between the three states defined in section \ref{softvalence}, 
and the soft vacuum, will be kept. We omit further details and write the final results,
 
 \begin{equation}\label{eikohg}
\mathsf{H}_{g}=\int_{\Lambda}^{e^{\delta\mathsf{Y}}\Lambda}\frac{dk^{+}}{2\pi}\int\frac{d^{2}\mathbf{k}}{(2\pi)^{2}}\frac{g\mathbf{k}^{i}}{\sqrt{2}|k^{+}|^{3/2}}\left[a_{i}^{a\dagger}(k^{+},\, \mathbf{k})\rho^{a}(-\mathbf{k})+a_{i}^{a}(k^{+},\, \mathbf{k})\rho^{a}(\mathbf{k})\right],
 \end{equation}

   \begin{eqnarray}
\mathsf{H}_{g\: qq}&&=\sum_{\lambda_{1},\,\lambda_{2}=\pm\frac{1}{2}}\int_{\Lambda}^{e^{\delta\mathsf{Y}}\Lambda}\frac{dk^{+}}{2\pi}\frac{dp^{+}}{2\pi}dq^{+}\,\int\frac{d^{2}\mathbf{k}}{(2\pi)^{2}}\frac{d^{2}\mathbf{p}}{(2\pi)^{2}}d^{2}\mathbf{q}\,\frac{gt_{\alpha\beta}^{a}}{2\sqrt{2k^{+}}}\,\delta^{(3)}(k-p-q)\,\Gamma_{\lambda_{1}\lambda_{2}}^{i}\nonumber\\
&&\times \,(a_{i}^{a}(k^{+},\, \mathbf{k})\, b_{\lambda_{1}}^{\alpha\dagger}(p^{+},\, \mathbf{p})\, d_{\lambda_{2}}^{\beta\dagger}(q^{+},\, \mathbf{q})\,+\, h.c.),\end{eqnarray}
 with
  \begin{equation}
 \Gamma_{\lambda_{1}\lambda_{2}}^{i}=\chi_{\lambda_{1}}^{\dagger}\left[\frac{2\mathbf{k}^{i}}{k^{+}}-\frac{\sigma\cdot \mathbf{p}}{p^{+}}\sigma^{i}-\sigma^{i}\frac{\sigma\cdot \mathbf{q}}{q^{+}}\right]\chi_{\lambda_{2}},
 \end{equation}

   \begin{equation}\begin{split}
\mathsf{H}_{g\: gg}=&-\int_{\Lambda}^{e^{\delta\mathsf{Y}}\Lambda}\frac{dk^{+}}{2\pi}\,\frac{dp^{+}}{2\pi}\, dq^{+}\,\int\frac{d^{2}\mathbf{k}}{(2\pi)^{2}}\,\frac{d^{2}\mathbf{p}}{(2\pi)^{2}}\, d^{2}\mathbf{q}\,\frac{igf^{abc}}{2\sqrt{2k^{+}p^{+}q^{+}}}\\
&\times\left[\left(\mathbf{q}^{i}-\frac{q^{+}}{p^{+}+q^{+}}\mathbf{k}^{i}\right)a_{i}^{a}(k)a_{j}^{b\dagger}(p)a_{j}^{c\dagger}(q)\delta^{(3)}(-k+p+q)\right.\\
&+\left.\left(\mathbf{q}^{i}+\frac{p^{+}+q^{+}}{q^{+}-p^{+}}\mathbf{k}^{i}\right)a_{i}^{a\dagger}(k)a_{j}^{b\dagger}(p)a_{j}^{c}(q)\delta^{(3)}(k+p-q),\,+\,h.c.\right]\\
 \end{split}\end{equation}

    \begin{eqnarray}
\mathsf{H}_{gg-inst}&&=\int_{\Lambda}^{e^{\delta\mathsf{Y}}\Lambda}\frac{dp^{+}}{2\pi}\frac{dq^{+}}{2\pi}\int\frac{d^{2}\mathbf{p}}{(2\pi)^{2}}\frac{d^{2}\mathbf{q}}{(2\pi)^{2}}\frac{ig^{2}f^{abc}}{\sqrt{p^{+}q^{+}}}\,\left(\frac{q^{+}\rho^{a}(-\mathbf{p}-\mathbf{q})}{2(p^{+}+q^{+})^{2}}\, a_{i}^{b\dagger}(p^{+},\,\mathbf{p})\, a_{i}^{c\dagger}(q^{+},\,\mathbf{q})\right.\nonumber\\
&&\left.-\, h.c.\,-\,\frac{(p^{+}+q^{+})\rho^{a}(-\mathbf{p}+\mathbf{q})}{2(p^{+}-q^{+})^{2}}\, a_{i}^{b\dagger}(p^{+},\,\mathbf{p})\, a_{i}^{c}(q^{+},\,\mathbf{q})\right),\end{eqnarray}

      \begin{equation}\begin{split}\label{qqinst}
\mathsf{H}_{qq-inst}&=\sum_{\lambda_{1},\,\lambda_{2}}\int_{\Lambda}^{e^{\delta\mathsf{Y}}\Lambda}\frac{dp^{+}}{2\pi}\frac{dq^{+}}{2\pi}\int\frac{d^{2}\mathbf{p}}{(2\pi)^{2}}\frac{d^{2}\mathbf{q}}{(2\pi)^{2}}\frac{g^{2}t_{\alpha\beta}^{a}\rho^{a}(-\mathbf{p}-\mathbf{q})\chi_{\lambda_{1}}^{\dagger}\chi_{\lambda_{2}}}{(p^{+}+q^{+})^{2}}\\
&\times\left[b_{\lambda_{1}}^{\alpha\dagger}(p^{+},\mathbf{p})\,d_{\lambda_{2}}^{\beta\dagger}(q^{+},\mathbf{q})\,+\,h.c.\right]\\
\end{split}\end{equation}

 \subsection{Eikonal Scattering}\label{eiksc}
 
At very high collision energies,  a fast projectile parton moves along its straight-line classical trajectory and the only scattering effect is the eikonal 
phase factor acquired along its propagation path.  We assume this approximation to be valid for all partons with longitudinal momenta above $\Lambda$.
The $\hat{S}$ matrix operator is diagonal in the coordinate space:
\begin{equation}\label{gscat}
\hat{S}\, a_{i}^{a\dagger}(x^{+},\, \mathbf{x})\left|0\right\rangle \:=\: S_{A}^{ab}(\mathbf{x})\, a_{i}^{b\dagger}(x^{+},\, \mathbf{x})\left|0\right\rangle ,
   \end{equation}
 \begin{equation}\begin{split}\label{qscat}
\hat{S} \,b_{i}^{\alpha\dagger}(x^{+},\mathbf{x})\left|0\right\rangle =S^{\beta\alpha}(\mathbf{x})\, b_{i}^{\beta\dagger}(x^{+},\mathbf{x})\left|0\right\rangle,
\,\,\,\,\hat{S}\, d_{i}^{\alpha\dagger}(x^{+},\mathbf{x})\left|0\right\rangle =S^{\alpha\beta}(\mathbf{x})\, d_{i}^{\beta\dagger}(x^{+},\mathbf{x})\left|0\right\rangle.
 \end{split}\end{equation}
$S_{A}^{ab}(\mathbf{x})$ and $S^{\alpha\beta}(\mathbf{x})$ are Wilson lines in the adjoint and fundamental representations,
\begin{equation}\begin{split}\label{SA}
&S_{A}^{ab}(\mathbf{x})\,=\,\left[\mathcal{P}\: exp\left\{ ig\int dx^{+}\, T^{c}A^{-c}(x^{+},\, \mathbf{x})\right\} \right]^{ab},
    \end{split}\end{equation}
   \begin{equation}\begin{split}\label{S}&S^{\alpha\beta}(\mathbf{x})\,=\,\left[\mathcal{P}\: exp\left\{ ig\int dx^{+}\, t^{c}A^{-c}(x^{+},\, \mathbf{x})\right\} \right]^{\alpha\beta}.
    \end{split}\end{equation}
    
The valence part scattering: $\hat S|v\rangle =|S v\rangle$.    
The operator $\rho$ gets affected by the scattering as well. As explained in \cite{KLremarks}, it is possible to express the action of $\rho$ on the 
state $|S v\rangle$ as a functional Lie derivative:
  \begin{equation}\begin{split}\label{rhoJ}
  &\rho_{g}^{a}(\mathbf{x})\,\hat{S}\left|\upsilon\right\rangle \,=\, J_{R,\, adj}^{a}(\mathbf{x})\left|\hat{S}\upsilon\right\rangle ,\qquad\qquad\qquad\hat{S}\rho_{g}^{a}(\mathbf{x})\,\left|\upsilon\right\rangle \,=\, J_{L,\, adj}^{a}(\mathbf{x})\,\left|\hat{S}\upsilon\right\rangle ,
 \end{split}\end{equation}\\
which act as right and left rotations on the Wilson lines $S$,
\begin{equation}\begin{split}&J_{R,\, adj}^{a}(\mathbf{x})\,\equiv\,-tr\left[S_{A}(\mathbf{x})T^{a}\frac{\delta}{\delta S_{A}^{\dagger}(\mathbf{x})}\right]\,;\\
&J_{L,\, adj}^{a}(\mathbf{x})\,\equiv\,[S_{A}(\mathbf{x})\, J_{R}(\mathbf{x})]^{a}\,=\,-tr\left[T^{a}S_{A}(\mathbf{x})\frac{\delta}{\delta S_{A}^{\dagger}(\mathbf{x})}\right].\\
    \end{split}\end{equation}
In a similar manner we can also express $\rho_{q\bar{q}}$:
\begin{equation}
\rho_{q\bar{q}}^{a}(\mathbf{x})\,\hat{S}\,\left|\upsilon\right\rangle \,=\, J_{R,\, F}^{a}(\mathbf{x})\,\left|\hat{S}\upsilon\right\rangle ,\qquad\qquad\qquad\hat{S}\,\rho_{q\bar{q}}(\mathbf{x})\,\left|\upsilon\right\rangle \,=\, J_{L,\, F}^{a}(\mathbf{x})\,\left|\hat{S}\upsilon\right\rangle . \end{equation}\\
with
\begin{equation}J_{R,\, F}^{a}(\mathbf{x})\,\equiv\, tr\left[\frac{\delta}{\delta S^{T}(\mathbf{x})}S(\mathbf{x})t^{a}\right]\,-\, tr\left[\frac{\delta}{\delta S^{*}(\mathbf{x})}t^{a}S^{\dagger}(\mathbf{x})\right],
      \end{equation}
      
 \begin{equation}J_{L,\, F}^{a}(\mathbf{x})\,\equiv\, tr\left[\frac{\delta}{\delta S^{T}(\mathbf{x})}t^{a}S(\mathbf{x})\right]\,-\, tr\left[\frac{\delta}{\delta S^{*}(\mathbf{x})}S^{\dagger}(\mathbf{x})t^{a}\right].
      \end{equation}
The following relations will be useful for the NLO calculations:
\begin{equation}\begin{split}\label{relrr}
&\rho^{a}(\mathbf{y})\,\rho^{b}(\mathbf{x})\,\hat{S}\,\left|\upsilon\right\rangle \,=\, J_{R}^{b}(\mathbf{x})\, J_{R}^{a}(\mathbf{y})\,\left|\hat{S}\upsilon\right\rangle, \\
&\rho^{a}(\mathbf{y})\,\hat{S}\,\rho^{b}(\mathbf{x})\,\left|\upsilon\right\rangle \,=\, J_{L}^{b}(\mathbf{x})\, J_{R}^{a}(\mathbf{y})\,\left|\hat{S}\upsilon\right\rangle, \\
&\hat{S}\,\rho^{a}(\mathbf{y})\,\rho^{b}(\mathbf{x})\,\left|\upsilon\right\rangle =\, J_{L}^{a}(\mathbf{y})\, J_{L}^{b}(\mathbf{x})\,\left|\hat{S}\upsilon\right\rangle. \\
 \end{split}\end{equation}
 The algebra of the operators $J_{L/R}^{a}(\mathbf{x})\equiv J_{L/R,\,F}^{a}(\mathbf{x})+J_{L/R,\, adj}^{a}(\mathbf{x})$ follows from that of $\rho$:
\begin{equation}\begin{split}\label{algJ}
&\left[J_{L}^{a}(\mathbf{x}),\, J_{L}^{b}(\mathbf{y})\right]=-if^{abc}J_{L}^{c}(\mathbf{x})\delta(\mathbf{x}-\mathbf{y});\quad\left[J_{R}^{a}(\mathbf{x}),\, J_{R}^{b}(\mathbf{y})\right]=if^{abc}J_{R}^{c}(\mathbf{x})\delta(\mathbf{x}-\mathbf{y}),
\end{split}\end{equation}
while $\left[J_{L}^{a}(\mathbf{x}),\, J_{R}^{b}(\mathbf{y})\right]=0$. As mentioned above, $J_{L/R}$ act on $S$ as left and right rotations:
\begin{equation}
\label{alg2}
 \left[J_{L}^{c}(\mathbf{x}),\, S_{A}^{ab}(\mathbf{z})\right]=-if^{cad}S_{A}^{db}(\mathbf{z})\delta(\mathbf{x}-\mathbf{z});\quad\left[J_{R}^{c}(\mathbf{x}),\, S_{A}^{ab}(\mathbf{z})\right]=if^{cbd}S_{A}^{ad}(\mathbf{z})\delta(\mathbf{x}-\mathbf{z}).
  \end{equation}
These algebra will be used in section \ref{reduclo}, where we subtract the $(LO)^{2}$ effects.

 \subsection{LO JIMWLK Hamiltonian}\label{losect}

\begin{figure}[hb]
  \centering
\includegraphics[scale=1.1]{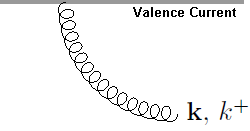}
  \caption{Soft gluon emission with momentum $k\,=\,(\mathbf{k},\,k^{+})$ from a valence current $\rho$ at order $g$. The emission vertex is defined by $\left|\psi^{LO}\right\rangle$.\label{curv}}
\end{figure}

To warm up, we will follow the approach of \cite{kovnerbasic} and derive the LO JIMWLK Hamiltonian using the LCWF formalism. At leading order, the only relevant process is one soft gluon emission by valence current, which is displayed in Fig.\ref{curv}.
\\
The soft part of the wave-function is given by the first order perturbation theory:
 \begin{equation}\label{lowxp}
\left|\psi^{LO}\right\rangle \,=\,\mathcal{N}^{LO}\left|0\right\rangle \,-\,\left|i\right\rangle \,\frac{\left\langle i\left|H_{int}\right|0\right\rangle }{E_{i}}, \end{equation}\\
 where $\mathcal{N}^{LO}$ is determined by the normalisation condition, $\left\langle \psi^{LO}|\psi^{LO}\right\rangle =1$:
   \begin{equation}\label{nrlo}
 \mathcal{N}^{LO}\,\equiv\,1-\frac{\left|\left\langle i\left|H_{int}\right|0\right\rangle \right|^{2}}{2E_{i}^{2}}.
 \end{equation}\\
At order $g$, there is only one state, $\left|g_{i}^{a}(k)\right\rangle $ that contributes to (\ref{lowxp}). The only non-vanishing matrix-element is that of $H_{g}$, (\ref{eikohg}):
\begin{equation}\label{g}
\left\langle g_{i}^{a}(k)\left|\, H_{g}\,\right|0\right\rangle =\frac{g\mathbf{k}^{i}\rho^{a}(-\mathbf{k})}{4\pi^{3/2}|k^{+}|^{3/2}}.
 \end{equation}
Therefore, (\ref{lowxp}) becomes:
\begin{equation}
\left|\psi^{LO}\right\rangle \,=\,\mathcal{N}^{LO}\left|0\right\rangle \,-\,\int_{\Lambda}^{e^{\delta\mathsf{Y}}\Lambda}dk^{+}\int d^{2}\mathbf{k}\,\frac{g\, \mathbf{k}^{i}}{2\pi^{3/2}\sqrt{k^{+}}\mathbf{k}^{2}}\,\rho^{a}(-\mathbf{k})\left|g_{i}^{a}(k)\right\rangle .
  \end{equation}
 
We will proceed by calculating the Fourier transformation of $\left|\psi^{LO}\right\rangle$. The coordinate space representation will make it possible to compute the scattered wave function, since, as explained in the previous section, the $S$-matrix is diagonal in the $x$-space. The charge densities $\rho^{a}(-\mathbf{k})$ are Fourier transformed according to (\ref{tran.rho}), and the transformation to the coordinates space is done using (\ref{fourier.1}). The wave function becomes:
\begin{equation}\begin{split}\label{lowav}
&\left|\psi^{LO}\right\rangle \,=\,\mathcal{N}^{LO}\left|0\right\rangle \,+\,\int_{\Lambda}^{e^{\delta\mathsf{Y}}\Lambda}\,\frac{dk^{+}}{\sqrt{k^{+}}}\int_{\mathbf{x},\, \mathbf{z}}\,\frac{i\, g\, X^{i}}{2\pi^{3/2}X^{2}}\,\rho^{a}(\mathbf{x})\left|g_{i}^{a}(k^{+},\, \mathbf{z})\right\rangle ,
  \end{split}\end{equation}
where $X^{i}\,\equiv\, \mathbf{x}^{i}-\mathbf{z}^{i}$. From (\ref{nrlo}):
    \begin{equation}\begin{split}&\mathcal{N}^{LO}\,=\,1\,-\,\frac{g^{2}}{8\pi^{3}}\int_{\Lambda}^{e^{\delta\mathsf{Y}}\Lambda}\,\frac{dk^{+}}{k^{+}}\int_{\mathbf{x},\, \mathbf{y},\, \mathbf{z}}\frac{X\cdot Y}{X^{2}Y^{2}}\,\rho^{a}(\mathbf{y})\,\rho^{a}(\mathbf{x})\,+\,\mathcal{O}(g^{4}),
   \end{split}\end{equation}
where $Y^{i}\,\equiv\, \mathbf{y}^{i}-\mathbf{z}^{i}$. The scattered wave function reads:
\begin{equation}\begin{split}\hat{S}\left|\psi^{LO}\right\rangle &=\left(1-\frac{g^{2}}{8\pi^{3}}\int_{\Lambda}^{e^{\delta\mathsf{Y}}\Lambda}\,\frac{dk^{+}}{k^{+}}\int_{\mathbf{x},\, \mathbf{y},\, \mathbf{z}}\frac{X\cdot Y}{X^{2}Y^{2}}\,\hat{S}\,\rho^{a}(\mathbf{y})\,\rho^{a}(\mathbf{x})\right)\left|0\right\rangle \\
&\,+\,\int_{\Lambda}^{e^{\delta\mathsf{Y}}\Lambda}\,\frac{dk^{+}}{\sqrt{k^{+}}}\,\int_{\mathbf{x},\, \mathbf{z}}\,\frac{i\, g\, X^{i}}{2\pi^{3/2}X^{2}}\,\hat{S}\,\rho^{a}(\mathbf{x})\left|g{}_{i}^{a}(k^{+},\mathbf{z})\right\rangle. 
\end{split}\end{equation}

\begin{figure}
  \centering
\includegraphics[scale=0.68]{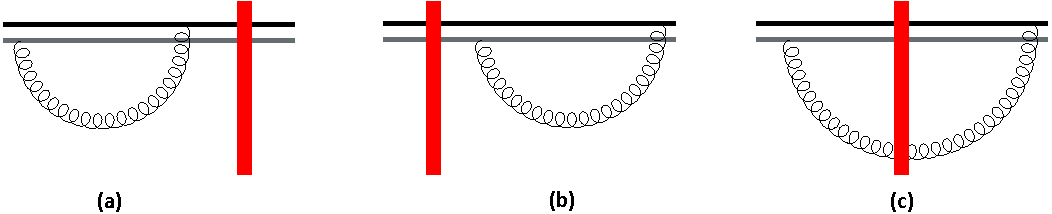}
  \caption{Figures $(a)$ and $(b)$ correspond to the contributions from the normalisation of the wave function, while the figure $(c)$ illustrates the real gluon scattering contribution. The instantaneous shockwave interaction between the projectile and the target is represented by the thick (red) line. Each operator $J_{L}/J_{R}$ in the JIMWLK Hamiltonian correspond to the emission vertice of a soft gluon from the valence current.\label{lorep}}
\end{figure}

By inserting the wave functions before and after the scattering process in (\ref{Sigma}) and using the relations (\ref{relrr}), we arrive at:
\begin{eqnarray}\label{siglo}
\Sigma^{LO}&\equiv&\left\langle \psi^{LO}\right|\,\hat{S}\,-\,1\,\left|\psi^{LO}\right\rangle \\
&=&-\frac{\alpha_{s}}{2\pi^{2}}\,\delta\mathsf{Y}\int_{\mathbf{x},\, \mathbf{y},\, \mathbf{z}}\,\frac{X\cdot Y}{X^{2}Y^{2}}\,\left[\, J_{L}^{a}(\mathbf{x})\, J_{L}^{a}(\mathbf{y})\,+\, J_{R}^{a}(\mathbf{x})\, J_{R}^{a}(\mathbf{y})\,-\,2J_{L}^{a}(\mathbf{x})\, S_{A}^{ab}(\mathbf{z})\, J_{R}^{b}(\mathbf{y})\,\right].
\nonumber \end{eqnarray}
Following the definition (\ref{defH}), the LO Hamiltonian is $H^{LO}\,\equiv\,-\frac{\delta\Sigma^{LO}}{\delta\mathsf{Y}}$, which gives (see Fig.\ref{lorep} for schematic illustration of the LO Hamiltonian)
  \begin{equation}H^{LO}_{JIMWLK}
  =\,\frac{\alpha_{s}}{2\pi^{2}}\int_{\mathbf{x},\, \mathbf{y},\, \mathbf{z}}\frac{X\cdot Y}{X^{2}Y^{2}}\,\left[\, J_{L}^{a}(\mathbf{x})\, J_{L}^{a}(\mathbf{y})\,+\, J_{R}^{a}(\mathbf{x})\, J_{R}^{a}(\mathbf{y})\,-\,2J_{L}^{a}(\mathbf{x})\, S_{A}^{ab}(\mathbf{z})\, J_{R}^{b}(\mathbf{y})\,\right].
  \end{equation}
  The LO Hamiltonian is invariant under $SU_{L}(N_{c})\times SU_{R}(N_{c})$ rotations, which reflect gauge invariance of scattering
amplitudes. In addition, the Hamiltonian is invariant under $Z_{2}$ transformation, $S\rightarrow S^{\dagger}$ together with $J_{L}\rightarrow-J_{R}$, which in \cite{KLreggeon}  was identified as signature, and the charge conjugation symmetry $S\rightarrow S^{*}$.

 \subsection{NLO JIMWLK Hamiltonian}\label{sdfla}
The NLO JIMWLK Hamiltonian deduced in \cite{nlojimwlk} is:
\begin{eqnarray}\label{ham}
&&H_{JIMWLK}^{NLO}=\int_{\mathbf{x}, \mathbf{y}, \mathbf{z}}K_{JSJ}(\mathbf{x}, \mathbf{y},\mathbf{z})\left[J_{L}^{a}(\mathbf{x}) J_{L}^{a}(\mathbf{y})+J_{R}^{a}(\mathbf{x})J_{R}^{a}(\mathbf{y})-2J_{L}^{a}(\mathbf{x}) S_{A}^{ab}(\mathbf{z})J_{R}^{b}(\mathbf{y})\right]\ \nonumber\\
&&+\int_{\mathbf{x},\, \mathbf{y},\mathbf{z}, \mathbf{z}^{\prime}}\, K_{JSSJ}(\mathbf{x},\mathbf{y},\mathbf{z}, \mathbf{z}^{\prime})\left[f^{abc} f^{def}J_{L}^{a}(\mathbf{x})S_{A}^{be}(\mathbf{z})S_{A}^{cf}(\mathbf{z}^{\prime}) J_{R}^{d}(\mathbf{y})\right.
\left.-N_{c}J_{L}^{a}(\mathbf{x})S_{A}^{ab}(\mathbf{z})J_{R}^{b}(\mathbf{y})\right]\nonumber\\
&&+\int_{\mathbf{x},\, \mathbf{y},\, \mathbf{z},\, \mathbf{z}^{\prime}}K_{q\bar{q}}(\mathbf{x},\, \mathbf{y},\, \mathbf{z},\, \mathbf{z}^{\prime})\left[2\, J_{L}^{a}(\mathbf{x})\, tr[S^{\dagger}(\mathbf{z})\, t^{a}\, S(\mathbf{z}^{\prime})t^{b}]\, J_{R}^{b}(\mathbf{y})\,-\, J_{L}^{a}(\mathbf{x})\, S_{A}^{ab}(\mathbf{z})\, J_{R}^{b}(\mathbf{y})\right]\nonumber\\
&&+\int_{\mathbf{w}\,,\mathbf{x},\, \mathbf{y},\, \mathbf{z},\, \mathbf{z}^{\prime}}K_{JJSSJ}(\mathbf{w}\,,\mathbf{x},\, \mathbf{y},\, \mathbf{z},\, \mathbf{z}^{\prime})\, f^{acb}\,\Big[J_{L}^{d}(\mathbf{x})\, J_{L}^{e}(\mathbf{y})\, S_{A}^{dc}(\mathbf{z})\, S_{A}^{eb}(\mathbf{z}^{\prime})\, J_{R}^{a}(\mathbf{w})\,\nonumber\\
&&-\, J_{L}^{a}(\mathbf{w})\, S_{A}^{cd}(\mathbf{z})\, S_{A}^{be}(\mathbf{z}^{\prime})\, J_{R}^{d}(\mathbf{x})\, J_{R}^{e}(\mathbf{y})\,+\,\frac{1}{3}(J_{L}^{c}(\mathbf{x})\, J_{L}^{b}(\mathbf{y})\, J_{L}^{a}(\mathbf{w})\,-\, J_{R}^{c}(\mathbf{x})\, J_{R}^{b}(\mathbf{y})\, J_{R}^{a}(\mathbf{w}))\,\Big]\,\nonumber\\
&&+\int_{\mathbf{w}\,,\mathbf{x},\, \mathbf{y},\, \mathbf{z}}\, K_{JJSJ}(\mathbf{w}\,,\mathbf{x},\, \mathbf{y},\, \mathbf{z})\, f^{bde}\,\Big[J_{L}^{d}(\mathbf{x})\, J_{L}^{e}(\mathbf{y})\, S_{A}^{ba}(\mathbf{z})\, J_{R}^{a}(\mathbf{w})\\
&&-\, J_{L}^{a}(\mathbf{w})\, S_{A}^{ab}(\mathbf{z})\, J_{R}^{d}(\mathbf{x})\, J_{R}^{e}(\mathbf{y})+\,\frac{1}{3}(J_{L}^{d}(\mathbf{x})\, J_{L}^{e}(\mathbf{y})\, J_{L}^{b}(\mathbf{w})\,-\, J_{R}^{d}(\mathbf{x})\, J_{R}^{e}(\mathbf{y})\, J_{R}^{b}(\mathbf{w}))\Big].\nonumber\end{eqnarray}
 It is important to stress that in (\ref{ham}) all the rotation operators $J_{L}$ and $J_{R}$ are assumed not to act on $S$ in the Hamiltonian itself. We now list the kernels:

\begin{equation}\begin{split}\label{JJSSJ}
&K_{JJSSJ}(\mathbf{w},\,\mathbf{x},\,\mathbf{y},\,\mathbf{z},\,\mathbf{z}^{\prime})=\,-\frac{i\alpha_{s}^{2}}{4\pi^{4}}\left(\frac{(Y^{\prime})^{j}X^{i}}{(Y^{\prime})^{2}X^{2}}\,-\,\frac{Y^{i}(X^{\prime})^{j}}{(X^{\prime})^{2}Y^{2}}\right)\\
&\quad\quad \quad\times\left(\frac{\delta^{ij}}{2Z^{2}}\,-\,\frac{(W^{\prime})^{j}Z^{i}}{(W^{\prime})^{2}Z^{2}}\,+\,\frac{W^{i}Z^{j}}{W^{2}Z^{2}}\,-\,\frac{W^{i}(W^{\prime})^{j}}{W^{2}(W^{\prime})^{2}}\right)\ln\left(\frac{W^{2}}{(W^{\prime})^{2}}\right),
\end{split}\end{equation}

$X\equiv \mathbf{x}-\mathbf{z},\: X^{\prime}\equiv \mathbf{x}-\mathbf{z}^{\prime},\: Y\equiv \mathbf{y}-\mathbf{z},\: Y^{\prime}\equiv \mathbf{y}-\mathbf{z}^{\prime},W\equiv \mathbf{w}-\mathbf{z}\:,W^{\prime}\equiv \mathbf{w}-\mathbf{z}^{\prime}\:,Z\equiv \mathbf{z}-\mathbf{z}^{\prime}$.

 \begin{equation}\begin{split}\label{JJSJ}
K_{JJSJ}(\mathbf{w},\, \mathbf{x},\, \mathbf{y},\, \mathbf{z})=-\frac{i\alpha_{s}^{2}}{4\pi^{3}}\left(\frac{X\cdot W}{X^{2}W^{2}}-\frac{Y\cdot W}{Y^{2}W^{2}}\right)\ln\left(\frac{Y^{2}}{(X-Y)^{2}}\right)\ln\left(\frac{X^{2}}{(X-Y)^{2}}\right),
  \end{split}\end{equation}
  
  \begin{eqnarray}\label{JSSJ}
&&K_{JSSJ}(\mathbf{x},\, \mathbf{y},\, \mathbf{z},\, \mathbf{z}^{\prime})\nonumber\\
&&=\frac{\alpha_{s}^{2}}{16\pi^{4}}\left[\frac{4}{Z^{4}}+\left\{ 2\frac{X^{2}(Y^{\prime})^{2}+(X^{\prime})^{2}Y^{2}-4(X-Y)^{2}Z^{2}}{Z^{4}(X^{2}(Y^{\prime})^{2}-(X^{\prime})^{2}Y^{2})}+\frac{(X-Y)^{4}}{X^{2}(Y^{\prime})^{2}-(X^{\prime})^{2}Y^{2}}\right.\right.\nonumber\\
&&\left.\times\left(\frac{1}{X^{2}(Y^{\prime})^{2}}+\frac{1}{Y^{2}(X^{\prime})^{2}}\right)+\frac{(X-Y)^{2}}{Z^{2}}\left(\frac{1}{X^{2}(Y^{\prime})^{2}}-\frac{1}{Y^{2}(X^{\prime})^{2}}\right)\right\} \ln\left(\frac{X^{2}(Y^{\prime})^{2}}{(X^{\prime})^{2}Y^{2}}\right)\nonumber\\
&&\left.-\frac{2I(\mathbf{x},\,\mathbf{z},\,\mathbf{z}^{\prime})}{Z^{2}}-\frac{2I(\mathbf{y},\,\mathbf{z},\,\mathbf{z}^{\prime})}{Z^{2}}\right]+\widetilde{K}(\mathbf{x},\,\mathbf{y},\,\mathbf{z},\,\mathbf{z}^{\prime}),\end{eqnarray}
where:
\begin{equation}\begin{split}
&I(\mathbf{x},\,\mathbf{z},\,\mathbf{z}^{\prime})\,\equiv\,\frac{1}{X^{2}-(X^{\prime})^{2}}\left(\frac{X^{2}+(X^{\prime})^{2}}{Z^{2}}\,-\,\frac{X\cdot X^{\prime}}{X^{2}}\,-\,\frac{X\cdot X^{\prime}}{(X^{\prime})^{2}}\,-\,2\right)\ln\left(\frac{X^{2}}{(X^{\prime})^{2}}\right)\\
&\quad\quad =\,\frac{1}{X^{2}-(X^{\prime})^{2}}\left(\frac{X^{2}+(X^{\prime})^{2}}{Z^{2}}\,+\,\frac{Z^{2}-X^{2}}{2(X^{\prime})^{2}}\,+\,\frac{Z^{2}-(X^{\prime})^{2}}{2X^{2}}\,-\,3\right)\ln\left(\frac{X^{2}}{(X^{\prime})^{2}}\right).\\
\end{split}\end{equation}
The terms which depend on one variable only, either $\mathbf{x}$ or $\mathbf{y}$, will not contribute when the
Hamiltonian is taken to act on gauge invariant operators.

   \begin{equation}\begin{split}\label{ktild1}
\widetilde{K}(\mathbf{x},\, \mathbf{y},\, \mathbf{z},\, \mathbf{z}^{\prime})&\equiv\,\frac{i}{2}\,\left(K_{JJSSJ}(\mathbf{x},\, \mathbf{x},\, \mathbf{y},\, \mathbf{z},\, \mathbf{z}^{\prime})-K_{JJSSJ}(\mathbf{y},\, \mathbf{x},\, \mathbf{y},\, \mathbf{z},\, \mathbf{z}^{\prime})\right.\\
&\left.-K_{JJSSJ}(\mathbf{x},\, \mathbf{y},\, \mathbf{x},\, \mathbf{z},\, \mathbf{z}^{\prime})+K_{JJSSJ}(\mathbf{y},\, \mathbf{y},\, \mathbf{x},\, \mathbf{z},\, \mathbf{z}^{\prime})\right).\\
   \end{split}\end{equation}
Explicitly:
\begin{eqnarray}\label{ktild2}
&&\widetilde{K}(\mathbf{x},\, \mathbf{y},\, \mathbf{z},\, \mathbf{z}^{\prime})=\frac{\alpha_{s}^{2}}{16\pi^{4}}\left(\frac{(Y^{\prime})^{2}}{(X^{\prime})^{2}Z^{2}Y^{2}}-\frac{Y^{2}}{Z^{2}X^{2}(Y^{\prime})^{2}}+\frac{1}{Z^{2}(Y^{\prime})^{2}}-\frac{1}{Z^{2}Y^{2}}+\frac{(X-Y)^{2}}{X^{2}Z^{2}Y^{2}}\right.\nonumber\\
&&\quad\quad \left.-\frac{(X-Y)^{2}}{(X^{\prime})^{2}Z^{2}(Y^{\prime})^{2}}+\frac{(X-Y)^{2}}{(X^{\prime})^{2}X^{2}(Y^{\prime})^{2}}-\frac{(X-Y)^{2}}{X^{2}(X^{\prime})^{2}Y^{2}}\right)\ln\left(\frac{X^{2}}{(X^{\prime})^{2}}\right)+\left(\mathbf{x}\leftrightarrow \mathbf{y}\right).\end{eqnarray}\\
    The following useful equality holds:
    
    \begin{equation}
    \int_{\mathbf{z}^{\prime}}\widetilde{K}(\mathbf{x},\, \mathbf{y},\, \mathbf{z},\, \mathbf{z}^{\prime})\,=\,-\frac{\alpha_{s}^{2}}{4\pi^{3}}\left(\frac{1}{X^{2}}-\frac{1}{Y^{2}}\right)\ln\left(\frac{Y^{2}}{(X-Y)^{2}}\right)\ln\left(\frac{X^{2}}{(X-Y)^{2}}\right).
 \end{equation}
The kernel $K_{JSJ}$ reads\footnote{The term $2b(\gamma-\ln\,2)$ was missing in our previous results  for $K_{JSJ}$ \cite{nlojimwlk,nlojimwlk2}, even though it should be
there as is clear from \cite{neqfo2}.}:
\begin{eqnarray}
&&K_{JSJ}(\mathbf{x},\,\mathbf{y},\,\mathbf{z})=\,-\frac{\alpha_{s}^{2}}{16\pi^{3}}\left(\frac{(X-Y)^{2}}{X^{2}Y^{2}}\left[b\ln\left((X-Y)^{2}\mu_{\overline{MS}}^{2}\right)-b\frac{X^{2}-Y^{2}}{(X-Y)^{2}}\ln\left(\frac{X^{2}}{Y^{2}}\right)\right.\right.\nonumber\\
&&\left.+2b\left(\gamma-\ln2\right)+\left(\frac{67}{9}-\frac{\pi^{2}}{3}\right)N_{c}-\frac{10}{9}N_{f}\right]-\left[\frac{1}{X^{2}}+\frac{1}{Y^{2}}\right]\left[2b\left(\gamma-\ln2\right)+\left(\frac{67}{9}-\frac{\pi^{2}}{3}\right)N_{c}\right.\nonumber\\
&&\left.\left.-\frac{10}{9}N_{f}\right]-\frac{b}{X^{2}}\ln X^{2}\mu_{\overline{MS}}^{2}-\frac{b}{Y^{2}}\ln Y^{2}\mu_{\overline{MS}}^{2}\right)-\frac{N_{c}}{2}\int_{\mathbf{z}^{\prime}}\widetilde{K}(\mathbf{x},\,\mathbf{y},\,\mathbf{z},\,\mathbf{z}^{\prime}).\end{eqnarray}
Here $\mu_{\overline{MS}}^{2}$ is the normalisation point in the $\overline{MS}$ scheme and $b$ is the first coefficient of the $\beta$-function:
\begin{equation}\label{beta}
b\,\equiv\,\frac{11N_{c}\,-\,2N_{f}}{3}\,.
 \end{equation}
An alternative representation for $K_{JSJ}$:
   \begin{eqnarray}\label{JSJ}
&&K_{JSJ}(\mathbf{x},\, \mathbf{y},\, \mathbf{z})\nonumber\\
&&=\frac{\alpha_{s}^{2}}{8\pi^{3}}\left[b\left(\frac{1}{2X^{2}}\ln\left(Y^{2}\mu_{\overline{MS}}^{2}\right)+\frac{1}{2Y^{2}}\ln\left(X^{2}\mu_{\overline{MS}}^{2}\right)-\frac{(X-Y)^{2}}{2X^{2}Y^{2}}\ln\left((X-Y)^{2}\mu_{\overline{MS}}^{2}\right)\right)\right.\nonumber\\
&&\left.+\frac{X\cdot Y}{X^{2}Y^{2}}\left[2b\left(\gamma-\ln2\right)+\left(\frac{67}{9}-\frac{\pi^{2}}{3}\right)N_{c}-\frac{10}{9}N_{f}\right]\right]-\frac{N_{c}}{2}\int_{\mathbf{z}^{\prime}}\,\widetilde{K}(\mathbf{x},\,\mathbf{y},\,\mathbf{z},\,\mathbf{z}^{\prime}).\end{eqnarray}
The quark sector:
\begin{eqnarray}\label{kqq}
K_{q\bar{q}}(\mathbf{x},\,\mathbf{y},\,\mathbf{z},\,\mathbf{z}^{\prime})&&=\frac{\alpha_{s}^{2}N_{f}}{8\pi^{4}}\left(\frac{2}{Z^{4}}-\frac{(X^{\prime})^{2}Y^{2}+(Y^{\prime})^{2}X^{2}-(X-Y)^{2}Z^{2}}{Z^{4}\left(X^{2}(Y^{\prime})^{2}-(X^{\prime})^{2}Y^{2}\right)}\ln\left(\frac{X^{2}(Y^{\prime})^{2}}{(X^{\prime})^{2}Y^{2}}\right)\right.\nonumber\\
&&\left.-\frac{I_{f}(\mathbf{x},\,\mathbf{z},\,\mathbf{z}^{\prime})}{Z^{2}}-\frac{I_{f}(\mathbf{y},\,\mathbf{z},\,\mathbf{z}^{\prime})}{Z^{2}}\right),\end{eqnarray}

\begin{equation}\begin{split}&I_{f}(\mathbf{x},\, \mathbf{z},\, \mathbf{z}^{\prime})\,\equiv\,\frac{2}{Z^{2}}-\frac{2X\cdot X^{\prime}}{Z^{2}(X^{2}-(X^{\prime})^{2})}\ln\left(\frac{X^{2}}{(X^{\prime})^{2}}\right).\\
\end{split}\end{equation}
The kernels $K_{JJSSJ}$ and $K_{JJSJ}$ are anti-symmetric under $\mathbf{x}$ and $\mathbf{y}$ replacements, while the kernels $K_{JSJ}$, $K_{JSSJ}$, and $K_{q\bar{q}}$ are symmetric. Additional properties of the kernels and relations among them can be found in Appendix \ref{kerproper}.

 \section{The Light Cone Wave Function at NLO}\label{wfcalc}

In this section, we compute the LCWF $|\psi\rangle$ perturbatively in $H_{int}$, up to order $g^3$.  Normalisation  of the LCWF is done to order $g^4$.
Apart of the usual technical aspects, such as divergencies,  related to loop calculations, there are two unfamiliar subtleties. The first one originates
from the fact that the matrix elements of $H_{int}$ (\ref{hint}) over the soft states defined in section {\ref{states} 
are non-commuting operators on the valence Hilbert space. We thus have to revise the standard perturbation theory in order to account  
for non-commutativity of the  matrix  elements, carefully tracing their ordering.  This is done in Appendix \ref{wavexp}.  
The second subtlety is related to the Born-Oppenheimer approximation and its applicability. It leads to  a non-trivial condition 
on the phase of the LCWF, which otherwise cannot be determined from the perturbation theory. 
Summary of our results is presented in section \ref{finres}.

 \subsection{Third Order Perturbation Theory}
 
General perturbative expression for the wave function before normalisation, correct up to order $g^{3}$, is computed in Appendix \ref{wavexp}. After simplifying notations ($\left|n^{(i)}\right\rangle \longrightarrow\left|i\right\rangle $) and also normalising the state when dividing by its norm:
\begin{eqnarray}\label{LCWF}
\left|\psi^{NLO}\right\rangle &&=\,\mathcal{N}^{NLO}\left|0\right\rangle \,-\,\left|i\right\rangle \,\frac{\left\langle i\left|H_{int}\right|0\right\rangle }{E_{i}}\,+\,\left|i\right\rangle \,\frac{\left\langle i\left|H_{int}\right|j\right\rangle \,\left\langle j\left|H_{int}\right|0\right\rangle }{E_{i}\, E_{j}}\,\\
&&-\,\left|i\right\rangle \,\frac{\left\langle i\left|H_{int}\right|k\right\rangle \,\left\langle k\left|H_{int}\right|j\right\rangle \,\left\langle j\left|H_{int}\right|0\right\rangle }{E_{i}\, E_{j}\, E_{k}}+\,\left|i\right\rangle \,\frac{\left|\left\langle j\left|H_{int}\right|0\right\rangle \right|^{2}\left\langle i\left|H_{int}\right|0\right\rangle \,(E_{i}\,+\,2E_{j})}{2E_{i}^{2}\, E_{j}^{2}}.\nonumber\end{eqnarray}
Compared to Appendix \ref{wavexp}, the extra term $\left|i\right\rangle \,\frac{\left|\left\langle j\left|H_{int}\right|0\right\rangle \right|^{2}\left\langle i\left|H_{int}\right|0\right\rangle }{E_{i}^{2}\, E_{j}}$is a contribution to the
norm from the first order perturbation theory. Up to phase, $\mathcal{N}^{NLO}$ is determined  from  normalisation of the state:
\begin{eqnarray}\label{normaa}
\left\Vert \mathcal{N}^{NLO}\right\Vert &&\equiv\,1\,-\,\frac{\left|\left\langle i\left|H_{int}\right|0\right\rangle \right|^{2}}{2E_{i}^{2}}\,+\,\frac{\left\langle 0\left|H_{int}\right|i\right\rangle \,\left\langle i\left|H_{int}\right|j\right\rangle \,\left\langle j\left|H_{int}\right|0\right\rangle }{2E_{i}\, E_{j}}\left(\frac{1}{E_{i}}\,+\,\frac{1}{E_{j}}\right)\nonumber\\
&&-\,\frac{\left\langle 0\left|H_{int}\right|i\right\rangle \,\left\langle i\left|H_{int}\right|j\right\rangle \,\left\langle j\left|H_{int}\right|k\right\rangle \,\left\langle k\left|H_{int}\right|0\right\rangle }{2E_{i}\, E_{j}\, E_{k}}\left(\frac{1}{E_{i}}\,+\,\frac{1}{E_{j}}\,+\,\frac{1}{E_{k}}\right)\,\\
&&+\,\frac{3\left|\left\langle j\left|H_{int}\right|0\right\rangle \right|^{2}\,\left|\left\langle i\left|H_{int}\right|0\right\rangle \right|^{2}}{8E_{i}^{2}\, E_{j}^{2}}.\nonumber\end{eqnarray}
Summation over repeated indices $i$, $j$ and $k$  is over the eigenstates of free Hamiltonian  $H_0$, 
excluding the vacuum state $\left|0\right\rangle$.  The  ones that are 
relevant for the NLO calculation are defined in (\ref{gsta}), (\ref{ggst}), and (\ref{qqst}). $E_{i}$ denotes the energy of the respective eigenstate, and $H_{int}$ is defined by (\ref{hint}). 
(\ref{LCWF}) is correct even when the matrix
elements are operator valued and do not commute with each other. Notice that the overall phase of the wave function is not determined by the expansion (\ref{normaa}). We will discuss below how this phase can be nevertheless uniquely fixed.

Our objective  is to compute $\left|\psi^{NLO}\right\rangle $. It is convenient to split the wave function according to its soft component content:
 \begin{equation}\label{olLCWF}
\left|\psi^{NLO}\right\rangle \,=\,\mathcal{N}^{NLO}\left|0\right\rangle \,+\,\left|\psi_{g\:\rho}^{LO}\right\rangle \,+\,\sum_{i=1}^{i=2}\left|\psi_{q\bar{q}}^{i}\right\rangle \,+\,\sum_{i=1}^{i=3}\left|\psi_{gg}^{i}\right\rangle \,+\,\sum_{i=1}^{i=8}\left|\psi_{g}^{i}\right\rangle ,
 \end{equation}
   where the subscripts $g$, $gg$ and $q\bar{q}$ correspond to the soft particle content in each state. $\left|\psi_{g\:\rho}^{LO}\right\rangle$ is the LO contribution (see section \ref{losect}):
  \begin{equation}\begin{split}\label{lotra}
\left|\psi_{g\:\rho}^{LO}\right\rangle =-\int_{\Lambda}^{e^{\delta\mathsf{Y}}\Lambda}dk^{+}\int\frac{d^{2}\mathbf{k}}{(2\pi)^{2}}\frac{g\mathbf{k}^{i}\rho^{a}(-\mathbf{k})}{\sqrt{\pi k^{+}}\mathbf{k}^{2}}\left|g_{i}^{a}(k)\right\rangle 
=\int_{\Lambda}^{e^{\delta\mathsf{Y}}\Lambda}\frac{dk^{+}}{\sqrt{k^{+}}}\int_{\mathbf{x},\mathbf{z}}\frac{ig\rho^{a}(\mathbf{x})X^{i}}{2\pi^{3/2}X^{2}}\left|g_{i}^{a}(k^{+},\mathbf{z})\right\rangle.
\end{split}\end{equation} 

 The following states (integration over the momenta and summation over the quantum numbers of each particle are assumed) contribute and will be computed one by one below.\\
 Order $g^{2}$ states:
 
     \begin{equation}\label{qq1}
  \hspace{0.2 cm} \left|\psi_{q\bar{q}}^{1}\right\rangle \,\equiv\,\left|\overline{q}\, q\right\rangle \left\langle \overline{q}\, q\left|H_{gqq}\right|g\right\rangle \left\langle g\left|H_{g}\right|0\right\rangle /E_{q\bar{q}}\, E_{g},
   \end{equation}
  \begin{equation}\label{qq2}
\hspace{-1 cm}\left|\psi_{q\bar{q}}^{2}\right\rangle \,\equiv\,-\left|\overline{q}\, q\right\rangle \left\langle \overline{q}\, q\left|H_{qq-inst}\right|0\right\rangle /E_{q\bar{q}},
  \end{equation} 
   \begin{equation}\label{agg1}
\left|\psi_{gg}^{1}\right\rangle \,\equiv\,\left|g\, g\right\rangle \left\langle g\, g\left|H_{g}\right|g\right\rangle \left\langle g\left|H_{g}\right|0\right\rangle /E_{gg}\, E_{g},
  \end{equation}
   \begin{equation}\label{agg2}
\hspace{0.3 cm}\left|\psi_{gg}^{2}\right\rangle \,\equiv\,\left|g\, g\right\rangle \left\langle g\, g\left|H_{ggg}\right|g\right\rangle \left\langle g\left|H_{g}\right|0\right\rangle /E_{gg}\, E_{g},
\end{equation}
  \begin{equation}\label{agg3}
\hspace{-0.9 cm}\left|\psi_{gg}^{3}\right\rangle \,\equiv\,-\left|g\, g\right\rangle \left\langle g\, g\left|H_{gg-inst}\right|0\right\rangle /E_{gg}.
   \end{equation}
Order $g^{3}$ states:
      \begin{equation}\label{ag1}
\hspace{-0.7 cm}\left|\psi_{g}^{1}\right\rangle \,\equiv\,-\left|g\right\rangle \left\langle g\left|H_{gqq}\right|\overline{q}\, q\right\rangle \left\langle \overline{q}\, q\left|H_{gqq}\right|g_{1}\right\rangle \left\langle g_{1}\left|H_{g}\right|0\right\rangle /E_{g}\, E_{q\bar{q}}\, E_{g_{1}},
\end{equation}
  \begin{equation}\label{ag2}
  \hspace{-0.6 cm}\left|\psi_{g}^{2}\right\rangle \,\equiv\,-\left|g\right\rangle \left\langle g\left|H_{ggg}\right|g\, g\right\rangle \left\langle g\, g\left|H_{ggg}\right|g_{1}\right\rangle \left\langle g_{1}\left|H_{g}\right|0\right\rangle /E_{g}\, E_{gg}\, E_{g_{1}},
\end{equation}
  \begin{equation}\label{ag3}
  \hspace{-3.4 cm}\left|\psi_{g}^{3}\right\rangle \,\equiv\,\left|g\right\rangle \left\langle g\left|H_{gg-inst}\right|g_{1}\right\rangle \left\langle g_{1}\left|H_{g}\right|0\right\rangle /E_{g}\, E_{g_{1}},
\end{equation}
  \begin{equation}\label{ag4}
\hspace{-3.2 cm}\left|\psi_{g}^{4}\right\rangle  \,\equiv\,\left|g\right\rangle \left\langle g\left|H_{g}\right|g\, g\right\rangle \left\langle g\, g\left|H_{gg-inst}\right|0\right\rangle /E_{g}\, E_{gg},
\end{equation}
 \begin{equation}\label{ag5}
  \hspace{-1.2 cm}\left|\psi_{g}^{5}\right\rangle  \,\equiv\,-\left|g\right\rangle \left\langle g\left|H_{g}\right|g\, g\right\rangle \left\langle g\, g\left|H_{g}\right|g_{1}\right\rangle \left\langle g_{1}\left|H_{g}\right|0\right\rangle /E_{g}\, E_{gg}\, E_{g_{1}},
  \end{equation}
  \begin{equation}\label{ag6}
 \hspace{-0.9 cm}\left|\psi_{g}^{6}\right\rangle \,\equiv\,-\left|g\right\rangle \left\langle g\left|H_{ggg}\right|g\, g\right\rangle \left\langle g\, g\left|H_{g}\right|g_{1}\right\rangle \left\langle g_{1}\left|H_{g}\right|0\right\rangle /E_{g}\, E_{gg}\, E_{g_{1}},
\end{equation}
  \begin{equation}\label{ag7}
\hspace{-0.9 cm}\left|\psi_{g}^{7}\right\rangle \,\equiv\,-\left|g\right\rangle \left\langle g\left|H_{g}\right|g\, g\right\rangle \left\langle g\, g\left|H_{ggg}\right|g_{1}\right\rangle \left\langle g_{1}\left|H_{g}\right|0\right\rangle /E_{g}\, E_{gg}\, E_{g_{1}},
\end{equation}
  \begin{equation}\label{ag8}
\hspace{-2 cm}\left|\psi_{g}^{8}\right\rangle \,\equiv\,\left|g\right\rangle \left\langle g\left|H_{g}\right|0\right\rangle \left|\left\langle g_{1}\left|H_{g}\right|0\right\rangle \right|^{2}(2E_{g_{1}}+E_{g})/2E_{g}^{2}\, E_{g_{1}}^{2}.
\end{equation}
The following contributions vanish:
      \begin{equation}
-\left|g\right\rangle \left\langle g\left|H_{ggg}\right|g\, g\right\rangle \left\langle g\, g\left|H_{gg-inst}\right|g_{1}\right\rangle \left\langle g_{1}\left|H_{g}\right|0\right\rangle /E_{g}\, E_{gg}\, E_{g_{1}}\,=\,0,
\end{equation}
  \begin{equation}
-\left|g\right\rangle \left\langle g\left|H_{gg-inst}\right|g\, g\right\rangle \left\langle g\, g\left|H_{ggg}\right|g_{1}\right\rangle \left\langle g_{1}\left|H_{g}\right|0\right\rangle /E_{g}\, E_{gg}\, E_{g_{1}}\,=\,0,
\end{equation}
  \begin{equation}
\left|g\right\rangle \left\langle g\left|H_{gggg}\right|g_{1}\right\rangle \left\langle g_{1}\left|H_{g}\right|0\right\rangle /E_{g}\, E_{\bar{g}_{1}}\,=\,0.
 \end{equation}
 
     \subsection{Matrix Elements}
We write down all the matrix elements that are relevant for the NLO calculation. These expressions are computed based on (\ref{eikohg}) $-$ (\ref{qqinst}). At NLO there are two matrix elements for the soft-soft interactions and four matrix elements for the soft-valence interactions:\\
 $\bullet$ \quad\textit{\textbf{Gluon splits into quark and anti-quark pair}}
\begin{equation}\begin{split}\label{gqq1}
     \left\langle \overline{q}_{\lambda_{2}}^{\beta}(q)\, q_{\lambda_{1}}^{\alpha}(p)\left|H_{gqq}\right|g_{i}^{a}(k)\right\rangle
     =\frac{gt_{\alpha\beta}^{a}}{8\pi^{3/2}\sqrt{k^{+}}}\chi_{\lambda_{1}}^{\dagger}\left[\frac{2\mathbf{k}^{i}}{k^{+}}-\frac{\sigma\cdot \mathbf{p}}{p^{+}}\sigma^{i}-\sigma^{i}\frac{\sigma\cdot \mathbf{q}}{q^{+}}\right]\chi_{\lambda_{2}}\delta^{(3)}(k-p-q).
 \end{split}\end{equation}\\
 $\bullet$ \quad\textit{\textbf{Triple gluon interaction}}
\begin{eqnarray}\label{ggg_split}
&&\left\langle g_{l}^{c}(q)\, g_{n}^{b}(p)\left|H_{ggg}\right|g_{m}^{a}(k)\right\rangle =\frac{igf^{abc}\delta^{(3)}(k-p-q)}{8\pi^{3/2}\sqrt{k^{+}p^{+}q^{+}}}\left[\left(\mathbf{p}^{m}-\mathbf{q}^{m}+\frac{q^{+}-p^{+}}{k^{+}}\mathbf{k}^{m}\right)\delta_{nl}\right.\nonumber\\
&&\quad\quad\quad \left.+\left(\mathbf{k}^{n}+\mathbf{q}^{n}-\frac{k^{+}+q^{+}}{p^{+}}\mathbf{p}^{n}\right)\delta_{ml}+\left(\frac{k^{+}+p^{+}}{q^{+}}\mathbf{q}^{l}-\mathbf{p}^{l}-\mathbf{k}^{l}\right)\delta_{mn}\right].\end{eqnarray}\\
$\bullet$ \quad\textit{\textbf{Valence adds an extra gluon to an existing one}}
\begin{equation}\label{g2}
\left\langle g_{l}^{c}(q)\, g_{j}^{b}(p)\left|H_{g}\right|g_{i}^{a}(k)\right\rangle \,=\,\delta^{ac}\delta_{il}\delta^{(3)}(k-q)\,\frac{g\mathbf{p}^{j}\rho^{b}(-\mathbf{p})}{4\pi^{3/2}|p^{+}|^{3/2}}\,+\,\delta^{ab}\delta_{ij}\delta^{(3)}(k-p)\,\frac{g\mathbf{q}^{l}\rho^{c}(-\mathbf{q})}{4\pi^{3/2}|q^{+}|^{3/2}}.
 \end{equation}
\\
 $\bullet$ \quad\textit{\textbf{Valence instantaneously creates a quark and anti-quark pair}}
\begin{equation}\label{qq}
\left\langle \overline{q}_{\lambda_{2}}^{\beta}(q)\, q_{\lambda_{1}}^{\alpha}(p)\left|H_{qq-inst}\right|0\right\rangle \,=\,\frac{g^{2}t_{\alpha\beta}^{a}\rho^{a}(-\mathbf{p}-\mathbf{q})}{8\pi^{3}(p^{+}+q^{+})^{2}}\chi_{\lambda_{1}}^{\dagger}\chi_{\lambda_{2}}.
 \end{equation}
 \\
  $\bullet$ \quad\textit{\textbf{Valence instantaneously creates two gluons}}
\begin{equation}\label{gg1}
\left\langle g_{j}^{c}(q)\, g_{i}^{b}(p)\left|H_{gg-inst}\right|0\right\rangle \,=\,\frac{ig^{2}f^{abc}(q^{+}-p^{+})\delta_{ij}\rho^{a}(-\mathbf{p}-\mathbf{q})}{2(2\pi)^{3}\sqrt{p^{+}q^{+}}(p^{+}+q^{+})^{2}}.
 \end{equation}
 \\
   $\bullet$ \quad\textit{\textbf{Valence instantaneously interacts with a soft gluon}}
\begin{equation}\label{gg2}
\left\langle g_{j}^{c}(q)\left|H_{gg-inst}\right|g_{i}^{b}(p)\right\rangle \,=\,\frac{ig^{2}f^{abc}(p^{+}+q^{+})\delta_{ij}\rho^{a}(\mathbf{p}-\mathbf{q})}{2(2\pi)^{3}\sqrt{p^{+}q^{+}}(p^{+}-q^{+})^{2}}.
 \end{equation}

\subsection{Technical Aspects of the Calculation}

Before diving into the calculation of the LCWF (\ref{LCWF}) in the next subsection, let us first introduce several important technical aspect,
which are used throughout this calculation.\\ \\
 $\bullet$ \textit{\textbf{The Decomposition Procedure}}
 
 As explained in the Introduction, the result of the calculation of the LCWF will have the form (\ref{psi}).  Many of the terms we are to compute will
 have a product of  more than one $\rho$ operator. These operators do not commute and their mutual ordering is very important. (\ref{LCWF})
 will be later used to compute the NLO JIMWLK Hamiltonian. In order to compare  with section \ref{sdfla}, we would have to represent 
 our result as in (\ref{ham}),  including the symmetry properties for  the kernels (the ones mentioned at the end of section \ref{sdfla}).  These 
 symmetry properties do not come automatically in our calculation and we would have to manipulate our final results.
 It turns out that in order to achieve this, it suffices to symmetrise any product of two $\rho$ operators in the LCWF
via the following identity: 
\begin{equation}\begin{split}\label{decompose}
&\rho^{b}(-\mathbf{k}+\mathbf{p})\,\rho^{a}(-\mathbf{p})\,=\,\frac{1}{2}\left[\rho^{b}(-\mathbf{k}+\mathbf{p}),\,\rho^{a}(-\mathbf{p})\right]\,+\,\frac{1}{2}\left\{ \rho^{b}(-\mathbf{k}+\mathbf{p}),\,\rho^{a}(-\mathbf{p})\right\} \\
&=\,\frac{i}{2}f^{bad}\rho^{d}(-\mathbf{k})\,+\,\frac{1}{2}\left\{ \rho^{b}(-\mathbf{k}+\mathbf{p}),\,\rho^{a}(-\mathbf{p})\right\}. 
\end{split}\end{equation}
The replacement (\ref{decompose}) will be referred to as  \textit{``decomposition procedure''}.  
  The physics behind this decomposition is that it explicitly separates between contributions from two emitters being in different  transverse
 points in coordinate space from that when both emitters are in the  same point. The latter is then equivalent to a contribution  from a single emitter, in 
 accordance to the colour algebra.
After implementing the decomposition  procedure where applicable, 
each component of the LCWF will receive an extra label in accord to the number of $\rho$ operators it contains.
For example, in section \ref{compqq} we show that  $\left|\psi_{q\bar{q}}^{1}\right\rangle$ contains only one $\rho$ operator, 
and  this component will be denoted thereafter by $\left|\psi_{q\bar{q}\:\rho}^{1}\right\rangle$.
\\
\\
 $\bullet$ \textit{\textbf{Regularisation of Divergent Integrals}}
 
 There are two types of divergencies that appear throughout the calculation. These are the usual divergencies in loop calculations. First, we will face divergencies that come from the longitudinal phase space integrals of the type $\int dk^+/k^+$, which correspond to  emission of soft partons, just as in the case of LO calculation.   These divergencies are regularised by confining the soft modes to live within the window $\Lambda<k^+< \Lambda e^{\delta Y}$. 
When computing the Hamiltonian in the next section, we will also encounter this longitudinal divergence "squared".  Physically, such double
 divergence corresponds to independent (uncorrelated) emission of two soft gluons, which is accounted for by iteration of the LO Hamiltonian. 
 
  $\Lambda$ acts as an IR cutoff which will be set to zero (and $\delta Y$ to infinity) whenever it is safe to do so,  without stating it explicitly. 
 

UV divergences in transverse space integrals will be treated via  dimensional regularisation, switching to 
$d\,\equiv\,2\,-\,\epsilon$ dimensions, 
\begin{equation}\label{dimregm}
\int\frac{d^{2}\mathbf{p}}{(2\pi)^{2}}\,\longrightarrow\,\mu^{\epsilon}\int\frac{d^{d}\mathbf{p}}{(2\pi)^{d}}.
 \end{equation}
In addition, we will be frequently  using the identity:
\begin{equation}\begin{split}\label{por}
\int d^{d}\mathbf{p}\, F(\mathbf{p}^{2})\, \mathbf{p}^{i}\, \mathbf{p}^{j}\,=\,\frac{\delta^{ij}}{d}\int d^{d}\mathbf{p}\, F(\mathbf{p}^{2})\, \mathbf{p}^{2}.
  \end{split}\end{equation}
A couple of relevant integrals in dimensional regularisation are quoted in Appendix \ref{inttab}.  
As usual, (\ref{gamexp}) will be used for $\epsilon$ expansion.
We will retain the divergent $\frac{1}{\epsilon}$ terms  at the level of the wave function, as by itself the wave function is not a physical observable. 
Coupling constant renormalisation is done for physical observables, such as $\Sigma$ (or the JIMWLK Hamiltonian).  This will be implemented
 in the  $\overline{MS}$ scheme as a default action without any further explanations. 
  The calculation also involves overlapping singularities when both the longitudinal and transverse integrals diverge. Those are treated 
 similarly as above.
 
In addition to the above discussed UV divergencies, 
there are also potential IR divergencies at small transverse momenta. Those, however, do not emerge in our calculation explicitly 
because they always get regularised by the charge density operators $\rho$.  Yet, this depends on a non-specified IR behaviour of the valence degrees of freedom and a priori is not guaranteed to lead to a safe IR.  Particularly the IR divergencies  resurface when the resulting JIMWLK  Hamiltonian is taken to act on a gauge non-invariant operator, such as in the problem of gluon reggeization. The gluon trajectory is then found to be IR divergent even at LO.


 $\bullet$ \textit{\textbf{Phase of the LCWF}}

The perturbation theory cannot fix the phase of the LCWF, that is the phase of ${\cal N}^{NLO}$. Normally, the phase is unobservable 
as it cancels in physical observables. However,  this is not the case at hand, as we are not computing the full wave function but only its soft component. The normalisation coefficient ${\cal N}^{NLO}$ is operator valued on the valence Hilbert space and thus
its phase can make a non-trivial contribution to the observables.  

How can we find this phase which is clearly beyond the perturbation theory? We recall that our formalism is based on the Born-Oppenheimer adiabatic approximation\footnote{We would like to specially thank Alex Kovner 
who significantly helped us to establish the argument.}. 
Obviously, there is some  freedom in the phase, as, in principle,
it can be always absorbed into the valence part of the LCWF.   Yet, the factorisation  (\ref{psisplit}) assumes that
\begin{equation}\label{phase}
\langle v|\otimes \langle \psi |\,H_V\, |\psi \rangle \otimes |v\rangle\,\simeq\, 
   \langle v|\,H_V\,|v \rangle \,
~~~~or~~~~~
 \left\langle \psi \right|\,H_V\,\left|\psi \right\rangle_{soft} \,\simeq\,0\,.
 \end{equation}
where $H_V$ is  a Hamiltonian for the valence degrees of freedom. The second expectation value is taken over the soft modes only.  
The condition (\ref{phase}) means that the dynamics of the 
soft modes does not significantly affect that of the valence. This assumption is implicit in the way the high energy factorisation is set,
eliminating the freedom in the phase.

While the explicit form of $H_V$ is not known, the condition  (\ref{phase}) presumably should be valid for any operator on the valence 
Hilbert space. Particularly, $H_V$  can be thought as been constructed from some powers of $\rho$, with
 (\ref{phase}) implying  that they  commute with $|\psi\rangle $ on the valence Hilbert space. Commuting one $\rho$ with 
$|\psi\rangle$  is equivalent to taking derivative  $\delta/\delta\rho$. We thus require the following condition to hold
\begin{equation}\label{condphase}
\left\langle \psi\right|\,\frac{\delta}{\delta\rho^{d}(\mathbf{w})}\,\left|\psi\right\rangle \,=\,0\,.
 \end{equation}
 Notice that (\ref{condphase}) is  the Berry connection, which we require to vanish.
 It is easy to check that at LO the condition is fulfilled by $|\psi^{LO}\rangle $ automatically. At NLO, this condition becomes 
 a non-trivial constraint on the phase, which is worked out in  Appendix \ref{wfphase}.


\subsection{Computation of the NLO Wave Function}\label{eval}
In this section we  compute the wave function (\ref{LCWF}). For each of the contributions (\ref{qq1}) - (\ref{ag8}), a separate subsection is devoted. Lengthy computations are available in Appendix \ref{supsect3}. The reader who is not interested in these details can proceed directly to the final results in section \ref{finres}.

\subsubsection{Quark anti$-$Quark State}\label{compqq}

\subsubsection*{Computation of $\left|\psi_{q\bar{q}}^{1}\right\rangle $}
$\left|\psi_{q\bar{q}\:\rho}^{1}\right\rangle$ is defined in (\ref{qq1}) (Fig. \ref{quarfig}a).
 \begin{figure}[!h]
$\qquad\quad\qquad$ \includegraphics[scale=0.66]{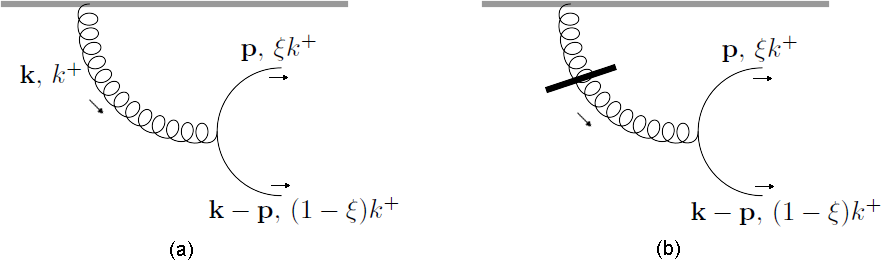}
  \caption{Quark and anti-quark emission diagrams.  The bold black line over the gluon in $(b)$ marks the instantaneous interaction. $(a)$ - $\left|\psi_{q\bar{q}}^{1}\right\rangle$, $(b)$ - $\left|\psi_{q\bar{q}}^{2}\right\rangle$. 
 \label{quarfig}}
\end{figure}
\begin{equation}\begin{split}\label{qor}
\left|\psi_{q\bar{q}}^{1}\right\rangle & \equiv\,\sum_{\lambda_{1},\lambda_{2},f}\int_{\Lambda}^{e^{\delta\mathsf{Y}}\Lambda}dk^{+}\, dp^{+}\, dq^{+}\,\int d^{2}\mathbf{k}\, d^{2}\mathbf{p}\, d^{2}\mathbf{q}\\
&\times\left|\bar{q}_{\lambda_{2}}^{\beta,\, f}(q)\, q_{\lambda_{1}}^{\alpha,\, f}(p)\right\rangle \frac{\left\langle \bar{q}_{\lambda_{2}}^{\beta,\, f}(q)\, q_{\lambda_{1}}^{\alpha,\, f}(p)\left|H_{gqq}\right|g_{i}^{a}(k)\right\rangle \left\langle g_{i}^{a}(k)\left|H_{g}\right|0\right\rangle }{E_{q\bar{q}}(p,\, q)\, E_{g}(k)}.
\end{split}\end{equation}
 The physics content of (\ref{qor}) becomes clear by reading the matrix elements  from the right to left:  the first one corresponds to  initial emission of a soft gluon from the valence current; then this soft gluon subsequently splits to  a quark and anti-quark pair.
The momentum conservation  $\delta^{(3)}(q-k+p)$, which is explicit in the second matrix element (\ref{gqq1}), removes $q$ integral and effectively turns it
into $k-p$.  The process  is illustrated in Fig. \ref{quarfig}a. 


After inserting the relevant matrix elements, (\ref{g}) and (\ref{gqq1}), and integrating over $q$:
\begin{equation}\begin{split}\label{qq_cont1}
&\left|\psi_{q\bar{q}\:\rho}^{1}\right\rangle \,\equiv\,\left|\psi_{q\bar{q}}^{1}\right\rangle \,=\,\sum_{\lambda_{1},\lambda_{2},f}\int_{\Lambda}^{e^{\delta\mathsf{Y}}\Lambda}dk^{+}\,\int_{0}^{1}d\xi\,\int d^{2}\mathbf{k}\, d^{2}\mathbf{p}\:\frac{4\xi(1-\xi)}{\left((1-\xi)\mathbf{p}^{2}+\xi(\mathbf{k}-\mathbf{p})^{2}\right)\mathbf{k}^{2}}\\
&\times\left(\frac{gt_{\alpha\beta}^{a}}{8\pi^{3/2}}\chi_{\lambda_{1}}^{\dagger}\left[2\mathbf{k}^{i}-\frac{\sigma\cdot\mathbf{p}}{\xi}\sigma^{i}-\sigma^{i}\frac{\sigma\cdot(\mathbf{k}-\mathbf{p})}{1-\xi}\right]\chi_{\lambda_{2}}\right)\left(\frac{g\rho^{a}(-\mathbf{k})\mathbf{k}^{i}}{4\pi^{3/2}}\right)\\
&\times\left|\bar{q}_{\lambda_{2}}^{\beta,\,f}((1-\xi)k^{+},\,\mathbf{k}-\mathbf{p})\,q_{\lambda_{1}}^{\alpha,\,f}(\xi k^{+},\,\mathbf{p})\right\rangle .
\end{split}\end{equation} 
The variable $\xi$ is defined as:
\begin{equation}\label{chanvar}
\xi\,\equiv\,\frac{p^{+}}{k^{+}}.
 \end{equation}
This definition will be used repeatedly  throughout this paper.
    \subsubsection*{Computation of $\left|\psi_{q\bar{q}}^{2}\right\rangle $}
$\left|\psi_{q\bar{q}\:\rho}^{2}\right\rangle$ is defined in (\ref{qq2}) (Fig. \ref{quarfig}b).
\begin{equation}\begin{split}
\left|\psi_{q\bar{q}}^{2}\right\rangle & \equiv\,-\sum_{\lambda_{1},\lambda_{2},f}\int_{\Lambda}^{e^{\delta\mathsf{Y}}\Lambda}dk^{+}\,\int_{0}^{k^{+}}dp^{+}\,\int d^{2}\mathbf{k}\, d^{2}\mathbf{p}\:\\
&\times\left|\bar{q}_{\lambda_{2}}^{\beta,\, f}(k-p)\, q_{\lambda_{1}}^{\alpha,\, f}(p)\right\rangle \frac{\left\langle \bar{q}_{\lambda_{2}}^{\beta,\, f}(k-p)\, q_{\lambda_{1}}^{\alpha,\, f}(p)\left|H_{qq-inst}\right|0\right\rangle }{E_{q\bar{q}}(p,\, k-p)}.
\end{split}\end{equation} 

Inserting the matrix element (\ref{qq}) and changing variables according to (\ref{chanvar}):
\begin{equation}\begin{split}\label{qq_cont2}
\left|\psi_{q\bar{q}\:\rho}^{2}\right\rangle &\equiv\,\left|\psi_{q\bar{q}}^{2}\right\rangle =\,-\sum_{\lambda_{1},\lambda_{2},f}\int_{\Lambda}^{e^{\delta\mathsf{Y}}\Lambda}dk^{+}\,\int_{0}^{1}d\xi\,\int d^{2}\mathbf{k}\, d^{2}\mathbf{p}\:\\
&\times\frac{g^{2}t_{\alpha\beta}^{a}\xi(1-\xi)\rho^{a}(-\mathbf{k})\chi_{\lambda_{1}}^{\dagger}\chi_{\lambda_{2}}}{4\pi^{3}\left((1-\xi)\mathbf{p}^{2}+\xi(\mathbf{k}-\mathbf{p})^{2}\right)}\left|\bar{q}_{\lambda_{2}}^{\beta,\,f}((1-\xi)k^{+},\,\mathbf{k}-\mathbf{p})\,q_{\lambda_{1}}^{\alpha,\,f}(\xi k^{+},\,\mathbf{p})\right\rangle .\\
\end{split}\end{equation}

\subsubsection{Two Gluon State}

\subsubsection*{Computation of $\left|\psi_{gg}^{1}\right\rangle$}
$\left|\psi_{gg}^{1}\right\rangle$ is defined in (\ref{agg1}) (Fig. \ref{ggfigo}a).
    \begin{figure}[!ht]
   $\:$ \includegraphics[scale=0.79]{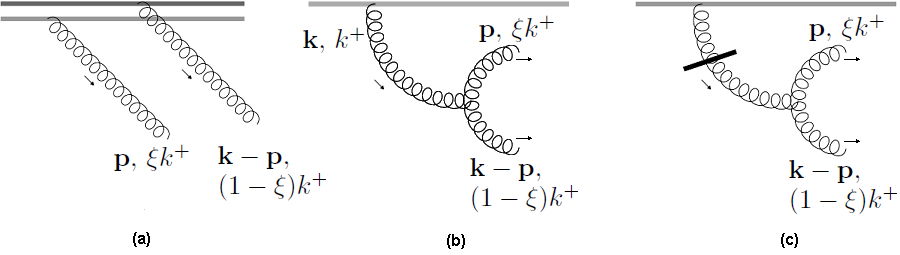}
  \caption{Two gluon producation diagrams. (a) - $\left|\psi_{gg}^{1}\right\rangle$, (b) - $\left|\psi_{gg}^{2}\right\rangle$, (c) - $\left|\psi_{gg}^{3}\right\rangle$.\label{ggfigo}}
\end{figure}
\begin{equation}\begin{split}\label{ggone}
\left|\psi_{gg}^{1}\right\rangle &\equiv\,\frac{1}{2}\int_{\Lambda}^{e^{\delta\mathsf{Y}}\Lambda}dq^{+}\, dk^{+}\,\int_{\Lambda}^{k^{+}-\Lambda}dp^{+}\,\int\, d^{2}\mathbf{q}\, d^{2}\mathbf{k}\, d^{2}\mathbf{p}\\
&\times\left|g_{j}^{b}(k-p)\, g_{l}^{c}(p)\right\rangle \frac{\left\langle g_{j}^{b}(k-p)\, g_{l}^{c}(p)\left|H_{g}\right|g_{i}^{a}(q)\right\rangle \left\langle g_{i}^{a}(q)\left|H_{g}\right|0\right\rangle }{E_{gg}(p,\, k-p)\, E_{g}(q)}.
\end{split}\end{equation}
Here $\frac{1}{2}$ is to avoid double-counting in two-gluon states. Inserting the matrix elements, (\ref{g}) and (\ref{g2}):
\begin{eqnarray}\label{afsda1}
\left|\psi_{gg}^{1}\right\rangle& =&\frac{1}{2}\int_{\Lambda}^{e^{\delta\mathsf{Y}}\Lambda}dk^{+}\, dq^{+}\,\int_{\Lambda}^{k^{+}-\Lambda}dp^{+}\,\int d^{2}\mathbf{k}\, d^{2}\mathbf{p}\, d^{2}\mathbf{q}\,\frac{1}{\frac{\mathbf{q}^{2}}{2q^{+}}\left(\frac{\mathbf{p}^{2}}{2p^{+}}+\frac{(\mathbf{k}-\mathbf{p})^{2}}{2(k^{+}-p^{+})}\right)}\nonumber \\
&\times&\left(\delta^{ac}\delta_{il}\delta^{(3)}(p-q)\,\frac{g\left(\mathbf{k}^{j}-\mathbf{p}^{j}\right)\rho^{b}(-\mathbf{k}+\mathbf{p})}{4\pi^{3/2}|k^{+}-p^{+}|^{3/2}}+\delta^{ab}\delta_{ji}\delta^{(3)}(k-p-q)\,\frac{g\mathbf{p}^{l}\rho^{c}(-\mathbf{p})}{4\pi^{3/2}|p^{+}|^{3/2}}\right)\nonumber \\
&\times&\left(\frac{g\mathbf{q}^{i}\rho^{a}(-\mathbf{q})}{4\pi^{3/2}|q^{+}|^{3/2}}\right)\left|g_{j}^{b}(k-p)\, g_{l}^{c}(p)\right\rangle. 
\end{eqnarray} 
 The physics content of (\ref{ggone}) is quite obvious and  corresponds to subsequent emissions of two soft gluons. 
The momentum conservation  imposed via the delta functions  turns the momentum  $q$ into either $k$ or
$k-p$, which  is illustrated in Fig. \ref{ggfigo}a.  After integration over  $q$, (\ref{afsda1})  becomes:
\begin{equation}\label{11}\begin{split}
\left|\psi_{gg}^{1}\right\rangle &=\int_{\Lambda}^{e^{\delta\mathsf{Y}}\Lambda}dk^{+}\,\int d^{2}\mathbf{k}\, d^{2}\mathbf{p}\,\int_{\frac{\Lambda}{k^{+}}}^{1-\frac{\Lambda}{k^{+}}}d\xi\,\frac{g^{2}\mathbf{p}^{j}\left(\mathbf{k}^{i}-\mathbf{p}^{i}\right)}{8\pi^{3}\left(\xi(\mathbf{k}-\mathbf{p})^{2}+(1-\xi)\mathbf{p}^{2}\right)}\\
&\times\left(\frac{\rho^{a}(-\mathbf{k}+\mathbf{p})\rho^{b}(-\mathbf{p})\sqrt{\xi}}{\mathbf{p}^{2}\sqrt{1-\xi}}\,+\,\frac{\rho^{b}(-\mathbf{p})\rho^{a}(-\mathbf{k}+\mathbf{p})\sqrt{1-\xi}}{(\mathbf{k}-\mathbf{p})^{2}\sqrt{\xi}}\right)\,\\
&\times\left|g_{i}^{a}((1-\xi)k^{+},\,\mathbf{k}-\mathbf{p})\,g_{j}^{b}(\xi k^{+},\,\mathbf{p})\right\rangle .\\
\end{split}\end{equation}
Here $\xi\,\equiv\,\frac{p^{+}}{k^{+}}$ as already defined  in (\ref{chanvar}).

The result (\ref{11}) can be decomposed according to (\ref{decompose}), and represented as contributions with one and two $\rho$ operators:
\begin{equation}
\left|\psi_{gg}^{1}\right\rangle \,\equiv\,\left|\psi_{gg\:\rho}^{1}\right\rangle \,+\,\left|\psi_{gg\:\rho\rho}\right\rangle,
\end{equation} 
with
\begin{eqnarray}\label{contt}
\left|\psi_{gg\:\rho}^{1}\right\rangle &\equiv&\int_{\Lambda}^{e^{\delta\mathsf{Y}}\Lambda}dk^{+}\,\int d^{2}\mathbf{k}\, d^{2}\mathbf{p}\,\int_{\frac{\Lambda}{k^{+}}}^{1-\frac{\Lambda}{k^{+}}}d\xi\,\frac{-ig^{2}f^{abc}\mathbf{p}^{i}\left(\mathbf{k}^{j}-\mathbf{p}^{j}\right)\rho^{c}(-\mathbf{k})\sqrt{\xi(1-\xi)}}{16\pi^{3}\left(\xi(\mathbf{k}-\mathbf{p})^{2}+(1-\xi)\mathbf{p}^{2}\right)}\nonumber\\
&\times&\left(\frac{1}{(1-\xi)\mathbf{p}^{2}}-\frac{1}{\xi(\mathbf{k}-\mathbf{p})^{2}}\right)\left|g_{j}^{b}((1-\xi)k^{+},\,\mathbf{k}-\mathbf{p})\,g_{i}^{a}(\xi k^{+},\,\mathbf{p})\right\rangle ,\end{eqnarray}
and
\begin{eqnarray}\label{ggrhorho}
\left|\psi_{gg\:\rho\rho}\right\rangle &\equiv&\int_{\Lambda}^{e^{\delta\mathsf{Y}}\Lambda}dk^{+}\,\int d^{2}\mathbf{k}\, d^{2}\mathbf{p}\,\int_{\frac{\Lambda}{k^{+}}}^{1-\frac{\Lambda}{k^{+}}}d\xi\,\\
&\times&\frac{g^{2}\mathbf{p}^{i}\left(\mathbf{k}^{j}-\mathbf{p}^{j}\right)\left\{ \rho^{a}(-\mathbf{p}),\,\rho^{b}(-\mathbf{k}+\mathbf{p})\right\} }{16\pi^{3}\mathbf{p}^{2}(\mathbf{k}-\mathbf{p})^{2}\sqrt{\xi(1-\xi)}}\left|g_{j}^{b}((1-\xi)k^{+},\,\mathbf{k}-\mathbf{p})\,g_{i}^{a}(\xi k^{+},\,\mathbf{p})\right\rangle .\nonumber
\end{eqnarray}

\subsubsection*{Computation of $\left|\psi_{gg}^{2}\right\rangle$}
$\left|\psi_{gg}^{2}\right\rangle$ is defined in (\ref{agg2}) (Fig. \ref{ggfigo}b).
\begin{equation}\begin{split}\label{ggtwo}
\left|\psi_{gg}^{2}\right\rangle &\equiv\,\frac{1}{2}\int_{\Lambda}^{e^{\delta\mathsf{Y}}\Lambda}dk^{+}\, dp^{+}\, dq^{+}\,\int d^{2}\mathbf{k}\, d^{2}\mathbf{p}\, d^{2}\mathbf{q}\\
&\times\left|g_{j}^{b}(q)\, g_{l}^{c}(p)\right\rangle \frac{\left\langle g_{j}^{b}(q)\, g_{l}^{c}(p)\left|H_{ggg}\right|g_{i}^{a}(k)\right\rangle \left\langle g_{i}^{a}(k)\left|H_{g}\right|0\right\rangle }{E_{gg}(p,\, q)\, E_{g}(k)}.\\
\end{split}\end{equation}
The condition $q=k-p$ is enforced via momentum conservation, which is  reflected on Fig. \ref{ggfigo}b. 
By inserting the relevant matrix elements, (\ref{g}) and (\ref{ggg_split}), and  integrating over $q$:
\begin{eqnarray}\label{ggmid}
\left|\psi_{gg}^{2}\right\rangle 
&=&\int_{\Lambda}^{e^{\delta\mathsf{Y}}\Lambda}dk^{+}\int_{\Lambda}^{k^{+}}dp^{+}\int d^{2}\mathbf{k}\,d^{2}\mathbf{p}\:\frac{igf^{acb}}{16\pi^{3/2}\left(\frac{\mathbf{p}^{2}}{2p^{+}}+\frac{(\mathbf{k}-\mathbf{p})^{2}}{2(k^{+}-p^{+})}\right)\left(\frac{\mathbf{k}^{2}}{2k^{+}}\right)\sqrt{k^{+}p^{+}(k^{+}-p^{+})}}\nonumber\\
&\times&\left(\left[2\mathbf{p}^{i}-\frac{2p^{+}}{k^{+}}\mathbf{k}^{i}\right]\delta_{jl}+\left[\frac{k^{+}+p^{+}}{k^{+}-p^{+}}(\mathbf{k}^{j}-\mathbf{p}^{j})-\mathbf{k}^{j}-\mathbf{p}^{j}\right]\delta_{il}+\left[2\mathbf{k}^{l}-\frac{2k^{+}}{p^{+}}\mathbf{p}^{l}\right]\delta_{ij}\right)\nonumber\\
&\times&\left(\frac{g\rho^{a}(-\mathbf{k})\mathbf{k}^{i}}{4\pi^{3/2}|k^{+}|^{3/2}}\right)\left|g_{j}^{b}(k-p)\, g_{l}^{c}(p)\right\rangle. \end{eqnarray} \\
By introducing new variables according to (\ref{chanvar}) and
\begin{equation}\label{chanvar2}
\mathbf{p}\,=\,\xi \mathbf{k}\,+\,\widetilde{\mathbf{p}},
 \end{equation}
%
(\ref{ggmid}) becomes:

\begin{equation}\begin{split}\label{contgg}
\left|\psi_{gg\:\rho}^{2}\right\rangle &\equiv\,\left|\psi_{gg}^{2}\right\rangle\, =\,-\int_{\Lambda}^{e^{\delta\mathsf{Y}}\Lambda}dk^{+}\,\int d^{2}\mathbf{k}\, d^{2}\widetilde{\mathbf{p}}\,\int_{\frac{\Lambda}{k^{+}}}^{1-\frac{\Lambda}{k^{+}}}d\xi\,\frac{ig^{2}f^{abc}\rho^{a}(-\mathbf{k})\sqrt{\xi(1-\xi)}\mathbf{k}^{i}}{8\pi^{3}\left(\mathbf{k}^{2}\xi(1-\xi)+\widetilde{\mathbf{p}}^{2}\right)\mathbf{k}^{2}}\\
&\times\left(\widetilde{\mathbf{p}}^{i}\delta_{jl}-\frac{1}{1-\xi}\widetilde{\mathbf{p}}^{j}\delta_{il}-\frac{1}{\xi}\widetilde{\mathbf{p}}^{l}\delta_{ij}\right)\left|g_{j}^{b}((1-\xi)k^{+},\,(1-\xi)\mathbf{k}-\widetilde{\mathbf{p}})\,g_{l}^{c}(\xi k^{+},\,\xi\mathbf{k}+\widetilde{\mathbf{p}})\right\rangle .\\
 \end{split}\end{equation}

\subsubsection*{Computation of $\left|\psi_{gg}^{3}\right\rangle$}
$\left|\psi_{gg}^{3}\right\rangle$ is defined in (\ref{agg3}) (Fig. \ref{ggfigo}c.).
\begin{equation}\begin{split}\label{ggthree}
\left|\psi_{gg}^{3}\right\rangle &\equiv\,-\frac{1}{2}\int_{\Lambda}^{e^{\delta\mathsf{Y}}\Lambda}dk^{+}\,\int_{\Lambda}^{k^{+}-\Lambda}dp^{+}\,\int d^{2}\mathbf{k}\, d^{2}\mathbf{p}\,\\
&\times\left|g_{i}^{b}(k-p)\, g_{j}^{c}(p)\right\rangle \frac{\left\langle g_{i}^{b}(k-p)\, g_{j}^{c}(p)\left|H_{gg-inst}\right|0\right\rangle }{E_{gg}(k-p,\, p)}.\\
\end{split}\end{equation}
After inserting the relevant matrix element (\ref{qq}) and changing variables according to (\ref{chanvar}) and (\ref{chanvar2}), we can write the last result as:

\begin{eqnarray}\label{contgg3}
&&\left|\psi_{gg\:\rho}^{3}\right\rangle \,\equiv\,\left|\psi_{gg}^{3}\right\rangle \,=\,\int_{\Lambda}^{e^{\delta\mathsf{Y}}\Lambda}dk^{+}\,\int d^{2}\mathbf{k}\,d^{2}\widetilde{\mathbf{p}}\,\int_{\frac{\Lambda}{k^{+}}}^{1-\frac{\Lambda}{k^{+}}}d\xi\,\frac{ig^{2}f^{abc}\rho^{a}(-\mathbf{k})(1-2\xi)\sqrt{\xi(1-\xi)}}{16\pi^{3}\left(\mathbf{k}^{2}\xi(1-\xi)+\widetilde{\mathbf{p}}^{2}\right)}\delta_{ij}\nonumber\\
&&\times\left|g_{i}^{b}((1-\xi)k^{+},\,(1-\xi)\mathbf{k}-\widetilde{\mathbf{p}})\,g_{j}^{c}(\xi k^{+},\,\xi\mathbf{k}+\widetilde{\mathbf{p}})\right\rangle .\end{eqnarray}

While focusing on two-parton production,
 we have not encountered any divergencies. This is obviously because we have not yet ran into any loop integrations. At this stage we can only anticipate 
their appearance   in the next section, where the overlaps of two wavefunctions are computed. Particularly, it is pretty obvious that  $\psi_{qq}^1$ 
 will lead to  the logarithmic  singularity in longitudinal momenta integration squared, which corresponds  to second iteration of the
 LO JIMWLK Hamiltonian. Somewhat less obvious, but $\psi_{qq}^3$ also contributes to square of the LO. This is due to the kinematical region where
 one of the two gluons is much softer than another. In the next subsection, we will encounter both longitudinal and UV divergencies in the LCWF itself.
The divergencies will be present in  most of the loop corrections to the single gluon state.

\subsubsection{One Gluon State}

\subsubsection*{Computation of $\left|\psi_{g}^{1}\right\rangle$}
$\left|\psi_{g}^{1}\right\rangle$ is defined in (\ref{ag1}) (Fig. \ref{loopfig}a.).
\begin{eqnarray}\label{gone}
\left|\psi_{g}^{1}\right\rangle &\equiv&-\int_{\Lambda}^{e^{\delta\mathsf{Y}}\Lambda}dk^{+}\, dp^{+}\, dq^{+}\, dr^{+}\,\int d^{2}\mathbf{k}\, d^{2}\mathbf{p}\, d^{2}\mathbf{q}\, d^{2}\mathbf{r}\\
&\times&\left|g_{j}^{d}(r)\right\rangle \frac{\left\langle g_{j}^{d}(r)\left|H_{gqq}\right|q_{\lambda_{1}}^{\alpha,f}(p)\,\bar{q}_{\lambda_{2}}^{\beta,f}(q)\right\rangle \left\langle q_{\lambda_{1}}^{\alpha,f}(p)\,\bar{q}_{\lambda_{2}}^{\beta,f}(q)\left|H_{gqq}\right|g_{i}^{a}(k)\right\rangle \left\langle g_{i}^{a}(k)\left|H_{g}\right|0\right\rangle }{E_{g}(r)\, E_{q\bar{q}}(p,\, q)\, E_{g}(k)}.\nonumber\end{eqnarray}
    \begin{figure}[h]
 $\qquad\quad$ \includegraphics[scale=0.7]{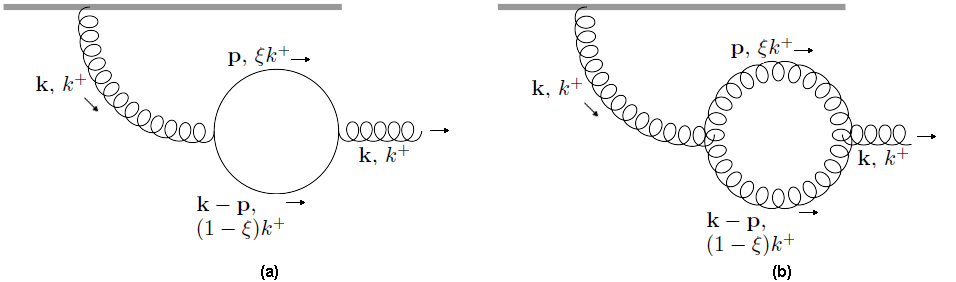}
  \caption{One gluon production. (a) - $\left|\psi_{g}^{1}\right\rangle$, (b) - $\left|\psi_{g}^{2}\right\rangle$. \label{loopfig}}
\end{figure}\\
The physics of  (\ref{gone}) is transparent from Fig. \ref{loopfig}a.  The figure reflects the momentum conservation, which eventually imposes 
 $q=k-p$ and $r=k$.} The rest of the calculation is available in Appendix \ref{swf4}. At the end of  calculation there, we arrive at the following result:
 \begin{eqnarray}\label{grho1}
\left|\psi_{g\:\rho}^{1}\right\rangle &\equiv&\left|\psi_{g}^{1}\right\rangle \\
&=&-\int_{\Lambda}^{e^{\delta\mathsf{Y}}\Lambda}dk^{+}\,\int d^{2}\mathbf{k}\,\frac{g^{3}\, N_{f}\,\,\rho^{a}(-\mathbf{k})\, \mathbf{k}^{i}}{32\pi^{7/2}\sqrt{k^{+}}\mathbf{k}^{2}}\left(\frac{2}{3}\left[-\frac{2}{\epsilon}+\ln\left(\frac{\mathbf{k}^{2}}{\mu_{\overline{MS}}^{2}}\right)\right]-\frac{10}{9}\right)\left|g_{i}^{a}(k)\right\rangle, \nonumber
\end{eqnarray}
where we have introduced $\mu_{\overline{MS}}^{2}\equiv4\pi e^{-\gamma}\mu^{2}$. { $2/\epsilon$ reflects  dimensional regularisation 
used to regulate the usual UV divergence of the quark loop integral. The renormalisation procedure will be applied in the next section, where we compute the physical Hamiltonian: the $2/\epsilon$ singularity will be then absorbed into redefinition of the coupling constant.

\subsubsection*{Computation of $\left|\psi_{g}^{2}\right\rangle$}
$\left|\psi_{g}^{2}\right\rangle$ is defined in (\ref{ag2}) (Fig. \ref{loopfig}b). 
\begin{equation}\begin{split}\label{gtwo}
\left|\psi_{g}^{2}\right\rangle &\equiv\,-\frac{1}{2}\int_{\Lambda}^{e^{\delta\mathsf{Y}}\Lambda}dk^{+}\, dp^{+}\, dq^{+}\, dr^{+}\int d^{2}\mathbf{k}\, d^{2}\mathbf{p}\, d^{2}\mathbf{q}\, d^{2}\mathbf{r}\\
&\times\left|g_{l}^{d}(r)\right\rangle \frac{\left\langle g_{l}^{d}(r)\left|H_{ggg}\right|g_{k}^{c}(q)\, g_{j}^{b}(p)\right\rangle \left\langle g_{k}^{c}(q)\, g_{j}^{b}(p)\left|H_{ggg}\right|g_{i}^{a}(k)\right\rangle \left\langle g_{i}^{a}(k)\left|H_{g}\right|0\right\rangle }{E_{g}(r)\, E_{gg}(p,\, q)\, E_{g}(k)},\\
\end{split}\end{equation}
Just as in the previous calculation and as  demonstrated on Fig. \ref{loopfig}b, 
both $q$ and $r$  become $q=k-p$ and $r=k$ after realisation of the 
momentum conservation imposing delta functions.  
The rest of the calculation is available in Appendix \ref{swf5}. The  result is:}
\begin{eqnarray}\label{grho2}
 \left|\psi_{g\:\rho}^{2}\right\rangle &\equiv&\left|\psi_{g}^{2}\right\rangle\,=\,\int_{\Lambda}^{e^{\delta\mathsf{Y}}\Lambda}dk^{+}\,\int d^{2}\mathbf{k}\,\frac{g^{3}N_{c}\rho^{a}(-\mathbf{k})\mathbf{k}^{i}}{32\pi^{7/2}\mathbf{k}^{2}\sqrt{k^{+}}}\\
&\times&\left(\left[\frac{11}{3}+4\ln\,\left(\frac{\Lambda}{k^{+}}\right)\right]\left[-\frac{2}{\epsilon}+\ln\left(\frac{\mathbf{k}^{2}}{\mu_{\overline{MS}}^{2}}\right)\right]+2\ln^{2}\,\left(\frac{\Lambda}{k^{+}}\right)-\frac{67}{9}+\frac{2\pi^{2}}{3}\right)\left|g_{i}^{a}(k)\right\rangle .\nonumber
\end{eqnarray}
{ Apart the terms which are easily recognised as the gluon loop contribution to the $\beta$-function, we notice additional terms and particularly the 
$\ln \Lambda/k^+$, which are potentially divergent in the $\Lambda\rightarrow 0$ limit.  These contributions arise from the phase space region where 
one of the gluons in the loop is much softer than another.  This is precisely the kinematical domain where the
triple gluon vertex $H_{ggg}$ admits eikonal approximation. It thus becomes clear that these logarithmic contributions correspond to iteration of the 
LO emission process. Indeed, in the next section we will demonstrate that all  the $\ln \Lambda/k^+$ terms eventually contribute $\delta Y^2$ to the evolution
identified with second iteration of the LO JIMWLK Hamiltonian.
}
  
\subsubsection*{Computation of $\left|\psi_{g}^{3}\right\rangle $}
$\left|\psi_{g}^{3}\right\rangle$ is defined in (\ref{ag3}). 
\begin{equation}\begin{split}\label{koa}
\left|\psi_{g}^{3}\right\rangle \,\equiv\,\int_{\Lambda}^{e^{\delta\mathsf{Y}}\Lambda}dk^{+}\,\int_{\Lambda}^{e^{\delta\mathsf{Y}}\Lambda}dp^{+}\,\int d^{2}\mathbf{k}\,d^{2}\mathbf{p}\:\left|g_{j}^{b}(k)\right\rangle \frac{\left\langle g_{j}^{b}(k)\left|H_{gg-inst}\right|g_{i}^{a}(p)\right\rangle \left\langle g_{i}^{a}(p)\left|H_{g}\right|0\right\rangle }{E_{g}(k)E_{g}(p)}.
\end{split}\end{equation}
Due to the instantaneous interaction, the gluon that was originally emitted can either gain or lose longitudinal momentum. 
{ The longitudinal integration over $p^{+}$ in (\ref{koa}) can be split into two intervals:
\begin{equation}
\int_{\Lambda}^{e^{\delta\mathsf{Y}}\Lambda}dp^{+}\,=\,\int_{\Lambda}^{k^{+}-\Lambda}dp^{+}\,+\,\int_{k^{+}-\Lambda}^{e^{\delta\mathsf{Y}}\Lambda}dp^{+}.\end{equation}
Accordingly,  we also write $\left|\psi_{g}^{3}\right\rangle$ as a sum of two contributions 
$\left|\psi_{g}^{3d}\right\rangle$ and $\left|\psi_{g}^{3u}\right\rangle$, ~
$\left|\psi_{g}^{3}\right\rangle =\left|\psi_{g}^{3d}\right\rangle +\left|\psi_{g}^{3u}\right\rangle $ where the upper script $u$ or $d$ represents the direction of the longitudinal momentum flow (corresponding to the integration intervals $[\Lambda,\,k^{+}-\Lambda]$ and $[k^{+}-\Lambda,\,e^{\delta\mathsf{Y}}\Lambda]$ respectively). 
  \begin{figure}[!ht]
  \includegraphics[scale=0.8]{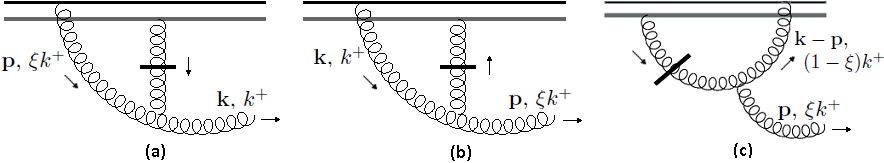}
  \caption{One gluon production via instantaneous interaction. (a) - $\left|\psi_{g}^{3d}\right\rangle$, (b) - $\left|\psi_{g}^{3u}\right\rangle$, (c) - $\left|\psi_{g}^{4}\right\rangle$.\label{instfig}}
\end{figure}
Let's start with $\left|\psi_{g}^{3d}\right\rangle$ defined by
\begin{equation}\begin{split}\label{gthreed}
&\left|\psi_{g}^{3d}\right\rangle \,\equiv\,\int_{\Lambda}^{e^{\delta\mathsf{Y}}\Lambda}dk^{+}\,\int_{\Lambda}^{k^{+}-\Lambda}dp^{+}\,\int d^{2}\mathbf{k}\, d^{2}\mathbf{p}\:\left|g_{j}^{b}(k)\right\rangle \frac{\left\langle g_{j}^{b}(k)\left|H_{gg-inst}\right|g_{i}^{a}(p)\right\rangle \left\langle g_{i}^{a}(p)\left|H_{g}\right|0\right\rangle }{E_{g}(k)E_{g}(p)}
  \end{split}\end{equation}
and  illustrated in Figure \ref{instfig}a,  where our standard relation ${p^{+}}=\xi\,{k^{+}}$ is also used.
The rest of the calculation is available in Appendix \ref{swf6}. The end result  integrated over $\xi$
 with the aid of (\ref{int.5}) reads:}
 \begin{equation}\begin{split}\label{grho3}
\left|\psi_{g\:\rho\rho}^{3d}\right\rangle &\equiv\,\left|\psi_{g}^{3d}\right\rangle \,=\,\int_{\Lambda}^{e^{\delta\mathsf{Y}}\Lambda}dk^{+}\,\int d^{2}\mathbf{k}\, d^{2}\mathbf{p}\,\frac{ig^{3}f^{abc}\mathbf{p}^{i}\left\{ \rho^{a}(-\mathbf{p}),\,\rho^{c}(-\mathbf{k}+\mathbf{p})\right\} }{16\pi^{9/2}\mathbf{k}^{2}\mathbf{p}^{2}\sqrt{k^{+}}}\\
&\times\left(\frac{k^{+}}{\Lambda}\,-\,1\,-\,\ln\left(\frac{\Lambda}{k^{+}}\right)\right)\left|g_{i}^{b}(k)\right\rangle .  \end{split}\end{equation}
{ We seem to again encounter the $\ln (k^+/\Lambda)$ type divergence. The latter is, however, fictitious and does not represent any physics. 
The divergence will cancel when we collect all similar contributions.

Next is $\left|\psi_{g}^{3u}\right\rangle$. We find it convenient to exchange the dummy momenta $k$ and $p$,  $k\leftrightarrow p$. 
The expression for $\left|\psi_{g}^{3u}\right\rangle$ is then given by
}
\begin{equation}\begin{split}\label{gthreeu}
&\left|\psi_{g}^{3u}\right\rangle \,\equiv\,\int_{\Lambda}^{e^{\delta\mathsf{Y}}\Lambda}dk^{+}\,\int_{\Lambda}^{k^{+}-\Lambda}dp^{+}\,\int d^{2}\mathbf{k}\, d^{2}\mathbf{p}\:\left|g_{j}^{b}(p)\right\rangle \frac{\left\langle g_{j}^{b}(p)\left|H_{gg-inst}\right|g_{i}^{a}(k)\right\rangle \left\langle g_{i}^{a}(k)\left|H_{g}\right|0\right\rangle }{E_{g}(p)E_{g}(k)}.
  \end{split}\end{equation}
{  and  illustrated in Figure \ref{instfig}b.}
The result obtained after inserting the relevant matrix elements, (\ref{g}) and (\ref{gg2}), and changing variables according to (\ref{chanvar}) 
becomes:
 \begin{equation}\begin{split}\label{sta3u}
&\left|\psi_{g}^{3u}\right\rangle =\int_{\Lambda}^{e^{\delta\mathsf{Y}}\Lambda}dk^{+}\,\int_{\Lambda}^{k^{+}-\Lambda}dp^{+}\,\int d^{2}\mathbf{k}\, d^{2}\mathbf{p}\:\frac{1}{\left(\frac{\mathbf{p}^{2}}{2p^{+}}\right)\left(\frac{\mathbf{k}^{2}}{2k^{+}}\right)}\\
&\times\left(\frac{ig^{2}f^{abc}(p^{+}+k^{+})\rho^{c}(-\mathbf{p}+\mathbf{k})\delta_{j}^{i}}{2(2\pi)^{3}\sqrt{k^{+}p^{+}}(k^{+}-p^{+})^{2}}\right)\left(\frac{g\rho^{a}(-\mathbf{k})\mathbf{k}^{i}}{4\pi^{3/2}|k^{+}|^{3/2}}\right)\left|g_{j}^{b}(p)\right\rangle .\\
\end{split}\end{equation}
Introducing $\xi$ for compactness the last result is:
  \begin{equation}\begin{split}\label{grhorho4}
\left|\psi_{g\:\rho\rho}^{3u}\right\rangle &\equiv\,\left|\psi_{g}^{3u}\right\rangle =\int_{\Lambda}^{e^{\delta\mathsf{Y}}\Lambda}dk^{+}\,\int d^{2}\mathbf{k}\, d^{2}\mathbf{p}\,\int_{\frac{\Lambda}{k^{+}}}^{1-\frac{\Lambda}{k^{+}}}d\xi\,\\
&\times\frac{ig^{3}f^{abc}(1+\xi)\sqrt{\xi}\left\{ \rho^{c}(-\mathbf{k}+\mathbf{p}),\,\rho^{a}(-\mathbf{p})\right\} \mathbf{p}^{i}}{32\pi^{9/2}(1-\xi)^{2}\mathbf{k}^{2}\mathbf{p}^{2}\sqrt{k^{+}}}\left|g_{i}^{b}(\xi k^{+},\,\mathbf{k})\right\rangle .
  \end{split}\end{equation}
 In principle, it is possible to compute one of the longitudinal integrals in (\ref{grhorho4}), but we have chosen to postpone this step until the final result gets assembled.
 
\subsubsection*{Computation of $\left|\psi_{g}^{4}\right\rangle $ and $\left|\psi_{g}^{5}\right\rangle $}

{ The contributions $\left|\psi_{g}^{4}\right\rangle$ and $\left|\psi_{g}^{5}\right\rangle$ have very similar integrand structures. 
This is the reason we founnd it convenient to combine the two.}
$\left|\psi_{g}^{4}\right\rangle$ is defined in (\ref{ag4}) (Fig. \ref{instfig}c), 
\begin{equation}\begin{split} \label{gfour}
\left|\psi_{g}^{4}\right\rangle &\equiv\,\frac{1}{2}\int_{\Lambda}^{e^{\delta\mathsf{Y}}\Lambda}\, dq^{+}\, dk^{+}\,\int_{\Lambda}^{k^{+}-\Lambda}\, dp^{+}\,\int\, d^{2}\mathbf{q}\, d^{2}\mathbf{k}\, d^{2}\mathbf{p}\\
&\times\left|g_{l}^{d}(p)\right\rangle \frac{\left\langle g_{l}^{d}(p)\left|H_{g}\right|g_{j}^{b}(q)\, g_{i}^{a}(k-q)\right\rangle \left\langle g_{j}^{b}(q)\, g_{i}^{a}(k-q)\left|H_{gg-inst}\right|0\right\rangle }{E_{g}(p)E_{gg}(q,\, k-q)}.
  \end{split}\end{equation}
{ The matrix element $\langle g_{l}^{d}(p)\left|H_{g}\right|g_{j}^{b}(q)\, g_{i}^{a}(k-q)\rangle$ contains a momentum conserving 
delta function which eventually sets $q=p$ (or $q=k-p$). Fig. \ref{instfig}c illustrates the process. 
$\left|\psi_{g}^{4}\right\rangle$ is computed  in Appendix \ref{swf8b}.}
  $\left|\psi_{g}^{5}\right\rangle$ is defined in (\ref{ag5}) (Fig. \ref{two_rhos}a), 
  \begin{equation}\begin{split}\label{gsix}
\left|\psi_{g}^{5}\right\rangle & \equiv\,-\frac{1}{2}\int_{\Lambda}^{e^{\delta\mathsf{Y}}\Lambda}dk^{+}\, dp^{+}\, dq^{+}\, dr^{+}\,\int d^{2}\mathbf{k}\, d^{2}\mathbf{p}\, d^{2}\mathbf{q}\, d^{2}\mathbf{r}\\
&\times\left|g_{m}^{d}(p)\right\rangle \frac{\left\langle g_{m}^{d}(p)\left|H_{g}\right|g_{k}^{c}(r)\, g_{j}^{b}(q)\right\rangle \left\langle g_{k}^{c}(r)\, g_{j}^{b}(q)\left|H_{ggg}\right|g_{i}^{a}(k)\right\rangle \left\langle g_{i}^{a}(k)\left|H_{g}\right|0\right\rangle }{E_{g}(k)\, E_{gg}(q,\, r)\, E_{g}(p)}.
\end{split}\end{equation}
{ 
The momentum conservation in  $\left|\psi_{g}^{5}\right\rangle$ imposes $q=p$ and $r=k-p$ (or $q=k-p$ and $r=p$), which is the case
displayed in Fig. \ref{two_rhos}a. The  calculation of $\left|\psi_{g}^{5}\right\rangle$ can be found  in Appendix \ref{swf8b}.

We now introduce  sum of $\left|\psi_{g}^{4}\right\rangle $ and $\left|\psi_{g}^{5}\right\rangle $,}
\begin{equation}\label{fouplfi}
\left|\psi_{g}^{4+5}\right\rangle \,\equiv\,\left|\psi_{g}^{5}\right\rangle +\left|\psi_{g}^{4}\right\rangle.
\end{equation}
Next,  $\left|\psi_{g}^{4+5}\right\rangle$ is split according to (\ref{decompose}) into two parts involving one and two $\rho$ operators:
\begin{equation}
 \left|\psi_{g}^{4+5}\right\rangle \,=\,\left|\psi_{g\:\rho}^{4+5}\right\rangle +\left|\psi_{g\:\rho\rho}^{4+5}\right\rangle.
\end{equation}
{ All the algebra is performed in Appendix \ref{swf8b}. Below, we  quote the final results only}.\\ \\
$\bullet$ \textit{\textbf{One $\rho$ part}}\\
     \begin{equation}\begin{split}\label{ffone}
&\left|\psi_{g\:\rho}^{4+5}\right\rangle \,\equiv\,\int_{\Lambda}^{e^{\delta\mathsf{Y}}\Lambda}dk^{+}\,\int d^{2}\mathbf{k}\,\int_{\frac{\Lambda}{k^{+}}}^{1-\frac{\Lambda}{k^{+}}}d\xi\,\frac{g^{3}N_{c}\rho^{a}(-\mathbf{k})\mathbf{k}^{j}}{32\pi^{7/2}\sqrt{\xi}(1-\xi)\mathbf{k}^{2}\sqrt{k^{+}}}\\
&\times\left(\left(\xi-2\right)\ln\left(\xi\right)+\left(\xi+1\right)\ln\left(1-\xi\right)+\left(\xi-2\right)\left[-\frac{2}{\epsilon}+\ln\left(\frac{\mathbf{k}^{2}}{\mu_{\overline{MS}}^{2}}\right)\right]\right)\left|g_{j}^{a}(\xi k^{+},\, \mathbf{k})\right\rangle. 
\end{split}\end{equation} \\
{ One of the longitudinal integrals, either $\xi$ or $k^+$, in (\ref{ffone}) could be computed, but we have again chosen to leave the expressions as is until the final result is assembled. Yet, we see that the longitudinal integral would result in a logarithmic divergence, although as usual regularised by $\Lambda$. 
The origin of this divergence can be traced to  $\left|\psi_{g}^{5}\right\rangle$ and particularly to the phase space region where $H_{ggg}$ admits eikonal approximation. Thus $\left|\psi_{g}^{5}\right\rangle$ has a piece, which can be interpreted as a soft gluon emission from a LO evolved LCWF.  In the next section,
such contributions will be identified with (LO)$^2$ terms.

Furthermore, $2/\epsilon$ points to  a regularised UV divergence.  Several diagrams below will also 
contribute such terms. Most, however, will cancel against each other. Some UV divergence will nevertheless survive and that is a feature of the LCWF at NLO. It will not, however, contribute any UV divergence at the level of the Hamiltonian.  There they will cancel against a subtraction term made of two produced gluons, which cross the shock wave   at the same transverse point.
}

 $\bullet$ \textit{\textbf{Two $\rho$ part}}\\
From (\ref{foplusfi}) (after the substitution $\mathbf{p}\leftrightarrow \mathbf{k}$):
\begin{eqnarray}\label{afdsp}
&&\left|\psi_{g\:\rho\rho}^{4+5}\right\rangle \equiv
\int_{\Lambda}^{e^{\delta\mathsf{Y}}\Lambda}dk^{+}\,\int d^{2}\mathbf{k}\, d^{2}\mathbf{p}\int_{\frac{\Lambda}{k^{+}}}^{1-\frac{\Lambda}{k^{+}}}d\xi\,\frac{ig^{3}f^{abc}\left\{ \rho^{c}(-\mathbf{k}+\mathbf{p}),\,\rho^{a}(-\mathbf{p})\right\} \xi^{3/2}}{32\pi^{9/2}\mathbf{k}^{2}\mathbf{p}^{2}\left(\xi(\mathbf{k}-\mathbf{p})^{2}+(1-\xi)\mathbf{k}^{2}\right)(1-\xi)\sqrt{k^{+}}}\nonumber\\
&&\times\left(\left[\frac{2}{\xi}\left(\mathbf{p}^{2}-\mathbf{k}\cdot \mathbf{p}\right)-\mathbf{p}^{2}+2\mathbf{k}\cdot \mathbf{p}\right]\mathbf{k}^{j}-\left[\mathbf{k}^{2}+\frac{1+\xi}{1-\xi}(\mathbf{k}-\mathbf{p})^{2}\right]\mathbf{p}^{j}\right)\left|g_{j}^{b}(\xi k^{+},\, \mathbf{k})\right\rangle.
\end{eqnarray}
A convenient way to represent (\ref{afdsp}):
 \begin{eqnarray}\label{sta6s}
\left|\psi_{g\:\rho\rho}^{4+5}\right\rangle &=&-\left|\psi_{g\:\rho\rho}^{3u}\right\rangle \nonumber\\
&&+\int_{\Lambda}^{e^{\delta\mathsf{Y}}\Lambda}dk^{+}\,\int d^{2}\mathbf{k}\, d^{2}\mathbf{p}\int_{\frac{\Lambda}{k^{+}}}^{1-\frac{\Lambda}{k^{+}}}d\xi\,\frac{ig^{3}f^{abc}\left\{ \rho^{c}(-\mathbf{k}+\mathbf{p}),\,\rho^{a}(-\mathbf{p})\right\} \xi^{3/2}}{32\pi^{9/2}\mathbf{k}^{2}\mathbf{p}^{2}\left(\xi(\mathbf{k}-\mathbf{p})^{2}+(1-\xi)\mathbf{k}^{2}\right)(1-\xi)\sqrt{k^{+}}}\nonumber\\
&&\times\left(\left[\frac{2}{\xi}\left(\mathbf{p}^{2}-\mathbf{k}\cdot \mathbf{p}\right)-\mathbf{p}^{2}+2\mathbf{k}\cdot \mathbf{p}\right]\mathbf{k}^{i}+\frac{1}{\xi}\mathbf{k}^{2}\mathbf{p}^{i}\right)\left|g_{i}^{b}(\xi k^{+},\, \mathbf{k})\right\rangle.\end{eqnarray}
{The advantage  of the last representation is obvious: in the next subsection we will add up $\left|\psi_{g\:\rho\rho}^{4+5}\right\rangle $ with
$\left|\psi_{g\:\rho\rho}^{3u}\right\rangle$ and the mutual cancelation becomes transparent via representation (\ref{sta6s}).}

\subsubsection*{Computation of $\left|\psi_{g}^{6}\right\rangle$}
\begin{figure}[!ht]
 $\qquad\quad$  \includegraphics[scale=0.8]{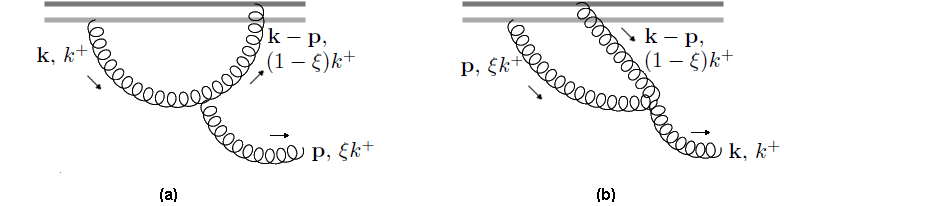}
  \caption{One gluon production diagrams involving the triple gluon vertex. (a) - $\left|\psi_{g}^{5}\right\rangle$, (b) - $\left|\psi_{g}^{6}\right\rangle$.\label{two_rhos}}
\end{figure}

$\left|\psi_{g}^{6}\right\rangle$ is defined in (\ref{ag6}) (Fig. \ref{two_rhos}b).
\begin{equation}\begin{split}\label{gfive}
\left|\psi_{g}^{6}\right\rangle &\equiv\,-\frac{1}{2}\int_{\Lambda}^{e^{\delta\mathsf{Y}}\Lambda}dk^{+}\, dp^{+}\, dq^{+}\, dr^{+}\,\int d^{2}\mathbf{k}\, d^{2}\mathbf{p}\, d^{2}\mathbf{q}\, d^{2}\mathbf{r}\\
&\times\left|g_{l}^{e}(k)\right\rangle \frac{\left\langle g_{l}^{e}(k)\left|H_{ggg}\right|g_{n}^{c}(r)\, g_{j}^{b}(p)\right\rangle \left\langle g_{n}^{c}(r)\, g_{j}^{b}(p)\left|H_{g}\right|g_{i}^{a}(q)\right\rangle \left\langle g_{i}^{a}(q)\left|H_{g}\right|0\right\rangle }{E_{g}(k)\, E_{gg}(p,\, r)\, E_{g}(q)}.
\end{split}\end{equation}
{ Just as in the previous calculations, the momentum conservation imposes $q=k$ and $r=k-p$. This is how this process is illustrated in
 Fig. \ref{two_rhos}b with our usual definition of $\xi$.} The rest of the calculation is available in Appendix \ref{swf8}, from which we quote:
 \begin{equation}\begin{split}\label{stasixsim}
 \left|\psi_{g}^{6}\right\rangle&=-\int_{\Lambda}^{e^{\delta\mathsf{Y}}\Lambda}dk^{+}\,\int d^{2}\mathbf{k}\, d^{2}\mathbf{p}\,\int_{\frac{\Lambda}{k^{+}}}^{1-\frac{\Lambda}{k^{+}}}d\xi\,\frac{ig^{3}f^{bad}\rho^{b}(-\mathbf{p})\rho^{a}(-\mathbf{k}+\mathbf{p})}{16\pi^{9/2}\xi(\mathbf{k}-\mathbf{p})^{2}\left((1-\xi)\mathbf{p}^{2}+\xi(\mathbf{k}-\mathbf{p})^{2}\right)\mathbf{k}^{2}\sqrt{k^{+}}}\\
&\hspace{-0.3cm}\times\left[\left(-2(1-\xi)\mathbf{k}\cdot \mathbf{p}+\frac{2(1-\xi^{2})}{\xi}\mathbf{p}^{2}\right)\mathbf{k}^{l}+\left(\mathbf{k}^{2}-\frac{1+\xi}{1-\xi}(\mathbf{k}-\mathbf{p})^{2}+\frac{\xi-2}{\xi}\mathbf{p}^{2}\right)\mathbf{p}^{l}\right]\left|g^{d}_{l}(k)\right\rangle .\\
\end{split}\end{equation}
The contributions with one and two $\rho$ operators can be isolated according to (\ref{decompose}):
\begin{equation}
\left|\psi_{g}^{6}\right\rangle \,=\,\left|\psi_{g\:\rho}^{6}\right\rangle \,+\,\left|\psi_{g\:\rho\rho}^{6}\right\rangle .
\end{equation}
$\bullet$ \textit{\textbf{One $\rho$ part}}\\
\begin{equation}\begin{split}\label{sixon}
\left|\psi_{g\:\rho}^{6}\right\rangle &\equiv\,-\int_{\Lambda}^{e^{\delta\mathsf{Y}}\Lambda}dk^{+}\,\int d^{2}\mathbf{k}\,\frac{g^{3}N_{c}\rho^{a}(-\mathbf{k})\mathbf{k}^{i}}{32\pi^{7/2}\mathbf{k}^{2}\sqrt{k^{+}}}\\
&\times\left(3\ln\left(\frac{\Lambda}{k^{+}}\right)\left[-\frac{2}{\epsilon}+\ln\left(\frac{\mathbf{k}^{2}}{\mu_{\overline{MS}}^{2}}\right)\right]+\frac{\pi^{2}}{3}+2\ln^{2}\left(\frac{\Lambda}{k^{+}}\right)\right)\left|g_{i}^{a}(k)\right\rangle .
\end{split}\end{equation}\\
  $\bullet$ \textit{\textbf{Two $\rho$ part}}\\
       \begin{eqnarray}\label{sixtw}
\left|\psi_{g\:\rho\rho}^{6}\right\rangle &\equiv&-\int_{\Lambda}^{e^{\delta\mathsf{Y}}\Lambda}dk^{+}\,\int d^{2}\mathbf{k}\, d^{2}\mathbf{p}\,\frac{ig^{3}f^{abc}\left\{ \rho^{b}(-\mathbf{p}),\,\rho^{c}(-\mathbf{k}+\mathbf{p})\right\} }{32\pi^{9/2}\mathbf{p}^{2}(\mathbf{k}-\mathbf{p})^{2}\mathbf{k}^{2}\sqrt{k^{+}}}\left[\ln\left(\frac{\Lambda}{k^{+}}\right)\mathbf{k}\cdot \mathbf{p}\mathbf{k}^{l}\right.\nonumber\\
&&\left.-\left(\ln\left(\frac{\Lambda}{k^{+}}\right)\mathbf{k}^{2}+2\left(\frac{k^{+}}{\Lambda}-1-\ln\left(\frac{\Lambda}{k^{+}}\right)\right)(\mathbf{k}-\mathbf{p})^{2}\right)\mathbf{p}^{l}\right]\left|g_{l}^{a}(k)\right\rangle .\end{eqnarray}
  \\

    \subsubsection*{Computation of $\left|\psi_{g}^{7}\right\rangle $}
    $\left|\psi_{g}^{7}\right\rangle $ is defined in  (\ref{ag7}) 
    \begin{figure}[hb]
    \includegraphics[scale=0.8]{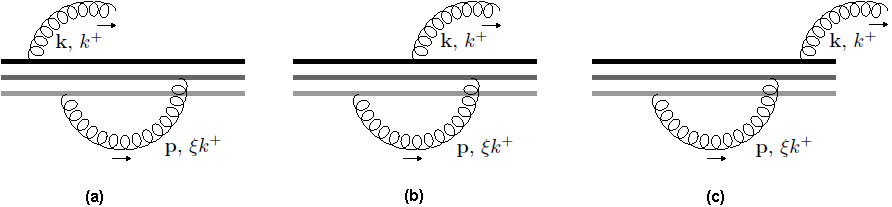}
  \caption{One gluon production diagrams with contribution to three $\rho$. (a,b) - $\left|\psi_{g}^{7}\right\rangle $, (c) - $\left|\psi_{g}^{8}\right\rangle$.\label{3rhofig}}
\end{figure}
\begin{equation}\begin{split}\label{gseven}
 \left|\psi_{g}^{7}\right\rangle  &\equiv\,-\frac{1}{2}\int_{\Lambda}^{e^{\delta\mathsf{Y}}\Lambda}dk^{+}\, dp^{+}\, dq^{+}\, dr^{+}\,\int d^{2}\mathbf{k}\, d^{2}\mathbf{p}\, d^{2}\mathbf{q}\, d^{2}\mathbf{r}\\
&\times\left|g_{l}^{d}(k)\right\rangle \frac{\left\langle g_{l}^{d}(k)\left|H_{g}\right|g_{j}^{b}(r)\, g_{k}^{c}(q)\right\rangle \left\langle g_{j}^{b}(r)\, g_{k}^{c}(q)\left|H_{g}\right|g_{i}^{a}(p)\right\rangle \left\langle g_{i}^{a}(p)\left|H_{g}\right|0\right\rangle }{E_{g}(k)\, E_{gg}(r,\, q)\, E_{g}(p)}.\\
\end{split}\end{equation}
{ $\left|\psi_{g}^{7}\right\rangle$ accounts for two distinct processes depicted, after the dummy momenta are integrated over, 
in Fig.\ref{3rhofig}a and Fig.\ref{3rhofig}b.  They correspond to two different light cone orderings in gluon emission/absorption process.
Intuitively, we expect $\left|\psi_{g}^{7}\right\rangle$  to mostly contribute to (LO)$^2$ effects. Indeed the structure of singularities, see below,
is consistent with our intuition. Particularly, we are to see a logarithmic divergence in the longitudinal integration, which can be identified 
as a virtual contribution at the LO, followed by  emission of another soft gluon.   
}
The calculation is available in Appendix \ref{swf9}, from which we quote the result:
\begin{equation}\begin{split}\label{sevsimto}
\left|\psi_{g}^{7}\right\rangle &=-\int_{\Lambda}^{e^{\delta\mathsf{Y}}\Lambda}dk^{+}\,\int d^{2}\mathbf{k}\, d^{2}\mathbf{p}\,\int_{\frac{\Lambda}{k^{+}}}^{\frac{e^{\delta\mathsf{Y}}\Lambda}{k^{+}}}d\xi\,\frac{g^{3}\mathbf{k}^{i}\rho^{b}(\mathbf{p})}{8\pi^{9/2}\sqrt{k^{+}}\xi\mathbf{k}^{2}}\\
&\times\left(\frac{\rho^{b}(-\mathbf{p})\rho^{a}(-\mathbf{k})}{\xi\mathbf{k}^{2}}+\frac{if^{abc}\rho^{c}(-\mathbf{k}-\mathbf{p})}{\xi\mathbf{k}^{2}+\mathbf{p}^{2}}\right)\left|g_{i}^{a}(k)\right\rangle ,\\
\end{split}\end{equation}
which can be written as a sum:
\begin{equation}\label{sevsplit}
\left|\psi_{g}^{7}\right\rangle \,=\,\left|\psi_{g\:\rho}^{7}\right\rangle \,+\,\left|\psi_{g\:\rho\rho}^{7}\right\rangle \,+\,\left|\psi_{g\:\rho\rho\rho}^{7}\right\rangle .
 \end{equation}\\
 $\bullet$ \textit{\textbf{One $\rho$ part}}\\
 \begin{equation}\begin{split}\label{sevone}
\left|\psi_{g\:\rho}^{7}\right\rangle &\equiv\,-\int_{\Lambda}^{e^{\delta\mathsf{Y}}\Lambda}dk^{+}\,\int d^{2}\mathbf{k}\,\frac{g^{3}N_{c}\rho^{a}(-\mathbf{k})\mathbf{k}^{i}}{32\pi^{7/2}\sqrt{k^{+}}\mathbf{k}^{2}}\\
&\times\left(2\delta\mathsf{Y}\left[-\frac{2}{\epsilon}+\ln\left(\frac{\mathbf{k}^{2}}{\mu_{\overline{MS}}^{2}}\right)\right]+\ln^{2}\left(\frac{\Lambda e^{\delta\mathsf{Y}}}{k^{+}}\right)-\ln^{2}\left(\frac{\Lambda}{k^{+}}\right)\right)\left|g_{i}^{a}(k)\right\rangle .
  \end{split}\end{equation}\\
$\bullet$ \textit{\textbf{Two $\rho$ part}}\\
  This part can be directly deduced from (\ref{sevsimto}):  
  \begin{equation}\label{sls}
\left|\psi_{g\:\rho\rho}^{7}\right\rangle \,\equiv\,-\int_{\Lambda}^{e^{\delta\mathsf{Y}}\Lambda}dk^{+}\,\int d^{2}\mathbf{k}\, d^{2}\mathbf{p}\,\int_{\frac{\Lambda}{k^{+}}}^{\frac{e^{\delta\mathsf{Y}}\Lambda}{k^{+}}}d\xi\,\frac{ig^{3}f^{abc}\mathbf{k}^{i}\left\{ \rho^{b}(-\mathbf{p}),\,\rho^{c}(-\mathbf{k}+\mathbf{p})\right\} }{16\pi^{9/2}\sqrt{k^{+}}\xi \mathbf{k}^{2}\left(\xi \mathbf{k}^{2}+\mathbf{p}^{2}\right)}\left|g_{i}^{a}(k)\right\rangle .
  \end{equation}
After integration over $\xi\,\equiv\,\frac{p^{+}}{k^{+}}$ according to (\ref{int.11as}):
\begin{equation}\begin{split}
\left|\psi_{g\:\rho\rho}^{7}\right\rangle &=\,\int_{\Lambda}^{e^{\delta\mathsf{Y}}\Lambda}dk^{+}\,\int d^{2}\mathbf{k}\, d^{2}\mathbf{p}\,\frac{ig^{3}f^{abc}\mathbf{k}^{i}\left\{ \rho^{b}(-\mathbf{p}),\,\rho^{c}(-\mathbf{k}+\mathbf{p})\right\} }{16\pi^{9/2}\sqrt{k^{+}}\mathbf{k}^{2}\mathbf{p}^{2}}\\
&\times\left(\ln\left(\frac{\Lambda}{k^{+}}\right)-\ln\left(\frac{\mathbf{p}^{2}}{\mathbf{k}^{2}}\right)\right)\left|g_{i}^{a}(k)\right\rangle .\\
\end{split}\end{equation}
Another useful representation is obtained by changing $p\,\rightarrow\, k\,-\, p$ in the second term,
  \begin{equation}\begin{split}
\left|\psi_{g\:\rho\rho}^{7}\right\rangle &=\,\int_{\Lambda}^{e^{\delta\mathsf{Y}}\Lambda}dk^{+}\,\int d^{2}\mathbf{k}\, d^{2}\mathbf{p}\,\frac{ig^{3}f^{abc}\mathbf{k}^{i}\left\{ \rho^{b}(-\mathbf{p}),\,\rho^{c}(-\mathbf{k}+\mathbf{p})\right\} }{16\pi^{9/2}\sqrt{k^{+}}\mathbf{k}^{2}}\\
&\times\left(\frac{1}{\mathbf{p}^{2}}\ln\left(\frac{\Lambda}{k^{+}}\right)+\frac{1}{(\mathbf{k}-\mathbf{p})^{2}}\ln\left(\frac{(\mathbf{k}-\mathbf{p})^{2}}{\mathbf{k}^{2}}\right)\right)\left|g_{i}^{a}(k)\right\rangle .
\end{split}\end{equation}\\ 
  $\bullet$ \textit{\textbf{Three $\rho$ part}}\\
  This part can be directly deduced from (\ref{sevsimto}):
\begin{equation}
\left|\psi_{g\:\rho\rho\rho}^{7}\right\rangle \,\equiv\,-\int_{\Lambda}^{e^{\delta\mathsf{Y}}\Lambda}dk^{+}\,\int d^{2}\mathbf{k}\, d^{2}\mathbf{p}\,\int_{\frac{\Lambda}{k^{+}}}^{\frac{e^{\delta\mathsf{Y}}\Lambda}{k^{+}}}d\xi\,\frac{g^{3}\mathbf{k}^{i}\rho^{b}(\mathbf{p})\rho^{b}(-\mathbf{p})\rho^{a}(-\mathbf{k})}{8\pi^{9/2}\sqrt{k^{+}}\xi^{2}\mathbf{k}^{4}}\left|g_{i}^{a}(k)\right\rangle .
 \end{equation}
 After integration over $\xi$:
 \begin{equation}\begin{split}\label{threerhos}
&\left|\psi_{g\:\rho\rho\rho}^{7}\right\rangle \,\equiv\,\int_{\Lambda}^{e^{\delta\mathsf{Y}}\Lambda}dk^{+}\,\int d^{2}\mathbf{k}\, d^{2}\mathbf{p}\,\frac{g^{3}\mathbf{k}^{i}\rho^{b}(\mathbf{p})\rho^{b}(-\mathbf{p})\rho^{a}(-\mathbf{k})\sqrt{k^{+}}}{8\pi^{9/2}\mathbf{k}^{4}\Lambda}\left(\frac{1}{e^{\delta\mathsf{Y}}}-1\right)\left|g_{i}^{a}(k)\right\rangle. 
\end{split}\end{equation}
{ From the general structure of the NLO JIMWLK Hamiltonian, we know that $\left|\psi_{g\:\rho\rho\rho}^{7}\right\rangle$ is not expected to 
contribute to it. At the level of Hamiltonian, $\left|\psi_{g\:\rho\rho\rho}^{7}\right\rangle$ gives rise to  $\rho^4$ terms, which are not present
in the NLO JIMWLK Hamiltonian.  Consistency demands that all such terms are expected to be absorbed into (LO)$^2$ Hamiltonian.  After a trivial
cancelation against $\left|\psi_{g}^{8}\right\rangle$, which is to be computed next, a combined contribution is  indeed found of the right form.
}
 
\subsubsection*{Computation of $\left|\psi_{g}^{8}\right\rangle$}
$\left|\psi_{g}^{8}\right\rangle$ is defined in (\ref{ag8}) (Fig. \ref{3rhofig}c):
\begin{equation}\begin{split}\label{geight}
&\left|\psi_{g}^{8}\right\rangle \,\equiv\,\int_{\Lambda}^{e^{\delta\mathsf{Y}}\Lambda}dk^{+}\, dp^{+}\,\int d^{2}\mathbf{k}\, d^{2}\mathbf{p}\:\left|g_{i}^{a}(k)\right\rangle \frac{\left|\left\langle g_{j}^{b}(p)\left|H_{g}\right|0\right\rangle \right|^{2}\left\langle g_{i}^{a}(k)\left|H_{g}\right|0\right\rangle (2E_{g}(p)+E_{g}(k))}{2E_{g}^{2}(k)E_{g}^{2}(p)}.
     \end{split}\end{equation}
After inserting the relevant matrix element (\ref{g}) and changing variables according to (\ref{chanvar}), we arrive at:
     \begin{equation}\begin{split}
   &  \left|\psi_{g}^{8}\right\rangle =\int_{\Lambda}^{e^{\delta\mathsf{Y}}\Lambda}dk^{+}\,\int d^{2}\mathbf{k}\, d^{2}\mathbf{p}\,\int_{\frac{\Lambda}{k^{+}}}^{\frac{e^{\delta\mathsf{Y}}\Lambda}{k^{+}}}d\xi\,\frac{g^{3}\mathbf{k}^{i}\rho^{b}(\mathbf{p})\rho^{b}(-\mathbf{p})\rho^{a}(-\mathbf{k})}{16\pi^{9/2}\sqrt{k^{+}}\xi^{2}\mathbf{k}^{2}}\,\left(\frac{\xi}{\mathbf{p}^{2}}\,+\,\frac{2}{\mathbf{k}^{2}}\right)\,\left|g_{i}^{a}(k)\right\rangle. 
     \end{split}\end{equation}
     After integration over $\xi$:
    \begin{eqnarray}\label{rrrow}
 && \left|\psi_{g\:\rho\rho\rho}^{8}\right\rangle \equiv \left|\psi_{g}^{8}\right\rangle \\
  &&\quad
  =\int_{\Lambda}^{e^{\delta\mathsf{Y}}\Lambda}dk^{+}\,\int d^{2}\mathbf{k}\, d^{2}\mathbf{p}\,\frac{g^{3}\mathbf{k}^{i}\rho^{b}(\mathbf{p})\rho^{b}(-\mathbf{p})\rho^{a}(-\mathbf{k})}{16\pi^{9/2}\sqrt{k^{+}}\mathbf{k}^{2}}\,\left(\frac{\delta\mathsf{Y}}{\mathbf{p}^{2}}\,-\,\left(\frac{1}{e^{\delta\mathsf{Y}}}-1\right)\frac{2k^{+}}{\Lambda \mathbf{k}^{2}}\right)\,\left|g_{i}^{a}(k)\right\rangle\nonumber .\end{eqnarray}

\subsection{The Final Result}\label{finres}


In this section we will assemble together all the different contributions that were computed in section \ref{eval}. It turns out to be useful to replace the representation of the LCFW as in (\ref{replo}) by a new representation schematically outlined in (\ref{psi}), in which the classification of the contributions will be made according to the soft particle content ($q\bar{q}$, $gg$ or $g$) and the number of valence current operators ($\rho$, $\rho\rho$ or $\rho\rho\rho$). This information will be indicated by a subscript. For example, the state $\left|\psi_{gg\:\rho\rho}\right\rangle $ contains two soft gluons and two $\rho$ operators. The NLO wave function is decomposed as follows:
\begin{equation}\label{expenw}
\left|\psi^{NLO}\right\rangle=\mathcal{N}^{NLO}\left|0\right\rangle +\left|\psi_{g\:\rho}^{LO}\right\rangle +\left|\psi_{q\overline{q}\:\rho}\right\rangle 
+\left|\psi_{gg\:\rho}\right\rangle +\left|\psi_{gg\:\rho\rho}\right\rangle +\left|\psi_{g\:\rho}\right\rangle +\left|\psi_{g\:\rho\rho}\right\rangle +
\left|\psi_{g\:\rho\rho\rho}\right\rangle .
\end{equation}
 $\left|\psi_{g\:\rho}^{LO}\right\rangle $ was defined in (\ref{lotra}). Additional states are:
   \begin{equation}\label{quuar}
 \left|\psi_{q\bar{q}\:\rho}\right\rangle \,\equiv\,\left|\psi_{q\bar{q}\:\rho}^{1}\right\rangle \,+\,\left|\psi_{q\bar{q}\:\rho}^{2}\right\rangle, 
  \end{equation}
  \begin{equation}\label{ggrar}
 \left|\psi_{gg\:\rho}\right\rangle \,\equiv\,\left|\psi_{gg\:\rho}^{1}\right\rangle \,+\,\left|\psi_{gg\:\rho}^{2}\right\rangle \,+\,\left|\psi_{gg\:\rho}^{3}\right\rangle, 
   \end{equation} 
  \begin{equation}\label{grar}
\left|\psi_{g\:\rho}\right\rangle \,\equiv\,\left|\psi_{g\:\rho}^{1}\right\rangle \,+\,\left|\psi_{g\:\rho}^{2}\right\rangle \,+\,\left|\psi_{g\:\rho}^{4+5}\right\rangle \,+\,\left|\psi_{g\:\rho}^{6}\right\rangle \,+\,\left|\psi_{g\:\rho}^{7}\right\rangle ,
  \end{equation}
\begin{equation}\label{grrar}
\left|\psi_{g\:\rho\rho}\right\rangle \,\equiv\,\left|\psi_{g\:\rho\rho}^{3u}\right\rangle \,+\,\left|\psi_{g\:\rho\rho}^{3d}\right\rangle \,+\,\left|\psi_{g\:\rho\rho}^{4+5}\right\rangle \,+\,\left|\psi_{g\:\rho\rho}^{6}\right\rangle \,+\,\left|\psi_{g\:\rho\rho}^{7}\right\rangle ,
  \end{equation}
  \begin{equation}\label{grrrar}
 \left|\psi_{g\:\rho\rho\rho}\right\rangle \,\equiv\,\left|\psi_{g\:\rho\rho\rho}^{7}\right\rangle \,+\,\left|\psi_{g\:\rho\rho\rho}^{8}\right\rangle .
  \end{equation}
  
\noindent$\bullet$ \textit{\textbf{The quark and anti-quark state}}\\
Inserting (\ref{qq_cont1}) and (\ref{qq_cont2}) to (\ref{quuar}), and changing the measure according to (\ref{dimregm}), yeilds:
 \begin{equation}\begin{split}\label{stateqq}
\left|\psi_{q\bar{q}\:\rho}\right\rangle &=\sum_{\lambda_{1},\lambda_{2},f}\int_{\Lambda}^{e^{\delta\mathsf{Y}}\Lambda}dk^{+}\,\int_{0}^{1}d\xi\,\int\frac{d^{2}\mathbf{k}}{(2\pi)^{2}}\,\frac{d^{2}\mathbf{p}}{(2\pi)^{2}}\:\frac{2\pi g^{2}t_{\alpha\beta}^{a}\rho^{a}(-\mathbf{k})\xi(1-\xi)}{(1-\xi)\mathbf{p}^{2}+\xi(\mathbf{k}-\mathbf{p})^{2}}\\
&\times\chi_{\lambda_{1}}^{\dagger}\left(\frac{\mathbf{k}^{i}}{\mathbf{k}^{2}}\left[2\mathbf{k}^{i}-\frac{\sigma\cdot \mathbf{p}}{\xi}\sigma^{i}-\sigma^{i}\frac{\sigma\cdot(\mathbf{k}-\mathbf{p})}{1-\xi}\right]-2\right)\chi_{\lambda_{2}}\left|\bar{q}_{\lambda_{2}}^{\beta,\, f}(k-p)\, q_{\lambda_{1}}^{\alpha,\, f}(p)\right\rangle ,
\end{split}\end{equation}
where $p^{+}=\xi k^{+}$. Another representation is obtained after introducing the variable $\widetilde{\mathbf{p}}$ according to (\ref{chanvar2}):
  \begin{eqnarray}
 \left|\psi_{q\bar{q}\:\rho}\right\rangle &=&-\sum_{\lambda_{1},\lambda_{2},f}\int_{\Lambda}^{e^{\delta\mathsf{Y}}\Lambda}dk^{+}\,\int_{0}^{1}d\xi\,\int\frac{d^{2}\mathbf{k}}{(2\pi)^{2}}\,\frac{d^{2}\mathbf{p}}{(2\pi)^{2}}\:\frac{2\pi g^{2}t_{\alpha\beta}^{a}\rho^{a}(-\mathbf{k})}{\xi(1-\xi)\mathbf{k}^{2}+\widetilde{\mathbf{p}}^{2}}\nonumber\\
&\times&\chi_{\lambda_{1}}^{\dagger}\left(\frac{\mathbf{k}^{i}\widetilde{\mathbf{p}}^{j}}{\mathbf{k}^{2}}\left[(1-2\xi)\delta^{ij}-i\varepsilon^{ijk}\sigma^{k}\right]+2\xi(1-\xi)\right)\chi_{\lambda_{2}}\\
&\times&\left|\bar{q}_{\lambda_{2}}^{\beta,\, f}((1-\xi)k^{+},\,(1-\xi)\mathbf{k}-\widetilde{\mathbf{p}})\, q_{\lambda_{1}}^{\alpha,\, f}(\xi k^{+},\,\xi\mathbf{k}+\widetilde{\mathbf{p}})\right\rangle .\nonumber\end{eqnarray}
Transforming to coordinate space using (\ref{fourier.3}) and (\ref{fourier.4}),
\begin{eqnarray}\label{qqtri}
&&\left|\psi_{q\bar{q}\:\rho}\right\rangle =-\sum_{\lambda_{1},\lambda_{2},f}\int_{\mathbf{x},\,\mathbf{z},\,\mathbf{z}^{\prime}}\,\int_{\Lambda}^{e^{\delta\mathsf{Y}}\Lambda}dk^{+}\,\int_{0}^{1}d\xi\:\frac{g^{2}t_{\alpha\beta}^{a}\rho^{a}(\mathbf{x})}{8\pi^{3}\left((1-\xi)(X^{\prime})^{2}+\xi X^{2}\right)}\chi_{\lambda_{1}}^{\dagger}\left(\frac{(X^{\prime}-\xi Z)^{i}Z^{j}}{Z^{2}}\right.\nonumber\\
&&\times\left[(1-2\xi)\delta^{ij}-i\varepsilon^{ijk}\sigma^{k}\right]+2\xi(1-\xi)\bigg)\chi_{\lambda_{2}}\left|\bar{q}_{\lambda_{2}}^{\beta,\, f}((1-\xi)k^{+},\,\mathbf{z}^{\prime})\, q_{\lambda_{1}}^{\alpha,\, f}(\xi k^{+},\,\mathbf{z})\right\rangle .\end{eqnarray} 

\noindent$\bullet$ \textit{\textbf{The two gluon state with one $\rho$}}  \\
    By inserting (\ref{contt}), (\ref{contgg}), and (\ref{contgg3}) to (\ref{ggrar}) we arrive at:
 \begin{eqnarray}
 \left|\psi_{gg\:\rho}\right\rangle &=& \int_{\Lambda}^{e^{\delta\mathsf{Y}}\Lambda}dk^{+}\,\int d^{2}\mathbf{k}\, d^{2}\widetilde{\mathbf{p}}\,\int_{\frac{\Lambda}{k^{+}}}^{1-\frac{\Lambda}{k^{+}}}d\xi\,\frac{ig^{2}f^{abc}\rho^{a}(-\mathbf{k})\sqrt{\xi(1-\xi)}}{8\pi^{3}\left(\mathbf{k}^{2}\xi(1-\xi)+\widetilde{\mathbf{p}}^{2}\right)\mathbf{k}^{2}}\nonumber\\
&\times&\left(\frac{\mathbf{k}^{2}\left(\xi \mathbf{k}^{i}+\widetilde{\mathbf{p}}^{i}\right)\left((1-\xi)\mathbf{k}^{j}-\widetilde{\mathbf{p}}^{j}\right)\left(\xi((1-\xi)\mathbf{k}-\widetilde{\mathbf{p}})^{2}-(1-\xi)(\xi \mathbf{k}+\widetilde{\mathbf{p}})^{2}\right)}{2\xi(1-\xi)(\xi \mathbf{k}+\widetilde{\mathbf{p}})^{2}((1-\xi)\mathbf{k}-\widetilde{\mathbf{p}})^{2}}\right.\nonumber\\
&&\left.+\frac{\delta^{jl}}{2}\left((1-2\xi)\mathbf{k}^{2}-2\mathbf{k}\cdot\widetilde{\mathbf{p}}\right)+\frac{1}{\xi}\mathbf{k}^{j}\widetilde{\mathbf{p}}^{l}+\frac{1}{1-\xi}\mathbf{k}^{l}\widetilde{\mathbf{p}}^{j}\right)\\
&\times&\left|g_{j}^{b}((1-\xi)k^{+},\,(1-\xi)\mathbf{k}-\widetilde{\mathbf{p}})\, g_{l}^{c}(\xi k^{+},\,\xi\mathbf{k}+\widetilde{\mathbf{p}})\right\rangle .\nonumber \end{eqnarray}
Transforming to coordinate space using (\ref{fourier.3}) and (\ref{fourier.4}):
 \begin{eqnarray}\label{gg_rho}
   \left|\psi_{gg\:\rho}\right\rangle &=&-\int_{\mathbf{x},\, \mathbf{z},\, \mathbf{z}^{\prime}}\int_{\Lambda}^{e^{\delta\mathsf{Y}}\Lambda}dk^{+}\,\int_{\frac{\Lambda}{k^{+}}}^{1-\frac{\Lambda}{k^{+}}}d\xi\,\frac{ig^{2}f^{abc}\rho^{a}(\mathbf{x})\sqrt{\xi(1-\xi)}}{8\pi^{3}\left((1-\xi)(X^{\prime})^{2}+\xi X^{2}\right)}\\
&\times&\left(\frac{\delta^{jl}}{2Z^{2}}\left(X^{2}-(X^{\prime})^{2}\right)+\frac{(X^{\prime})^{j}Z^{l}}{\xi Z^{2}}+\frac{X^{l}Z^{j}}{(1-\xi)Z^{2}}+\frac{X^{l}(X^{\prime})^{j}}{2\xi X^{2}}-\frac{X^{l}(X^{\prime})^{j}}{2(1-\xi)(X^{\prime})^{2}}\right)\nonumber\\
&\times&\left|g_{l}^{b}(\xi k^{+},\,\mathbf{z})\, g_{j}^{c}((1-\xi)k^{+},\,\mathbf{z}^{\prime})\right\rangle .\nonumber\end{eqnarray}
   
\noindent$\bullet$ \textit{\textbf{The two gluon state with two $\rho$}}  \\
  This result was shown in (\ref{ggrhorho}):
  \begin{equation}\begin{split}
\left|\psi_{gg\:\rho\rho}\right\rangle &=\int_{\Lambda}^{e^{\delta\mathsf{Y}}\Lambda}dk^{+}\,\int d^{2}\mathbf{k}\, d^{2}\mathbf{p}\,\int_{\frac{\Lambda}{k^{+}}}^{1-\frac{\Lambda}{k^{+}}}d\xi\,\\
&\times\frac{g^{2}\mathbf{p}^{i}\left(\mathbf{k}^{j}-\mathbf{p}^{j}\right)\left\{ \rho^{a}(-\mathbf{p}),\,\,\rho^{b}(-\mathbf{k}+\mathbf{p})\right\} }{16\pi^{3}\mathbf{p}^{2}(\mathbf{k}-\mathbf{p})^{2}\sqrt{\xi(1-\xi)}}\left|g_{j}^{b}(k-p)\, g_{i}^{a}(p)\right\rangle .\\
  \end{split}\end{equation}
After transformation to coordinate space using (\ref{fourier.1}):
      \begin{equation}\begin{split}\label{gg_rhorho}
\left|\psi_{gg\:\rho\rho}\right\rangle &=\,-\int_{\mathbf{x},\,\mathbf{y},\,\mathbf{z},\,\mathbf{z}^{\prime}}\int_{\Lambda}^{e^{\delta\mathsf{Y}}\Lambda}dk^{+}\,\int_{\frac{\Lambda}{k^{+}}}^{1-\frac{\Lambda}{k^{+}}}d\xi\\
&\times\frac{g^{2}\left\{ \rho^{a}(\mathbf{y}),\,\rho^{b}(\mathbf{x})\right\} Y^{i}(X^{\prime})^{j}}{16\pi^{3}\sqrt{\xi(1-\xi)}Y^{2}(X^{\prime})^{2}}\left|g_{j}^{b}((1-\xi)k^{+},\mathbf{z}^{\prime})\, g_{i}^{a}(\xi k^{+},\mathbf{z})\right\rangle .
 \end{split}\end{equation}

\noindent$\bullet$ \textit{\textbf{The one gluon state with one $\rho$}}\\
  By inserting (\ref{grho1}), (\ref{grho2}), (\ref{ffone}), (\ref{sixon}), and (\ref{sevone}) to (\ref{grar}) we arrive at:
     \begin{eqnarray}
\left|\psi_{g\:\rho}\right\rangle &=&\int_{\Lambda}^{e^{\delta\mathsf{Y}}\Lambda}dk^{+}\,\int d^{2}\mathbf{k}\,\frac{g^{3}\rho^{a}(-\mathbf{k})\mathbf{k}^{i}}{32\pi^{7/2}\mathbf{k}^{2}\sqrt{k^{+}}}\left[\left(\left[b+N_{c}\ln\,\left(\frac{\Lambda}{k^{+}}\right)-2N_{c}\delta\mathsf{Y}\right]\right.\right.\nonumber\\
&\times&\left[-\frac{2}{\epsilon}+\ln\left(\frac{\mathbf{k}^{2}}{\mu_{\overline{MS}}^{2}}\right)\right]-N_{c}\ln^{2}\left(\frac{\Lambda e^{\delta\mathsf{Y}}}{k^{+}}\right)+N_{c}\ln^{2}\left(\frac{\Lambda}{k^{+}}\right)-\left(\frac{67}{9}-\frac{\pi^{2}}{3}\right)N_{c}\nonumber \\
&&\left.+\frac{10}{9}N_{f}\right)\left|g_{i}^{a}(k)\right\rangle +\int_{\frac{\Lambda}{k^{+}}}^{1-\frac{\Lambda}{k^{+}}}d\xi\,\frac{N_{c}}{\sqrt{\xi}(1-\xi)}\bigg(\left(\xi-2\right)\ln\left(\xi\right)+\left(\xi+1\right)\ln\left(1-\xi\right)\nonumber\\
&&\left.-\left(\xi-2\right)\left[-\frac{2}{\epsilon}+\ln\left(\frac{\mathbf{k}^{2}}{\mu_{\overline{MS}}^{2}}\right)\right]\bigg)\left|g_{i}^{a}(\xi k^{+},\,\mathbf{k})\right\rangle \right].\end{eqnarray}
where the parameter $b$ was defined in (\ref{beta}) and $\mu_{\overline{MS}}^{2}\equiv4\pi e^{-\gamma}\mu^{2}$. The transformation to coordinate space is possible using (\ref{fourier.1}) and (\ref{fourier.2}):
       \begin{eqnarray}\label{g_rho}
\left|\psi_{g\:\rho}\right\rangle &=&-\int_{\mathbf{x},\,\mathbf{z}}\int_{\Lambda}^{e^{\delta\mathsf{Y}}\Lambda}dk^{+}\,\frac{ig^{3}\rho^{a}(\mathbf{x})X^{i}}{32\pi^{7/2}X^{2}\sqrt{k^{+}}}\left[\left(\left[b+N_{c}\ln\,\left(\frac{\Lambda}{k^{+}}\right)-2N_{c}\delta\mathsf{Y}\right]\left[-\frac{2}{\epsilon}-2\gamma\right.\right.\right.\nonumber\\
&&\left.-\ln\left(\frac{X^{2}\mu_{\overline{MS}}^{2}}{4}\right)\right]-N_{c}\ln^{2}\left(\frac{\Lambda e^{\delta\mathsf{Y}}}{k^{+}}\right)+N_{c}\ln^{2}\left(\frac{\Lambda}{k^{+}}\right)-\left(\frac{67}{9}-\frac{\pi^{2}}{3}\right)N_{c}\nonumber\\
&&\left.+\frac{10}{9}N_{f}\right)\left|g_{i}^{a}(k^{+},\,\mathbf{z})\right\rangle +\int_{\frac{\Lambda}{k^{+}}}^{1-\frac{\Lambda}{k^{+}}}d\xi\,\frac{N_{c}}{\sqrt{\xi}(1-\xi)}\bigg(\left(\xi-2\right)\ln\left(\xi\right)+\left(\xi+1\right)\ln\left(1-\xi\right)\nonumber \\
&&\left.-\left(\xi-2\right)\left[-\frac{2}{\epsilon}-2\gamma-\ln\left(\frac{X^{2}\mu_{\overline{MS}}^{2}}{4}\right)\right]\bigg)\left|g_{i}^{a}(\xi k^{+},\,\mathbf{z})\right\rangle \right].\end{eqnarray}
In section 4, we will use this form of $\left|\psi_{g\:\rho}\right\rangle$. However, after defining a new variable $\bar{k}^{+}=\xi k^{+}$, one of the longitudinal integrals can be evaluated. For completeness of the presentation, we also quote this result:
  \begin{eqnarray}
&&\hspace{-0.3cm}\left|\psi_{g\:\rho}\right\rangle =-\int_{\mathbf{x},\mathbf{z}}\int_{\Lambda}^{e^{\delta\mathsf{Y}}\Lambda}dk^{+}\,\frac{ig^{3}\rho^{a}(\mathbf{x})X^{i}}{32\pi^{7/2}X^{2}\sqrt{k^{+}}}\left\{ \left[b+N_{c}\ln\left(\frac{\Lambda}{k^{+}}\right)+N_{c}\ln\left(\frac{\Lambda e^{\delta\mathsf{Y}}}{k^{+}}-1\right)-N_{c}\delta\mathsf{Y}\right]\right.\nonumber\\
&&\hspace{-0.3cm}\times\left[-\frac{2}{\epsilon}-2\gamma-\ln\left(\frac{X^{2}\mu_{\overline{MS}}^{2}}{4}\right)\right]+\frac{10}{9}N_{f}+N_{c}\left[\frac{\pi^{2}}{3}-\frac{67}{9}+\ln^{2}\left(1-\frac{k^{+}}{\Lambda e^{\delta\mathsf{Y}}}\right)+\ln\left(1-\frac{k^{+}}{\Lambda e^{\delta\mathsf{Y}}}\right)\right.\nonumber\\
&&\hspace{-0.3cm}\left.\left.\times\ln\left(\frac{\Lambda e^{\delta\mathsf{Y}}}{k^{+}}\right)\right]\right\} \left|g_{i}^{a}(k^{+},\,\mathbf{z})\right\rangle\end{eqnarray}
{ $\left|\psi_{g\:\rho}\right\rangle$ appears to be one of the most non-trivial  parts of our calculation, which to our knowledge has not been 
computed before. It has a mixture of various terms, including singular ones.  It is hard to give interpretation to every term at the level of the LCWF.
We only comment here that $\left|\psi_{g\:\rho}\right\rangle$ contributes both  to NLO JIMWLK Hamiltonian (the kernel $K_{JSJ}$) and  second
 iteration of the LO one.    $\left|\psi_{g\:\rho}\right\rangle$ is also a source of the $\beta$-function of the running coupling.
 }
 

\noindent$\bullet$ \textit{\textbf{The one gluon state with two $\rho$}} \\
   By inserting (\ref{grho3}), (\ref{grhorho4}), (\ref{sta6s}), (\ref{sixtw}), and (\ref{sls}) to (\ref{grrar}) we arrive at:
 \begin{eqnarray}\label{g_rhorho}
\left|\psi_{g\:\rho\rho}\right\rangle &\equiv&\int_{\Lambda}^{e^{\delta\mathsf{Y}}\Lambda}dk^{+}\,\int d^{2}\mathbf{k}\, d^{2}\mathbf{p}\,\frac{ig^{3}f^{abc}\left\{ \rho^{b}(-\mathbf{p}),\,\rho^{c}(-\mathbf{k}+\mathbf{p})\right\} }{32\pi^{9/2}\mathbf{p}^{2}\mathbf{k}^{2}\sqrt{k^{+}}}\left\{ \left(\left[\left(2-\frac{\mathbf{k}\cdot\mathbf{p}}{(\mathbf{k}-\mathbf{p})^{2}}\right)\right.\right.\right.\nonumber\\
&&\left.\left.\times\ln\left(\frac{\Lambda}{k^{+}}\right)+\frac{2\mathbf{p}^{2}}{(\mathbf{k}-\mathbf{p})^{2}}\ln\left(\frac{(\mathbf{k}-\mathbf{p})^{2}}{\mathbf{k}^{2}}\right)\right]\mathbf{k}^{i}+\ln\left(\frac{\Lambda}{k^{+}}\right)\frac{\mathbf{k}^{2}}{(\mathbf{k}-\mathbf{p})^{2}}\mathbf{p}^{i}\right)\left|g_{i}^{a}(k)\right\rangle \nonumber \\
&&-\int_{\frac{\Lambda}{k^{+}}}^{1-\frac{\Lambda}{k^{+}}}d\xi\,\frac{\xi^{3/2}}{\left(\xi(\mathbf{k}-\mathbf{p})^{2}+(1-\xi)\mathbf{k}^{2}\right)(1-\xi)}\left(\left[\frac{2}{\xi}\left(\mathbf{p}^{2}-\mathbf{k}\cdot\mathbf{p}\right)-\mathbf{p}^{2}+2\mathbf{k}\cdot\mathbf{p}\right]\mathbf{k}^{i}\right.\nonumber\\
&&\left.\left.+\frac{1}{\xi}\mathbf{k}^{2}\mathbf{p}^{i}\right)\left|g_{i}^{a}(\xi k^{+},\,\mathbf{k})\right\rangle \right\} .\end{eqnarray}
Again, despite the possibility to evaluate one of the integrals,  in section 4, we will use this form of $\left|\psi_{g\:\rho\rho}\right\rangle$. 
For completeness of presentation, we also quote the final result:
\begin{eqnarray}
&&\left|\psi_{g\:\rho\rho}\right\rangle \,\equiv\,\int_{\Lambda}^{e^{\delta\mathsf{Y}}\Lambda}dk^{+}\,\int d^{2}\mathbf{k}\, d^{2}\mathbf{p}\,\frac{ig^{3}f^{abc}\left\{ \rho^{b}(-\mathbf{p}),\,\rho^{c}(-\mathbf{k}+\mathbf{p})\right\} }{32\pi^{9/2}\mathbf{p}^{2}\mathbf{k}^{2}\sqrt{k^{+}}}\left\{ \left[\left(2-\frac{\mathbf{k}\cdot\mathbf{p}}{(\mathbf{k}-\mathbf{p})^{2}}\right)\right.\right.\nonumber\\
&&\left.\times\ln\left(\frac{\Lambda}{k^{+}}\right)+\frac{2\mathbf{p}^{2}}{(\mathbf{k}-\mathbf{p})^{2}}\ln\left(\frac{(\mathbf{k}-\mathbf{p})^{2}}{\mathbf{k}^{2}}\right)\right]\mathbf{k}^{i}+\ln\left(\frac{\Lambda}{k^{+}}\right)\frac{\mathbf{k}^{2}}{(\mathbf{k}-\mathbf{p})^{2}}\mathbf{p}^{i}\nonumber \\
&&+\frac{\left(2\mathbf{p}^{2}-2\mathbf{k}\cdot\mathbf{p}+\mathbf{k}^{2}\right)\mathbf{k}^{i}+\mathbf{k}^{2}\mathbf{p}^{i}}{(\mathbf{k}-\mathbf{p})^{2}}\ln\left(1+\left[\frac{e^{\delta\mathsf{Y}}\Lambda}{k^{+}}-1\right]\frac{\mathbf{k}^{2}}{(\mathbf{k}-\mathbf{p})^{2}}\right)-\frac{\mathbf{p}^{2}\mathbf{k}^{i}+\mathbf{k}^{2}\mathbf{p}^{i}}{(\mathbf{k}-\mathbf{p})^{2}}\nonumber\\
&&\left.\times\ln\left(e^{\delta\mathsf{Y}}-\frac{k^{+}}{\Lambda}\right)-\mathbf{k}^{i}\ln\left(\frac{e^{\delta\mathsf{Y}}\Lambda}{k^{+}}\right)\right\} \left|g_{i}^{a}(k)\right\rangle \end{eqnarray}
We did not manage to compute a coordinate representation  for this component. 
Together with $\left|\psi_{g\:\rho}\right\rangle$, $\left|\psi_{g\:\rho\rho}\right\rangle$ constitutes one of the main new non-trivial result. 
In the next section, we will demonstrate how $\left|\psi_{g\:\rho\rho}\right\rangle$ contributes both to NLO JIMWLK Hamiltonian and the LO iteration. 
Particularly $\left|\psi_{g\:\rho\rho}\right\rangle$ will give rise to the NLO kernel $K_{JJSJ}$. The kernel $K_{JJSJ}$ is 
related to the kernel $K_{JJSSJ}$ through the equation (\ref{jjsjiden}). In its turn, the kernel $K_{JJSSJ}$ will be obtained from 
$\left|\psi_{gg\:\rho\rho}\right\rangle$ (\ref{ggrhorho}).  We notice that while there is a relation between the kernels, it does not seem to be possible 
to establish any interesting  relation at the level of the LCWF components. 

\noindent$\bullet$ \textit{\textbf{The one gluon state with three $\rho$}}  \\
Substituting (\ref{threerhos}) and (\ref{rrrow}) in (\ref{grrrar}), we arrive at:
   \begin{equation}\begin{split}\label{3ro}
\left|\psi_{g\:\rho\rho\rho}\right\rangle \,=\,\delta\mathsf{Y}\,\int_{\Lambda}^{e^{\delta\mathsf{Y}}\Lambda}dk^{+}\,\int d^{2}\mathbf{k}\, d^{2}\mathbf{p}\,\frac{g^{3}\,\rho^{b}(\mathbf{p})\,\rho^{b}(-\mathbf{p})\,\rho^{a}(-\mathbf{k})\, \mathbf{k}^{i}}{16\pi^{9/2}\mathbf{k}^{2}\mathbf{p}^{2}\sqrt{k^{+}}}\,\left|g_{i}^{a}(k)\right\rangle .
  \end{split}\end{equation}
Inserting the identity $\int  d^{2}\mathbf{q}\frac{\mathbf{p}\cdot \mathbf{q}}{\mathbf{q}^{2}}\delta^{(2)}(\mathbf{q}-\mathbf{p})\,=\,1$ 
in (\ref{3ro}), the latter can  be straightforwardly written in  coordinate space:
   \begin{equation}\begin{split}\label{g_rhorhorho}
  \left|\psi_{g\:\rho\rho\rho}\right\rangle \,=\,-\,\delta\mathsf{Y}\int_{\mathbf{w},\, \mathbf{x},\, \mathbf{y},\, \mathbf{z},\, \mathbf{z}^{\prime}}\int_{\Lambda}^{e^{\delta\mathsf{Y}}\Lambda}dk^{+}\,\frac{ig^{3}\,\rho^{b}(\mathbf{x})\,\rho^{b}(\mathbf{y})\,\rho^{a}(\mathbf{w})\, X\cdot Y\,(W^{\prime})^{i}}{16\pi^{9/2}X^{2}Y^{2}(W^{\prime})^{2}\sqrt{k^{+}}}\left|g_{i}^{a}(k^{+},\mathbf{z}^{\prime})\right\rangle .
    \end{split}\end{equation}
The  $\left|\psi_{g\:\rho\rho\rho}\right\rangle$ component of the LCWF is pretty transparent. It is already proportional to $\delta Y$ because of the virtual soft gluon, which we have integrated over. One can recognise in the integrand of (\ref{g_rhorhorho}) the kernel of the LO JIMWLK Hamiltonian. The extra soft gluon $g(k)$ in $\left|\psi_{g\:\rho\rho\rho}\right\rangle$ corresponds to another iteration of the kernel as will be demonstrated in the next section.
 
\noindent  $\bullet$ \textit{\textbf{Normalisation}}  \\
  The norm of the wave function can be represented as:
  \begin{equation}
  \mathcal{N}^{NLO}\,=\,\left\Vert \mathcal{N}^{NLO}\right\Vert \, e^{i\phi^{NLO}}.
 \end{equation}
  From  normalisation condition on the expansion (\ref{expenw}),
  \begin{eqnarray}\label{nlonor}
&&\left\Vert \mathcal{N}^{NLO}\right\Vert =\,1-\frac{1}{2}\left(\left\langle \psi_{g\:\rho}^{LO}\left|\psi_{g\:\rho}^{LO}\right.\right\rangle +\left\langle \psi_{q\overline{q}\:\rho}\left|\psi_{q\overline{q}\:\rho}\right.\right\rangle +\left\langle \psi_{g\:\rho}^{LO}\left|\psi_{g\:\rho}\right.\right\rangle +\left\langle \psi_{g\:\rho}\left|\psi_{g\:\rho}^{LO}\right.\right\rangle \right.\\
&&\quad\quad+\,\left\langle \psi_{g\:\rho}^{LO}\left|\psi_{g\:\rho\rho}\right.\right\rangle \,+\,\left\langle \psi_{g\:\rho\rho}\left|\psi_{g\:\rho}^{LO}\right.\right\rangle \,+\,\left\langle \psi_{gg\:\rho}\left|\psi_{gg\:\rho}\right.\right\rangle \,+\,\left\langle \psi_{gg\:\rho}\left|\psi_{gg\:\rho\rho}\right.\right\rangle \,+\,\left\langle \psi_{gg\:\rho\rho}\left|\psi_{gg\:\rho}\right.\right\rangle\nonumber \\
&&\quad\quad\left.\,+\,\left\langle \psi_{gg\:\rho\rho}\left|\psi_{gg\:\rho\rho}\right.\right\rangle \,+\,\left\langle \psi_{g\:\rho}^{LO}\left|\psi_{g\:\rho\rho\rho}\right.\right\rangle \,+\,\left\langle \psi_{g\:\rho\rho\rho}\left|\psi_{g\:\rho}^{LO}\right.\right\rangle \right)\,-\,\frac{1}{8}\left\langle \psi_{g\:\rho}^{LO}\left|\psi_{g\:\rho}^{LO}\right.\right\rangle \left\langle \psi_{g\:\rho}^{LO}\left|\psi_{g\:\rho}^{LO}\right.\right\rangle .\nonumber \end{eqnarray}
The computation relevant for $\left\Vert \mathcal{N}^{NLO}\right\Vert$ is postponed until section \ref{normcon}. Here we quote the final result:
\begin{eqnarray}
\left\Vert \mathcal{N}^{NLO}\right\Vert &=&1-\frac{\alpha_{s}}{2\pi^{2}}\,\delta\mathsf{Y}\,\int_{\mathbf{x},\, \mathbf{y},\, \mathbf{z},\, \mathbf{z}^{\prime}}\,\frac{X\cdot Y}{X^{2}Y^{2}}\rho^{a}(\mathbf{x})\,\rho^{a}(\mathbf{y})-\delta\mathsf{Y}\,\int_{\mathbf{x},\, \mathbf{y},\, \mathbf{z}}K_{JSJ}(\mathbf{x},\, \mathbf{y},\, \mathbf{z})\,\rho^{a}(\mathbf{x})\,\rho^{a}(\mathbf{y})\,\nonumber\\
&+&\frac{\alpha_{s}^{2}}{8\pi^{4}}\,\left(\delta\mathsf{Y}\right)^{2}\,\int_{\mathbf{w},\, \mathbf{v},\, \mathbf{x},\, \mathbf{y},\, \mathbf{z},\, \mathbf{z}^{\prime}}\,\frac{X\cdot Y\, W^{\prime}\cdot V^{\prime}}{X^{2}Y^{2}(W^{\prime})^{2}(V^{\prime})^{2}}\rho^{b}(\mathbf{w})\,\rho^{b}(\mathbf{v})\rho^{a}(\mathbf{x})\,\rho^{a}(\mathbf{y}).\end{eqnarray}
where $K_{JSJ}(\mathbf{x},\, \mathbf{y},\, \mathbf{z})$ is defined in (\ref{JSJ}). 
The first non-trivial term is a LO contribution, which is of order $\alpha_s$.  The second  term is an $\alpha_s^2$ correction and, because of its
linear dependence on $\delta Y$, it will be contributing to virtual terms in the NLO JIMWLK Hamiltonian. The last term will contribute 
to subtraction of (LO)$^2$ terms in the evolution.

The  phase of the LCWF is computed in Appendix \ref{wfphase} from the condition (\ref{condphase}):  
\begin{equation}\begin{split}
i\phi^{NLO}=\,-\delta\mathsf{Y}\,\int_{\mathbf{w},\, \mathbf{x},\, \mathbf{y},\, \mathbf{z},\, \mathbf{z}^{\prime}}\, K_{JJSSJ}(\mathbf{w},\, \mathbf{x},\, \mathbf{y},\, \mathbf{z},\, \mathbf{z}^{\prime})f^{acb}\rho^{a}(\mathbf{x})\,\rho^{b}(\mathbf{y})\,\rho^{c}(\mathbf{w}),
 \end{split}\end{equation}
where $K_{JJSSJ}$ is defined in (\ref{JJSSJ}). Up to order $\alpha_s^2$,
\begin{equation}\label{nphase}
\mathcal{N}^{NLO}\,=\,\left\Vert \mathcal{N}^{NLO}\right\Vert \,+\,i\phi^{NLO}.
\end{equation}

\section{The Next to Leading Order JIMWLK Hamiltonian}\label{kerderi}
The aim of this section is to compute $\Sigma$
 (\ref{Sigma}), that is the $\hat S$-matrix expectation value  in the LCWF (\ref{expenw}),  which was calculated in the previous section.   Each soft quark or gluon in the LCWF will  contribute one eikonal factor $S$ defined respectively in (\ref{S}) and (\ref{SA}). The $\rho$ factors
 in the ket/bra states will be converted into left/right rotation operators $J_{L/R}$ according to (\ref{rhoJ}).
 There are nine  overlaps that  emerge in this computation, which we classify according to the number of $J$ operators and the Wilson lines $S$ that they contain:
 \begin{eqnarray}\label{basicrel}
&&\Sigma\,=\,\left\langle \psi^{NLO}\right|\,\hat{S}\,-\,1\,\left|\psi^{NLO}\right\rangle \,\\
&&=\,\Sigma^{LO}\,+\,\Sigma_{q\bar{q}}\,+\,\Sigma_{JJSSJ}\,+\,\Sigma_{JSSJ}\,+\,\Sigma_{JJSJ}\,+\,\Sigma_{JSJ}\,+\,\Sigma_{JJSSJJ}\,+\,\Sigma_{JJJSJ}\,+\,\Sigma_{virtual}.\nonumber\end{eqnarray}
$\Sigma^{LO}$ was  defined in (\ref{siglo}). The other entrees (all of order $\alpha_{s}^{2}$) are defined below:
\begin{equation}\label{sigmaqq}
\hspace{-4.4 cm}\Sigma_{q\bar{q}}\,\equiv\,\left\langle \psi_{q\bar{q}\:\rho}\right|\,\hat{S}\,\left|\psi_{q\bar{q}\:\rho}\right\rangle ,
 \end{equation}
\begin{equation}\label{sigmaJJSSJ}
\hspace{-0.2 cm}\Sigma_{JJSSJ}\,\equiv\,\left\langle \psi_{gg\:\rho}\right|\,\hat{S}\,\left|\psi_{gg\:\rho\rho}\right\rangle +\left\langle \psi_{gg\:\rho\rho}\right|\,\hat{S}\,\left|\psi_{gg\:\rho}\right\rangle, \\
\end{equation}
\begin{equation}\label{sigmaJSSJ}
\hspace{-3.8 cm}\Sigma_{JSSJ}\,\equiv\,\left\langle \psi_{gg\:\rho}\right|\,\hat{S}\,\left|\psi_{gg\:\rho}\right\rangle, \\ 
\end{equation}
\begin{equation}\label{sigmaJJSJ}
\hspace{-0.8 cm}\Sigma_{JJSJ}\,\equiv\,\left\langle \psi_{g\:\rho\rho}\right|\,\hat{S}\,\left|\psi_{g\:\rho}^{LO}\right\rangle +\left\langle \psi_{g\:\rho}^{LO}\right|\,\hat{S}\,\left|\psi_{g\:\rho\rho}\right\rangle,
\end{equation}
\begin{equation}\label{sigmaJSJ}
\hspace{-1.2 cm}\Sigma_{JSJ}\,\equiv\,\left\langle \psi_{g\:\rho}\right|\,\hat{S}\,\left|\psi_{g\:\rho}^{LO}\right\rangle +\left\langle \psi_{g\:\rho}^{LO}\right|\,\hat{S}\,\left|\psi_{g\:\rho}\right\rangle ,
 \end{equation}
\begin{equation}\label{sigmaJJSSJJ}
\hspace{-3.1 cm}\Sigma_{JJSSJJ}\,\equiv\,\left\langle \psi_{gg\:\rho\rho}\right|\,\hat{S}\,\left|\psi_{gg\:\rho\rho}\right\rangle,
\end{equation}
\begin{equation}\label{sigmaJJSJJ}
\hspace{-0.1 cm}\Sigma_{JJJSJ}\,\equiv\,\left\langle \psi_{g\:\rho\rho\rho}\right|\,\hat{S}\,\left|\psi_{g\:\rho}^{LO}\right\rangle \,+\,\left\langle \psi_{g\:\rho}^{LO}\right|\,\hat{S}\,\left|\psi_{g\:\rho\rho\rho}\right\rangle ,
\end{equation}
\begin{equation}\begin{split}\label{signor}
\hspace{-1.4 cm}\Sigma_{virtual}\,\equiv\,\left\langle 0\right|\left(\mathcal{N}^{NLO}\right)^{\dagger}\,\hat{S}\,\mathcal{N}^{NLO}\left|0\right\rangle -1.
  \end{split}\end{equation}
For each of these contributions a dedicated subsection will be devoted. Some details of the calculations are provided in Appendix \ref{supsect4}.
 
As anticipated, we will find not only contributions linear in $\delta\mathsf{Y}$, but also contributions of order $(\delta\mathsf{Y})^{2}$. To make  separation between these very different types of contributions crystal clear,  we will introduce an additional label (upper script) $-$ NLO or $(\delta\mathsf{Y})^{2}$ respectively. Each $\Sigma_{...}$ defined in (\ref{sigmaqq}) $-$ (\ref{signor}) is decomposed according to:
   \begin{equation}
 \Sigma_{...}\,=\,\Sigma^{NLO}_{...}(\alpha_{s}^{2},\,\delta\mathsf{Y})\,+\,\Sigma_{...}^{(\delta\mathsf{Y})^{2}}(\alpha_{s}^{2},\,(\delta\mathsf{Y})^{2}).
  \end{equation}
At the end, in section \ref{subterms}, we will assemble together all the contributions at NLO and write down the NLO JIMWLK Hamiltonian 
defined in (\ref{defH}).

Finally, in section \ref{reduclo} we will collect all the $(\delta\mathsf{Y})^{2}$ contributions, and show that they correspond to contributions which are generated from the LO Hamiltonian applied twice. Based on the expansion (\ref{opr}), we will show that the anticipated relation:
 \begin{equation}\begin{split}\label{lolo}
 \Sigma^{(\delta\mathsf{Y})^{2}}\,=\,\frac{1}{2}\,(\delta Y\,H_{JIMWLK}^{LO})^{2}
  \end{split}\end{equation}
  holds indeed. This relation serves as a strong consistency check on our calculation.

\subsection{Computation of $\Sigma_{q\bar{q}}$}\label{qqsec}
\begin{figure}[!h]
  \centering
\includegraphics[scale=0.7]{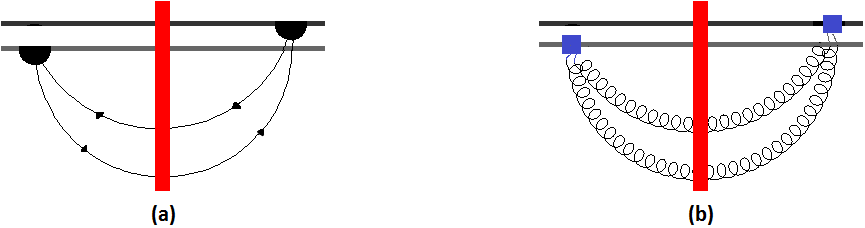}
  \caption{(a) - The diagram for $\Sigma_{q\bar{q}}$, (b) - The diagram for $\Sigma_{JSSJ}$. The black blub denotes the effective emission vertex, which is defined by $\left|\psi_{q\bar{q}\:\rho}\right\rangle$. The effective two gluon emission vertex corresponds to $\left|\psi_{gg\: \rho}\right\rangle $.\label{kerqq}}
\end{figure}

$\Sigma_{q\bar{q}}$ is defined in (\ref{sigmaqq}) as an overlap between the incoming state $\left|\psi_{q\overline{q}\:\rho}\right\rangle $ (\ref{stateqq}), before and after passing through the shockwave (Fig.\ref{kerqq}a).
\\
Representing the operators $\rho$ in terms of the operators $J_{L}$ and $J_{R}$ as explained in section \ref{eiksc}, and replacing the products of Pauli matrices  using the identity $\sigma^{i}\sigma^{j}=i\varepsilon^{ijk}\sigma^{k}+\delta^{ij}I$,

\begin{eqnarray}\label{starqq}
\Sigma_{q\bar{q}}&=&\int_{\mathbf{x},\,\mathbf{y},\,\mathbf{z},\,\mathbf{z}^{\prime}}\int_{\Lambda}^{e^{\delta\mathsf{Y}}\Lambda}dk^{+}\,\int_{0}^{1}d\xi\,\int d^{2}\mathbf{k}\, d^{2}\mathbf{p}\, d^{2}\mathbf{u}\, d^{2}\mathbf{v}\: e^{-i\mathbf{v}\cdot Z+i\mathbf{u}\cdot Y^{\prime}+i\mathbf{p}\cdot Z-i\mathbf{k}\cdot X^{\prime}}\nonumber\\
&\times&\frac{g^{4}N_{f}\, J_{L}^{a}(\mathbf{x})\, Tr\left[S^{\dagger}(\mathbf{z})t^{a}S(\mathbf{z}^{\prime})t^{b}\right]\, J_{R}^{b}(\mathbf{y})}{512\pi^{10}\left((1-\xi)\mathbf{p}^{2}+\xi(\mathbf{k}-\mathbf{p})^{2}\right)\left((1-\xi)\mathbf{v}^{2}+\xi(\mathbf{u}-\mathbf{v})^{2}\right)k^{+}}\nonumber\\
&\times&\left(\frac{(1-2\xi)^{2}\left(\mathbf{p}^{i}-\xi \mathbf{k}^{i}\right)\left(\mathbf{v}^{j}-\xi \mathbf{u}^{j}\right)+\varepsilon^{im}\varepsilon^{js}\left(\mathbf{p}^{m}-\xi \mathbf{k}^{m}\right)\left(\mathbf{v}^{s}-\xi \mathbf{u}^{s}\right)}{\mathbf{k}^{2}\mathbf{u}^{2}}\mathbf{k}^{i}\mathbf{u}^{j}\right.\nonumber\\
&&\left.-2\xi(1-\xi)(1-2\xi)\left(\frac{\mathbf{k}\cdot(\xi\mathbf{k}-\mathbf{p})}{\mathbf{k}^{2}}+\frac{\mathbf{u}\cdot(\xi\mathbf{u}-\mathbf{v})}{\mathbf{u}^{2}}\right)+4\xi^{2}(1-\xi)^{2}\right).\end{eqnarray}
We have used the fact that the scattering matrix is diagonal in coordinate space:
 \begin{eqnarray}\label{qq.scat}
&&\left\langle \bar{q}_{\lambda_{4}}^{\delta,\,\bar{f}}(u-v)\, q_{\lambda_{3}}^{\gamma,\,\bar{f}}(v)\left|\,\hat{S}\,\right|\bar{q}_{\lambda_{2}}^{\beta,\, f}(k-p)\, q_{\lambda_{1}}^{\alpha,\, f}(p)\right\rangle \\ 
&&\quad=\frac{N_{f}\delta_{\lambda_{1}\lambda_{3}}\delta_{\lambda_{2}\lambda_{4}}}{(2\pi)^{4}k^{+}}\int_{\mathbf{z},\, \mathbf{z}^{\prime}}S^{\delta\alpha}(\mathbf{z})S^{\dagger\beta\gamma}(\mathbf{z}^{\prime})e^{-i\mathbf{v}\cdot(\mathbf{z}-\mathbf{z}^{\prime})+i\mathbf{p}\cdot(\mathbf{z}-\mathbf{z}^{\prime})-i\mathbf{u}\cdot \mathbf{z}^{\prime}+i\mathbf{k}\cdot \mathbf{z}^{\prime}}\,\delta(u^{+}-k^{+})\,\delta(\xi-\vartheta),
\nonumber \end{eqnarray}
Here $\xi\,\equiv\,\frac{p^{+}}{k^{+}}$ and $\vartheta\,\equiv\,\frac{v^{+}}{u^{+}}$. Equation (\ref{starqq}) coincides with eq. $(14)$ of \cite{Weigertrun}, after  the identity  (\ref{epsi}) and a shift in momenta are applied. The rest of the computation can be found in Appendix \ref{sfk1}. 
After integration over $\xi$ in (\ref{siqq3}), using (\ref{int.2}), (\ref{int.3}), and (\ref{int.4}),

\begin{eqnarray}\label{resqq}
&&\Sigma_{q\bar{q}}\,=\,\delta\mathsf{Y}\int_{\mathbf{x},\mathbf{y},\mathbf{z},\mathbf{z}^{\prime}}\frac{g^{4}N_{f}\, J_{L}^{a}(\mathbf{x})\, Tr\left[S^{\dagger}(\mathbf{z})t^{a}S(\mathbf{z}^{\prime})t^{b}\right]\, J_{R}^{b}(\mathbf{y})}{64\pi^{6}Z^{4}}\left(2-\frac{Z^{2}}{(X^{\prime})^{2}-X^{2}}\ln\left(\frac{X^{2}}{(X^{\prime})^{2}}\right)\right.\nonumber\\
&&-\frac{Z^{2}}{(Y^{\prime})^{2}-Y^{2}}\ln\left(\frac{Y^{2}}{(Y^{\prime})^{2}}\right)-\frac{2X^{2}(X^{\prime})^{2}\left((Y^{\prime})^{2}-Y^{2}\right)}{\left((X^{\prime})^{2}-X^{2}\right)\left((X^{\prime})^{2}Y^{2}-X^{2}(Y^{\prime})^{2}\right)}\ln\left(\frac{X^{2}}{(X^{\prime})^{2}}\right)\\
&&\left.+\frac{2Y^{2}(Y^{\prime})^{2}\left((X^{\prime})^{2}-X^{2}\right)}{\left((Y^{\prime})^{2}-Y^{2}\right)\left((X^{\prime})^{2}Y^{2}-X^{2}(Y^{\prime})^{2}\right)}\ln\left(\frac{Y^{2}}{(Y^{\prime})^{2}}\right)-\frac{(X-Y)^{2}Z^{2}}{(X^{\prime})^{2}Y^{2}-X^{2}(Y^{\prime})^{2}}\ln\left(\frac{(X^{\prime})^{2}Y^{2}}{X^{2}(Y^{\prime})^{2}}\right)\right).\nonumber\end{eqnarray}
The result contains terms of order $\delta\mathsf{Y}$ only. To indicate this, we add the upper script $NLO$, so that $\Sigma_{q\bar{q}}^{NLO}\,\equiv\,\Sigma_{q\bar{q}}$. (\ref{resqq}) can be equivalently written as:
  \begin{equation}\begin{split}\label{sigqq}
  \Sigma_{q\bar{q}}^{NLO}\,=\,-\,\delta\mathsf{Y}\int_{\mathbf{x},\, \mathbf{y},\, \mathbf{z},\, \mathbf{z}^{\prime}}K_{q\bar{q}}(\mathbf{x},\, \mathbf{y},\, \mathbf{z},\, \mathbf{z}^{\prime})\left[2J_{L}^{a}(\mathbf{x})\, tr\left[S^{\dagger}(\mathbf{z})t^{a}S(\mathbf{z}^{\prime})t^{b}\right]\, J_{R}^{b}(\mathbf{y})\right],
   \end{split}\end{equation}
   where $K_{q\bar{q}}(\mathbf{x},\, \mathbf{y},\, \mathbf{z},\, \mathbf{z}^{\prime})$ is defined in  (\ref{kqq}).

\subsection{Computation of $\Sigma_{JSSJ}$}

$\Sigma_{JSSJ}$ is defined in (\ref{sigmaJSSJ}) as an overlap of $\left|\psi_{gg\:\rho}\right\rangle $, (\ref{gg_rho}), before and after passing through the shockwave. The relevant diagrams appears in Fig.\ref{kerqq}b. 

 \begin{eqnarray}\label{starjssj}
\Sigma_{JSSJ}
&&=\int_{\mathbf{x},\, \mathbf{y},\, \mathbf{z},\, \mathbf{z}^{\prime}}\int_{\Lambda}^{e^{\delta\mathsf{Y}}\Lambda}dk^{+}\int_{\frac{\Lambda}{k^{+}}}^{1-\frac{\Lambda}{k^{+}}}d\xi\,\frac{g^{4}f^{abc}f^{def}\, J_{L}^{a}(\mathbf{x})\, S_{A}^{be}(\mathbf{z})\, S_{A}^{cf}(\mathbf{z}^{\prime})\, J_{R}^{d}(\mathbf{y})\,\xi(1-\xi)}{32\pi^{6}k^{+}\left((1-\xi)(X^{\prime})^{2}+\xi X^{2}\right)\left((1-\xi)(Y^{\prime})^{2}+\xi Y^{2}\right)}\nonumber\\
&&\times\left(\delta^{jl}\frac{X^{2}-(X^{\prime})^{2}}{2Z^{2}}+\frac{1}{\xi}\left[\frac{(X^{\prime})^{j}Z^{l}}{Z^{2}}+\frac{X^{l}(X^{\prime})^{j}}{2X^{2}}\right]+\frac{1}{1-\xi}\left[\frac{X^{l}Z^{j}}{Z^{2}}-\frac{X^{l}(X^{\prime})^{j}}{2(X^{\prime})^{2}}\right]\right)\nonumber\\
&&\times\left(\delta^{jl}\frac{Y^{2}-(Y^{\prime})^{2}}{2Z^{2}}+\frac{1}{\xi}\left[\frac{(Y^{\prime})^{j}Z^{l}}{Z^{2}}+\frac{Y^{l}(Y^{\prime})^{j}}{2Y^{2}}\right]+\frac{1}{1-\xi}\left[\frac{Y^{l}Z^{j}}{Z^{2}}-\frac{Y^{l}(Y^{\prime})^{j}}{2(Y^{\prime})^{2}}\right]\right).\end{eqnarray}\\
where the following matrix element was used:
\begin{eqnarray}\label{matss}
&&\left\langle g_{m}^{e}(u^{+}\,-\,v^{+},\,\mathbf{z}^{\prime})\,g_{n}^{f}(v^{+},\,\mathbf{z})\,\right|\,\hat{S}\,\left|\,g_{j}^{b}(k^{+}\,-\,p^{+},\,\mathbf{\overline{z}^{\prime}})\,g_{l}^{c}(p^{+},\,\mathbf{\overline{z}})\right\rangle  \nonumber\\
&&=\,\delta(u^{+}\,-\,k^{+})\,\left(\,S_{A}^{bf}(\mathbf{z})\,S_{A}^{ce}(\mathbf{z}^{\prime})\,\delta(k^{+}\,-\,p^{+}\,-\,v^{+})\,\delta^{(2)}(\mathbf{z}\,-\,\mathbf{\overline{z}^{\prime}})\,\delta^{(2)}(\mathbf{z}^{\prime}\,-\,\mathbf{\overline{z}})\,\delta_{nj}\,\delta_{lm}\right.\nonumber\\
&&\left.+\,S_{A}^{be}(\mathbf{z})\,S_{A}^{cf}(\mathbf{z}^{\prime})\,\delta(p^{+}-v^{+})\,\delta^{(2)}(\mathbf{z}^{\prime}-\mathbf{\overline{z}^{\prime}})\,\delta^{(2)}(\mathbf{z}-\mathbf{\overline{z}})\,\delta_{mj}\,\delta_{nl}\,\right).\end{eqnarray}
The remaining  computation is moved to Appendix \ref{sfk3}. The result is:
\begin{equation}\label{sigijssj}
\Sigma_{JSSJ}\,=\,\Sigma_{JSSJ}^{NLO}\,+\,\Sigma_{JSSJ}^{(\delta\mathsf{Y})^{2}},
  \end{equation}
with
 \begin{equation}\begin{split}\label{ajssjsig}
&\Sigma_{JSSJ}^{NLO}\,=\,-\,\delta\mathsf{Y}\int_{\mathbf{x},\, \mathbf{y},\, \mathbf{z},\, \mathbf{z}^{\prime}}K_{JSSJ}(\mathbf{x},\, \mathbf{y},\, \mathbf{z},\, \mathbf{z}^{\prime})\, f^{abc}\, f^{def}\, S_{A}^{be}(\mathbf{z})\, S_{A}^{cf}(\mathbf{z}^{\prime})\, J_{L}^{a}(\mathbf{x})\, J_{R}^{d}(\mathbf{y}),
 \end{split}\end{equation}
where $K_{JSSJ}(\mathbf{x},\, \mathbf{y},\, \mathbf{z},\, \mathbf{z}^{\prime})$ is defined in (\ref{JSSJ}), and 

 \begin{eqnarray}\label{losq_JSSJ}
\Sigma_{JSSJ}^{(\delta\mathsf{Y})^{2}}&=&\frac{\alpha_{s}^{2}}{4\pi^{4}}\,\left(\delta\mathsf{Y}\right)^{2}\,\int_{\mathbf{x},\,\mathbf{y},\,\mathbf{z},\,\mathbf{z}^{\prime}}\left(\frac{2X\cdot Y}{Z^{2}X^{2}Y^{2}}-\frac{X\cdot Y\, Z\cdot Y^{\prime}}{X^{2}Y^{2}Z^{2}(Y^{\prime})^{2}}-\frac{X\cdot Y\, X^{\prime}\cdot Z}{X^{2}Y^{2}(X^{\prime})^{2}Z^{2}}\right.\nonumber\\
&&\left.+\frac{X\cdot Y\, X^{\prime}\cdot Y^{\prime}}{2X^{2}Y^{2}(X^{\prime})^{2}(Y^{\prime})^{2}}\right)\, f^{abc}\, f^{def}\, S_{A}^{be}(\mathbf{z})\, S_{A}^{cf}(\mathbf{z}^{\prime})\, J_{L}^{a}(\mathbf{x})\, J_{R}^{d}(\mathbf{y}).\end{eqnarray}

 \subsection{Computation of $\Sigma_{JJSSJ}$}\label{sejjssjoo}
\begin{figure}[!h]
  \centering
\includegraphics[scale=0.6]{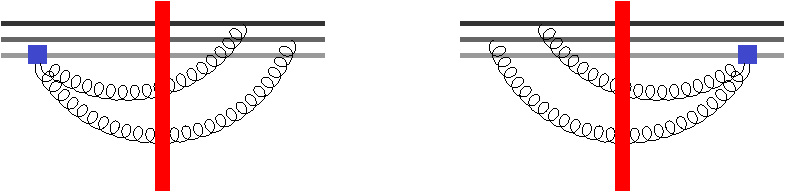}
  \caption{The diagrams for $\Sigma_{JJSSJ}$. Effective two gluon emission vertex corresponds to $\left|\psi_{gg\: \rho}\right\rangle $.\label{figjjssj}}
\end{figure}

 $\Sigma_{JJSSJ}$ is defined in (\ref{sigmaJJSSJ}) as an $S$-matrix overlap between $\left|\psi_{gg\:\rho}\right\rangle $ (\ref{gg_rho}) and $\left|\psi_{gg\:\rho\rho}\right\rangle $ (\ref{gg_rhorho}). The relevant diagrams appear in Fig.\ref{figjjssj}. With the aid of (\ref{matss}),

\begin{eqnarray}\label{jjssjsta}
\Sigma_{JJSSJ}&=&\frac{ig^{4}f^{acb}}{64\pi^{6}}\int_{\mathbf{w},\,\mathbf{x},\,\mathbf{y},\,\mathbf{z},\,\mathbf{z}^{\prime}}\int_{\Lambda}^{e^{\delta\mathsf{Y}}\Lambda}\frac{dk^{+}}{k^{+}}\,\int_{\frac{\Lambda}{k^{+}}}^{1-\frac{\Lambda}{k^{+}}}d\xi\,\\
&\times&\frac{1}{\left((1-\xi)(W^{\prime})^{2}\,+\,\xi W^{2}\right)}\,\left(\frac{(Y^{\prime})^{j}X^{i}}{(Y^{\prime})^{2}X^{2}}\,-\,\frac{Y^{i}(X^{\prime})^{j}}{(X^{\prime})^{2}Y^{2}}\right)\nonumber\\
&\times&\left(\frac{\delta^{ij}}{2Z^{2}}\left(W^{2}-(W^{\prime})^{2}\right)\,-\,\frac{(W^{\prime})^{j}Z^{i}}{\xi Z^{2}}\,+\,\frac{W^{i}Z^{j}}{(1-\xi)Z^{2}}\,+\,\frac{W^{i}(W^{\prime})^{j}}{2\xi W^{2}}\,-\,\frac{W^{i}(W^{\prime})^{j}}{2(1-\xi)(W^{\prime})^{2}}\right)\nonumber\\
&\times&\left[J_{L}^{d}(\mathbf{x})\, J_{L}^{e}(\mathbf{y})\, S_{A}^{dc}(\mathbf{z})\, S_{A}^{eb}(\mathbf{z}^{\prime})\, J_{R}^{a}(\mathbf{w})\,-\, J_{L}^{a}(\mathbf{w})\, S_{A}^{cd}(\mathbf{z})\, S_{A}^{be}(\mathbf{z}^{\prime})\, J_{R}^{d}(\mathbf{x})\, J_{R}^{e}(\mathbf{y})\right],\nonumber\end{eqnarray}\
Notice that the minus sign between the terms in the last line in (\ref{jjssjsta}) is due to  complex conjugation. The integration over $\xi$ is done using (\ref{int.2}), (\ref{heaso1}) and (\ref{heaso2}):

\begin{eqnarray}\label{pods}
&&\Sigma_{JJSSJ}=\frac{ig^{4}f^{acb}}{64\pi^{6}}\int_{\mathbf{w},\,\mathbf{x},\,\mathbf{y},\,\mathbf{z},\,\mathbf{z}^{\prime}}\int_{\Lambda}^{e^{\delta\mathsf{Y}}\Lambda}\frac{dk^{+}}{k^{+}}\,\left(\frac{(Y^{\prime})^{j}X^{i}}{(Y^{\prime})^{2}X^{2}}\,-\,\frac{Y^{i}(X^{\prime})^{j}}{(X^{\prime})^{2}Y^{2}}\right)\\
&&\times\left[\left(\frac{\delta^{ij}}{2Z^{2}}-\frac{(W^{\prime})^{j}Z^{i}}{(W^{\prime})^{2}Z^{2}}+\frac{W^{i}Z^{j}}{W^{2}Z^{2}}-\frac{W^{i}(W^{\prime})^{j}}{W^{2}(W^{\prime})^{2}}\right)\ln\left(\frac{W^{2}}{(W^{\prime})^{2}}\right)\,-\,\left(\frac{(W^{\prime})^{j}Z^{i}}{(W^{\prime})^{2}Z^{2}}+\frac{W^{i}Z^{j}}{W^{2}Z^{2}}\right)\right.\nonumber\\
&&\left.\times\ln\left(\frac{\Lambda}{k^{+}}\right)\right]\left[J_{L}^{d}(\mathbf{x})\, J_{L}^{e}(\mathbf{y})\, S_{A}^{dc}(\mathbf{z})\, S_{A}^{eb}(\mathbf{z}^{\prime})\, J_{R}^{a}(\mathbf{w})-J_{L}^{a}(\mathbf{w})\, S_{A}^{cd}(\mathbf{z})\, S_{A}^{be}(\mathbf{z}^{\prime})\, J_{R}^{d}(\mathbf{x})\, J_{R}^{e}(\mathbf{y})\right].\nonumber\end{eqnarray}
 It is important to notice that $J_{L}$ and $J_{R}$ in (\ref{pods}) by construction do not act on $S_{A}$ inside (\ref{pods}).
 As anticipated, the result contains both terms proportional to $\delta\mathsf{Y}$ and $(\delta\mathsf{Y})^{2}$, which are split as:
 \begin{equation}\label{si}
\Sigma_{JJSSJ}\,=\,\Sigma_{JJSSJ}^{NLO}\,+\,\Sigma_{JJSSJ}^{(\delta\mathsf{Y})^{2}}.
 \end{equation}
$\Sigma_{JJSSJ}^{(\delta\mathsf{Y})^{2}}$ emerges from the term $\ln\left(\frac{\Lambda}{k^{+}}\right)$, which becomes proportional to $(\delta\mathsf{Y})^{2}$ after integration over $k^{+}$. $\Sigma_{JJSSJ}^{NLO}$ reads:
\begin{equation}\begin{split}\label{jjssjsig}
&\Sigma_{JJSSJ}^{NLO}\,=\,-\,\delta\mathsf{Y}\int_{\mathbf{w},\, \mathbf{x},\, \mathbf{y},\, \mathbf{z},\, \mathbf{z}^{\prime}}\, K_{JJSSJ}(\mathbf{w},\, \mathbf{x},\, \mathbf{y},\, \mathbf{z},\, \mathbf{z}^{\prime})\\
&\times f^{acb}\left[J_{L}^{d}(\mathbf{x})\, J_{L}^{e}(\mathbf{y})\, S_{A}^{dc}(\mathbf{z})\, S_{A}^{eb}(\mathbf{z}^{\prime})\, J_{R}^{a}(\mathbf{w})-J_{L}^{a}(\mathbf{w})\, S_{A}^{cd}(\mathbf{z})\, S_{A}^{be}(\mathbf{z}^{\prime})\, J_{R}^{d}(\mathbf{x})\, J_{R}^{e}(\mathbf{y})\right].
\end{split}\end{equation}
where $K_{JJSSJ}(\mathbf{w},\, \mathbf{x},\, \mathbf{y},\, \mathbf{z},\, \mathbf{z}^{\prime})$ is defined in (\ref{JJSSJ}).

  \begin{equation}\begin{split}\label{losq_JJSSJ}
 &\Sigma_{JJSSJ}^{(\delta\mathsf{Y})^{2}}\,=\,\frac{i\alpha_{s}^{2}}{8\pi^{4}}\left(\delta\mathsf{Y}\right)^{2}\int_{\mathbf{w},\, \mathbf{x},\, \mathbf{y},\, \mathbf{z},\, \mathbf{z}^{\prime}}\left(\frac{(Y^{\prime})^{j}X^{i}}{(Y^{\prime})^{2}X^{2}}-\frac{Y^{i}(X^{\prime})^{j}}{(X^{\prime})^{2}Y^{2}}\right)\,\left(\frac{(W^{\prime})^{j}Z^{i}}{(W^{\prime})^{2}Z^{2}}+\frac{W^{i}Z^{j}}{W^{2}Z^{2}}\right)\\
&\times\, f^{acb}\,\left[J_{L}^{d}(\mathbf{x})\, J_{L}^{e}(\mathbf{y})\, S_{A}^{dc}(\mathbf{z})\, S_{A}^{eb}(\mathbf{z}^{\prime})\, J_{R}^{a}(\mathbf{w})\,-\, J_{L}^{a}(\mathbf{w})\, S_{A}^{cd}(\mathbf{z})\, S_{A}^{be}(\mathbf{z}^{\prime})\, J_{R}^{d}(\mathbf{x})\, J_{R}^{e}(\mathbf{y})\right].
  \end{split}\end{equation}

    \subsection{Computation of $\Sigma_{JJSJ}$}
\begin{figure}[!h]
  \centering
  \includegraphics[scale=0.6]{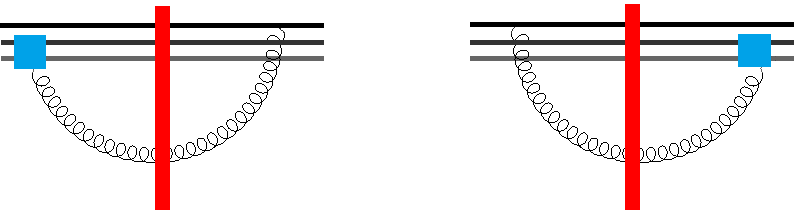}
  \caption{The diagrams for $\Sigma_{JJSJ}$. Effective one gluon emission vertex corresponds to $\left|\psi_{g\: \rho\rho}\right\rangle $.\label{figjjsj}}
\end{figure}

     $\Sigma_{JJSJ}$ is defined in (\ref{sigmaJJSJ}) as an $S$-matrix element between the state $\left|\psi_{g\:\rho\rho}\right\rangle $, (\ref{g_rhorho}), and the state $\left|\psi_{g\:\rho}^{LO}\right\rangle $, (\ref{lotra}). The relevant diagrams appear in Fig.\ref{figjjsj}.
 \begin{eqnarray}
&&\Sigma_{JJSJ}=-\frac{g^{4}f^{bde}}{128\pi^{7}}\int_{\Lambda}^{e^{\delta\mathsf{Y}}\Lambda}\frac{dk^{+}}{k^{+}}\int_{\mathbf{w},\,\mathbf{x},\,\mathbf{y},\,\mathbf{z}}\,\frac{W^{i}}{W^{2}}\left[J_{L}^{d}(\mathbf{x})\, J_{L}^{e}(\mathbf{y})\, S_{A}^{ba}(\mathbf{z})\, J_{R}^{a}(\mathbf{w})\right.\nonumber\\
&&\left.-J_{L}^{a}(\mathbf{w})\, S_{A}^{ab}(\mathbf{z})\, J_{R}^{d}(\mathbf{x})\, J_{R}^{e}(\mathbf{y})\right]\int d^{2}\mathbf{k}\, d^{2}\mathbf{p}\: e^{-i\mathbf{k}\cdot X+i\mathbf{p}\cdot\left(X-Y\right)}\frac{1}{\mathbf{p}^{2}\mathbf{k}^{2}}\\
&&\left\{ \left(\left[\left(2-\frac{\mathbf{k}\cdot\mathbf{p}}{(\mathbf{k}-\mathbf{p})^{2}}\right)\ln\left(\frac{\Lambda}{k^{+}}\right)+\frac{2\mathbf{p}^{2}}{(\mathbf{k}-\mathbf{p})^{2}}\ln\left(\frac{(\mathbf{k}-\mathbf{p})^{2}}{\mathbf{k}^{2}}\right)\right]\mathbf{k}^{i}+\ln\left(\frac{\Lambda}{k^{+}}\right)\frac{\mathbf{k}^{2}}{(\mathbf{k}-\mathbf{p})^{2}}\mathbf{p}^{i}\right)\right.\nonumber\\
&&\left.-\int_{\frac{\Lambda}{k^{+}}}^{1-\frac{\Lambda}{k^{+}}}d\xi\,\frac{\left[2\left(\mathbf{p}^{2}-\mathbf{k}\cdot\mathbf{p}\right)-\xi\mathbf{p}^{2}+2\xi\mathbf{k}\cdot\mathbf{p}\right]\mathbf{k}^{i}+\mathbf{k}^{2}\mathbf{p}^{i}}{\left(\xi(\mathbf{k}-\mathbf{p})^{2}+(1-\xi)\mathbf{k}^{2}\right)(1-\xi)}\right\} .\nonumber\end{eqnarray}
After the $\xi$ integral  is evaluated using (\ref{jjsjino}) and (\ref{jjsjino2}):
 \begin{eqnarray}\label{sojjsj}
&&\Sigma_{JJSJ}=-\frac{g^{4}f^{bde}}{128\pi^{7}}\int_{\Lambda}^{e^{\delta\mathsf{Y}}\Lambda}\frac{dk^{+}}{k^{+}}\int_{\mathbf{w},\,\mathbf{x},\,\mathbf{y},\,\mathbf{z}}\,\frac{W^{i}}{W^{2}}\left[J_{L}^{d}(\mathbf{x})\, J_{L}^{e}(\mathbf{y})\, S_{A}^{ba}(\mathbf{z})\, J_{R}^{a}(\mathbf{w})\right.\nonumber\\
&&\left.-J_{L}^{a}(\mathbf{w})\, S_{A}^{ab}(\mathbf{z})\, J_{R}^{d}(\mathbf{x})\, J_{R}^{e}(\mathbf{y})\right]\left(\int d^{2}\mathbf{k}\, d^{2}\mathbf{p}\: e^{-i\mathbf{k}\cdot X+i\mathbf{p}\cdot\left(X-Y\right)}\left[\left(\frac{\mathbf{p}^{i}}{\mathbf{p}^{2}(\mathbf{k}-\mathbf{p})^{2}}-\frac{\mathbf{k}^{i}}{\mathbf{k}^{2}(\mathbf{k}-\mathbf{p})^{2}}\right.\right.\right.\nonumber\\
&&\quad\quad\quad\left.\left.+\frac{\mathbf{k}^{i}}{\mathbf{k}^{2}\mathbf{p}^{2}}\right)\ln\left(\frac{(\mathbf{k}-\mathbf{p})^{2}}{\mathbf{k}^{2}}\right)-\left(\frac{2\mathbf{k}^{i}}{\mathbf{k}^{2}\mathbf{p}^{2}}\,+\,\frac{\mathbf{p}\cdot(\mathbf{p}-\mathbf{k})\,\mathbf{k}^{i}}{\mathbf{k}^{2}\mathbf{p}^{2}(\mathbf{k}-\mathbf{p})^{2}}\,+\,\frac{2\mathbf{p}^{i}}{\mathbf{p}^{2}(\mathbf{k}-\mathbf{p})^{2}}\right)\ln\left(\frac{k^{+}}{\Lambda}\right)\right]\nonumber\\
&&\quad\quad\quad\,-\,(X\leftrightarrow Y)\bigg).\end{eqnarray}
As usual, this expression can be split according to:
\begin{equation}\label{splitjjsj}
\Sigma_{JJSJ}\,=\,\Sigma_{JJSJ}^{NLO}\,+\,\Sigma_{JJSJ}^{(\delta\mathsf{Y})^{2}}.
 \end{equation}
The remaining computation is available in Appendix \ref{sfk5}. The results:
 \begin{equation}\begin{split}\label{jjsjlab}
\Sigma_{JJSJ}^{NLO}&=\,-\delta\mathsf{Y}\int_{\mathbf{w},\, \mathbf{x},\, \mathbf{y},\, \mathbf{z}}\, K_{JJSJ}(\mathbf{w},\, \mathbf{x},\, \mathbf{y},\, \mathbf{z})\, f^{bde}\left[J_{L}^{d}(\mathbf{x})\, J_{L}^{e}(\mathbf{y})\, S_{A}^{ba}(\mathbf{z})\, J_{R}^{a}(\mathbf{w})\right.\\
&\left.-J_{L}^{a}(\mathbf{w})\, S_{A}^{ab}(\mathbf{z})\, J_{R}^{d}(\mathbf{x})\, J_{R}^{e}(\mathbf{y})\right].\\
 \end{split}\end{equation}
  With $K_{JJSJ}(\mathbf{w},\, \mathbf{x},\, \mathbf{y},\, \mathbf{z})$ defined in (\ref{JJSJ}). In addition:
 \begin{eqnarray}\label{losqjjsjne}
&&\Sigma_{JJSJ}^{(\delta\mathsf{Y})^{2}}\\
&&\quad=\frac{i\alpha_{s}^{2}f^{bde}}{4\pi^{4}}(\delta\mathsf{Y})^{2}\int_{\mathbf{w},\,\mathbf{x},\,\mathbf{y},\,\mathbf{z},\,\mathbf{z}^{\prime}}\,\left[J_{L}^{d}(\mathbf{x})\, J_{L}^{e}(\mathbf{y})\, S_{A}^{ba}(\mathbf{z}^{\prime})\, J_{R}^{a}(\mathbf{w})-J_{L}^{a}(\mathbf{w})\, S_{A}^{ab}(\mathbf{z}^{\prime})\, J_{R}^{d}(\mathbf{x})\, J_{R}^{e}(\mathbf{y})\right]\nonumber\\
&&\quad\times\left[\frac{X\cdot Y\, Y^{\prime}\cdot W^{\prime}}{X^{2}Y^{2}(Y^{\prime})^{2}(W^{\prime})^{2}}-\frac{X\cdot Y\, X^{\prime}\cdot W^{\prime}}{X^{2}Y^{2}(X^{\prime})^{2}(W^{\prime})^{2}}-\frac{X\cdot Z\, Y^{\prime}\cdot W^{\prime}}{X^{2}Z^{2}(Y^{\prime})^{2}(W^{\prime})^{2}}+\frac{Y\cdot Z\, X^{\prime}\cdot W^{\prime}}{Y^{2}Z^{2}(X^{\prime})^{2}(W^{\prime})^{2}}\right].\nonumber\end{eqnarray}

      \subsection{Computation of $\Sigma_{JSJ}$}\label{jsjsec}
\begin{figure}[!h]
  \centering
  \includegraphics[scale=0.6]{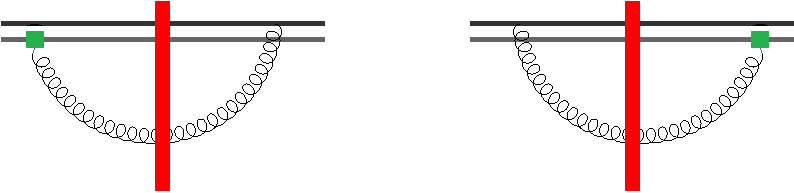}
  \caption{The diagrams for $\Sigma_{JSJ}$. Effective one gluon emission vertex corresponds to $\left|\psi_{g\: \rho}\right\rangle $.\label{figjsj1}}
\end{figure}

 $\Sigma_{JSJ}$ is defined in (\ref{sigmaJSJ}) as an $S$-matrix element between the state $\left|\psi_{g\:\rho}\right\rangle $, (\ref{g_rho}), and the state $\left|\psi_{g\:\rho}^{LO}\right\rangle $, (\ref{lotra}). The relevant diagrams appears in Fig.\ref{figjsj1}. 
 \begin{eqnarray}
&&\Sigma_{JSJ}\,=\,-\frac{g^{4}}{64\pi^{5}}\int_{\mathbf{x},\,\mathbf{y},\,\mathbf{z}}\,\int_{\Lambda}^{e^{\delta\mathsf{Y}}\Lambda}\, dk^{+}\frac{X\cdot Y}{X^{2}Y^{2}k^{+}}\bigg\{\left(\left[b-N_{c}\ln\,\left(\frac{k^{+}}{\Lambda}\right)-2N_{c}\delta\mathsf{Y}\right]\right.\nonumber\\
&&\left.\times\left[-2\gamma-\ln\left(\frac{X^{2}\mu_{\overline{MS}}^{2}}{4}\right)\right]-N_{c}\ln^{2}\left(\frac{\Lambda e^{\delta\mathsf{Y}}}{k^{+}}\right)+N_{c}\ln^{2}\left(\frac{\Lambda}{k^{+}}\right)-\left(\frac{67}{9}-\frac{\pi^{2}}{3}\right)N_{c}+\frac{10}{9}N_{f}\right)\nonumber\\
&&+\int_{\frac{\Lambda}{k^{+}}}^{1-\frac{\Lambda}{k^{+}}}d\xi\,\frac{N_{c}}{\xi(1-\xi)}\bigg(\left(\xi-2\right)\ln\left(\xi\right)+\left(\xi+1\right)\ln\left(1-\xi\right)\\
&&-\left(\xi-2\right)\left[-2\gamma-\ln\left(\frac{X^{2}\mu_{\overline{MS}}^{2}}{4}\right)\right]\bigg)+(X\leftrightarrow Y)\bigg\}\, J_{L}^{a}(\mathbf{x})\, S_{A}^{ab}(\mathbf{z})\, J_{R}^{b}(\mathbf{y}).\nonumber\end{eqnarray}
Using (\ref{int.9}), (\ref{int.10}) and (\ref{int.11}) we can integrate over $\xi$:
 \begin{eqnarray}
&&\int_{\frac{\Lambda}{k^{+}}}^{1-\frac{\Lambda}{k^{+}}}d\xi\,\frac{N_{c}}{\xi(1-\xi)}\bigg(\left(\xi-2\right)\ln\left(\xi\right)+\left(\xi+1\right)\ln\left(1-\xi\right)\\
&&-\left(\xi-2\right)\left[-2\gamma-\ln\left(\frac{X^{2}\mu_{\overline{MS}}^{2}}{4}\right)\right]\bigg)\,=\,-3N_{c}\ln\,\left(\frac{\Lambda}{k^{+}}\right)\,\left[-2\gamma-\ln\left(\frac{X^{2}\mu_{\overline{MS}}^{2}}{4}\right)\right].\nonumber\end{eqnarray}\\
 and obtain:
   \begin{eqnarray}\label{sigjsj}
&&\Sigma_{JSJ}\,=\,-\,\frac{g^{4}}{32\pi^{5}}\,\int_{\mathbf{x},\,\mathbf{y},\,\mathbf{z}}\,\int_{\Lambda}^{e^{\delta\mathsf{Y}}\Lambda}dk^{+}\,\frac{X\cdot Y}{X^{2}Y^{2}k^{+}}\left(\left(b+2N_{c}\ln\,\left(\frac{k^{+}}{\Lambda}\right)-2N_{c}\delta\mathsf{Y}\right)\right.\nonumber\\
&&\times\left(-2\gamma-\frac{1}{2}\ln\left(\frac{X^{2}\mu_{\overline{MS}}^{2}}{4}\right)-\frac{1}{2}\ln\left(\frac{Y^{2}\mu_{\overline{MS}}^{2}}{4}\right)\right)-N_{c}\ln^{2}\left(\frac{\Lambda e^{\delta\mathsf{Y}}}{k^{+}}\right)+N_{c}\ln^{2}\left(\frac{\Lambda}{k^{+}}\right)\nonumber\\
&&\left.-\left(\frac{67}{9}-\frac{\pi^{2}}{3}\right)N_{c}+\frac{10}{9}N_{f}\right)\, J_{L}^{a}(\mathbf{x})\, S_{A}^{ab}(\mathbf{z})\, J_{R}^{b}(\mathbf{y}).\end{eqnarray}
Now,  after integration over $k^{+}$, $\Sigma_{JSJ}$ is split according to:
\begin{equation}
\Sigma_{JSJ}\,=\,\Sigma_{JSJ}^{NLO}\,+\,\Sigma_{JSJ}^{(\delta\mathsf{Y})^{2}},
 \end{equation}
 with
  \begin{equation}\begin{split}\label{jsjsig}
  \Sigma_{JSJ}^{NLO}\,=\,-\,\delta\mathsf{Y}\int_{\mathbf{x},\, \mathbf{y},\, \mathbf{z}}\, K_{JSJ}^{\prime}(\mathbf{x},\, \mathbf{y},\, \mathbf{z})\,\left[\,-\,2\, J_{L}^{a}(\mathbf{x})\, S_{A}^{ab}(\mathbf{z})\, J_{R}^{b}(\mathbf{y})\right],
  \end{split}\end{equation}
       where
   \begin{equation}\begin{split} \label{kprime}
K_{JSJ}^{\prime}(\mathbf{x},\,\mathbf{y},\,\mathbf{z})&\equiv\,\frac{\alpha_{s}^{2}X\cdot Y}{4\pi^{3}X^{2}Y^{2}}\,\left(b\,\left(\,\frac{1}{2}\ln\left(\frac{X^{2}\mu_{\overline{MS}}^{2}}{4}\right)\,+\,\frac{1}{2}\ln\left(\frac{Y^{2}\mu_{\overline{MS}}^{2}}{4}\right)\,+\,2\gamma\right)\right.\\
&\left.+\,\left(\frac{67}{9}\,-\,\frac{\pi^{2}}{3}\right)N_{c}\,-\,\frac{10}{9}N_{f}\right).
  \end{split}\end{equation}
 The result for $K_{JSJ}^{\prime}$ can be identified (up to $2\gamma-\ln\,4$) in equation $(87)$ of \cite{BC}. The relation between $K_{JSJ}^{\prime}(\mathbf{x},\, \mathbf{y},\, \mathbf{z})$ and $K_{JSJ}(\mathbf{x},\, \mathbf{y},\, \mathbf{z})$ is given below, in (\ref{jsjsub}). Also,

  \begin{equation}\begin{split}
\Sigma_{JSJ}^{(\delta\mathsf{Y})^{2}}&=\,-\frac{\alpha_{s}^{2}N_{c}}{4\pi^{3}}\,\left(\delta\mathsf{Y}\right)^{2}\int_{\mathbf{x},\,\mathbf{y},\,\mathbf{z}}\,\frac{X\cdot Y}{X^{2}Y^{2}}\left[\frac{4}{\epsilon}+4\gamma+\ln\left(\frac{X^{2}\mu_{\overline{MS}}^{2}}{4}\right)+\ln\left(\frac{Y^{2}\mu_{\overline{MS}}^{2}}{4}\right)\right]\\
&\times\, J_{L}^{a}(\mathbf{x})\, S_{A}^{ab}(\mathbf{z})\, J_{R}^{b}(\mathbf{y}),\\
\end{split}\end{equation}
 which can be equivalently written as:
  \begin{equation}\begin{split}\label{fekal}
\Sigma_{JSJ}^{(\delta\mathsf{Y})^{2}}&=\,-\frac{\alpha_{s}^{2}N_{c}}{4\pi^{4}}\,(\delta\mathsf{Y})^{2}\int_{\mathbf{x},\,\mathbf{y},\,\mathbf{z},\,\mathbf{z}^{\prime}}\,\left(\frac{2X\cdot Y}{Z^{2}X^{2}Y^{2}}-\frac{X\cdot Y\, X^{\prime}\cdot Z}{X^{2}Y^{2}(X^{\prime})^{2}Z^{2}}-\frac{X\cdot Y\, Y^{\prime}\cdot Z}{X^{2}Y^{2}(Y^{\prime})^{2}Z^{2}}\right)\,\\
&\times J_{L}^{a}(\mathbf{x})\, S_{A}^{ab}(\mathbf{z})\, J_{R}^{b}(\mathbf{y}).
\end{split}\end{equation}

 \subsection{Computation of $\Sigma_{JJSSJJ}$ and $\Sigma_{JJJSJ}$}
\begin{figure}[!h]
  \centering
  \includegraphics[scale=0.6]{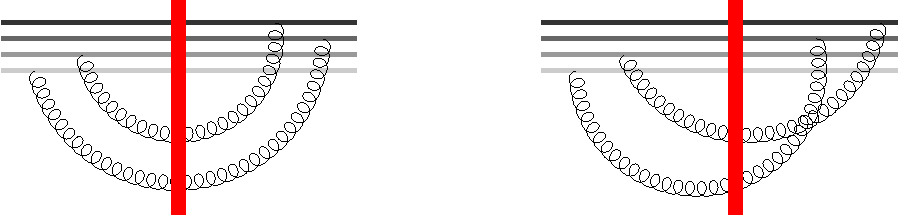}
  \caption{The diagrams for $\Sigma_{JJSSJJ}$. \label{figjsj}}
\end{figure}

\begin{figure}
  \centering
\includegraphics[scale=0.45]{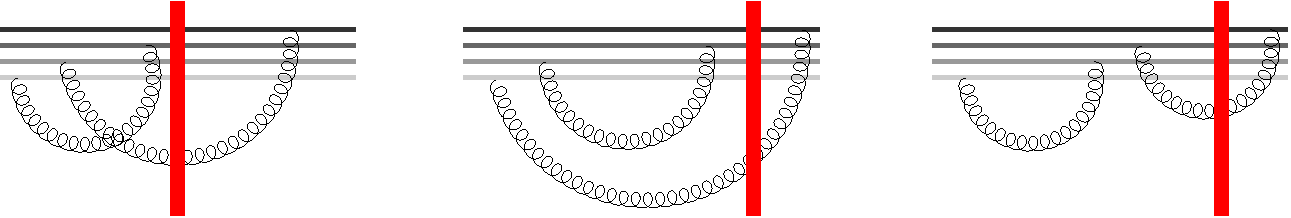}
  \caption{The diagrams for $\Sigma_{JJJSJ}$.\label{jjsjjdi}}
\end{figure}

$\Sigma_{JJSSJJ}$ is defined in (\ref{sigmaJJSSJJ}). This contribution is driven by $\left|\psi_{gg\:\rho\rho}\right\rangle$, (\ref{gg_rhorho}), interfering with itself before and after passing though the shockwave, as shown in Fig.\ref{figjsj}.
\begin{eqnarray}\label{starjjssjj}
\Sigma_{JJSSJJ}&=&\int_{\mathbf{w},\,\mathbf{v},\,\mathbf{x},\,\mathbf{y},\,\mathbf{z},\,\mathbf{z}^{\prime},\,\overline{\mathbf{z}},\,\overline{\mathbf{z}}^{\prime}}\int_{\Lambda}^{e^{\delta\mathsf{Y}}\Lambda}dk^{+}\, d\bar{k}^{+}\,\int_{\frac{\Lambda}{k^{+}}}^{1-\frac{\Lambda}{k^{+}}}d\xi\, d\vartheta\,\nonumber\\
&\times&\frac{g^{4}\, X^{i}\,(W^{\prime})^{j}\,\bar{Y}^{k}\cdot(\bar{V}^{\prime})^{l}}{256\pi^{6}\sqrt{\xi(1-\xi)\vartheta(1-\vartheta)}X^{2}\bar{Y}^{2}(W^{\prime})^{2}(\bar{V}^{\prime})^{2}}\left\langle g_{l}^{c}((1-\vartheta)\bar{k}^{+},\,\overline{\mathbf{z}}^{\prime})\, g_{k}^{d}(\vartheta\bar{k}^{+},\,\overline{\mathbf{z}})\right|\nonumber \\
&\times&\left\{ \rho^{c}(\mathbf{v}),\,\rho^{d}(\mathbf{y})\right\} \,\hat{S}\,\left\{ \rho^{b}(\mathbf{w}),\,\rho^{a}(\mathbf{x})\right\} \left|g_{j}^{b}((1-\xi)k^{+},\,\mathbf{z}^{\prime})\, g_{i}^{a}(\xi k^{+},\,\mathbf{z})\right\rangle ,
 \end{eqnarray}
 where  $\bar{Y}=\mathbf{y}-\mathbf{\bar{z}}$ and $\bar{V}^{\prime}=\mathbf{v}-\mathbf{\bar{z}^{\prime}}$. The rest of this calculation is available in Appendix \ref{sijjjj}. The result after integration over $k^{+}$ is:
 
 \begin{eqnarray}\label{losjjssjj}
\Sigma_{JJSSJJ}^{(\delta\mathsf{Y})^{2}}
&=&\frac{\alpha_{s}^{2}}{2\pi^{4}}\,(\delta\mathsf{Y})^{2}\int_{\mathbf{w},\, \mathbf{v},\, \mathbf{x},\, \mathbf{y},\, \mathbf{z},\, \mathbf{z}^{\prime}}\frac{X\cdot Y\, W^{\prime}\cdot V^{\prime}}{X^{2}Y^{2}(W^{\prime})^{2}(V^{\prime})^{2}}J_{L}^{a}(\mathbf{x})J_{L}^{b}(\mathbf{w})
S_{A}^{ad}(\mathbf{z})S_{A}^{bc}(\mathbf{z}^{\prime})J_{R}^{d}(\mathbf{y})J_{R}^{c}(\mathbf{v})\nonumber\\
&-&\frac{\alpha_{s}^{2}}{8\pi^{4}}\,\left(\delta\mathsf{Y}\right)^{2}\,\int_{\mathbf{x},\, \mathbf{y},\, \mathbf{z},\, \mathbf{z}^{\prime}}\frac{X\cdot Y\, X^{\prime}\cdot Y^{\prime}}{X^{2}Y^{2}(X^{\prime})^{2}(Y^{\prime})^{2}} f^{abc} f^{def} S_{A}^{be}(\mathbf{z}) S_{A}^{cf}(\mathbf{z}^{\prime})
J_{L}^{a}(\mathbf{x}) J_{R}^{d}(\mathbf{y}).\end{eqnarray}
    Now let us compute $\Sigma_{JJJSJ}$ as defined in (\ref{sigmaJJSJJ}). The contribution  comes from the $S$-matrix element 
    between the state $\left|\psi_{g\:\rho\rho\rho}\right\rangle $ (\ref{g_rhorhorho})  and the state $\left|\psi_{g\:\rho}^{LO}\right\rangle $
     (\ref{lotra}), as shown in Fig.\ref{jjsjjdi}.
  \begin{equation}\begin{split}
  \Sigma_{JJJSJ}&=\,-\frac{\alpha_{s}^{2}}{2\pi^{4}}\left(\delta\mathsf{Y}\right)^{2}\,\int_{\mathbf{w},\, \mathbf{v},\, \mathbf{x},\, \mathbf{y},\, \mathbf{z},\, \mathbf{z}^{\prime}}\,\frac{X\cdot Y\, W^{\prime}\cdot V^{\prime}}{X^{2}Y^{2}(W^{\prime})^{2}(V^{\prime})^{2}}\\
&\times\left[S_{A}^{cd}(\mathbf{z}^{\prime})\, J_{L}^{a}(\mathbf{x})\, J_{L}^{a}(\mathbf{y})\, J_{L}^{c}(\mathbf{w})\, J_{R}^{d}(\mathbf{v})+S_{A}^{cd}(\mathbf{z}^{\prime})\, J_{L}^{c}(\mathbf{v})\, J_{R}^{a}(\mathbf{x})\, J_{R}^{a}(\mathbf{y})\, J_{R}^{d}(\mathbf{w})\right].\\
\end{split}\end{equation}
Which we write as:
  \begin{equation}\begin{split}\label{matjjsjj}
&\Sigma_{JJJSJ}=\,-\frac{\alpha_{s}^{2}}{2\pi^{4}}\left(\delta\mathsf{Y}\right)^{2}\,\int_{\mathbf{w},\,\mathbf{v},\,\mathbf{x},\,\mathbf{y},\,\mathbf{z},\,\mathbf{z}^{\prime}}\,\frac{X\cdot Y\,W^{\prime}\cdot V^{\prime}}{X^{2}Y^{2}(W^{\prime})^{2}(V^{\prime})^{2}}\\
&\times\left[S_{A}^{cd}(\mathbf{z}^{\prime})\,J_{L}^{c}(\mathbf{w})\,J_{L}^{a}(\mathbf{x})\,J_{L}^{a}(\mathbf{y})\,J_{R}^{d}(\mathbf{v})+S_{A}^{cd}(\mathbf{z}^{\prime})\,J_{L}^{c}(\mathbf{v})\,J_{R}^{d}(\mathbf{w})\,J_{R}^{a}(\mathbf{x})\,J_{R}^{a}(\mathbf{y})\right]\\
&-\frac{if^{abc}\alpha_{s}^{2}}{4\pi^{4}}\left(\delta\mathsf{Y}\right)^{2}\,\int_{\mathbf{w},\,\mathbf{x},\,\mathbf{y},\,\mathbf{z},\,\mathbf{z}^{\prime}}\left(\frac{X\cdot Y\,X^{\prime}\cdot W^{\prime}}{X^{2}Y^{2}(X^{\prime})^{2}(W^{\prime})^{2}}-\frac{X\cdot Y\,Y^{\prime}\cdot W^{\prime}}{X^{2}Y^{2}(Y^{\prime})^{2}(W^{\prime})^{2}}\right)\\
&\times\left[S_{A}^{cd}(\mathbf{z}^{\prime})\,J_{L}^{b}(\mathbf{x})\,J_{L}^{a}(\mathbf{y})\,J_{R}^{d}(\mathbf{w})-S_{A}^{dc}(\mathbf{z}^{\prime})\,J_{L}^{d}(\mathbf{w})J_{R}^{a}(\mathbf{x})\,J_{R}^{b}(\mathbf{y})\right]\\
\end{split}\end{equation}

 \subsection{Virtual Contributions}\label{normcon}
 
 $\Sigma_{virtual}$ is defined in (\ref{signor}). This contribution comes from the scattering of the vacuum component $\mathcal{N}^{NLO}\left|0\right\rangle $ of the LCWF.  To save space, we will not illustrate this contribution, but, quite obviously, these are all the graphs with no soft partons 
crossing  the shockwave, similarly to Fig. (\ref{lorep}a,\ref{lorep}b).  
With (\ref{nphase}) and (\ref{nlonor}) in mind, and counting powers of $g$,  
 \begin{eqnarray}\label{signor2}
\Sigma_{virtual}&=&\left\langle 0\right|\left\Vert \mathcal{N}^{NLO}\right\Vert \,\hat{S}\,\left|0\right\rangle +\left\langle 0\right|\,\hat{S}\,\left\Vert \mathcal{N}^{NLO}\right\Vert \left|0\right\rangle +\frac{1}{4}\left\langle 0\right|\,\left\langle \psi_{g\:\rho}^{LO}\left|\psi_{g\:\rho}^{LO}\right.\right\rangle \,\hat{S}\,\left\langle \psi_{g\:\rho}^{LO}\left|\psi_{g\:\rho}^{LO}\right.\right\rangle \left|0\right\rangle \nonumber \\
&+&\langle 0|\,\hat S\, i\phi^{NLO}\,|0\rangle \,-\, \langle 0|\,i\phi^{NLO\,\dagger}\, \hat S\,|0\rangle \,-\,2.
\end{eqnarray}
To organise the computation, we introduce the following decomposition for  
$\Sigma_{virtual}$,
 \begin{equation}\begin{split}\label{sigivi}
\Sigma_{virtual}\,=\,\Sigma_{v}^{LO}\,+\,\Sigma_{v\, JJ}\,+\,\Sigma_{v\, JJJ}\,+\,\Sigma_{v\, JJJJ}\,+\,\Sigma_{\phi}.
 \end{split}\end{equation}
 where,  after inserting $||\mathcal{N}^{NLO}||$, as defined in equation (\ref{nlonor}), 
  into (\ref{signor2}), 
   \begin{equation}\begin{split}
&\Sigma_{v}^{LO}\,\equiv\,-\,\frac{1}{2}\,\left\langle 0\right|\left\langle \psi_{g\:\rho}^{LO}\left|\psi_{g\:\rho}^{LO}\right.\right\rangle \,\hat{S}\,\left|0\right\rangle \,-\,\frac{1}{2}\left\langle 0\right|\,\hat{S}\,\left\langle \psi_{g\:\rho}^{LO}\left|\psi_{g\:\rho}^{LO}\right.\right\rangle \left|0\right\rangle. \\
 \end{split}\end{equation}
 The contribution with two $J$ operators
        \begin{equation}
\hspace{-8cm}       \Sigma_{v\, JJ}\,\equiv\,\Sigma_{v\, q\bar{q}}\,+\,\Sigma_{v\, J\cancel{S}J}\,+\,\Sigma_{v\, J\cancel{S}\cancel{S}J},
 \end{equation}
     \begin{equation}\begin{split}\label{qqnorm}
&\Sigma_{v\, q\bar{q}}\,\equiv\,-\,\frac{1}{2}\,\left\langle 0\right|\left\langle \psi_{q\bar{q}\:\rho}\left|\psi_{q\bar{q}\:\rho}\right.\right\rangle \,\hat{S}\,\left|0\right\rangle \,-\,\frac{1}{2}\,\left\langle 0\right|\,\hat{S}\,\left\langle \psi_{q\bar{q}\:\rho}\left|\psi_{q\bar{q}\:\rho}\right.\right\rangle \left|0\right\rangle.\\
 \end{split}\end{equation}
            \begin{equation}\begin{split}\label{jsjno}
  \hspace{-2cm}      \Sigma_{v\, J\cancel{S}J}\equiv&-\,\frac{1}{2}\left\langle 0\right|\left(\left\langle \psi_{g\:\rho}^{LO}\left|\psi_{g\:\rho}\right.\right\rangle +\left\langle \psi_{g\:\rho}\left|\psi_{g\:\rho}^{LO}\right.\right\rangle \right)\,\hat{S}\,\left|0\right\rangle \,\\
 &-\,\frac{1}{2}\left\langle 0\right|\,\hat{S}\,\left(\left\langle \psi_{g\:\rho}^{LO}\left|\psi_{g\:\rho}\right.\right\rangle +\left\langle \psi_{g\:\rho}\left|\psi_{g\:\rho}^{LO}\right.\right\rangle \right)\left|0\right\rangle ,\\
 \end{split}\end{equation}
 \begin{equation}\begin{split}\label{jssjnorm}
   &\Sigma_{v\, J\cancel{S}\cancel{S}J}\,\equiv\,-\,\frac{1}{2}\left\langle 0\right|\left\langle \psi_{gg\:\rho}\left|\psi_{gg\:\rho}\right.\right\rangle \,\hat{S}\,\left|0\right\rangle \,-\,\frac{1}{2}\left\langle 0\right|\,\hat{S}\,\left\langle \psi_{gg\:\rho}\left|\psi_{gg\:\rho}\right.\right\rangle \left|0\right\rangle.\\
 \end{split}\end{equation}
 The contribution with three $J$ operators
 \begin{equation}
\hspace{-8cm}\Sigma_{v\, JJJ}\,\equiv\,\Sigma_{v\, JJ\cancel{S}J}\,+\,\Sigma_{v\, JJ\cancel{S}\cancel{S}J}.
 \end{equation}       
 \begin{equation}\begin{split}\label{norm.JJSSJ}
\Sigma_{v\, JJ\cancel{S}\cancel{S}J}\,\equiv&-\,\frac{1}{2}\left\langle 0\right|\left(\left\langle \psi_{gg\:\rho\rho}\left|\psi_{gg\:\rho}\right.\right\rangle +\left\langle \psi_{gg\:\rho}\left|\psi_{gg\:\rho\rho}\right.\right\rangle \right)\,\hat{S}\,\left|0\right\rangle \,\\
&-\,\frac{1}{2}\left\langle 0\right|\,\hat{S}\,\left(\left\langle \psi_{gg\:\rho\rho}\left|\psi_{gg\:\rho}\right.\right\rangle +\left\langle \psi_{gg\:\rho}\left|\psi_{gg\:\rho\rho}\right.\right\rangle \right)\left|0\right\rangle ,\\
 \end{split}\end{equation}
   \begin{equation}\begin{split}\label{norm.JJSJ}
\Sigma_{v\, JJ\cancel{S}J}\,\equiv&-\,\frac{1}{2}\left\langle 0\right|\left(\left\langle \psi_{g\:\rho}^{LO}\left|\psi_{g\:\rho\rho}\right.\right\rangle +\left\langle \psi_{g\:\rho\rho}\left|\psi_{g\:\rho}^{LO}\right.\right\rangle \right)\,\hat{S}\,\left|0\right\rangle \,\\
&-\,\frac{1}{2}\left\langle 0\right|\,\hat{S}\,\left(\left\langle \psi_{g\:\rho}^{LO}\left|\psi_{g\:\rho\rho}\right.\right\rangle +\left\langle \psi_{g\:\rho\rho}\left|\psi_{g\:\rho}^{LO}\right.\right\rangle \right)\left|0\right\rangle .
 \end{split}\end{equation}
 The contribution with four $J$ operators
   \begin{eqnarray}\label{norJJJJ}
\Sigma_{v\, JJJJ}&&\equiv\,\left\langle 0\right|\left(-\frac{1}{2}\left\langle \psi_{g\:\rho}^{LO}\left|\psi_{g\:\rho\rho\rho}\right.\right\rangle -\frac{1}{2}\left\langle \psi_{g\:\rho\rho\rho}\left|\psi_{g\:\rho}^{LO}\right.\right\rangle -\frac{1}{2}\left\langle \psi_{gg\:\rho\rho}\left|\psi_{gg\:\rho\rho}\right.\right\rangle \right.\nonumber\\
&&\left.-\frac{1}{8}\left\langle \psi_{g\:\rho}^{LO}\left|\psi_{g\:\rho}^{LO}\right.\right\rangle \left\langle \psi_{g\:\rho}^{LO}\left|\psi_{g\:\rho}^{LO}\right.\right\rangle \right)\,\hat{S}\,\left|0\right\rangle +\,\left\langle 0\right|\,\hat{S}\,\left(-\frac{1}{2}\left\langle \psi_{g\:\rho}^{LO}\left|\psi_{g\:\rho\rho\rho}\right.\right\rangle \right.\nonumber\\
&&\left.-\frac{1}{2}\left\langle \psi_{g\:\rho\rho\rho}\left|\psi_{g\:\rho}^{LO}\right.\right\rangle -\frac{1}{2}\left\langle \psi_{gg\:\rho\rho}\left|\psi_{gg\:\rho\rho}\right.\right\rangle \,-\frac{1}{8}\left\langle \psi_{g\:\rho}^{LO}\left|\psi_{g\:\rho}^{LO}\right.\right\rangle \left\langle \psi_{g\:\rho}^{LO}\left|\psi_{g\:\rho}^{LO}\right.\right\rangle \right)\,\left|0\right\rangle \nonumber\\
&&\,+\,\frac{1}{4}\,\left\langle 0\right|\left\langle \psi_{g\:\rho}^{LO}\left|\psi_{g\:\rho}^{LO}\right.\right\rangle \,\hat{S}\,\left\langle \psi_{g\:\rho}^{LO}\left|\psi_{g\:\rho}^{LO}\right.\right\rangle \left|0\right\rangle .\end{eqnarray}\\\\
Finally, 
 \begin{equation}\label{sigmaphi}
\Sigma_{\phi}\,\equiv\,\langle 0|\,\hat S\, i\phi^{NLO}\,|0\rangle \,-\, \langle 0|\,i\phi^{NLO\,\dagger}\, \hat S\,|0\rangle .
 \end{equation}

In the rest of this section we present the results for all the virtual terms  defined in (\ref{qqnorm})$-$(\ref{sigmaphi}).
In fact, it is  straightforward  to deduce most of them by recycling our previous computations. We notice that
most of the overlaps that appear in (\ref{qqnorm})$-$(\ref{norJJJJ}) could be read off from the corresponding real terms, 
when  the limit $S\longrightarrow1$  is taken. In addition, this limit has to be taken with understanding that all the factors $\rho$ in
(\ref{qqnorm})$-$(\ref{norJJJJ}) (apart of the last line in (\ref{norJJJJ})) are placed either on the left or  right of $\hat S$. This basically implies 
 a replacement in the real terms  $J_L\longrightarrow J_R$  ($J_R\longrightarrow J_L$) 
 in the case  $\hat{S}$ is placed to the right (left) of the matrix elements. One should, however, be careful with the ordering of the operators. 
 The correct procedure involves first writing all the $J_R$ operators to the left of $J_L$ operators. Second is to convert the $J$ operators back into
 $\rho$ in accord to (\ref{relrr}). Finally, the string of $\rho$ gets converted again to a string of $J$ operators. 
\\ \\
$\bullet$ \textit{\textbf{Computation of $\Sigma_{v\, JJ}$}}\\
 The contribution (\ref{qqnorm}) can be deduced from (\ref{sigqq}); (\ref{jsjno}) from (\ref{kprime}) and (\ref{fekal});
(\ref{jssjnorm}) from (\ref{ajssjsig}) and (\ref{losq_JSSJ}).
 (\ref{qqnorm}) can be split similarly to how it was done for the real terms:
\begin{equation}
\Sigma_{v\, JJ}\,=\,\Sigma_{v\, JJ}^{NLO}\,+\,\Sigma_{v\, JJ}^{(\delta\mathsf{Y})^{2}},
 \end{equation}
 with
 \begin{equation}\label{dkladf}
\Sigma_{v\, JJ}^{(\delta\mathsf{Y})^{2}}\,=\,-\,\frac{\alpha_{s}^{2}N_{c}}{16\pi^{4}}\,\left(\delta\mathsf{Y}\right)^{2}\,\int_{\mathbf{x},\, \mathbf{y},\, \mathbf{z},\, \mathbf{z}^{\prime}}\frac{X\cdot Y\, X^{\prime}\cdot Y^{\prime}}{X^{2}Y^{2}(X^{\prime})^{2}(Y^{\prime})^{2}}\,\left[J_{L}^{a}(\mathbf{x})\, J_{L}^{a}(\mathbf{y})\,+\, J_{R}^{a}(\mathbf{x})\, J_{R}^{a}(\mathbf{y})\right],
 \end{equation}
 and
   \begin{equation}\begin{split}\label{jjcon}
\Sigma_{v\, JJ}^{NLO}\,=\,-\,\delta\mathsf{Y}\,\int_{\mathbf{x},\, \mathbf{y},\, \mathbf{z}}\, K_{JSJ}(\mathbf{x},\, \mathbf{y},\, \mathbf{z})\,\left[J_{L}^{a}(\mathbf{x})\, J_{L}^{a}(\mathbf{y})\,+\, J_{R}^{a}(\mathbf{x})\, J_{R}^{a}(\mathbf{y})\right],
 \end{split}\end{equation}
 where we have introduced the following kernel:
 \begin{equation}\label{jsjsub}
K_{JSJ}(\mathbf{x},\, \mathbf{y},\, \mathbf{z})\,\equiv\, K_{JSJ}^{\prime}(\mathbf{x},\, \mathbf{y},\, \mathbf{z})\,-\,\frac{1}{2}\,\int_{\mathbf{z}^{\prime}}K_{q\bar{q}}(\mathbf{x},\, \mathbf{y},\, \mathbf{z},\, \mathbf{z}^{\prime})-\frac{N_{c}}{2}\,\int_{\mathbf{z}^{\prime}}\, K_{JSSJ}(\mathbf{x},\, \mathbf{y},\, \mathbf{z},\, \mathbf{z}^{\prime}).
  \end{equation}
The $z^\prime$ integrals are divergent and have to be properly regularised.  The results are quoted in (\ref{to1}) and (\ref{to2}).
The kernel $K_{JSJ}$ in (\ref{jsjsub}) is the one defined in (\ref{JSJ}).
 \\  \\
        $\bullet$ \textit{\textbf{Computation of $\Sigma_{v\, JJJ}$}}\\
Taking the $S\rightarrow 1$ limit in  (\ref{jjssjsig}), (\ref{losq_JJSSJ}), (\ref{jjsjlab}), (\ref{losqjjsjne}),  and splitting (\ref{norm.JJSSJ}) as usual,
 \begin{equation}
 \Sigma_{v\, JJJ}\,=\,\Sigma_{v\, JJJ}^{NLO}\,+\,\Sigma_{v\, JJJ}^{(\delta\mathsf{Y})^{2}},
 \end{equation}
we obtain:
 \begin{eqnarray}\label{sigvirjjssj}
&&\hspace{-0.4cm}\Sigma_{v\, JJJ}^{NLO}\,=\,-\,\frac{1}{2}\,\delta\mathsf{Y}\,\int_{\mathbf{w}\,,\mathbf{x},\,\mathbf{y},\,\mathbf{z},\,\mathbf{z}^{\prime}}\left(K_{JJSSJ}(\mathbf{w}\,,\mathbf{x},\,\mathbf{y},\,\mathbf{z},\,\mathbf{z}^{\prime})\,+\, K_{JJSJ}(\mathbf{w}\,,\mathbf{x},\,\mathbf{y},\,\mathbf{z})\right)\, f^{acb}\\
&&\times\left[J_{L}^{c}(\mathbf{x}) J_{L}^{b}(\mathbf{y})\, J_{L}^{a}(\mathbf{w})- J_{L}^{a}(\mathbf{w}) J_{L}^{c}(\mathbf{x})J_{L}^{b}(\mathbf{y})+ J_{R}^{c}(\mathbf{x}) J_{R}^{b}(\mathbf{y}) J_{R}^{a}(\mathbf{w})\right.
\left.-J_{R}^{a}(\mathbf{w}) J_{R}^{c}(\mathbf{x})J_{R}^{b}(\mathbf{y})\right].\nonumber
\end{eqnarray}
By using the identity (\ref{jjsjiden}) along with
\begin{eqnarray}\label{identt}
&&f^{acb}\left[J_{L}^{c}(\mathbf{x})\, J_{L}^{b}(\mathbf{y})\, J_{L}^{a}(\mathbf{w})\,-\, J_{L}^{a}(\mathbf{w})\, J_{L}^{c}(\mathbf{x})\, J_{L}^{b}(\mathbf{y})\,+\, J_{R}^{c}(\mathbf{x})\, J_{R}^{b}(\mathbf{y})\, J_{R}^{a}(\mathbf{w})\,\right.\\
&&-\left.\, J_{R}^{a}(\mathbf{w})\, J_{R}^{c}(\mathbf{x})\, J_{R}^{b}(\mathbf{y})\right]\,=\, iN_{c}\left[J_{L}^{a}(\mathbf{x})\, J_{L}^{a}(\mathbf{y})\,-\, J_{R}^{a}(\mathbf{x})\, J_{R}^{a}(\mathbf{y})\right]\left[\delta(\mathbf{x}-\mathbf{w})\,-\,\delta(\mathbf{y}-\mathbf{w})\right],\nonumber
\end{eqnarray}
It is possible to show that:
\begin{equation}
\Sigma_{v\,JJJ}^{NLO}\,=\,0,
\end{equation}
and
\begin{equation}\begin{split}\label{sdaf}
&\Sigma_{v\:JJJ}^{(\delta\mathsf{Y})^{2}}=-\frac{N_{c}\alpha_{s}^{2}}{2\pi^{4}}(\delta\mathsf{Y})^{2}\int_{\mathbf{x},\,\mathbf{y},\,\mathbf{z},\,\mathbf{z}^{\prime}}\frac{(X^{\prime}-Y^{\prime})^{2}X\cdot Y}{2X^{2}Y^{2}(Y^{\prime})^{2}(X^{\prime})^{2}}\left[J_{L}^{a}(\mathbf{x})\,J_{L}^{a}(\mathbf{y})\,-\,J_{R}^{a}(\mathbf{x})\,J_{R}^{a}(\mathbf{y})\right].\\
 \end{split}\end{equation}
\\ 
  $\bullet$ \textit{\textbf{Computation of $\Sigma_{JJJJ}$}}\\
Taking the $S\rightarrow 1$ limit in (\ref{losjjssjj}) and (\ref{matjjsjj}), 
we arrive at the following expression for (\ref{norJJJJ}):
       \begin{eqnarray}\label{jjjjvirlosq}
&&\Sigma_{v\,JJJJ}^{(\delta\mathsf{Y})^{2}}\,=\,\frac{\alpha_{s}^{2}}{8\pi^{4}}\left(\delta\mathsf{Y}\right)^{2}\int_{\mathbf{w},\,\mathbf{v},\,\mathbf{x},\,\mathbf{y},\,\mathbf{z},\,\mathbf{z}^{\prime}}\frac{X\cdot Y\,W^{\prime}\cdot V^{\prime}}{X^{2}Y^{2}(W^{\prime})^{2}(V^{\prime})^{2}}\nonumber\\
&&\times\left[J_{R}^{c}(\mathbf{w})\,J_{R}^{c}(\mathbf{v})J_{R}^{a}(\mathbf{x})\,J_{R}^{a}(\mathbf{y})+J_{L}^{c}(\mathbf{w})\,J_{L}^{c}(\mathbf{v})J_{L}^{a}(\mathbf{x})\,J_{L}^{a}(\mathbf{y})+2J_{L}^{a}(\mathbf{x})\,J_{L}^{a}(\mathbf{y})J_{R}^{c}(\mathbf{w})\,J_{R}^{c}(\mathbf{v})\right]\nonumber\\
&&+\,\frac{\alpha_{s}^{2}N_{c}}{16\pi^{4}}\,\left(\delta\mathsf{Y}\right)^{2}\,\int_{\mathbf{x},\,\mathbf{y},\,\mathbf{z},\,\mathbf{z}^{\prime}}\frac{X\cdot Y\,X^{\prime}\cdot Y^{\prime}}{X^{2}Y^{2}(X^{\prime})^{2}(Y^{\prime})^{2}}\,\left[J_{R}^{a}(\mathbf{x})\,J_{R}^{a}(\mathbf{y})\,+\,J_{L}^{a}(\mathbf{x})\,J_{L}^{a}(\mathbf{y})\right]\nonumber\\
&&+\frac{N_{c}\alpha_{s}^{2}}{2\pi^{4}}(\delta\mathsf{Y})^{2}\int_{\mathbf{x},\,\mathbf{y},\,\mathbf{z},\,\mathbf{z}^{\prime}}\frac{(X^{\prime}-Y^{\prime})^{2}X\cdot Y}{2X^{2}Y^{2}(Y^{\prime})^{2}(X^{\prime})^{2}}\left[J_{L}^{a}(\mathbf{x})\,J_{L}^{a}(\mathbf{y})\,-\,J_{R}^{a}(\mathbf{x})\,J_{R}^{a}(\mathbf{y})\right].
\end{eqnarray}
\\ 
       $\bullet$ \textit{\textbf{Computation of $\Sigma_{\phi}$}}\\
$\Sigma_{\phi}$ is defined in (\ref{sigmaphi}) and computed  in Appendix \ref{wfphase} from the condition (\ref{condphase}), 
    \begin{equation}\begin{split}
\Sigma_{\phi}^{NLO}&=\,-\delta\mathsf{Y}\,\int_{\mathbf{w},\, \mathbf{x},\, \mathbf{y},\, \mathbf{z},\, \mathbf{z}^{\prime}}\, K_{JJSSJ}(\mathbf{w},\, \mathbf{x},\, \mathbf{y},\, \mathbf{z},\, \mathbf{z}^{\prime})\\
&\times f^{acb}\left[J_{L}^{a}(\mathbf{x})\, J_{L}^{b}(\mathbf{y})\, J_{L}^{c}(\mathbf{w})\,-\, J_{R}^{a}(\mathbf{x})\, J_{R}^{b}(\mathbf{y})\, J_{R}^{c}(\mathbf{w})\right].\\
 \end{split}\end{equation}
 With the help of identities (\ref{jjsjiden}) and (\ref{dien1}) it is possible to cast $\Sigma_{\phi}$ into the form that is straightforwardly comparable with
 (\ref{ham}),
     \begin{equation}\begin{split}\label{phasres}
\Sigma_{\phi}^{NLO}&=\,-\,\frac{1}{3}\,\delta\mathsf{Y}\,\int_{\mathbf{w}\,,\mathbf{x},\,\mathbf{y},\,\mathbf{z},\,\mathbf{z}^{\prime}}\left(K_{JJSSJ}(\mathbf{w}\,,\mathbf{x},\,\mathbf{y},\,\mathbf{z},\,\mathbf{z}^{\prime})+K_{JJSJ}(\mathbf{w}\,,\mathbf{x},\,\mathbf{y},\,\mathbf{z})\right)\,f^{acb}\\
&\times\,\left[J_{L}^{c}(\mathbf{x})\,J_{L}^{b}(\mathbf{y})\,J_{L}^{a}(\mathbf{w})\,-\,J_{R}^{c}(\mathbf{x})\,J_{R}^{b}(\mathbf{y})\,J_{R}^{a}(\mathbf{w})\right]\\
  \end{split}\end{equation} \\ 
      $\bullet$ \textit{\textbf{Assembling the virtual  contributions}}\\
$ \Sigma_{virtual}$  has contributions  proportional  to $\delta\mathsf{Y}$ and $(\delta\mathsf{Y})^{2}$ and we split them via our standard procedure:
      \begin{equation}
     \Sigma_{virtual}\,=\,\Sigma_{virtual}^{NLO}\,+\,\Sigma_{virtual}^{(\delta\mathsf{Y})^{2}}.
 \end{equation}
We define $\Sigma_{virtual}^{NLO}$ by adding together contributions (\ref{dkladf}), (\ref{jjjjvirlosq}) and (\ref{phasres}):
\begin{eqnarray}\label{sivirtu}
&&\Sigma_{virtual}^{NLO}\,\equiv\,\Sigma_{v\,JJJ}^{NLO}\,+\,\Sigma_{v\,JJ}^{NLO}\,+\,\Sigma_{\phi}^{NLO}=\,-\,\delta\mathsf{Y}\,\int_{\mathbf{x},\,\mathbf{y},\,\mathbf{z}}\,K_{JSJ}(\mathbf{x},\,\mathbf{y},\,\mathbf{z})\,\left[J_{L}^{a}(\mathbf{x})\,J_{L}^{a}(\mathbf{y})\right.\nonumber \\
&&\left.+\,J_{R}^{a}(\mathbf{x})\,J_{R}^{a}(\mathbf{y})\right]-\,\frac{1}{3}\,\delta\mathsf{Y}\,\int_{\mathbf{w}\,,\mathbf{x},\,\mathbf{y},\,\mathbf{z},\,\mathbf{z}^{\prime}}\left(K_{JJSSJ}(\mathbf{w}\,,\mathbf{x},\,\mathbf{y},\,\mathbf{z},\,\mathbf{z}^{\prime})+K_{JJSJ}(\mathbf{w}\,,\mathbf{x},\,\mathbf{y},\,\mathbf{z})\right)\,f^{acb}\nonumber \\
&&\times\,\left[J_{L}^{c}(\mathbf{x})\,J_{L}^{b}(\mathbf{y})\,J_{L}^{a}(\mathbf{w})\,-\,J_{R}^{c}(\mathbf{x})\,J_{R}^{b}(\mathbf{y})\,J_{R}^{a}(\mathbf{w})\right]
 \end{eqnarray}
We define $\Sigma_{virtual}^{(\delta\mathsf{Y})^{2}}$ by adding together the contributions (\ref{dkladf}), (\ref{sdaf}), and (\ref{jjjjvirlosq}):
 \begin{eqnarray}\label{jjjjlosq}
 \Sigma_{virtual}^{(\delta\mathsf{Y})^{2}}&&\equiv\Sigma_{v\,JJJJ}^{(\delta\mathsf{Y})^{2}}+\Sigma_{v\,JJJ}^{(\delta\mathsf{Y})^{2}}+\Sigma_{v\,JJ}^{(\delta\mathsf{Y})^{2}}
 =\frac{\alpha_{s}^{2}}{8\pi^{4}}\left(\delta\mathsf{Y}\right)^{2}\int_{\mathbf{w},\,\mathbf{v},\,\mathbf{x},\,\mathbf{y},\,\mathbf{z},\,\mathbf{z}^{\prime}}\frac{X\cdot Y\,W^{\prime}\cdot V^{\prime}}{X^{2}Y^{2}(W^{\prime})^{2}(V^{\prime})^{2}}\nonumber\\
&&\times\left[J_{L}^{c}(\mathbf{w})\, J_{L}^{c}(\mathbf{v})+J_{R}^{c}(\mathbf{w})\, J_{R}^{c}(\mathbf{v})\right]\left[J_{L}^{a}(\mathbf{x})\, J_{L}^{a}(\mathbf{y})+J_{R}^{a}(\mathbf{x})\, J_{R}^{a}(\mathbf{y})\right].
 \end{eqnarray}

 \subsection{The NLO JIMWLK Hamiltonian Assembled}\label{subterms}
 We define $ \Sigma^{NLO}$ as a sum of all the contributions of order $\delta\mathsf{Y}$:
   \begin{equation}\begin{split}\label{fullnlo}
\Sigma^{NLO}\,\equiv\,\Sigma_{JSJ}^{NLO}\,+\,\Sigma_{q\overline{q}}^{NLO}\,+\,\Sigma_{JSSJ}^{NLO}\,+\,\Sigma_{JJSJ}^{NLO}\,+\,\Sigma_{JJSSJ}^{NLO}\,+\,\Sigma_{virtual}^{NLO},
   \end{split}\end{equation}
with the terms computed in  (\ref{sigqq}), (\ref{jjssjsig}), (\ref{ajssjsig}), (\ref{jjsjlab}), (\ref{jsjsig}), (\ref{sivirtu}).
The integral over $z$ in $\Sigma_{JSJ}^{NLO}$ has UV divergence. Similar divergence is found in $\Sigma^{NLO}_{JSSJ}$  (\ref{JSSJ}) and $\Sigma^{NLO}_{q\bar{q}}$ (\ref{kqq}) when $\mathbf{z}\,\rightarrow\, \mathbf{z}^{\prime}$. 
 Following \cite{BC}, we subtract (and later add)  terms which correspond to the case in which both gluons or the quark and anti-quark pair
pass through the  shockwave at the same transverse point. This corresponds to single gluon scattering, thanks to the identities
$S_{A}^{ab}(\mathbf{z})=2tr\,\left[S^{\dagger}(\mathbf{z})t^{a}S(\mathbf{z})t^{b}\right]$ and $f^{abc}\, f^{def}\, S_{A}^{be}(\mathbf{z})\, S_{A}^{cf}(\mathbf{z})\,=\, N_{c}\, S_{A}^{ad}(\mathbf{z})$.

Accordingly, we introduce the subtraction terms
   \begin{equation}\begin{split}
\Sigma_{q\bar{q}}^{sub}\,\equiv\,-\,\delta\mathsf{Y}\,\int_{\mathbf{x},\, \mathbf{y},\, \mathbf{z},\, \mathbf{z}^{\prime}}K_{q\bar{q}}(\mathbf{x},\, \mathbf{y},\, \mathbf{z},\, \mathbf{z}^{\prime})\, J_{L}^{a}(\mathbf{x})\, S_{A}^{ab}(\mathbf{z})\, J_{R}^{b}(\mathbf{y}).
  \end{split}\end{equation}
    \begin{equation}\begin{split}
 \Sigma_{JSSJ}^{sub}\,\equiv\,-\,\delta\mathsf{Y}\,\int_{\mathbf{x},\, \mathbf{y},\, \mathbf{z},\, \mathbf{z}^{\prime}}\, K_{JSSJ}(\mathbf{x},\, \mathbf{y},\, \mathbf{z},\, \mathbf{z}^{\prime})\, N_{c}\, J_{L}^{a}(\mathbf{x})\, S_{A}^{ab}(\mathbf{z})\, J_{R}^{b}(\mathbf{y}).
  \end{split}\end{equation}\\
As was mentioned in the previous subsection, proper dimensional regularisation of $z^\prime$ integrals is needed, and the results for the integrations 
are quoted in   (\ref{to1}) and (\ref{to2}). With the subtraction terms added,
 \begin{equation}\begin{split}\label{enlo}
\Sigma^{NLO}&=\,(\Sigma_{JSJ}^{NLO}\,+\,\Sigma_{q\bar{q}}^{sub}\,+\,\Sigma_{JSSJ}^{sub})\,+\,(\Sigma_{JSSJ}^{NLO}\,-\,\Sigma_{JSSJ}^{sub})\,+\,(\Sigma_{q\bar{q}}^{NLO}\,-\,\Sigma_{q\bar{q}}^{sub})\,\\
&+\,\Sigma_{JJSJ}^{NLO}\,+\,\Sigma_{JJSSJ}^{NLO}\,+\,\Sigma_{virtual}^{NLO},
   \end{split}\end{equation}
Following   (\ref{jsjsub}),
     \begin{equation}\begin{split}\label{h4}
\Sigma_{JSJ}^{NLO}\,+\,\Sigma_{q\bar{q}}^{sub}\,+\,\Sigma_{JSSJ}^{sub}
=\,-\,\delta\mathsf{Y}\,\int_{\mathbf{x},\, \mathbf{y},\, \mathbf{z}}\, K_{JSJ}(\mathbf{x},\, \mathbf{y},\, \mathbf{z})\,\left[-\,2\, J_{L}^{a}(\mathbf{x})\, S_{A}^{ab}(\mathbf{z})\, J_{R}^{b}(\mathbf{y})\right],\\
    \end{split}\end{equation}
The UV divergence is canceled in $K_{JSJ}$. The NLO JIMWLK Hamiltonian is obtained from (\ref{enlo}) 
via the definition (\ref{defH}). The result is identical to the one quoted in (\ref{ham}).

\subsection{Reduction of the $(LO)^{2}$ Contribution}\label{reduclo}
In addition to the contributions to the NLO JIMWLK Hamiltonian we found terms proportional to $(\delta\mathsf{Y})^{2}$:
  \begin{equation}\label{deltay}
\Sigma^{(\delta\mathsf{Y})^{2}}\,\equiv\,\Sigma_{JSJ}^{(\delta\mathsf{Y})^{2}}\,+\,\Sigma_{JJSJ}^{(\delta\mathsf{Y})^{2}}\,+\,\Sigma_{JJJSJ}^{(\delta\mathsf{Y})^{2}}\,+\,\Sigma_{JSSJ}^{(\delta\mathsf{Y})^{2}}\,+\,\Sigma_{JJSSJ}^{(\delta\mathsf{Y})^{2}}\,+\,\Sigma_{JJSSJJ}^{(\delta\mathsf{Y})^{2}}\,+\,\Sigma_{virtual}^{(\delta\mathsf{Y})^{2}}.
  \end{equation}
Our objective in this section is to show that all these contributions are generated by second iteration of  the LO Hamiltonian $H^{LO}_{JIMWLK}$.
Let's first summarise all the $\left(\delta\mathsf{Y}\right)^{2}$ contributions that were found in subsections \ref{sejjssjoo} - \ref{normcon}.
From  (\ref{jjjjlosq}),
\begin{equation}\begin{split}\label{dely1}
 &\Sigma_{virtual}^{(\delta\mathsf{Y})^{2}}\,\equiv\,\frac{\alpha_{s}^{2}}{8\pi^{4}}\,\left(\delta\mathsf{Y}\right)^{2}\,\int_{\mathbf{w},\, \mathbf{v},\, \mathbf{x},\, \mathbf{y},\, \mathbf{z},\, \mathbf{z}^{\prime}}\,\frac{X\cdot Y\, W^{\prime}\cdot V^{\prime}}{X^{2}Y^{2}(W^{\prime})^{2}(V^{\prime})^{2}}\\
&\times\left[J_{L}^{b}(\mathbf{w})\, J_{L}^{b}(\mathbf{v})+J_{R}^{b}(\mathbf{w})\, J_{R}^{b}(\mathbf{v})\right]\left[J_{L}^{a}(\mathbf{x})\, J_{L}^{a}(\mathbf{y})+J_{R}^{a}(\mathbf{x})\, J_{R}^{a}(\mathbf{y})\right].\\
\end{split}\end{equation}
Adding together the contribution with one $S$ factor, (\ref{losqjjsjne}), (\ref{fekal}) and (\ref{matjjsjj}):
    \begin{eqnarray}\label{dely2}
&&\Sigma_{JSJ}^{(\delta\mathsf{Y})^{2}}\,+\,\Sigma_{JJSJ}^{(\delta\mathsf{Y})^{2}}\,+\,\Sigma_{JJJSJ}^{(\delta\mathsf{Y})^{2}}\,=\,-\frac{\alpha_{s}^{2}}{2\pi^{4}}\left(\delta\mathsf{Y}\right)^{2}\left(\,\int_{\mathbf{w},\, \mathbf{v},\, \mathbf{x},\, \mathbf{y},\, \mathbf{z},\, \mathbf{z}^{\prime}}\,\frac{X\cdot Y\, W^{\prime}\cdot V^{\prime}}{X^{2}Y^{2}(W^{\prime})^{2}(V^{\prime})^{2}}\right.\nonumber\\
&&\times\left[S_{A}^{cd}(\mathbf{z}^{\prime})\,J_{L}^{c}(\mathbf{w})\,J_{L}^{a}(\mathbf{x})\,J_{L}^{a}(\mathbf{y})\,J_{R}^{d}(\mathbf{v})+S_{A}^{cd}(\mathbf{z}^{\prime})\,J_{L}^{c}(\mathbf{v})\,J_{R}^{d}(\mathbf{w})\,J_{R}^{a}(\mathbf{x})\,J_{R}^{a}(\mathbf{y})\right]\nonumber\\
&&+\,\frac{if^{abc}}{2}\,\int_{\mathbf{w},\,\mathbf{x},\,\mathbf{y},\,\mathbf{z},\,\mathbf{z}^{\prime}}\,\left(\frac{X\cdot Z\,Y^{\prime}\cdot W^{\prime}}{X^{2}Z^{2}(Y^{\prime})^{2}(W^{\prime})^{2}}-\frac{Y\cdot Z\,X^{\prime}\cdot W^{\prime}}{Y^{2}Z^{2}(X^{\prime})^{2}(W^{\prime})^{2}}\right)\\
&&\times\left[S_{A}^{cd}(\mathbf{z}^{\prime})\,J_{L}^{a}(\mathbf{x})\,J_{L}^{b}(\mathbf{y})\,J_{R}^{d}(\mathbf{w})\,-\,S_{A}^{dc}(\mathbf{z}^{\prime})\,J_{L}^{d}(\mathbf{w})\,J_{R}^{a}(\mathbf{x})\,J_{R}^{b}(\mathbf{y})\right]\nonumber\\
&&\left.+\,\frac{N_{c}}{2}\,\int_{\mathbf{x},\, \mathbf{y},\, \mathbf{z},\, \mathbf{z}^{\prime}}\left(\frac{2X\cdot Y}{Z^{2}X^{2}Y^{2}}-\frac{X\cdot Y\, Z\cdot Y^{\prime}}{X^{2}Y^{2}Z^{2}(Y^{\prime})^{2}}-\frac{X\cdot Y\, X^{\prime}\cdot Z}{X^{2}Y^{2}(X^{\prime})^{2}Z^{2}}\right)\, S_{A}^{ab}(\mathbf{z})\, J_{L}^{a}(\mathbf{x})\, J_{R}^{b}(\mathbf{y})\right).\nonumber\end{eqnarray}
Adding together the contributions which contain two $S$ factors, (\ref{losq_JJSSJ}), (\ref{losq_JSSJ}), and (\ref{losjjssjj}):
   \begin{eqnarray}\label{dely3}
&&\Sigma_{JSSJ}^{(\delta\mathsf{Y})^{2}}\,+\,\Sigma_{JJSSJ}^{(\delta\mathsf{Y})^{2}}\,+\,\Sigma_{JJSSJJ}^{(\delta\mathsf{Y})^{2}}=\,\frac{\alpha_{s}^{2}}{2\pi^{4}}\left(\delta\mathsf{Y}\right)^{2}\,\left(\int_{\mathbf{w},\,\mathbf{v},\,\mathbf{x},\,\mathbf{y},\,\mathbf{z},\,\mathbf{z}^{\prime}}\,\frac{X\cdot Y\, W^{\prime}\cdot V^{\prime}}{X^{2}Y^{2}(W^{\prime})^{2}(V^{\prime})^{2}}\,\right.\nonumber\\
&&\times S_{A}^{cd}(\mathbf{z}^{\prime})\, S_{A}^{ab}(\mathbf{z})\, J_{L}^{a}(\mathbf{x})\, J_{L}^{c}(\mathbf{w})\, J_{R}^{b}(\mathbf{y})\, J_{R}^{d}(\mathbf{v})+\,\frac{if^{abc}}{2}\,\int_{\mathbf{w},\,\mathbf{x},\,\mathbf{y},\,\mathbf{z},\,\mathbf{z}^{\prime}}\left(\frac{X\cdot Z\, Y^{\prime}\cdot W^{\prime}}{X^{2}Z^{2}(Y^{\prime})^{2}(W^{\prime})^{2}}\,\right.\nonumber\\
&&\left.-\,\frac{Y\cdot Z\, X^{\prime}\cdot W^{\prime}}{Y^{2}Z^{2}(X^{\prime})^{2}(W^{\prime})^{2}}\right)\,\left[S_{A}^{db}(\mathbf{z})\, S_{A}^{ea}(\mathbf{z}^{\prime})\, J_{L}^{d}(\mathbf{x})\, J_{L}^{e}(\mathbf{y})\, J_{R}^{c}(\mathbf{w})\,\right.\\
&&\left.-\, S_{A}^{ae}(\mathbf{z}^{\prime})\, S_{A}^{bd}(\mathbf{z})\, J_{L}^{c}(\mathbf{w})\, J_{R}^{d}(\mathbf{x})\, J_{R}^{e}(\mathbf{y})\right]+\,\frac{1}{2}\,\int_{\mathbf{x},\,\mathbf{y},\,\mathbf{z},\,\mathbf{z}^{\prime}}\left(\frac{2X\cdot Y}{Z^{2}X^{2}Y^{2}}\right.\nonumber\\
&&\left.-\frac{X\cdot Y\, Z\cdot Y^{\prime}}{X^{2}Y^{2}Z^{2}(Y^{\prime})^{2}}-\frac{X\cdot Y\, X^{\prime}\cdot Z}{X^{2}Y^{2}(X^{\prime})^{2}Z^{2}}\right)\, f^{abc}\, f^{def}\, S_{A}^{be}(\mathbf{z})\, S_{A}^{cf}(\mathbf{z}^{\prime})\, J_{L}^{a}(\mathbf{x})\, J_{R}^{d}(\mathbf{y})\bigg).\nonumber\end{eqnarray}
Now let us write down the result that is obtained by applying $H^{LO}$ twice:
\begin{equation}\begin{split}&(\delta\mathsf{Y})^{2}(H_{JIMWLK}^{LO})^{2}\,=\,(\Sigma^{LO})^{2}=\frac{\alpha_{s}^{2}}{4\pi^{4}}\left(\delta\mathsf{Y}\right)^{2}\,\int_{\mathbf{w},\, \mathbf{v},\, \mathbf{x},\, \mathbf{y},\, \mathbf{z},\, \mathbf{z}^{\prime}}\,\frac{X\cdot Y\, W^{\prime}\cdot V^{\prime}}{X^{2}Y^{2}(W^{\prime})^{2}(V^{\prime})^{2}}\\
&\times\left[J_{L}^{a}(\mathbf{x})\, J_{L}^{a}(\mathbf{y})\,+\, J_{R}^{a}(\mathbf{x})\, J_{R}^{a}(\mathbf{y})\,-\,2J_{L}^{a}(\mathbf{x})\, S_{A}^{ab}(\mathbf{z})\, J_{R}^{b}(\mathbf{y})\right]\\
&\times\left[J_{L}^{c}(\mathbf{w})\, J_{L}^{c}(\mathbf{v})\,+\, J_{R}^{c}(\mathbf{w})\, J_{R}^{c}(\mathbf{v})\,-\,2J_{L}^{c}(\mathbf{w})\, S_{A}^{cd}(\mathbf{z}^{\prime})\, J_{R}^{d}(\mathbf{v})\right].\\
\end{split}\end{equation}
$(\Sigma^{LO})^{2}$ can also be split in accord with the number of $S$ factors:
\begin{equation}\begin{split}\label{nodes}
(\Sigma^{LO})^{2}\,=\,\Sigma_{JJJJ}^{(LO)^{2}}\,+\,\Sigma_{JJJSJ}^{(LO)^{2}}\,+\,\Sigma_{JJSSJJ}^{(LO)^{2}},
 \end{split}\end{equation}
 with 
\begin{equation}\begin{split}\label{los1}
&\Sigma_{JJJJ}^{(LO)^{2}}\,\equiv\,\frac{\alpha_{s}^{2}}{4\pi^{4}}\left(\delta\mathsf{Y}\right)^{2}\,\int_{\mathbf{w},\, \mathbf{v},\, \mathbf{x},\, \mathbf{y},\, \mathbf{z},\, \mathbf{z}^{\prime}}\,\frac{X\cdot Y\, W^{\prime}\cdot V^{\prime}}{X^{2}Y^{2}(W^{\prime})^{2}(V^{\prime})^{2}}\\
&\times\left[J_{L}^{a}(\mathbf{x})\, J_{L}^{a}(\mathbf{y})+J_{R}^{a}(\mathbf{x})\, J_{R}^{a}(\mathbf{y})\right]\left[J_{L}^{c}(\mathbf{w})\, J_{L}^{c}(\mathbf{v})+J_{R}^{c}(\mathbf{w})\, J_{R}^{c}(\mathbf{v})\right],\\
 \end{split}\end{equation}
 \begin{equation}\begin{split}\label{losa2}
&\Sigma_{JJJSJ}^{(LO)^{2}}\,\equiv\,-\frac{\alpha_{s}^{2}}{2\pi^{4}}\left(\delta\mathsf{Y}\right)^{2}\,\int_{\mathbf{w},\,\mathbf{v},\,\mathbf{x},\,\mathbf{y},\,\mathbf{z},\,\mathbf{z}^{\prime}}\,\frac{X\cdot Y\,W^{\prime}\cdot V^{\prime}}{X^{2}Y^{2}(W^{\prime})^{2}(V^{\prime})^{2}}\big(\,\left[J_{L}^{a}(\mathbf{x})\,J_{L}^{a}(\mathbf{y})+J_{R}^{a}(\mathbf{x})\,J_{R}^{a}(\mathbf{y})\right]\\
&\times\,\left[S_{A}^{cd}(\mathbf{z}^{\prime})\,J_{L}^{c}(\mathbf{w})\,J_{R}^{d}(\mathbf{v})\right]+\,\left[S_{A}^{ab}(\mathbf{z})\,J_{L}^{a}(\mathbf{x})\,J_{R}^{b}(\mathbf{y})\right]\,\left[J_{L}^{c}(\mathbf{w})\,J_{L}^{c}(\mathbf{v})+J_{R}^{c}(\mathbf{w})\,J_{R}^{c}(\mathbf{v})\right]\big),\\
\end{split}\end{equation}
 \begin{equation}\begin{split}\label{losa3}
&\Sigma_{JJSSJJ}^{(LO)^{2}}\,\equiv\,\frac{\alpha_{s}^{2}}{\pi^{4}}\left(\delta\mathsf{Y}\right)^{2}\,\int_{\mathbf{w},\, \mathbf{v},\, \mathbf{x},\, \mathbf{y},\, \mathbf{z},\, \mathbf{z}^{\prime}}\,\frac{X\cdot Y\, W^{\prime}\cdot V^{\prime}}{X^{2}Y^{2}(W^{\prime})^{2}(V^{\prime})^{2}}\\
&\times\,\left[S_{A}^{ab}(\mathbf{z})\, J_{L}^{a}(\mathbf{x})\, J_{R}^{b}(\mathbf{y})\right]\,\left[S_{A}^{cd}(\mathbf{z}^{\prime})\, J_{L}^{c}(\mathbf{w})\, J_{R}^{d}(\mathbf{v})\right].\\
\end{split}\end{equation}
In order to compare with $\Sigma^{(\delta Y)^2}$, we have to drag all the Wilson lines $S$ to the left of all $J$.
This is done using the algebra  (\ref{algJ}) and (\ref{alg2}), which generate terms with lesser number of $J$ operators.
  The procedure is applied both to  
$\Sigma_{JJJSJ}^{(LO)^{2}}$ and $\Sigma_{JJSSJJ}^{(LO)^{2}}$.
By comparing (\ref{dely1}) with (\ref{los1}), (\ref{dely2}) with (\ref{losa2}), and (\ref{dely3}) with (\ref{losa3}), we arrive at the following equivalence relations:
  \begin{equation}\begin{split}
&\Sigma_{JJJSJ}^{(\delta\mathsf{Y})^{2}}\,+\,\Sigma_{JJSJ}^{(\delta\mathsf{Y})^{2}}\,+\,\Sigma_{JSJ}^{(\delta\mathsf{Y})^{2}}\,=\,\frac{1}{2}\Sigma_{JJJSJ}^{(LO)^{2}}\:;\\
&\Sigma_{JJSSJJ}^{(\delta\mathsf{Y})^{2}}\,+\,\Sigma_{JJSSJ}^{(\delta\mathsf{Y})^{2}}\,+\,\Sigma_{JSSJ}^{(\delta\mathsf{Y})^{2}}\,=\,\frac{1}{2}\Sigma_{JJSSJJ}^{(LO)^{2}}\:;\\
&\Sigma_{virtual}^{(\delta\mathsf{Y})^{2}}\,=\,\frac{1}{2}\Sigma_{JJJJ}^{(LO)^{2}}.\\
 \end{split}\end{equation}
This establishes (\ref{lolo}).

\acknowledgments
 We are most grateful to  Ian Balitsky, Guillaume Beuf, and Alex Kovner, who have contributed to our work a great deal of inspiration and wisdom, and without whose support our progress  would be noticeably slower.
 We also benefited from discussions with Nestor Armesto,  Simon Caron-Huot, Andrey Grabovsky, and Yura Kovchegov.
Y.M. thanks the Physics Department of the University of Connecticut for their hospitality during the times when part of this work was done. This work was supported by the ISRAELI SCIENCE FOUNDATION grant \#87277111, BSF grant \#012124, the People Program (Marie Curie Actions) of the European Union's Seventh Framework under REA grant agreement \#318921.

\appendix

\section{From QCD Lagrangian to Light Cone QCD Hamiltonian}\label{hamder}
The QCD Lagrangian:
\begin{equation}\label{QCD}
\mathcal{L}_{QCD}=-\frac{1}{4}F^{a\mu\nu}F_{\mu\nu}^{a}+\sum_{f}\,\overline{\psi}_{f}(x)\,(i\gamma^{\mu}D_{\mu}-m_{f})\,\psi_{f}(x).
 \end{equation}
The tensor $F_{\mu\nu}$ defined as $F_{\mu\nu}\equiv t^{a}F_{\mu\nu}^{a}=-\frac{i}{g}\left[D_{\mu},\, D_{v}\right]$, 
satisfies the following equations of motion:
\begin{equation}\label{glueq}
D_{\mu}F^{a\mu\nu}=\partial_{\mu}F^{a\mu\nu}-gf^{abc}A_{\mu}^{b}F^{c\mu\nu}=g\overline{\psi}\gamma^{\nu}t^{a}\psi\equiv-gJ^{a\nu}.
 \end{equation}
The canonical momenta:
\begin{equation}(\Pi_{A})^{a\mu}(x)\,\equiv\,\frac{\delta\mathcal{L}_{QCD}}{\delta(\partial^{-}A_{\mu}^{a})}=\frac{1}{2}F^{a\mu+},\qquad\qquad(\Pi_{\psi})(x)\,\equiv\,\frac{\delta\mathcal{L}_{QCD}}{\delta(\partial^{-}\psi)}=\frac{i}{2}\overline{\psi}\gamma^{+}.
\end{equation}
Neglecting the quark masses, the Dirac equation becomes (after multiplication by $\gamma^{0}$):
\begin{equation}\label{dir_lc}
(i\gamma^{0}\gamma^{+}D_{+}+i\gamma^{0}\gamma^{-}D_{-}+i\alpha^{i}D_{i})\psi=0.
\end{equation}
with $\alpha^{i}\equiv\gamma^{0}\gamma^{i}$. Introduce projecting operators $\Lambda_{\pm}=\Lambda^{\pm}=\frac{1}{2}\gamma^{0}\gamma_{\pm}$, where $\left(\Lambda_{\pm}\right)^{2}=\Lambda_{\pm}$ and $\Lambda_{+}\Lambda_{-}=\Lambda_{-}\Lambda_{+}=0$,  such that $\widetilde{\psi}_{\pm}\equiv\Lambda_{\pm}\psi$ and also $\psi=\widetilde{\psi}_{+}+\widetilde{\psi}_{-}$,
\begin{equation}
\psi=\left(\begin{array}{c}
\psi_{1}\\
\psi_{2}\\
\psi_{3}\\
\psi_{4}
\end{array}\right),\qquad\qquad\qquad\widetilde{\psi}_{+}=\left(\begin{array}{c}
0\\
\psi_{2}\\
\psi_{3}\\
0
\end{array}\right),\qquad\qquad\qquad\widetilde{\psi}_{-}=\left(\begin{array}{c}
\psi_{1}\\
0\\
0\\
\psi_{4}.
\end{array}\right)
 \end{equation}
Two-coulmn vector $\psi_{+}=\left(\begin{array}{c}
\psi_{2}\\
\psi_{3}
\end{array}\right)$  and $\psi_{-}=\left(\begin{array}{c}
\psi_{1}\\
\psi_{4}
\end{array}\right)$ are defined by keeping only the non-vanishing components. The equations of motion for $\psi_{\pm}$ can be found after multiplying equation (\ref{dir_lc}) by the projections $\Lambda_{\pm}$:
\begin{equation}(i\partial^{-}-gt^{a}A^{a-})\psi_{+}=-i\sigma^{i}D_{i}\psi_{-},\qquad\qquad(i\partial^{+}-gt^{a}A^{a+})\psi_{-}=-i\sigma^{i}D_{i}\psi_{+},
 \end{equation}
from which  $\psi_{-}$ is determined in terms of $\psi_{+}$. The following constraints are obtained from the $+$ component of (\ref{glueq}):\begin{equation}D_{\mu}F^{a\mu+}=-gJ^{a+},\qquad\qquad\qquad\qquad\partial^{+}A_{+}^{a}=-\frac{1}{\partial^{+}}(D_{i}^{ab}\partial^{+}A_{i}^{b}-gJ^{a+}).
\end{equation}
Eliminating $\psi_{-}$ and $A_{+}$, the LC Hamiltonian is:
\begin{eqnarray}
H_{LC\; QCD}&&=P_{+}=\int dx^{-}\, d^{2}\mathbf{x}\,\mathcal{H}_{QCD}=\int dx^{-}\, d^{2}\mathbf{x}\,\left((\Pi_{\psi})\partial^{-}\psi+(\Pi_{A})^{\mu}\partial^{-}A_{\mu}-\mathcal{L}_{QCD}\right)\nonumber\\
&&=\int dx^{-}\, d^{2}\mathbf{x}\,\left(\frac{i}{2}\overline{\psi}\gamma^{+}\partial^{-}\psi-F^{\mu+}F_{\mu+}+\frac{1}{4}F^{\mu\nu}F_{\mu\nu}-i\overline{\psi}\gamma^{\mu}D_{\mu}\psi\right).
\end{eqnarray}
(\ref{hzero}) and (\ref{hint}) are obtained using the following relations:
\begin{equation}F^{ij}F_{ij}\,=\,2\left((\partial_{i}A_{j}^{a})(\partial_{i}A_{j}^{a})\,-\,(\partial_{i}A_{j}^{a})(\partial_{j}A_{i}^{a})\right)\,-\,4gf^{abc}A_{i}^{b}A_{j}^{c}\partial_{i}A_{j}^{a}\,+\, g^{2}f^{abc}f^{ade}A_{i}^{b}A_{j}^{c}A_{i}^{d}A_{j}^{e}.
\end{equation}
\begin{eqnarray}
&&\Pi^{a}(x^{-},\, \mathbf{x})\Pi^{a}(x^{-},\, \mathbf{x})\\
&&\quad=\,\left(\frac{1}{\partial^{+}}D_{i}\partial^{+}A_{i}\right)^{a}\left(\frac{1}{\partial^{+}}D_{j}\partial^{+}A_{j}\right)^{a}\,+\,2g\left(\frac{1}{\partial^{+}}D_{i}\partial^{+}A_{i}\right)^{a}\frac{1}{\partial^{+}}J^{a+}\,+\, g^{2}\frac{1}{\partial^{+}}J^{a+}\frac{1}{\partial^{+}}J^{a+}.\nonumber
\end{eqnarray}

 \section{Deriving the Eikonal Approximation for LC Hamiltonian}\label{sec2}
In order to introduce the eikonal approximation for the LC QCD Hamiltonian, we split the gluon and quark fields into  the soft and valence modes: 
\begin{equation}\label{Am}
\underline{A}_{i}^{a}(x)=\int_{\Lambda}^{e^{\delta\mathsf{Y}}\Lambda}\frac{dk^{+}}{2\pi}\int\frac{d^{2}\mathbf{k}}{(2\pi)^{2}}\frac{1}{\sqrt{2k^{+}}}\left(a_{i}^{a}(k^{+},\mathbf{k})e^{-ik\cdot x}+a_{i}^{a\dagger}(k^{+},\mathbf{k})e^{ik\cdot x}\right),
 \end{equation}
and
\begin{equation}\label{Ap}
\overline{A}_{i}^{a}(x)=\int_{e^{\delta\mathsf{Y}}\Lambda}^{\infty}\frac{dk^{+}}{2\pi}\int\frac{d^{2}\mathbf{k}}{(2\pi)^{2}}\frac{1}{\sqrt{2k^{+}}}\left(a_{i}^{a}(k^{+},\mathbf{k})e^{-ik\cdot x}+a_{i}^{a\dagger}(k^{+},\mathbf{k})e^{ik\cdot x}\right).
 \end{equation}
 Similarly for the quark field:
  \begin{equation}
\underline{\psi}_{+}^{\alpha}(x)=\sum_{\lambda=\pm\frac{1}{2}}\chi_{\lambda}\int_{\Lambda}^{e^{\delta\mathsf{Y}}\Lambda}\frac{dk^{+}}{2\pi}\int\frac{d^{2}\mathbf{k}}{(2\pi)^{2}}\frac{1}{\sqrt{2}}\left(b_{\lambda}^{\alpha}(k^{+},\mathbf{k})e^{-ik\cdot x}+d_{\lambda}^{\alpha\dagger}(k^{+},\mathbf{k})e^{ik\cdot x}\right),
  \end{equation}
 and
  \begin{equation}
 \overline{\psi}_{+}^{\alpha}(x)=\sum_{\lambda=\pm\frac{1}{2}}\chi_{\lambda}\int_{e^{\delta\mathsf{Y}}\Lambda}^{\infty}\frac{dk^{+}}{2\pi}\int\frac{d^{2}\mathbf{k}}{(2\pi)^{2}}\frac{1}{\sqrt{2}}\left(b_{\lambda}^{\alpha}(k^{+},\mathbf{k})e^{-ik\cdot x}+d_{\lambda}^{\alpha\dagger}(k^{+},\mathbf{k})e^{ik\cdot x}\right).
  \end{equation}
After inserting $A_{i}^{a}(x)=\underline{A}_{i}^{a}(x)+\overline{A}_{i}^{a}(x)$ and $\psi_{+}^{\alpha}(x)=\underline{\psi}_{+}^{\alpha}(x)+\overline{\psi}_{+}^{\alpha}(x)$ to $H_{int}$ defined in (\ref{hint}), we define the following components of the interaction Hamiltonian:
 \begin{equation}\label{Hg}
H_{g}\,\equiv\,-g\int dx^{-}\, d^{2}\mathbf{x}\,(\partial_{i}\underline{A}_{i}^{a})\left(f^{abc}\frac{1}{\partial^{+}}(\overline{A}_{j}^{b}\partial^{+}\overline{A}_{j}^{c})+2\frac{1}{\partial^{+}}(\overline{\psi}_{+}^{\dagger}t^{a}\overline{\psi}_{+})\right),
  \end{equation}
 \begin{equation}\begin{split}\label{gqq} 
&H_{gqq}\,\equiv\,-g\int dx^{-}\, d^{2}\mathbf{x}\,\left(2(\partial_{i}\underline{A}_{i}^{a})\frac{1}{\partial^{+}}(\underline{\psi}_{+}^{\dagger}t^{a}\underline{\psi}_{+})+\underline{\psi}_{+}^{\dagger}t^{a}(\sigma_{i}\partial_{i})\frac{1}{\partial^{+}}(\sigma_{j}\underline{A}_{j}^{a}\underline{\psi}_{+})\right.\\
&\left.+\underline{\psi}_{+}^{\dagger}t^{a}(\sigma_{i}\underline{A}_{i}^{a})\frac{1}{\partial^{+}}(\sigma_{j}\partial_{j}\underline{\psi}_{+})\right),\\
 \end{split}\end{equation}
\begin{equation}\label{hggg}
H_{ggg}\,\equiv\,-gf^{abc}\int dx^{-}\, d^{2}\mathbf{x}\,\left((\partial_{i}\underline{A}_{j}^{a})\underline{A}_{i}^{b}\underline{A}_{j}^{c}+(\partial_{i}\underline{A}_{i}^{a})\frac{1}{\partial^{+}}(\underline{A}_{j}^{b}\partial^{+}\underline{A}_{j}^{c})\right),
 \end{equation}
  \begin{equation}
  \label{qqi} H_{qq-inst}\,\equiv\,2g^{2}\int dx^{-}\, d^{2}\mathbf{x}\,\frac{1}{\partial^{+}}(\underline{\psi}_{+}^{\dagger}t^{a}\underline{\psi}_{+})\left(f^{abc}\frac{1}{\partial^{+}}(\overline{A}_{j}^{b}\partial^{+}\overline{A}_{j}^{c})+2\frac{1}{\partial^{+}}(\overline{\psi}_{+}^{\dagger}t^{a}\overline{\psi}_{+})\right),
 \end{equation}
  \begin{equation}\begin{split}\label{ggi}
&H_{gg-inst}\,\equiv\,g^{2}f^{abc}\int dx^{-}\, d^{2}\mathbf{x}\,\frac{1}{\partial^{+}}(\underline{A}_{i}^{b}\partial^{+}\underline{A}_{i}^{c})\left(f^{ade}\frac{1}{\partial^{+}}(\overline{A}_{j}^{d}\partial^{+}\overline{A}_{j}^{e})+2\frac{1}{\partial^{+}}(\overline{\psi}_{+}^{\dagger}t^{a}\overline{\psi}_{+})\right),
 \end{split}\end{equation}
 \begin{equation}H_{gggg}\,\equiv\,\frac{g^{2}}{4}f^{abc}f^{ade}\int dx^{-}\, d^{2}\mathbf{x}\,\left(\underline{A}_{i}^{b}\underline{A}_{j}^{c}\underline{A}_{i}^{d}\underline{A}_{j}^{e}-2\frac{1}{\partial^{+}}(\underline{A}_{i}^{b}\partial^{+}\underline{A}_{i}^{c})\frac{1}{\partial^{+}}(\underline{A}_{j}^{d}\partial^{+}\underline{A}_{j}^{e})\right),
 \end{equation}
 \begin{equation}H_{ggqq}\,\equiv\,-ig^{2}\int dx^{-}\, d^{2}\mathbf{x}\,\underline{\psi}_{+}^{\dagger}t^{a}t^{b}\sigma_{i}\underline{A}_{i}^{a}\frac{1}{\partial^{+}}(\sigma_{j}\underline{A}_{j}^{b}\underline{\psi}_{+}),
 \end{equation}
 
   \begin{equation}\begin{split}\label{gq}
 &H_{gq}\,\equiv\,-g\int dx^{-}\, d^{2}\mathbf{x}\,\left(\overline{\psi}_{+}^{\dagger}t^{a}(\sigma_{i}\partial_{i})\frac{1}{\partial^{+}}(\sigma_{j}\underline{A}_{j}^{a}\underline{\psi}_{+})+\overline{\psi}_{+}^{\dagger}t^{a}(\sigma_{i}\underline{A}_{i}^{a})\frac{1}{\partial^{+}}(\sigma_{j}\partial_{j}\underline{\psi}_{+})\right.\\
&\left.+ig\overline{\psi}_{+}^{\dagger}t^{a}t^{b}\sigma_{i}\underline{A}_{i}^{a}\frac{1}{\partial^{+}}(\sigma_{j}\underline{A}_{j}^{b}\underline{\psi}_{+})\right).\\
  \end{split}\end{equation}
{ The eikonal approximation means that  all  contributions  involving division by large (valence) longitudinal momenta are neglected, such as $\frac{1}{\partial^{+}}(\underline{\psi}_{+}^{\dagger}t^{a}\overline{\psi}_{+})\sim\frac{1}{\partial^{+}}(\underline{A}_{j}^{b}\partial^{+}\overline{A}_{j}^{c})\sim0$. Furthermore, 
we only keep terms which are relevant for our NLO calculation, the ones that contain either interaction between soft modes  or interaction between soft and 
valence modes.   Some soft interaction terms that would contribute starting from NNLO only, such as instantaneous gluon-quark interaction term, 
as well as  the valence  Hamiltonian  are not written down explicitly.}

\section{Integrals and Fourier Transformations}\label{inttab}
In the following Appendix we assemble all the integrals that were used throughout our calculations in this paper.  
 \begin{equation}\label{int.1}
\int_{0}^{1}d\xi\,(4\xi^{2}-4\xi+2)\left(\ln\left(\xi(1-\xi)\right)-1\right)=-\frac{38}{9}.
 \end{equation}
 \begin{equation}\label{int.2}
 \int_{0}^{1}d\xi\,\frac{1}{(1-\xi)(X^{\prime})^{2}+\xi X^{2}}=\frac{1}{X^{2}-(X^{\prime})^{2}}\ln\left(\frac{X^{2}}{(X^{\prime})^{2}}\right).
\end{equation}
    \begin{equation}\label{int.3}
\int_{0}^{1}d\xi\,\frac{1}{\left((1-\xi)(X^{\prime})^{2}+\xi X^{2}\right)\left((1-\xi)(Y^{\prime})^{2}+\xi Y^{2}\right)}=\frac{1}{(X^{\prime})^{2}Y^{2}-X^{2}(Y^{\prime})^{2}}\ln\left(\frac{(X^{\prime})^{2}Y^{2}}{X^{2}(Y^{\prime})^{2}}\right).
\end{equation}
   \begin{eqnarray}\label{int.4}
&&\int_{0}^{1}d\xi\,\frac{\xi(1-\xi)}{\left((1-\xi)(X^{\prime})^{2}+\xi X^{2}\right)\left((1-\xi)(Y^{\prime})^{2}+\xi Y^{2}\right)}\,=\,-\,\frac{1}{\left((X^{\prime})^{2}-X^{2}\right)\left((Y^{\prime})^{2}-Y^{2}\right)}\nonumber\\
&&+\,\frac{(X^{\prime})^{2}X^{2}}{((X^{\prime})^{2}-X^{2})^{2}\left((X^{\prime})^{2}Y^{2}-X^{2}(Y^{\prime})^{2}\right)}\ln\left(\frac{X^{2}}{(X^{\prime})^{2}}\right)\\
&&+\,\frac{(Y^{\prime})^{2}Y^{2}}{((Y^{\prime})^{2}-Y)^{2}\left((X^{\prime})^{2}Y^{2}-X^{2}(Y^{\prime})^{2}\right)}\ln\left(\frac{Y^{2}}{(Y^{\prime})^{2}}\right).\nonumber\end{eqnarray}

  The integrals below are evaluated under the assumption $\Lambda\rightarrow0\,$ and $\delta\mathsf{Y}\rightarrow\infty$. While explicitly divergent 
  terms will be retain,   we will take the limit  in all the non-divergent terms without further notification.
  \begin{equation}\label{heaso1}
  \int_{\frac{\Lambda}{k^{+}}}^{1-\frac{\Lambda}{k^{+}}}d\xi\,\frac{1}{\xi\left((1-\xi)(W^{\prime})^{2}\,+\,\xi W^{2}\right)}\,=\,-\,\frac{1}{(W^{\prime})^{2}}\left(\ln\left(\frac{\Lambda}{k^{+}}\right)\,+\,\ln\left(\frac{W^{2}}{(W^{\prime})^{2}}\right)\right).
 \end{equation}
\begin{equation}\label{heaso2}
\int_{\frac{\Lambda}{k^{+}}}^{1-\frac{\Lambda}{k^{+}}}d\xi\,\frac{1}{(1-\xi)\left((1-\xi)(W^{\prime})^{2}\,+\,\xi W^{2}\right)}\,=\,-\,\frac{1}{W^{2}}\left(\ln\left(\frac{\Lambda}{k^{+}}\right)\,+\,\ln\left(\frac{W^{2}}{(W^{\prime})^{2}}\right)\right).
 \end{equation}
  \begin{equation}\begin{split}\label{int.5}
&\int_{\frac{\Lambda}{k^{+}}}^{1-\frac{\Lambda}{k^{+}}}d\xi\,\frac{1+\xi}{\xi(1-\xi)^{2}}\,=\,\frac{2k^{+}}{\Lambda}-2-2\ln\left(\frac{\Lambda}{k^{+}}\right).\\
 \end{split}\end{equation}
 \begin{equation}\begin{split}\label{int.6}
&\int_{\frac{\Lambda}{k^{+}}}^{1-\frac{\Lambda}{k^{+}}}d\xi\,\frac{1+\xi}{(1-\xi)^{2}}\,=\,\frac{2k^{+}}{\Lambda}-2+\ln\left(\frac{\Lambda}{k^{+}}\right).\\
 \end{split}\end{equation}
    \begin{equation}\begin{split}\label{int.7}
 \int_{\frac{\Lambda}{k^{+}}}^{\frac{e^{\delta\mathsf{Y}}\Lambda}{k^{+}}}d\xi\,\frac{1}{\xi}\ln\xi=\frac{1}{2}\ln^{2}\left(\frac{e^{\delta\mathsf{Y}}\Lambda}{k^{+}}\right)-\frac{1}{2}\ln^{2}\left(\frac{\Lambda}{k^{+}}\right).
 \end{split}\end{equation}
 \begin{equation}\begin{split}\label{int.8}
 &\int_{\frac{\Lambda}{k^{+}}}^{1-\frac{\Lambda}{k^{+}}}d\xi\,\left(\xi(1-\xi)-2+\frac{1}{\xi(1-\xi)}\right)\ln\left(\frac{\mathbf{k}^{2}\xi(1-\xi)}{\mu_{\overline{MS}}^{2}}\right)\\
&=\left[-\frac{11}{6}-2\ln\,\left(\frac{\Lambda}{k^{+}}\right)\right]\ln\left(\frac{\mathbf{k}^{2}}{\mu_{\overline{MS}}^{2}}\right)-\ln^{2}\left(\frac{\Lambda}{k^{+}}\right)+\frac{67}{18}-\frac{\pi^{2}}{3}.\\
\end{split}\end{equation}
  \begin{equation}\begin{split}\label{int.9}
 &\int_{\frac{\Lambda}{k^{+}}}^{1-\frac{\Lambda}{k^{+}}}d\xi\,\frac{(\xi-2)\ln\left(\xi\right)+(1+\xi)\ln\left(1-\xi\right)}{\xi(1-\xi)}\,=\,0.\\
  \end{split}\end{equation}
 \begin{equation}\begin{split}\label{int.9a}
 \int_{\frac{\Lambda}{k^{+}}}^{1-\frac{\Lambda}{k^{+}}}d\xi\,\frac{(\xi-2)\ln\xi-(1+\xi)\ln(1-\xi)}{\xi(1-\xi)}\,=\,\frac{\pi^{2}}{3}\,+\,2\ln^{2}\left(\frac{\Lambda}{k^{+}}\right).
 \end{split}\end{equation}
  \begin{equation}\label{int.10}
\int_{\frac{\Lambda}{k^{+}}}^{1-\frac{\Lambda}{k^{+}}}d\xi\,\frac{1+\xi}{\xi(1-\xi)}\,=\,-\int_{\frac{\Lambda}{k^{+}}}^{1-\frac{\Lambda}{k^{+}}}d\xi\,\frac{\xi-2}{\xi(1-\xi)}\,=\,-3\ln\left(\frac{\Lambda}{k^{+}}\right).
  \end{equation}
 \begin{equation}\label{int.11ab}
\int_{\frac{\Lambda}{k^{+}}}^{1-\frac{\Lambda}{k^{+}}}d\xi\,\frac{\xi-2}{\xi^{2}(1-\xi)}\,=\,2\left(1-\frac{k^{+}}{\Lambda}+\ln\left(\frac{\Lambda}{k^{+}}\right)\right).
  \end{equation}
  \begin{equation}\label{int.11}
\int_{\Lambda}^{\Lambda e^{\delta\mathsf{Y}}}\frac{dk^{+}}{k^{+}}\:\left(\ln^{2}\left(\frac{\Lambda e^{\delta\mathsf{Y}}}{k^{+}}\right)\,-\,\ln^{2}\left(\frac{\Lambda}{k^{+}}\right)\right)=0.
 \end{equation}
\begin{equation}\begin{split}
&\int_{1-\frac{\Lambda}{k^{+}}}^{\frac{e^{\delta Y}\Lambda}{k^{+}}}\frac{b+a\xi}{\xi(1-\xi)\left(c+d(\xi-1)\right)}\, d\xi\,=\,\left(\frac{a}{c}-\frac{bd}{c(c-d)}\right)\left[\ln\left(c-d+\frac{e^{\delta\mathsf{Y}}\Lambda}{k^{+}}d\right)\right.\\
&\left.-\ln\left(c-\frac{\Lambda}{k^{+}}d\right)\right]-\frac{a+b}{c}\left[\ln\left(1-\frac{e^{\delta\mathsf{Y}}\Lambda}{k^{+}}\right)-\ln\left(\frac{\Lambda}{k^{+}}\right)\right]+\frac{b}{c-d}\ln\left(\frac{e^{\delta\mathsf{Y}}\Lambda}{k^{+}}\right).\\
\end{split}\end{equation}
 \begin{equation}\label{int.11as}
 \int_{\frac{\Lambda}{k^{+}}}^{\frac{e^{\delta\mathsf{Y}}\Lambda}{k^{+}}}d\xi\,\frac{1}{\xi\left(\xi+\frac{\mathbf{p}^{2}}{\mathbf{k}^{2}}\right)}\,=\,\frac{\mathbf{k}^{2}}{\mathbf{p}^{2}}\left(\ln\left(\frac{\Lambda}{k^{+}}\right)\,-\,\ln\left(\frac{\mathbf{p}^{2}}{\mathbf{k}^{2}}\right)\right).
 \end{equation}
    \begin{equation}\begin{split}\label{jjsjino}
&\int_{\frac{\Lambda}{k^{+}}}^{1-\frac{\Lambda}{k^{+}}}d\xi\,\frac{\xi}{\left(\xi(\mathbf{k}-\mathbf{p})^{2}+(1-\xi)\mathbf{k}^{2}\right)(1-\xi)}\\
&=\,\frac{\mathbf{k}^{2}}{(\mathbf{k}-\mathbf{p})^{2}\left(\mathbf{k}^{2}-(\mathbf{k}-\mathbf{p})^{2}\right)}\ln\left(\frac{(\mathbf{k}-\mathbf{p})^{2}}{\mathbf{k}^{2}}\right)-\frac{1}{(\mathbf{k}-\mathbf{p})^{2}}\ln\left(\frac{\Lambda}{k^{+}}\right).\\
\end{split}\end{equation}
   \begin{equation}\label{jjsjino2}
\int_{\frac{\Lambda}{k^{+}}}^{1-\frac{\Lambda}{k^{+}}}d\xi\,\frac{1}{\left(\xi(\mathbf{k}-\mathbf{p})^{2}+(1-\xi)\mathbf{k}^{2}\right)(1-\xi)}\,=\,\frac{1}{(\mathbf{k}-\mathbf{p})^{2}}\ln\left(\frac{(\mathbf{k}-\mathbf{p})^{2}}{\mathbf{k}^{2}}\right)-\frac{1}{(\mathbf{k}-\mathbf{p})^{2}}\ln\left(\frac{\Lambda}{k^{+}}\right).
 \end{equation}
  $\bullet$ \quad\textit{\textbf{Fourier transforms}}\\
Here we present a list of the integrals necessary to perform the Fourier transformations during the computations. A comprehensive list of Fourier transforms can be found in \cite{oldgrab}.
  \begin{equation}
\frac{1}{(2\pi)^{3}}\int dx^{-}\, d^{2}\mathbf{x}\, e^{i(k-p)x}\,=\,\delta^{(3)}(k-p).
\end{equation}
 \begin{equation}\begin{split}\label{fourier.1}
 \int\frac{d^{2}\mathbf{k}}{2\pi}\,\frac{\mathbf{k}^{i}}{\mathbf{k}^{2}}e^{i\mathbf{k}\cdot X}=\frac{iX^{i}}{X^{2}}.
 \end{split}\end{equation}
  \begin{equation}\begin{split} \label{fourier.2}
  \int\frac{d^{2}\mathbf{k}}{2\pi}\,\frac{\mathbf{k}^{i}}{\mathbf{k}^{2}}\ln\left(\frac{\mathbf{k}^{2}}{\mu^{2}}\right)e^{i\mathbf{k}\cdot X}=\frac{iX^{i}}{X^{2}}\left(-2\gamma-\ln\left(\frac{X^{2}\mu^{2}}{4}\right)\right).
 \end{split}\end{equation}
    \begin{equation}\begin{split} \label{fourier.3}
    \int\frac{d^{2}\mathbf{k}}{2\pi}\frac{d^{2}\mathbf{p}}{2\pi}\,\frac{1}{\xi(1-\xi)\mathbf{k}^{2}+\mathbf{p}^{2}}e^{i\mathbf{k}\cdot X+i\mathbf{p}\cdot Z}=\frac{1}{X^{2}+\xi(1-\xi)Z^{2}}.
 \end{split}\end{equation}
  \begin{equation}\begin{split} \label{fourier.4}
  \int\frac{d^{2}\mathbf{k}}{2\pi}\frac{d^{2}\mathbf{p}}{2\pi}\,\frac{\mathbf{k}^{i}\mathbf{p}^{j}}{\mathbf{k}^{2}\left(a\mathbf{k}^{2}+b\mathbf{p}^{2}\right)}e^{i\mathbf{k}\cdot X+i\mathbf{p}\cdot Z}=-\frac{X^{i}Z^{j}}{Z^{2}\left(bX^{2}+aZ^{2}\right)}.
 \end{split}\end{equation}
  \begin{equation}\label{balitz}
\int\frac{d^{2}k}{2\pi}\,\frac{d^{2}p}{2\pi}\,\frac{\mathbf{k}\cdot \mathbf{p}}{\mathbf{k}^{2}\mathbf{p}^{2}}\ln\left(\frac{(\mathbf{k}-\mathbf{p})^{2}}{\mu_{\overline{MS}}^{2}}\right)e^{i\mathbf{k}\cdot X-i\mathbf{p}\cdot Y}\,=\,\frac{X\cdot Y}{X^{2}Y^{2}}\left(\ln\left(\frac{(X-Y)^{2}}{X^{2}Y^{2}\mu_{\overline{MS}}^{2}}\right)-2\gamma+\log4\right).
  \end{equation}
 \begin{equation}\label{oiam}
 \int\frac{d^{2}\mathbf{k}}{2\pi}\frac{d^{2}\mathbf{p}}{2\pi}\,\frac{1}{\mathbf{p}^{2}}\ln\left(\frac{(\mathbf{k}-\mathbf{p})^{2}}{\mathbf{k}^{2}}\right)e^{-i\mathbf{k}\cdot X+i\mathbf{p}\cdot Y}\,=\,\frac{1}{X^{2}}\ln\left(\frac{(X-Y)^{2}}{Y^{2}}\right).
  \end{equation}
\footnote{We thank Ian Balitsky for sharing this integral with us.}    
\vspace{-0.7cm}
\begin{eqnarray}\label{ian}
&&\int\frac{d^{2}\mathbf{k}}{2\pi}\,\frac{d^{2}\mathbf{p}}{2\pi}\left(\frac{\mathbf{k}^{i}-\mathbf{p}^{i}}{\mathbf{p}^{2}(\mathbf{k}-\mathbf{p})^{2}}-\frac{\mathbf{k}^{i}}{\mathbf{k}^{2}\mathbf{p}^{2}}+\frac{\mathbf{k}^{i}}{\mathbf{k}^{2}(\mathbf{k}-\mathbf{p})^{2}}\right)\ln\left(\frac{\mathbf{p}^{2}}{\mathbf{k}^{2}}\right)e^{-i\mathbf{k}\cdot Y-i\mathbf{p}\cdot(X-Y)}-(X\leftrightarrow Y)\nonumber \\
&&=-\frac{i}{2}\left(\frac{X^{i}}{X^{2}}-\frac{Y^{i}}{Y^{2}}\right)\ln\left(\frac{X^{2}}{(X-Y)^{2}}\right)\ln\left(\frac{Y^{2}}{(X-Y)^{2}}\right).
\end{eqnarray}
  \begin{equation}\label{frfr1}
 \int d^{2}\mathbf{k}\, d^{2}\mathbf{p}\: e^{-i(\mathbf{k}-\mathbf{p})\cdot \mathbf{y}-i\mathbf{p}\cdot \mathbf{x}+i\mathbf{k}\cdot \mathbf{z}}\frac{\mathbf{k}^{i}}{\mathbf{k}^{2}\mathbf{p}^{2}}\,=\,2i\pi\int_{\mathbf{z}^{\prime}}\frac{X^{\prime}\cdot Y^{\prime}\, Y^{i}}{(X^{\prime})^{2}(Y^{\prime})^{2}Y^{2}}.
 \end{equation}
 \begin{equation}\label{frfr2}
\int d^{2}\mathbf{k}\, d^{2}\mathbf{p}\, e^{-i(\mathbf{k}-\mathbf{p})\cdot \mathbf{x}-i\mathbf{p}\cdot \mathbf{y}+i\mathbf{k}\cdot \mathbf{z}}\,\frac{\mathbf{p}^{i}}{\mathbf{p}^{2}(\mathbf{k}-\mathbf{p})^{2}}\,=\,2i\pi\int_{\mathbf{z}^{\prime}}\frac{X^{\prime}\cdot Z\, Y^{i}}{(X^{\prime})^{2}Z^{2}Y^{2}}.
 \end{equation}
 \begin{equation}\label{frfr3}
\int d^{2}\mathbf{k}\, d^{2}\mathbf{p}\, e^{-i(\mathbf{k}-\mathbf{p})\cdot \mathbf{x}-i\mathbf{p}\cdot \mathbf{y}+i\mathbf{k}\cdot \mathbf{z}}\,\frac{\mathbf{p}\cdot(\mathbf{p}-\mathbf{k})\, \mathbf{k}^{i}}{\mathbf{k}^{2}\mathbf{p}^{2}(\mathbf{k}-\mathbf{p})^{2}}\,=\,2i\pi\int_{\mathbf{z}^{\prime}}\frac{X^{\prime}\cdot Y^{\prime}\, Z^{i}}{(X^{\prime})^{2}(Y^{\prime})^{2}Z^{2}}.
 \end{equation}
 $\bullet$ \quad\textit{\textbf{Dimensional Regularization}}\\
\begin{equation}\label{dim1}
\int\frac{d^{d}\mathbf{k}}{(2\pi)^{d}}\frac{\mathbf{k}^{2}}{\left(\mathbf{k}^{2}+\Delta\right)^{n}}=\frac{d}{2(4\pi)^{d/2}}\frac{\Gamma\left(n-1-\frac{d}{2}\right)}{\Gamma\left(n\right)}\left(\frac{1}{\Delta}\right)^{n-1-\frac{d}{2}}.
  \end{equation}
  \begin{equation}\label{misdim}
 \int\frac{d^{d}\mathbf{k}}{(2\pi)^{d}}\frac{1}{\left(\mathbf{k}^{2}+\Delta\right)^{n}}\,=\,\frac{1}{(4\pi)^{d/2}}\frac{\Gamma\left(n-\frac{d}{2}\right)}{\Gamma\left(n\right)}\left(\frac{1}{\Delta}\right)^{n-\frac{d}{2}}.
  \end{equation}
 $\Gamma$ function expansion:
 \begin{equation}\label{gamexp}
\Gamma(-n+\epsilon)=\frac{(-1)^{n}}{n!}\left[\frac{1}{\epsilon}+\sum_{k=1}^{k=n}\frac{1}{k}-\gamma\right]+\mathcal{O}(\epsilon).
  \end{equation}

    \section{Properties of the NLO JIMWLK Kernels}\label{kerproper}
Here we quote from \cite{nlojimwlk} some useful properties of the kernels defined in section \ref{sdfla}:
\begin{equation}\begin{split}\label{dien1}
\int_{\mathbf{z}^{\prime}}K_{JJSSJ}(\mathbf{y},\, \mathbf{x},\, \mathbf{y},\, \mathbf{z},\, \mathbf{z}^{\prime})=0.
   \end{split}\end{equation}
\begin{equation}\begin{split}
\int_{\mathbf{z}}K_{JJSJ}(\mathbf{y},\, \mathbf{x},\, \mathbf{y},\, \mathbf{z})=\int_{\mathbf{z},\mathbf{z}^{\prime}}K_{JJSSJ}(\mathbf{y},\, \mathbf{y},\, \mathbf{x},\, \mathbf{z},\, \mathbf{z}^{\prime})=0.
   \end{split}\end{equation}
\begin{equation}\begin{split}
\int_{\mathbf{z},\mathbf{z}^{\prime}}\widetilde{K}(\mathbf{x},\, \mathbf{y},\, \mathbf{z},\, \mathbf{z}^{\prime})=i\int_{\mathbf{z}}\left[K_{JJSJ}(\mathbf{y},\, \mathbf{x},\, \mathbf{y},\, \mathbf{z})+K_{JJSJ}(\mathbf{x},\, \mathbf{y},\, \mathbf{x},\, \mathbf{z})\right]=0.
   \end{split}\end{equation}
      \begin{equation}\begin{split}\label{jjsjiden}
       K_{JJSJ}(\mathbf{w},\, \mathbf{x},\, \mathbf{y},\, \mathbf{z})=\int_{\mathbf{z}^{\prime}}\left[K_{JJSSJ}(\mathbf{y},\, \mathbf{w},\, \mathbf{x},\, \mathbf{z},\, \mathbf{z}^{\prime})-K_{JJSSJ}(\mathbf{x},\, \mathbf{w},\, \mathbf{y},\, \mathbf{z},\, \mathbf{z}^{\prime})\right].
  \end{split}\end{equation}
Throughout the calculation, we encounter the kernels $K_{qq}$ and $K_{JSSJ}$ integrated over $z^\prime$. These integrals are UV divergent. In order to evaluate them, they have to be written in their momentum space representation and dimensionally regularised.
From \cite{Weigertrun} and using (\ref{balitz}), (\ref{oiam}),
     \begin{eqnarray}\label{to1}
&&\int_{\mathbf{z}^{\prime}}K_{q\bar{q}}(\mathbf{x},\,\mathbf{y},\,\mathbf{z},\,\mathbf{z}^{\prime})\,=\,-\frac{\alpha_{s}^{2}N_{f}}{12\pi^{3}}\left(\frac{1}{X^{2}}\ln\left(X^{2}\mu_{\overline{MS}}^{2}\right)\,+\,\frac{1}{Y^{2}}\ln\left(Y^{2}\mu_{\overline{MS}}^{2}\right)\right.\nonumber\\
&&\left.+\,\frac{(X-Y)^{2}}{X^{2}Y^{2}}\ln\left(\frac{(X-Y)^{2}}{X^{2}Y^{2}\mu_{\overline{MS}}^{2}}\right)\,+\,\frac{X\cdot Y}{X^{2}Y^{2}}\left(\frac{10}{3}+4\gamma-2\ln4\right)\right).\end{eqnarray}
From \cite{BC},
   \begin{eqnarray}\label{to2}
&&\int_{\mathbf{z}^{\prime}}K_{JSSJ}(\mathbf{x},\,\mathbf{y},\,\mathbf{z},\,\mathbf{z}^{\prime})\,=\,\frac{\alpha_{s}^{2}}{4\pi^{3}}\left(\frac{11}{6X^{2}}\ln\left(X^{2}\mu_{\overline{MS}}^{2}\right)\,+\,\frac{11}{6Y^{2}}\ln\left(Y^{2}\mu_{\overline{MS}}^{2}\right)\,+\,\frac{11(X-Y)^{2}}{6X^{2}Y^{2}}\right.\nonumber\\
&&\left.\times\ln\left(\frac{(X-Y)^{2}}{X^{2}Y^{2}\mu_{\overline{MS}}^{2}}\right)\,+\,\frac{X\cdot Y}{X^{2}Y^{2}}\left(\frac{67}{9}-\frac{\pi^{2}}{3}+\frac{11}{3}\left[2\gamma-\ln4\right]\right)\right)\,+\,\int_{\mathbf{z}^{\prime}}\,\widetilde{K}(\mathbf{x},\,\mathbf{y},\,\mathbf{z},\,\mathbf{z}^{\prime}).\end{eqnarray}

\section{Perturbation Theory with Non-commutative Matrix Elements}\label{wavexp}
In this Appendix we briefly sketch the derivation of the wave function up to order $g^{3}$. 
Our prime goal is to provide a derivation, which would be
valid for operator valued non-commuting matrix elements. 
We begin with an unperturbed Hamiltonian $H_{0}$. $E_{n}^{(0)}$ and $\left|n^{(0)}\right\rangle $ denote the eigenvalues and the eigenstates of $H_{0}$. Let introduce a weak perturbation $H_{int}$. The energy levels and eigenstates of the perturbed Hamiltonian are $E_{n}$, and $\left|n\right\rangle$:
\begin{equation}\label{dwf1}
H_{0}\,\left|n^{(0)}\right\rangle \,=\, E_{n}^{(0)}\,\left|n^{(0)}\right\rangle \:;\qquad\qquad\qquad\left(H_{0}\,+\, H_{int}\right)\,\left|n\right\rangle \,=\, E_{n}\,\left|n\right\rangle .
  \end{equation}
  \begin{equation}\begin{split}\label{poas}
  &E_{n}\,=\, E_{n}^{(0)}\,+\, E_{n}^{(1)}\,+\, E_{n}^{(2)}\,+\,\ldots\qquad\qquad\left|n\right\rangle \,=\,\left|n^{(0)}\right\rangle \,+\,\left|n^{(1)}\right\rangle \,+\,\left|n^{(2)}\right\rangle \,+\,\ldots
 \end{split}\end{equation}
 Where $\left|n^{(i)}\right\rangle $ are orthogonal states, corresponding to $i$-th order in perturbation theory. By inserting (\ref{poas}) in (\ref{dwf1}), expanding and equating terms in the same order of $g$, we arrive at:
  \begin{eqnarray}\label{waex}
&&\left|n\right\rangle \,=\,\left|n^{(0)}\right\rangle \,-\,\sum_{i\neq0}\frac{1}{E_{i}^{(0)}\,-\, E_{0}^{(0)}\,-\, E_{i}^{(1)}\,-\, E_{i}^{(2)}}\,\left|n^{(i)}\right\rangle \left\langle n^{(i)}\left|H_{int}\right|n^{(0)}\right\rangle\nonumber\\ 
&&+\,\sum_{i,j\neq0}\,\frac{1}{E_{i}^{(0)}-E_{0}^{(0)}-E_{i}^{(1)}}\,\left|n^{(i)}\right\rangle \,\left\langle n^{(i)}\left|H_{int}\right|n^{(j)}\right\rangle \,\frac{1}{E_{j}^{(0)}-E_{0}^{(0)}-E_{j}^{(1)}}\,\left\langle n^{(j)}\left|H_{int}\right|n^{(0)}\right\rangle \nonumber\\
&&-\,\sum_{i,j,k\neq0}\,\frac{1}{E_{i}^{(0)}\,-\, E_{0}^{(0)}}\left|n^{(i)}\right\rangle \,\left\langle n^{(i)}\left|H_{int}\right|n^{(j)}\right\rangle \,\frac{1}{E_{j}^{(0)}\,-\, E_{0}^{(0)}}\,\left\langle n^{(j)}\left|H_{int}\right|n^{(k)}\right\rangle \,\\
&&\times\frac{1}{E_{k}^{(0)}\,-\, E_{0}^{(0)}}\,\left\langle n^{(k)}\left|H_{int}\right|n^{(0)}\right\rangle ,\nonumber\end{eqnarray}
 with
\begin{equation}\label{edef}
E_{i}^{(1)}\,\equiv\,\left\langle n^{(i)}\right|H_{int}\left|n^{(i)}\right\rangle \qquad\qquad E_{i}^{(2)}\,\equiv\,-\sum_{i\neq0}\frac{\left|\left\langle n^{(i)}\left|H_{int}\right|0\right\rangle \right|^{2}}{E_{i}^{(0)}}.
  \end{equation}
 Let's focus on the vacuum state to be denoted $\left|\psi\right\rangle _{N}$ below.  We set  $E_{0}^{(0)}=0$. 
Having in mind $H_{int}$ that does not have a matrix element over the vacuum state, from (\ref{edef}), $E_{n}^{(1)}=0$.
The denominator in the first line of (\ref{waex}):
  \begin{equation}\begin{split}&\frac{1}{E_{i}^{(0)}\,-\, E_{0}^{(0)}\,-\, E_{i}^{(1)}\,-\, E_{i}^{(2)}}\,\approx\,\frac{1}{E_{i}^{(0)}}\,-\,\frac{1}{(E_{i}^{(0)})^{2}}\sum_{j\neq0}\frac{\left|\left\langle n^{(j)}\left|H_{int}\right|0\right\rangle \right|^{2}}{E_{j}^{(0)}}.
  \end{split}\end{equation}
Substituting  the last relation in (\ref{waex}), we finally get:
\begin{equation}\begin{split}\left|\psi\right\rangle _{N}&=\,\left|0\right\rangle \,-\,\left|n^{(i)}\right\rangle \frac{\left\langle n^{(i)}\left|H_{int}\right|0\right\rangle }{E_{i}^{(0)}}\,+\,\left|n^{(i)}\right\rangle \frac{\left\langle n^{(i)}\left|H_{int}\right|n^{(j)}\right\rangle \,\left\langle n^{(j)}\left|H_{int}\right|0\right\rangle }{E_{i}^{(0)}E_{j}^{(0)}}\\
&-\,\left|n^{(i)}\right\rangle \frac{\left\langle n^{(i)}\left|H_{int}\right|n^{(j)}\right\rangle \,\left\langle n^{(j)}\left|H_{int}\right|n^{(k)}\right\rangle \,\left\langle n^{(k)}\left|H_{int}\right|0\right\rangle }{E_{i}^{(0)}E_{j}^{(0)}E_{k}^{(0)}}\,\\
&+\,\left|n^{(i)}\right\rangle \frac{\left|\left\langle n^{(j)}\left|H_{int}\right|0\right\rangle \right|^{2}\,\left\langle n^{(i)}\left|H_{int}\right|0\right\rangle }{(E_{i}^{(0)})^{2}E_{j}^{(0)}}.
\end{split}\end{equation}

\section{The Phase of the Wave Function}\label{wfphase}
The condition (\ref{condphase})  imposed in order to find the phase  reads
\begin{equation}\label{condit}
\left\langle \psi^{NLO}\right|\,\frac{\delta}{\delta\rho^{d}(\mathbf{w})}\,\left|\psi^{NLO}\right\rangle =0.
\end{equation}
It is trivial to check that the LO LCWF satisfies (\ref{condit}). This happens automatically thanks to  cancelation between 
contributions of normalisation  against diagonal contribution from one gluon component. Similar cancelations happen when
 $\left|\psi^{NLO}\right\rangle $  (\ref{expenw}) is inserted in (\ref{condit}).  What we are left with are the "interference" contributions
 between different components of the LCWF and the phase:
 \begin{equation}\begin{split}
&\left\langle \psi^{NLO}\right|\,\frac{\delta}{\delta\rho^{d}(\mathbf{w})}\,\left|\psi^{NLO}\right\rangle \,=\,\left\langle \psi_{gg\:\rho}\right|\,\frac{\delta}{\delta\rho^{d}(\mathbf{w})}\,\left|\psi_{gg\:\rho\rho}\right\rangle \,+\,\left\langle \psi_{gg\:\rho\rho}\right|\,\frac{\delta}{\delta\rho^{d}(\mathbf{w})}\,\left|\psi_{gg\:\rho}\right\rangle \\
&\,+\,\left\langle \psi_{g\:\rho\rho}\right|\,\frac{\delta}{\delta\rho^{d}(\mathbf{w})}\,\left|\psi_{g\:\rho}^{LO}\right\rangle \,+\,\left\langle \psi_{g\:\rho}^{LO}\right|\,\frac{\delta}{\delta\rho^{d}(\mathbf{w})}\,\left|\psi_{g\:\rho\rho}\right\rangle \,+\,\left\langle 0\right|\,\frac{\delta}{\delta\rho^{d}(\mathbf{w})}\,i\phi^{NLO}\left|0\right\rangle\,=\,0 \\\
  \end{split}\end{equation}
After insertion of expressions (\ref{lotra}), (\ref{gg_rho}), (\ref{gg_rhorho}), (\ref{g_rho}), (\ref{g_rhorho}), the condition (\ref{condit}) becomes:
\begin{equation}\begin{split}\label{crite}
&-\,\delta\mathsf{Y}\,\int_{\mathbf{x},\, \mathbf{y},\, \mathbf{z},\, \mathbf{z}^{\prime}}f^{dcb}\rho^{b}(\mathbf{y})\rho^{c}(\mathbf{x})[-K_{JJSSJ}(\mathbf{w},\, \mathbf{x},\, \mathbf{y},\, \mathbf{z},\, \mathbf{z}^{\prime})+K_{JJSSJ}(\mathbf{x},\, \mathbf{w},\, \mathbf{y},\, \mathbf{z},\, \mathbf{z}^{\prime})\\
&-K_{JJSSJ}(\mathbf{y},\,\mathbf{w},\,\mathbf{x},\,\mathbf{z},\,\mathbf{z}^{\prime})]\,+\,\left\langle 0\right|\,\frac{\delta}{\delta\rho^{d}(\mathbf{w})}\,i\phi^{NLO}\left|0\right\rangle \,=\,0.\\
 \end{split}\end{equation}
 In  (\ref{crite}), the kernels  $K_{JJSSJ}$ and $K_{JJSJ}$ originate from the overlaps  
 $\left\langle \psi_{gg\:\rho\rho}|\psi_{gg\:\rho}\right\rangle$  and
 $\langle \psi_{g\:\rho\rho}|\psi_{g\:\rho}^{LO}\rangle$. The computation can be easily traced to a similar computation of $\Sigma_{vJJJ}$ 
 in section \ref{normcon}.
 
It is easy to see that the following ansatz for $i\phi^{NLO}$ solves (\ref{crite}),
\begin{equation}
i\phi^{NLO}\,=\,-\delta\mathsf{Y}\,\int_{\mathbf{v}\,,\mathbf{x},\,\mathbf{y},\,\mathbf{z},\,\mathbf{z}^{\prime}}K_{JJSSJ}(\mathbf{v},\,\mathbf{x},\,\mathbf{y},\,\mathbf{z},\,\mathbf{z}^{\prime})f^{abc}\rho^{b}(\mathbf{y})\rho^{a}(\mathbf{x})\rho^{c}(\mathbf{v}).
 \end{equation}

\section{Supplementary for Section 3}\label{supsect3}
\subsection{Supplement for Computation of $\left|\psi_{g}^{1}\right\rangle$} \label{swf4}
$\left|\psi_{g}^{1}\right\rangle$ is defined in (\ref{gone}). The calculation in this section is identical to \cite{Weigertrun}. After inserting the relevant matrix elements, (\ref{g}) and (\ref{gqq1}):
\begin{eqnarray}\label{insq}
&&\left|\psi_{g}^{1}\right\rangle =-\sum_{f,\lambda_{1},\lambda_{2}}\int_{\Lambda}^{e^{\delta\mathsf{Y}}\Lambda}dk^{+}\,\int_{0}^{k^{+}}dp^{+}\,\int d^{2}\mathbf{k}\, d^{2}\mathbf{p}\,\frac{g^{3}\rho^{a}(-\mathbf{k})\, tr[t^{a}\, t^{d}]\, \mathbf{k}^{i}}{32\pi^{9/2}\left(\frac{\mathbf{p}^{2}}{p^{+}}+\frac{(\mathbf{k}-\mathbf{p})^{2}}{(k^{+}-p^{+})}\right)\mathbf{k}^{4}\sqrt{k^{+}}}\nonumber\\
&&\times\chi_{\lambda_{2}}^{\dagger}\left(\frac{2\mathbf{k}^{j}}{k^{+}}-\sigma^{j}\frac{\sigma\cdot\mathbf{p}}{p^{+}}-\frac{\sigma\cdot(\mathbf{k}-\mathbf{p})}{k^{+}-p^{+}}\sigma^{j}\right)\chi_{\lambda_{1}}\\
&&\times\chi_{\lambda_{1}}^{\dagger}\left(\frac{2\mathbf{k}^{i}}{k^{+}}-\frac{\sigma\cdot\mathbf{p}}{p^{+}}\sigma^{i}-\sigma^{i}\frac{\sigma\cdot(\mathbf{k}-\mathbf{p})}{k^{+}-p^{+}}\right)\chi_{\lambda_{2}}\left|g_{j}^{d}(k)\right\rangle .\nonumber\end{eqnarray}
By using
\begin{equation}
\sum_{\lambda}\chi_{\lambda}\chi_{\lambda}^{\dagger}=\left(\begin{array}{cc}
1 & 0\\
0 & 1
\end{array}\right)=I\:,\qquad\qquad\sigma^{i}\sigma^{j}=i\varepsilon^{ij}\sigma^{3}+\delta^{ij}I\:,\qquad\qquad tr\left[t^{a}t^{d}\right]\,=\,\frac{1}{2}\delta^{ad}\:,
 \end{equation}
and
\begin{eqnarray}\label{simpql}
&&\sum_{\lambda}\chi_{\lambda}^{\dagger}\left(\frac{2\mathbf{k}^{j}}{k^{+}}-\sigma^{j}\frac{\sigma\cdot \mathbf{p}}{p^{+}}-\frac{\sigma\cdot(\mathbf{k}-\mathbf{p})}{k^{+}-p^{+}}\sigma^{j}\right)\left(\frac{2\mathbf{k}^{i}}{k^{+}}-\sigma^{i}\frac{\sigma\cdot(\mathbf{k}-\mathbf{p})}{k^{+}-p^{+}}-\frac{\sigma\cdot \mathbf{p}}{p^{+}}\sigma^{i}\right)\chi_{\lambda}\nonumber\\
&&=\sum_{\lambda}\chi_{\lambda}^{\dagger}\left(4\frac{\mathbf{k}^{i}\mathbf{k}^{j}}{(k^{+})^{2}}-2\frac{\mathbf{k}^{i}\mathbf{p}^{j}+\mathbf{k}^{j}\mathbf{p}^{i}}{k^{+}p^{+}}-2\frac{\mathbf{k}^{i}(\mathbf{k}^{j}-\mathbf{p}^{j})+\mathbf{k}^{j}(\mathbf{k}^{i}-\mathbf{p}^{i})}{k^{+}(k^{+}-p^{+})}+\frac{\mathbf{p}^{i}\mathbf{p}^{j}+\varepsilon^{il}\varepsilon^{jk}\mathbf{p}^{k}\mathbf{p}^{l}}{(p^{+})^{2}}\right.\nonumber\\
&&+\frac{\mathbf{p}^{i}(\mathbf{k}^{j}-\mathbf{p}^{j})+\mathbf{p}^{j}(\mathbf{k}^{i}-\mathbf{p}^{i})-\varepsilon^{il}\varepsilon^{jk}(\mathbf{k}^{k}-\mathbf{p}^{k})\mathbf{p}^{l}-\varepsilon^{il}\varepsilon^{jk}(\mathbf{k}^{l}-\mathbf{p}^{l})\mathbf{p}^{k}}{p^{+}(k^{+}-p^{+})}\nonumber\\
&&\left.+\frac{(\mathbf{k}^{i}-\mathbf{p}^{i})(\mathbf{k}^{j}-\mathbf{p}^{j})+\varepsilon^{il}\varepsilon^{jk}(\mathbf{k}^{i}-\mathbf{p}^{i})(\mathbf{k}^{j}-\mathbf{p}^{j})}{(k^{+}-p^{+})^{2}}\right)\chi_{\lambda}\\
&&=\frac{2\widetilde{\mathbf{p}}^{i}\widetilde{\mathbf{p}}^{j}}{(k^{+})^{2}}\left(\frac{1}{\xi^{2}}+\frac{1}{(1-\xi)^{2}}-\frac{2}{\xi(1-\xi)}\right)+\frac{2\varepsilon^{il}\varepsilon^{jk}\widetilde{\mathbf{p}}^{k}\widetilde{\mathbf{p}}^{l}}{(k^{+})^{2}}\left(\frac{1}{\xi^{2}}+\frac{1}{(1-\xi)^{2}}+\frac{2}{\xi(1-\xi)}\right)\nonumber\\
&&=2\frac{(4\xi^{2}-4\xi+1)\widetilde{\mathbf{p}}^{i}\widetilde{\mathbf{p}}^{j}+\varepsilon^{il}\varepsilon^{jk}\widetilde{\mathbf{p}}^{k}\widetilde{\mathbf{p}}^{l}}{\xi^{2}(1-\xi)^{2}(k^{+})^{2}}.\nonumber\end{eqnarray}
After change of variables according to (\ref{chanvar}) and (\ref{chanvar2}):
\begin{equation}\begin{split}\label{iqd}
&\left|\psi_{g}^{1}\right\rangle \,=\,-\int_{\Lambda}^{e^{\delta\mathsf{Y}}\Lambda}dk^{+}\,\int_{0}^{1}d\xi\,\int d^{2}\mathbf{k}\, d^{2}\widetilde{\mathbf{p}}\,\\
&\times\frac{g^{3}\, N_{f}\,\rho^{a}(-\mathbf{k})\,\left((4\xi^{2}-4\xi+1)\widetilde{\mathbf{p}}^{i}\widetilde{\mathbf{p}}^{j}+\varepsilon^{il}\varepsilon^{jk}\widetilde{\mathbf{p}}^{k}\widetilde{\mathbf{p}}^{l}\right)\mathbf{k}^{i}}{32\pi^{9/2}\xi(1-\xi)\mathbf{k}^{4}\left(\xi(1-\xi)\mathbf{k}^{2}+\widetilde{\mathbf{p}}^{2}\right)\sqrt{k^{+}}}\left|g_{j}^{a}(k)\right\rangle .\\
\end{split}\end{equation}
The UV divergent $p$-integral is regularised via dimensional regularisation according to (\ref{dimregm}). Then, using (\ref{por}) together with the following identity:
\begin{equation}\begin{split}\label{epsi}
\varepsilon^{il}\varepsilon^{jk}=\delta^{ij}\delta^{kl}-\delta^{ik}\delta^{jl},
\end{split}\end{equation}
we arrive at:
\begin{eqnarray}\label{qdre}
&&\left|\psi_{g}^{1}\right\rangle \,\\
&&=\,-\mu^{2\epsilon}\int_{\Lambda}^{e^{\delta\mathsf{Y}}\Lambda}dk^{+}\,\int\frac{d^{d}\mathbf{k}}{(2\pi)^{d}}\,\frac{d^{d}\widetilde{\mathbf{p}}}{(2\pi)^{d}}\,\int_{0}^{1}d\xi\,\frac{g^{3}\, N_{f}\,\rho^{a}(-\mathbf{k})\,(4\xi^{2}-4\xi+d)\,\widetilde{\mathbf{p}}^{2}\, \mathbf{k}^{i}}{2\sqrt{\pi k^{+}}\mathbf{k}^{4}\left(\xi(1-\xi)\mathbf{k}^{2}+\widetilde{\mathbf{p}}^{2}\right)\xi(1-\xi)d}\left|g_{i}^{a}(k)\right\rangle. \nonumber\end{eqnarray}\\
Integration over $\widetilde{\mathbf{p}}$ is done with the aid of (\ref{dim1}):
\begin{equation}\begin{split}\label{qdreg}
&\left|\psi_{g}^{1}\right\rangle =-\mu^{2\epsilon}\int_{\Lambda}^{e^{\delta\mathsf{Y}}\Lambda}dk^{+}\,\int\frac{d^{d}\mathbf{k}}{(2\pi)^{d}}\,\int_{0}^{1}d\xi\,\frac{g^{3}\, N_{f}\,\rho^{a}(-\mathbf{k})\,(4\xi^{2}-4\xi+d)\, \mathbf{k}^{i}}{2\sqrt{\pi k^{+}}\mathbf{k}^{2}d}\\
&\times\frac{1}{(4\pi)^{d/2}}\frac{d}{2}\Gamma\left(-\frac{d}{2}\right)\left(\frac{1}{\xi(1-\xi)\mathbf{k}^{2}}\right)^{1-\frac{d}{2}}\left|g_{i}^{a}(k)\right\rangle .\\
\end{split}\end{equation}
After expanding with (\ref{gamexp}) and taking $\epsilon\,\rightarrow\,0$ limit, the result becomes:
\begin{equation}\begin{split}
&\left|\psi_{g}^{1}\right\rangle =-\int_{\Lambda}^{e^{\delta\mathsf{Y}}\Lambda}dk^{+}\,\int d^{2}\mathbf{k}\,\int_{0}^{1}d\xi\,\frac{g^{3}\, N_{f}\,\rho^{a}(-\mathbf{k})\, \mathbf{k}^{i}}{64\pi^{7/2}\sqrt{k^{+}}\mathbf{k}^{2}}\\
&\times\left((4\xi^{2}-4\xi+2)\left[-\frac{2}{\epsilon}+\ln\left(\frac{\xi(1-\xi)\mathbf{k}^{2}}{\mu^{2}}\right)+\gamma-\ln4\pi-1\right]+2\right)\left|g_{i}^{a}(k)\right\rangle .\\
\end{split}\end{equation}
Finally, integrating over $\xi$ according to (\ref{int.1}):
 \begin{eqnarray}\label{qdre2}
&&\left|\psi_{g}^{1}\right\rangle \\
&&=-\int_{\Lambda}^{e^{\delta\mathsf{Y}}\Lambda}dk^{+}\,\int d^{2}\mathbf{k}\,\frac{g^{3}\, N_{f}\,\rho^{a}(-\mathbf{k})\, \mathbf{k}^{i}}{64\pi^{7/2}\sqrt{k^{+}}\mathbf{k}^{2}}\left(\frac{4}{3}\left[-\frac{2}{\epsilon}+\ln\left(\frac{\mathbf{k}^{2}}{\mu^{2}}\right)+\gamma-\ln4\pi\right]-\frac{20}{9}\right)\left|g_{i}^{a}(k)\right\rangle .\nonumber\end{eqnarray}
which we can write as in (\ref{grho1}).

\subsection{Supplement for Computation of $\left|\psi_{g}^{2}\right\rangle$} \label{swf5}
$\left|\psi_{g}^{2}\right\rangle$ is defined in (\ref{gtwo}). After inserting the matrix elements, (\ref{g}) and (\ref{ggg_split}),
\begin{eqnarray}\label{glop}
&&\left|\psi_{g}^{2}\right\rangle =-\int_{\Lambda}^{e^{\delta\mathsf{Y}}\Lambda}dk^{+}\,\int_{\Lambda}^{k^{+}-\Lambda}dp^{+}\,\int d^{2}\mathbf{k}\, d^{2}\mathbf{p}\,\\
&&\times\frac{g^{2}f^{abc}f^{dbc}}{\left(\frac{\mathbf{k}^{2}}{2k^{+}}\right)\left(\frac{\mathbf{p}^{2}}{2p^{+}}+\frac{(\mathbf{k}-\mathbf{p})^{2}}{2(k^{+}-p^{+})}\right)\left(\frac{\mathbf{k}^{2}}{2k^{+}}\right)16(2\pi)^{3}k^{+}p^{+}(k^{+}-p^{+})}\nonumber\\
&&\times\left(\left[2\mathbf{p}^{l}-\frac{2p^{+}}{k^{+}}\mathbf{k}^{l}\right]\delta_{jk}+\left[2\mathbf{k}^{j}-\frac{2k^{+}}{p^{+}}\mathbf{p}^{j}\right]\delta_{lk}+\left[\frac{k^{+}+p^{+}}{k^{+}-p^{+}}(\mathbf{k}^{k}-\mathbf{p}^{k})-\mathbf{k}^{k}-\mathbf{p}^{k}\right]\delta_{lj}\right)\nonumber\\
&&\times\left(\left[2\mathbf{p}^{i}-\frac{2p^{+}}{k^{+}}\mathbf{k}^{i}\right]\delta_{jk}+\left[2\mathbf{k}^{j}-\frac{2k^{+}}{p^{+}}\mathbf{p}^{j}\right]\delta_{ik}+\left[\frac{k^{+}+p^{+}}{k^{+}-p^{+}}(\mathbf{k}^{k}-\mathbf{p}^{k})-\mathbf{k}^{k}-\mathbf{p}^{k}\right]\delta_{ij}\right)\nonumber\\
&&\times\left(\frac{g\rho^{a}(-\mathbf{k})\mathbf{k}^{i}}{4\pi^{3/2}|k^{+}|^{3/2}}\right)\left|g_{l}^{d}(k)\right\rangle  .\nonumber\end{eqnarray}
By changing variables according to (\ref{chanvar}) and (\ref{chanvar2}) we can write the last result as:
\begin{equation}\begin{split}
&\left|\psi_{g}^{2}\right\rangle =-\int_{\Lambda}^{e^{\delta\mathsf{Y}}\Lambda}dk^{+}\,\int d^{2}\mathbf{k}\, d^{2}\widetilde{\mathbf{p}}\int_{\frac{\Lambda}{k^{+}}}^{1-\frac{\Lambda}{k^{+}}}d\xi\,\frac{g^{3}f^{abc}f^{dbc}\rho^{a}(-\mathbf{k})\mathbf{k}^{i}}{8(2\pi)^{3}\pi^{3/2}\xi(1-\xi)\mathbf{k}^{4}\left(\mathbf{k}^{2}\xi(1-\xi)+\widetilde{\mathbf{p}}^{2}\right)\sqrt{k^{+}}}\\
&\times\left(2\widetilde{\mathbf{p}}^{l}\delta_{jk}-\frac{2}{\xi}\widetilde{\mathbf{p}}^{j}\delta_{lk}-\frac{2}{1-\xi}\widetilde{\mathbf{p}}^{k}\delta_{lj}\right)\left(2\widetilde{\mathbf{p}}^{i}\delta_{jk}-\frac{2}{\xi}\widetilde{\mathbf{p}}^{j}\delta_{ik}-\frac{2}{1-\xi}\widetilde{\mathbf{p}}^{k}\delta_{ij}\right)\left|g^{d}_{l}(k)\right\rangle .\\
\end{split}\end{equation}
After some algebra:
\begin{eqnarray}
&&\left|\psi_{g}^{2}\right\rangle =-\int_{\Lambda}^{e^{\delta\mathsf{Y}}\Lambda}dk^{+}\,\int d^{2}\mathbf{k}\, d^{2}\widetilde{\mathbf{p}}\int_{\frac{\Lambda}{k^{+}}}^{1-\frac{\Lambda}{k^{+}}}d\xi\,\frac{g^{3}N_{c}\rho^{a}(-\mathbf{k})\mathbf{k}^{i}\widetilde{\mathbf{p}}^{2}}{16\pi{}^{9/2}\mathbf{k}^{4}\left(\mathbf{k}^{2}\xi(1-\xi)+\widetilde{\mathbf{p}}^{2}\right)\sqrt{k^{+}}}\nonumber\\
&&\times\left(1+\frac{1}{\xi^{2}}+\frac{1}{(1-\xi)^{2}}\right)\left|g_{i}^{a}(k)\right\rangle .\end{eqnarray}

By replacing the measure according to (\ref{dimregm}):
\begin{eqnarray}\label{glop2}
&&\left|\psi_{g}^{2}\right\rangle =-\mu^{2\epsilon}\int_{\Lambda}^{e^{\delta\mathsf{Y}}\Lambda}dk^{+}\,\int\frac{d^{d}\mathbf{k}}{(2\pi)^{d}}\,\frac{d^{d}\widetilde{\mathbf{p}}}{(2\pi)^{d}}\,\int_{\frac{\Lambda}{k^{+}}}^{1-\frac{\Lambda}{k^{+}}}d\xi\,\frac{g^{3}N_{c}\rho^{a}(-\mathbf{k})\mathbf{k}^{i}\widetilde{\mathbf{p}}^{2}}{\sqrt{\pi}\mathbf{k}^{4}\left(\mathbf{k}^{2}\xi(1-\xi)+\widetilde{\mathbf{p}}^{2}\right)\sqrt{k^{+}}}\nonumber\\
&&\times\left(1+\frac{1}{\xi^{2}}+\frac{1}{(1-\xi)^{2}}\right)\left|g_{i}^{a}(k)\right\rangle .\end{eqnarray}
We can now compute the integration over $\mathbf{p}$ with the aid of the integral (\ref{dim1}):
\begin{eqnarray}\label{glop3}
&&\left|\psi_{g}^{2}\right\rangle =-\mu^{2\epsilon}\int_{\Lambda}^{e^{\delta\mathsf{Y}}\Lambda}dk^{+}\,\int\frac{d^{d}\mathbf{k}}{(2\pi)^{d}}\,\int_{\frac{\Lambda}{k^{+}}}^{1-\frac{\Lambda}{k^{+}}}d\xi\,\frac{g^{3}N_{c}\rho^{a}(-\mathbf{k})\mathbf{k}^{i}}{\sqrt{\pi}\mathbf{k}^{2}\sqrt{k^{+}}}\\
&&\times\left(\frac{d}{2(4\pi)^{d/2}}\Gamma\left(-\frac{d}{2}\right)\left(\xi(1-\xi)\mathbf{k}^{2}\right){}^{d/2-1}\right)\left(\xi(1-\xi)-2+\frac{1}{\xi(1-\xi)}\right)\left|g_{i}^{a}(k)\right\rangle .\nonumber \end{eqnarray}
After expanding with the aid of (\ref{gamexp}) taking the $\epsilon\,\rightarrow\,0$ limit and, the last result becomes:
\begin{eqnarray}\label{glop4}
&&\left|\psi_{g}^{2}\right\rangle \,=\,-\int_{\Lambda}^{e^{\delta\mathsf{Y}}\Lambda}dk^{+}\,\int d^{2}\mathbf{k}\,\int_{\frac{\Lambda}{k^{+}}}^{1-\frac{\Lambda}{k^{+}}}d\xi\,\frac{g^{3}N_{c}\rho^{a}(-\mathbf{k})\mathbf{k}^{i}}{16\pi^{7/2}\mathbf{k}^{2}\sqrt{k^{+}}}\left(\xi(1-\xi)-2+\frac{1}{\xi(1-\xi)}\right)\nonumber\\
&&\times\left(-\frac{2}{\epsilon}+\ln\,\left(\frac{\mathbf{k}^{2}\xi(1-\xi)}{\mu_{\overline{MS}}^{2}}\right)\right)\left|g_{i}^{a}(k)\right\rangle .\end{eqnarray}
The relevant integral for the integration over $\xi$ is (\ref{int.8}), the result after the integration is:

 \begin{eqnarray}
 \left|\psi_{g\:\rho}^{2}\right\rangle &\equiv&\left|\psi_{g}^{2}\right\rangle\,=\,\int_{\Lambda}^{e^{\delta\mathsf{Y}}\Lambda}dk^{+}\,\int d^{2}\mathbf{k}\,\frac{g^{3}N_{c}\rho^{a}(-\mathbf{k})\mathbf{k}^{i}}{32\pi^{7/2}\mathbf{k}^{2}\sqrt{k^{+}}}\\
&\times&\left(\left[\frac{11}{3}+4\ln\,\left(\frac{\Lambda}{k^{+}}\right)\right]\left[-\frac{2}{\epsilon}+\ln\left(\frac{\mathbf{k}^{2}}{\mu_{\overline{MS}}^{2}}\right)\right]+2\ln^{2}\,\left(\frac{\Lambda}{k^{+}}\right)-\frac{67}{9}+\frac{2\pi^{2}}{3}\right)\left|g_{i}^{a}(k)\right\rangle .\nonumber
\end{eqnarray}

\subsection{Supplement for Computation of $\left|\psi_{g}^{3}\right\rangle$} \label{swf6}
$\left|\psi_{g}^{3}\right\rangle$ is defined in (\ref{gthreed}). After inserting the matrix elements, (\ref{g}) and (\ref{gg2}):
 \begin{equation}\begin{split}\label{sta3d}
&\left|\psi_{g}^{3d}\right\rangle =\int_{\Lambda}^{e^{\delta\mathsf{Y}}\Lambda}dk^{+}\,\int_{\Lambda}^{k^{+}-\Lambda}dp^{+}\,\int d^{2}\mathbf{k}\, d^{2}\mathbf{p}\:\frac{1}{\left(\frac{\mathbf{p}^{2}}{2p^{+}}\right)\left(\frac{\mathbf{k}^{2}}{2k^{+}}\right)}\\
&\times\left(\frac{ig^{2}f^{abc}(p^{+}+k^{+})\rho^{c}(-\mathbf{k}+\mathbf{p})\delta_{j}^{i}}{2(2\pi)^{3}\sqrt{k^{+}p^{+}}(k^{+}-p^{+})^{2}}\right)\left(\frac{g\rho^{a}(-\mathbf{p})\mathbf{p}^{i}}{4\pi^{3/2}|p^{+}|^{3/2}}\right)\left|g_{j}^{b}(k)\right\rangle .\\
\end{split}\end{equation}
By changing variables according to (\ref{chanvar}), we arrive at: 
\begin{equation}\begin{split}
&\left|\psi_{g}^{3d}\right\rangle =\int_{\Lambda}^{e^{\delta\mathsf{Y}}\Lambda}dk^{+}\,\int d^{2}\mathbf{k}\, d^{2}\mathbf{p}\,\int_{\frac{\Lambda}{k^{+}}}^{1-\frac{\Lambda}{k^{+}}}d\xi\,\frac{ig^{3}f^{abc}(1+\xi)\rho^{c}(-\mathbf{k}+\mathbf{p})\rho^{a}(-\mathbf{p})\mathbf{p}^{i}}{16\pi^{9/2}(1-\xi)^{2}\xi \mathbf{k}^{2}\mathbf{p}^{2}\sqrt{k^{+}}}\left|g_{i}^{b}(k)\right\rangle. \\
  \end{split}\end{equation}
Separating the last contribution to contribution with one and two $\rho$ operators according to (\ref{decompose}), we notice that the contribution with one $\rho$ operator vanishes after integraion over $\mathbf{p}$. The two $\rho$ part reads:
 \begin{equation}\begin{split}\label{simp3d}
&\left|\psi_{g}^{3d}\right\rangle  =\int_{\Lambda}^{e^{\delta\mathsf{Y}}\Lambda}dk^{+}\,\int d^{2}\mathbf{k}\, d^{2}\mathbf{p}\,\int_{\frac{\Lambda}{k^{+}}}^{1-\frac{\Lambda}{k^{+}}}d\xi\,\frac{ig^{3}f^{abc}(1+\xi)\mathbf{p}^{i}\left\{ \rho^{c}(-\mathbf{k}+\mathbf{p}),\,\rho^{a}(-\mathbf{p})\right\} }{32\pi^{9/2}(1-\xi)^{2}\xi \mathbf{k}^{2}\mathbf{p}^{2}\sqrt{k^{+}}}\left|g_{i}^{b}(k)\right\rangle.  \\
  \end{split}\end{equation}
The relevant integral for the integration over $\xi$ is (\ref{int.5}), the result after the integration appears in (\ref{grho3}). 

\subsection{Supplement for Computation of $\left|\psi_{g}^{4}\right\rangle $ and $\left|\psi_{g}^{5}\right\rangle $} \label{swf8b}
$\left|\psi_{g}^{4}\right\rangle $ is defined in (\ref{gfour}). After inserting the matrix elements, (\ref{gg1}) and (\ref{g2}),
 \begin{equation}\begin{split}\label{sta4}
&\left|\psi_{g}^{4}\right\rangle \,=\,\frac{1}{2}\int_{\Lambda}^{e^{\delta\mathsf{Y}}\Lambda}dk^{+}\,\int_{\Lambda}^{k^{+}-\Lambda}dp^{+}\, dq^{+}\,\int d^{2}\mathbf{k}\, d^{2}\mathbf{p}\, d^{2}\mathbf{q}\frac{1}{\left(\frac{\mathbf{q}^{2}}{2q^{+}}+\frac{(\mathbf{k}-\mathbf{q})^{2}}{2(k^{+}-q^{+})}\right)\left(\frac{\mathbf{p}^{2}}{2p^{+}}\right)}\\
&\times\left(\delta^{dc}\delta_{il}\delta^{(3)}(k-p-q)\frac{g\mathbf{q}^{j}\rho^{b}(\mathbf{q})}{4\pi^{3/2}|q^{+}|^{3/2}}+\delta^{db}\delta_{lj}\delta^{(3)}(q-p)\frac{g(\mathbf{k}^{i}-\mathbf{q}^{i})\rho^{c}(\mathbf{k}-\mathbf{q})}{4\pi^{3/2}|k^{+}-q^{+}|^{3/2}}\right)\\
&\times\left(\frac{ig^{2}f^{abc}(k^{+}-2q^{+})\delta_{ij}\rho^{a}(-\mathbf{k})}{2(2\pi)^{3}\sqrt{(k^{+}-q^{+})q^{+}}(k^{+})^{2}}\right)\left|g_{l}^{d}(p)\right\rangle .\\
 \end{split}\end{equation}
   After integration over $q$:
  \begin{equation}\begin{split}
&\left|\psi_{g}^{4}\right\rangle =-\int_{\Lambda}^{e^{\delta\mathsf{Y}}\Lambda}dk^{+}\,\int_{\Lambda}^{k^{+}-\Lambda}dp^{+}\,\int d^{2}\mathbf{k}\, d^{2}\mathbf{p}\,\\
&\times\frac{ig^{3}f^{abc}(2p^{+}-k^{+})\rho^{c}(\mathbf{k}-\mathbf{p})\rho^{a}(-\mathbf{k})(\mathbf{k}^{i}-\mathbf{p}^{i})}{64\pi^{9/2}\left(\frac{\mathbf{p}^{2}}{2p^{+}}+\frac{(\mathbf{k}-\mathbf{p})^{2}}{2(k^{+}-p^{+})}\right)\left(\frac{\mathbf{p}^{2}}{2p^{+}}\right)(k^{+}(k^{+}-p^{+}))^{2}\sqrt{p^{+}}}\left|g_{i}^{b}(p)\right\rangle .
 \end{split}\end{equation}
After changing variables according to (\ref{chanvar}):
\begin{eqnarray}\label{psif}
\left|\psi_{g}^{4}\right\rangle &=&-\int_{\Lambda}^{e^{\delta\mathsf{Y}}\Lambda}dk^{+}\,\int d^{2}\mathbf{k}\, d^{2}\mathbf{p}\,\int_{\frac{\Lambda}{k^{+}}}^{1-\frac{\Lambda}{k^{+}}}d\xi\,\nonumber\\
&\times&\frac{ig^{3}f^{abc}\rho^{c}(\mathbf{k}-\mathbf{p})\rho^{a}(-\mathbf{k})(\mathbf{k}^{i}-\mathbf{p}^{i})(2\xi-1)\xi^{3/2}}{16\pi^{9/2}\left(\xi(\mathbf{k}-\mathbf{p})^{2}+(1-\xi)\mathbf{p}^{2}\right)\mathbf{p}^{2}(1-\xi)\sqrt{k^{+}}}\left|g_{i}^{b}(\xi k^{+},\mathbf{p})\right\rangle .\end{eqnarray}

Now let us work out the case of $\left|\psi_{g}^{5}\right\rangle$ as defined in (\ref{gsix}). By inserting the relevant matrix elements, (\ref{g}), (\ref{g2}) and (\ref{ggg_split}),

 \begin{eqnarray}\label{sta5}
&&\left|\psi_{g}^{5}\right\rangle =-\int_{\Lambda}^{e^{\delta\mathsf{Y}}\Lambda}dk^{+}\,dp^{+}\,\int_{\Lambda}^{k^{+}-\Lambda}dq^{+}\,\int d^{2}\mathbf{k}\,d^{2}\mathbf{p}\,d^{2}\mathbf{q}\frac{1}{\left(\frac{\mathbf{p}^{2}}{2p^{+}}\right)\left(\frac{(\mathbf{k}-\mathbf{q})^{2}}{2(k^{+}-q^{+})}+\frac{\mathbf{q}^{2}}{2q^{+}}\right)\left(\frac{\mathbf{k}^{2}}{2k^{+}}\right)}\nonumber\\
&&\times\left(\delta^{dc}\delta^{km}\delta^{(3)}(k-p-q)\frac{g\mathbf{q}^{j}\rho^{b}(\mathbf{q})}{4\pi^{3/2}|q^{+}|^{3/2}}+\delta^{db}\delta^{jm}\delta^{(3)}(p-q)\frac{g(\mathbf{k}^{k}-\mathbf{q}^{k})\rho^{c}(\mathbf{k}-\mathbf{q})}{4\pi^{3/2}|k^{+}-q^{+}|^{3/2}}\right)\nonumber\\
&&\times\left(\left[2\mathbf{q}^{i}-\frac{2q^{+}}{k^{+}}\mathbf{k}^{i}\right]\delta_{jk}+\left[2\mathbf{k}^{j}-\frac{2k^{+}}{q^{+}}\mathbf{q}^{j}\right]\delta_{ik}+\left[\frac{k^{+}+q^{+}}{k^{+}-q^{+}}(\mathbf{k}^{k}-\mathbf{q}^{k})-\mathbf{k}^{k}-\mathbf{q}^{k}\right]\delta_{ij}\right)\nonumber\\
&&\times\frac{igf^{abc}}{16\pi^{3/2}\sqrt{k^{+}q^{+}(k^{+}-q^{+})}}\left(\frac{g\mathbf{k}^{i}\rho^{a}(-\mathbf{k})}{4\pi^{3/2}|k^{+}|^{3/2}}\right)\left|g_{m}^{d}(p)\right\rangle .\end{eqnarray}

After performing the multiplications, and changing variables according to (\ref{chanvar}), we arrive at:
 \begin{eqnarray}\label{psi5mul}
&&\left|\psi_{g}^{5}\right\rangle =\int_{\Lambda}^{e^{\delta\mathsf{Y}}\Lambda}dk^{+}\,\int d^{2}\mathbf{k}\, d^{2}\mathbf{p}\int_{\frac{\Lambda}{k^{+}}}^{1-\frac{\Lambda}{k^{+}}}d\xi\,\frac{ig^{3}f^{abc}\rho^{c}(\mathbf{k}-\mathbf{p})\rho^{a}(-\mathbf{k})\xi^{3/2}}{16\pi^{9/2}\mathbf{k}^{2}\mathbf{p}^{2}\left(\xi(\mathbf{k}-\mathbf{p})^{2}+(1-\xi)\mathbf{p}^{2}\right)(1-\xi)\sqrt{k^{+}}}\nonumber\\
&&\times\bigg(\left[-2\mathbf{k}\cdot \mathbf{p}+2\xi \mathbf{k}^{2}\right](\mathbf{k}^{j}-\mathbf{p}^{j})+\left[-2\mathbf{k}^{j}+\frac{2}{\xi}\mathbf{p}^{j}\right]\mathbf{k}\cdot(\mathbf{k}-\mathbf{p})\nonumber\\
&&\left.+\left[\mathbf{k}^{2}-\mathbf{p}^{2}-\frac{1+\xi}{1-\xi}(\mathbf{k}-\mathbf{p})^{2}\right]\mathbf{k}^{j}\right)\left|g_{j}^{b}(\xi k^{+},\, \mathbf{p})\right\rangle. \end{eqnarray}
Equivalently, we can write the last result as:
\begin{equation}\begin{split}\label{psi5per}
 &\left|\psi_{g}^{5}\right\rangle =\int_{\Lambda}^{e^{\delta\mathsf{Y}}\Lambda}dk^{+}\,\int d^{2}\mathbf{k}\, d^{2}\mathbf{p}\int_{\frac{\Lambda}{k^{+}}}^{1-\frac{\Lambda}{k^{+}}}d\xi\,\frac{ig^{3}f^{abc}\rho^{c}(\mathbf{k}-\mathbf{p})\rho^{a}(-\mathbf{k})\xi^{3/2}}{16\pi^{9/2}\mathbf{k}^{2}\mathbf{p}^{2}\left(\xi(\mathbf{k}-\mathbf{p})^{2}+(1-\xi)\mathbf{p}^{2}\right)(1-\xi)\sqrt{k^{+}}}\\
&\times\left(\left[(2\xi-1)\mathbf{k}^{2}-\mathbf{p}^{2}-\frac{1+\xi}{1-\xi}(\mathbf{k}-\mathbf{p})^{2}\right]\mathbf{k}^{j}+\left[\frac{\xi-1}{\xi}\mathbf{k}\cdot \mathbf{p}+\frac{1-\xi^{2}}{\xi}\mathbf{k}^{2}\right]2\mathbf{p}^{j}\right)\left|g_{j}^{b}(\xi k^{+},\, \mathbf{p})\right\rangle.  
\end{split}\end{equation}
 By adding together $\left|\psi_{g}^{4}\right\rangle $ and $\left|\psi_{g}^{5}\right\rangle $, we arrive at:
\begin{eqnarray}\label{foplusfi}
&&\left|\psi_{g}^{4+5}\right\rangle \,\equiv\,\left|\psi_{g}^{4}\right\rangle \,+\,\left|\psi_{g}^{5}\right\rangle \,\\
&&=\int_{\Lambda}^{e^{\delta\mathsf{Y}}\Lambda}dk^{+}\,\int d^{2}\mathbf{k}\, d^{2}\mathbf{p}\,\int_{\frac{\Lambda}{k^{+}}}^{1-\frac{\Lambda}{k^{+}}}d\xi\,\frac{ig^{3}f^{abc}\rho^{c}(\mathbf{k}-\mathbf{p})\rho^{a}(-\mathbf{k})\xi^{3/2}}{16\pi^{9/2}\mathbf{k}^{2}\mathbf{p}^{2}\left(\xi(\mathbf{k}-\mathbf{p})^{2}+(1-\xi)\mathbf{p}^{2}\right)(1-\xi)\sqrt{k^{+}}}\nonumber\\
&&\times\left(\left[\frac{2}{\xi}\left(\mathbf{k}^{2}-\mathbf{k}\cdot \mathbf{p}\right)-\mathbf{k}^{2}+2\mathbf{k}\cdot \mathbf{p}\right]\mathbf{p}^{j}-\left[\mathbf{p}^{2}+\frac{1+\xi}{1-\xi}(\mathbf{k}-\mathbf{p})^{2}\right]\mathbf{k}^{j}\right)\left|g_{j}^{b}(\xi k^{+},\, \mathbf{p})\right\rangle. \nonumber\end{eqnarray}\\
from (\ref{foplusfi}) we can directly deduce the part which is proportional to two $\rho$ operators. Below, we will focus on the part which involve one $\rho$ operator only, which can be isolated via the prescription (\ref{decompose}).\\ \\
  $\bullet$ \textit{\textbf{One $\rho$ part}}\\
    Let us now focus on the contribution which involves only one $\rho$ operator:
\begin{eqnarray}
&&\left|\psi_{g}^{4+5}\right\rangle \,\\
&&=\,-\int_{\Lambda}^{e^{\delta\mathsf{Y}}\Lambda}dk^{+}\,\int d^{2}\mathbf{k}\, d^{2}\mathbf{p}\,\int_{\frac{\Lambda}{k^{+}}}^{1-\frac{\Lambda}{k^{+}}}d\xi\,\frac{g^{3}N_{c}\rho^{a}(-\mathbf{p})\xi^{3/2}}{32\pi^{9/2}\mathbf{k}^{2}\mathbf{p}^{2}\left(\xi(\mathbf{k}-\mathbf{p})^{2}+(1-\xi)\mathbf{p}^{2}\right)(1-\xi)\sqrt{k^{+}}}\nonumber\\
&&\times\left(\left[\frac{2}{\xi}\left(\mathbf{k}^{2}-\mathbf{k}\cdot \mathbf{p}\right)-\mathbf{k}^{2}+2\mathbf{k}\cdot \mathbf{p}\right]\mathbf{p}^{j}-\left[\mathbf{p}^{2}+\frac{1+\xi}{1-\xi}(\mathbf{k}-\mathbf{p})^{2}\right]\mathbf{k}^{j}\right)\left|g_{j}^{a}(\xi k^{+},\, \mathbf{p})\right\rangle.\nonumber\end{eqnarray}\\
We rewrite the denominator of $\left|\psi_{g}^{4+5}\right\rangle  $ by using Feynman parameter. We introduce the following definitions:
 \begin{equation}\label{alph}
\alpha\,\equiv\,1-x+x\xi\:,\qquad\qquad\Delta\equiv\frac{x(1-x+x\xi-x\xi^{2})}{\alpha^{2}}.
 \end{equation}
and perform the shift $\mathbf{k}\:\longrightarrow\: \mathbf{k}\,+\,\frac{x\xi}{\alpha}\mathbf{p}$. Then we arrive at:
\begin{eqnarray}
&&\left|\psi_{g}^{4+5}\right\rangle \nonumber\\
&&=\,-\int_{\Lambda}^{e^{\delta\mathsf{Y}}\Lambda}dk^{+}\,\int d^{2}\mathbf{k}\, d^{2}\mathbf{p}\,\int_{\frac{\Lambda}{k^{+}}}^{1-\frac{\Lambda}{k^{+}}}d\xi\,\int_{0}^{1}dx\,\frac{g^{3}N_{c}\rho^{a}(-\mathbf{p})\xi^{3/2}}{32\pi^{9/2}\sqrt{k^{+}}(1-\xi)\mathbf{p}^{2}\left(\mathbf{k}^{2}+\Delta \mathbf{p}^{2}\right)^{2}}\nonumber\\
&&\times\left(\left[\frac{2}{\xi}\left(\left(\mathbf{k}\,+\,\frac{x\xi}{\alpha}\mathbf{p}\right)^{2}-\left(\mathbf{k}\,+\,\frac{x\xi}{\alpha}\mathbf{p}\right)\cdot \mathbf{p}\right)-\left(\mathbf{k}\,+\,\frac{x\xi}{\alpha}\mathbf{p}\right)^{2}+2\left(\mathbf{k}\,+\,\frac{x\xi}{\alpha}\mathbf{p}\right)\cdot \mathbf{p}\right]\mathbf{p}^{j}\right.\nonumber\\
&&\left.-\left[\mathbf{p}^{2}+\frac{1+\xi}{1-\xi}\left(\mathbf{k}\,+\,\left(\frac{x\xi}{\alpha}-1\right)\mathbf{p}\right)^{2}\right]\left(\mathbf{k}^{j}\,+\,\frac{x\xi}{\alpha}\mathbf{p}^{j}\right)\right)\left|g_{j}^{a}(\xi k^{+},\, \mathbf{p})\right\rangle.\end{eqnarray}
 After changing the measure according to (\ref{dimregm}), and using the relation (\ref{por}):
 \begin{eqnarray}
&&\left|\psi_{g\:\rho}^{4+5}\right\rangle \,=\,-\mu^{2\epsilon}\int_{\Lambda}^{e^{\delta\mathsf{Y}}\Lambda}dk^{+}\,\int\frac{d^{d}\mathbf{k}}{(2\pi)^{d}}\,\frac{d^{d}\widetilde{\mathbf{p}}}{(2\pi)^{d}}\,\int_{\frac{\Lambda}{k^{+}}}^{1-\frac{\Lambda}{k^{+}}}d\xi\,\int_{0}^{1}dx\,\nonumber\\
&&\times\frac{g^{3}N_{c}\rho^{a}(-\mathbf{p})\xi^{3/2}\mathbf{p}^{j}}{2\sqrt{\pi k^{+}}(1-\xi)\alpha^{2}\mathbf{p}^{2}\left(\mathbf{k}^{2}+\Delta\mathbf{p}^{2}\right)^{2}}\left(\bigg[\left(\frac{2}{\xi}-1+\frac{2}{d}\frac{1+\xi}{1-\xi}\frac{1-x}{\alpha}-\frac{x\xi}{\alpha}\frac{1+\xi}{1-\xi}\right)\mathbf{k}^{2}\right.\nonumber\\
&&+\left(1-\frac{2}{\xi}-\frac{(1+\xi)(1-x)}{\alpha(1-\xi)}\right)\frac{x(1-x)\xi}{\alpha^{2}}\mathbf{p}^{2}\bigg]\mathbf{p}^{j}-\bigg[\bigg(1+\frac{1+\xi}{1-\xi}\frac{x-1}{\alpha}\\
&&\left.-\frac{2}{d}\left(\frac{2-\xi}{\xi}\frac{x\xi}{\alpha}-\frac{x\xi}{\alpha}\frac{1+\xi}{1-\xi}\frac{x-1}{\alpha}-\frac{1-\xi}{\xi}\right)\bigg)\mathbf{p}^{2}+\frac{1+\xi}{1-\xi}\mathbf{k}^{2}\bigg]\mathbf{k}^{j}\right)\left|g_{j}^{a}(\xi k^{+},\,\mathbf{p})\right\rangle .\nonumber\end{eqnarray}
After integration over $\mathbf{k}$ with the aid of (\ref{dim1}) and (\ref{misdim}):
\begin{eqnarray}
&&\left|\psi_{g\:\rho}^{4+5}\right\rangle \,=\,-\mu^{2\epsilon}\int_{\Lambda}^{e^{\delta\mathsf{Y}}\Lambda}dk^{+}\,\int\frac{d^{d}\mathbf{p}}{(2\pi)^{d}}\,\int_{\frac{\Lambda}{k^{+}}}^{1-\frac{\Lambda}{k^{+}}}d\xi\,\int_{0}^{1}dx\,\frac{g^{3}N_{c}\rho^{a}(-\mathbf{p})\xi^{3/2}\mathbf{p}^{j}}{2\sqrt{\pi}(1-\xi)\alpha^{2}\mathbf{p}^{2}\sqrt{k^{+}}}\nonumber\\
&&\times\left(\left[1-\frac{2}{\xi}-\frac{1+\xi}{1-\xi}\frac{1-x}{\alpha}\right]\frac{x(1-x)\xi}{\alpha^{2}}\mathbf{p}^{2}\frac{1}{(4\pi)^{d/2}}\Gamma\left(2-\frac{d}{2}\right)\left(\frac{1}{\Delta}\right)^{2-\frac{d}{2}}\right.\\
&&\left.+\left[\frac{2}{\xi}-1+\frac{2}{d}\frac{1+\xi}{1-\xi}\frac{1-x}{\alpha}-\frac{x\xi}{\alpha}\frac{1+\xi}{1-\xi}\right]\frac{1}{(4\pi)^{d/2}}\frac{d}{2}\Gamma\left(1-\frac{d}{2}\right)\left(\frac{1}{\Delta}\right)^{1-\frac{d}{2}}\right)\left|g_{j}^{a}(\xi k^{+},\, \mathbf{p})\right\rangle. \nonumber\end{eqnarray}
After taking the $\epsilon\,\rightarrow\,0$ limit and expanding with the aid of (\ref{gamexp}), the last result becomes:
  \begin{eqnarray}\label{fff}
&&\left|\psi_{g\:\rho}^{4+5}\right\rangle \,=\,-\int_{\Lambda}^{e^{\delta\mathsf{Y}}\Lambda}dk^{+}\,\int d^{2}\mathbf{p}\,\int_{\frac{\Lambda}{k^{+}}}^{1-\frac{\Lambda}{k^{+}}}d\xi\,\int_{0}^{1}dx\,\frac{g^{3}N_{c}\rho^{a}(-\mathbf{p})\xi^{3/2}\mathbf{p}^{j}}{32\pi^{7/2}(1-\xi)\alpha^{2}\mathbf{p}^{2}\sqrt{k^{+}}}\nonumber\\
&&\times\left(\left[1-\frac{2}{\xi}-\frac{1+\xi}{1-\xi}\frac{1-x}{\alpha}\right]\frac{(1-x)\xi}{1-x+x\xi-x\xi^{2}}-\frac{2}{\xi}+1+\frac{x\xi}{\alpha}\frac{1+\xi}{1-\xi}\right.\\
&&\left.-\left[\frac{2}{\xi}-1+\frac{1+\xi}{1-\xi}\frac{1-x-x\xi}{\alpha}\right]\left[-\frac{2}{\epsilon}+\ln\left(\frac{x(1-x+x\xi-x\xi^{2})\mathbf{p}^{2}}{\alpha^{2}\mu_{\overline{MS}}^{2}}\right)\right]\right)\left|g_{j}^{a}(\xi k^{+},\, \mathbf{p})\right\rangle. \nonumber\end{eqnarray}
The integration over $x$ can be performed using the following integral:
  \begin{eqnarray}\label{intx1}
&&\int_{0}^{1}dx\,\frac{1}{(1-x+x\xi)^{2}}\left(\left[1-\frac{2}{\xi}-\frac{1+\xi}{1-\xi}\frac{1-x}{1-x+x\xi}\right]\frac{(1-x)\xi}{1-x+x\xi-x\xi^{2}}-\frac{2}{\xi}+1\right.\nonumber\\
&&\left.+\frac{x\xi}{1-x+x\xi}\frac{1+\xi}{1-\xi}-\left[\frac{2}{\xi}-1+\frac{1+\xi}{1-\xi}\frac{1-x-x\xi}{1-x+x\xi}\right]\left[-\frac{2}{\epsilon}+\ln\left(\frac{x(1-x+x\xi-x\xi^{2})\mathbf{p}^{2}}{(1-x+x\xi)^{2}\mu_{\overline{MS}}^{2}}\right)\right]\right)\nonumber\\
&&=-\frac{1}{\xi^{2}}\left(\left(\xi-2\right)\ln\left(\xi\right)+\left(\xi+1\right)\ln\left(1-\xi\right)-\left(\xi-2\right)\left[-\frac{2}{\epsilon}+\ln\left(\frac{\mathbf{p}^{2}}{\mu_{\overline{MS}}^{2}}\right)\right]\right).\end{eqnarray}
The final result is quoted in (\ref{ffone}) after renaming $\mathbf{p}\leftrightarrow \mathbf{k}$.

\subsection{Supplement for Computation of $\left|\psi_{g}^{6}\right\rangle$} \label{swf8}
$\left|\psi_{g}^{6}\right\rangle$ is defined in (\ref{gfive}). By inserting the relevant matrix elements, (\ref{g}), (\ref{g2}) and (\ref{ggg_split}),  we arrive at:
\begin{eqnarray}\label{stasix}
&&\left|\psi_{g}^{6}\right\rangle \,=\,\int_{\Lambda}^{e^{\delta\mathsf{Y}}\Lambda}dk^{+}\,\int_{\frac{\Lambda}{k^{+}}}^{1-\frac{\Lambda}{k^{+}}}dp^{+}\, dq^{+}\int d^{2}\mathbf{k}\, d^{2}\mathbf{p}\, d^{2}\mathbf{q}\:\nonumber\\
&&\times\frac{igf^{bcd}}{\left(\frac{\mathbf{q}^{2}}{2q^{+}}\right)\left(\frac{\mathbf{p}^{2}}{2p^{+}}+\frac{(\mathbf{k}-\mathbf{p})^{2}}{2(k^{+}-p^{+})}\right)\left(\frac{\mathbf{k}^{2}}{2k^{+}}\right)16\pi^{3/2}\sqrt{k^{+}p^{+}(k^{+}-p^{+})}}\\
&&\times\left[\left(2\mathbf{p}^{l}-\frac{2p^{+}}{k^{+}}\mathbf{k}^{l}\right)\delta_{nj}+\left(2\mathbf{k}^{j}-\frac{2k^{+}}{p^{+}}\mathbf{p}^{j}\right)\delta_{ln}+\left(\frac{k^{+}+p^{+}}{k^{+}-p^{+}}(\mathbf{k}^{n}-\mathbf{p}^{n})-\mathbf{k}^{n}-\mathbf{p}^{n}\right)\delta_{lj}\right]\nonumber\\
&&\times\left(\delta^{ac}\delta^{ni}\delta^{(3)}(q-k+p)\frac{g\mathbf{p}^{j}\rho^{b}(-\mathbf{p})}{4\pi^{3/2}|p^{+}|^{3/2}}+\delta^{ab}\delta^{ni}\delta^{(3)}(q-p)\frac{g(\mathbf{k}^{n}-\mathbf{p}^{n})\rho^{c}(-\mathbf{k}+\mathbf{p})}{4\pi^{3/2}|k^{+}-p^{+}|^{3/2}}\right)\nonumber\\
&&\times\left(\frac{g\rho^{a}(-\mathbf{q})\mathbf{q}^{i}}{4\pi^{3/2}|q^{+}|^{3/2}}\right)\left|g_{l}^{d}(k)\right\rangle .\nonumber\end{eqnarray}
After simplifications the last result becomes:
 \begin{eqnarray}
 &&\left|\psi_{g}^{6}\right\rangle =\int_{\Lambda}^{e^{\delta\mathsf{Y}}\Lambda}dk^{+}\,\int d^{2}\mathbf{k}\, d^{2}\mathbf{p}\,\int_{\frac{\Lambda}{k^{+}}}^{1-\frac{\Lambda}{k^{+}}}d\xi\,\frac{ig^{3}f^{bad}\rho^{b}(-\mathbf{p})\rho^{a}(-\mathbf{k}+\mathbf{p})}{16\pi^{9/2}\xi(\mathbf{k}-\mathbf{p})^{2}\left((1-\xi)\mathbf{p}^{2}+\xi(\mathbf{k}-\mathbf{p})^{2}\right)\mathbf{k}^{2}\sqrt{k^{+}}}\nonumber\\
&&\times\left[\left(2\mathbf{p}^{l}-2\xi \mathbf{k}^{l}\right)\mathbf{p}\cdot(\mathbf{k}-\mathbf{p})+\left(2\mathbf{k}\cdot \mathbf{p}-\frac{2}{\xi}\mathbf{p}^{2}\right)(\mathbf{k}-\mathbf{p})^{l}\right.\\
&&\left.+\left(\frac{1+\xi}{1-\xi}(\mathbf{k}-\mathbf{p})^{2}-\mathbf{k}^{2}+\mathbf{p}^{2}\right)\mathbf{p}^{l}\right]\left|g_{l}^{d}(k)\right\rangle. \nonumber\end{eqnarray}
Which we can equivalently write as in (\ref{stasixsim}). We can now isolate the contribution to one and two $\rho$ via the prescription in (\ref{decompose}).
\\ \\
$\bullet$ \textit{\textbf{One $\rho$ part}}\\
  Let us now focus on the contribution which involves only one $\rho$ operator:
\begin{eqnarray}\label{6rhoi}
&&\left|\psi_{g\:\rho}^{6}\right\rangle\\
&&=\int_{\Lambda}^{e^{\delta\mathsf{Y}}\Lambda}dk^{+}\,\int d^{2}\mathbf{k}\, d^{2}\mathbf{p}\,\int_{\frac{\Lambda}{k^{+}}}^{1-\frac{\Lambda}{k^{+}}}d\xi\,\int_{0}^{1}dx\,\frac{g^{3}f^{bad}f^{bac}\rho^{c}(-\mathbf{k})}{32\pi^{9/2}\xi\left(\left(\mathbf{p}-\alpha \mathbf{k}\right)^{2}+\alpha x(1-\xi)\mathbf{k}^{2}\right)^{2}\mathbf{k}^{2}\sqrt{k^{+}}}\nonumber\\
&&\times\left[\left(-2(1-\xi)\mathbf{k}\cdot \mathbf{p}+\frac{2(1-\xi^{2})}{\xi}\mathbf{p}^{2}\right)\mathbf{k}^{l}+\left(\mathbf{k}^{2}-\frac{1+\xi}{1-\xi}(\mathbf{k}-\mathbf{p})^{2}+\frac{\xi-2}{\xi}\mathbf{p}^{2}\right)\mathbf{p}^{l}\right]\left|g^{d}_{l}(k)\right\rangle. \nonumber\end{eqnarray}\\
The denominators of the last expression can be rewritten by using Feynman parameter. We introduce again the variable $\alpha$ which was defined in (\ref{alph}). After shifting the momenta according to $\mathbf{p}\:\longrightarrow\: \mathbf{p}\,+\,\alpha \mathbf{k}$ we arrive at:
\begin{eqnarray}
&&\left|\psi_{g\:\rho}^{6}\right\rangle \,=\,\int_{\Lambda}^{e^{\delta\mathsf{Y}}\Lambda}dk^{+}\,\int d^{2}\mathbf{k}\, d^{2}\mathbf{p}\,\int_{\frac{\Lambda}{k^{+}}}^{1-\frac{\Lambda}{k^{+}}}d\xi\,\int_{0}^{1}dx\,\frac{g^{3}N_{c}\rho^{a}(-\mathbf{k})}{32\pi^{9/2}\xi\left(\mathbf{p}^{2}\,+\,\alpha x(1-\xi)\mathbf{k}^{2}\right)^{2}\mathbf{k}^{2}\sqrt{k^{+}}}\nonumber\\
&&\times\left(\left(-2(1-\xi)\mathbf{k}\cdot\left(\mathbf{p}+\alpha \mathbf{k}\right)+\frac{2(1-\xi^{2})}{\xi}\left(\mathbf{p}+\alpha \mathbf{k}\right)^{2}\right)\mathbf{k}^{l}\right.\\
&&\left.+\left(\mathbf{k}^{2}-\frac{1+\xi}{1-\xi}\left((1-\alpha)\mathbf{k}-\mathbf{p}\right)^{2}+\frac{\xi-2}{\xi}\left(\mathbf{p}+\alpha \mathbf{k}\right)^{2}\right)\left(\mathbf{p}^{l}+\alpha \mathbf{k}^{l}\right)\right)\left|g_{l}^{a}(k)\right\rangle. \nonumber\end{eqnarray}\\
Changing the measure according to (\ref{dimregm}) and using the relation (\ref{por}) we can rewrite the last result as:
  \begin{eqnarray}
&&\left|\psi_{g\:\rho}^{6}\right\rangle =\mu^{2\epsilon}\int_{\Lambda}^{e^{\delta\mathsf{Y}}\Lambda}dk^{+}\int\frac{d^{d}\mathbf{k}}{(2\pi)^{d}}\frac{d^{d}\mathbf{p}}{(2\pi)^{d}}\int_{\frac{\Lambda}{k^{+}}}^{1-\frac{\Lambda}{k^{+}}}d\xi\,\int_{0}^{1}dx\,\frac{g^{3}\,N_{c}\,\rho^{a}(-\mathbf{k})}{2\sqrt{\pi k^{+}}\xi\left(\mathbf{p}^{2}+\alpha x(1-\xi)\mathbf{k}^{2}\right)^{2}\mathbf{k}^{2}}\nonumber\\
&&\times\left(\left[1-\frac{(1-\alpha)^{2}(1+\xi)}{(1-\xi)}+\frac{2\alpha(1-\alpha)(1+\xi)}{(1-\xi)d}+\frac{(2+d)\alpha^{2}(\xi-2)+4\alpha(1-\xi^{2})}{\xi d}\right.\right.\nonumber\\
&&\left.-\frac{2\xi(1-\xi)}{\xi d}\right]\mathbf{k}^{2}\,\mathbf{p}^{l}+\left[\frac{\alpha(\xi-2)(d+2)+2d(1-\xi^{2})}{\xi d}+\frac{2(1-\alpha)(1+\xi)-\alpha(1+\xi)d}{(1-\xi)d}\right]\mathbf{p}^{2}\,\mathbf{k}^{l}\nonumber\\
&&+\left[\frac{\alpha^{3}(\xi-2)+2\alpha^{2}(1-\xi^{2})}{\xi}-\frac{\alpha(1-\alpha)^{2}(1+\xi)}{1-\xi}+\alpha(2\xi-1)\right]\mathbf{k}^{2}\,\mathbf{k}^{l}\\
&&\left.+\left[\frac{\xi-2}{\xi}-\frac{1+\xi}{1-\xi}\right]\mathbf{p}^{2}\,\mathbf{p}^{l}\right)\,\left|g_{l}^{a}(k)\right\rangle .\nonumber\end{eqnarray}
After integration over $\mathbf{p}$ by using (\ref{dim1}) and (\ref{misdim}):
\begin{eqnarray}
&&\left|\psi_{g\:\rho}^{6}\right\rangle \,=\,\mu^{2\epsilon}\int_{\Lambda}^{e^{\delta\mathsf{Y}}\Lambda}dk^{+}\,\int\frac{d^{d}\mathbf{k}}{(2\pi)^{d}}\,\int_{\frac{\Lambda}{k^{+}}}^{1-\frac{\Lambda}{k^{+}}}d\xi\,\int_{0}^{1}dx\,\frac{g^{3}\, N_{c}\,\rho^{a}(-\mathbf{k})\mathbf{k}^{l}}{2\sqrt{\pi k^{+}}\xi\mathbf{k}^{2}}\left(\frac{\Gamma\left(2-\frac{d}{2}\right)}{(4\pi)^{d/2}}\right.\nonumber\\
&&\times\left(\frac{1}{\alpha x(1-\xi)\mathbf{k}^{2}}\right)^{2-\frac{d}{2}}\left[\frac{\alpha^{3}(\xi-2)+2\alpha^{2}(1-\xi^{2})}{\xi}-\frac{\alpha(1-\alpha)^{2}(1+\xi)}{1-\xi}+(2\xi-1)\alpha\right]\mathbf{k}^{2}\nonumber\\
&&+\frac{\Gamma\left(1-\frac{d}{2}\right)}{2(4\pi)^{d/2}}\left(\frac{1}{\alpha x(1-\xi)\mathbf{k}^{2}}\right)^{1-\frac{d}{2}}\left[\frac{2(1-\xi^{2})d+\alpha(\xi-2)d+2\alpha(\xi-2)}{2\xi}\right.\\
&&\left.\left.+\frac{2(1-\alpha)(1+\xi)-\alpha(1+\xi)d}{2(1-\xi)}\right]\right)\left|g_{l}^{a}(k)\right\rangle .\nonumber\end{eqnarray}
After taking the $\epsilon\,\rightarrow\,0$ limit and expanding with the aid of (\ref{gamexp}), the last result becomes:
\begin{eqnarray}
&&\left|\psi_{g\:\rho}^{6}\right\rangle \,=\,\int_{\Lambda}^{e^{\delta\mathsf{Y}}\Lambda}dk^{+}\,\int d^{2}\mathbf{k}\,\int_{\frac{\Lambda}{k^{+}}}^{1-\frac{\Lambda}{k^{+}}}d\xi\,\int_{0}^{1}dx\,\frac{g^{3}\, N_{c}\,\rho^{a}(-\mathbf{k})\,\mathbf{k}^{l}}{32\pi^{7/2}\xi\mathbf{k}^{2}\sqrt{k^{+}}}\left(\frac{2(1-\xi^{2})+\alpha(\xi-2)}{\xi}\right.\nonumber\\
&&-\frac{\alpha(1+\xi)}{1-\xi}+\left[\frac{\alpha^{3}(\xi-2)+2\alpha^{2}(1-\xi^{2})}{\xi}-\frac{\alpha(1-\alpha)^{2}(1+\xi)}{1-\xi}+(2\xi-1)\alpha\right]\frac{1}{\alpha x(1-\xi)}\nonumber\\
&&-\left[\frac{2(1-\xi^{2})+2\alpha(\xi-2)}{\xi}+\frac{(1-\alpha)(1+\xi)-\alpha(1+\xi)}{1-\xi}\right]\\
&&\left.\times\left[-\frac{2}{\epsilon}+\ln\left(\frac{\alpha x(1-\xi)\mathbf{k}^{2}}{4\pi e^{-\gamma}\mu^{2}}\right)\right]\right)\left|g_{l}^{a}(k)\right\rangle .\nonumber\end{eqnarray}
After integration over $x$:
\begin{equation}\begin{split}\label{6rhof}
&\left|\psi_{g\:\rho}^{6}\right\rangle \,=\,-\int_{\Lambda}^{e^{\delta\mathsf{Y}}\Lambda}dk^{+}\,\int d^{2}\mathbf{k}\,\int_{\frac{\Lambda}{k^{+}}}^{1-\frac{\Lambda}{k^{+}}}d\xi\:\frac{g^{3}N_{c}\rho^{a}(-\mathbf{k})\mathbf{k}^{l}}{32\pi^{7/2}\xi(1-\xi)\mathbf{k}^{2}\sqrt{k^{+}}}\\
&\times\left((\xi-2)\ln\left(\xi\right)-(1+\xi)\ln\left(1-\xi\right)-(1+\xi)\left[-\frac{2}{\epsilon}+\ln\left(\frac{\mathbf{k}^{2}}{\mu_{\overline{MS}}^{2}}\right)\right]\right)\left|g_{l}^{a}(k)\right\rangle. \\
 \end{split}\end{equation}\\
 The result which is obtained after integration over $\xi$, using equations (\ref{int.9a}) and (\ref{int.10}), is:
 \begin{equation}\begin{split}
\left|\psi_{g\:\rho}^{6}\right\rangle &\equiv\,-\int_{\Lambda}^{e^{\delta\mathsf{Y}}\Lambda}dk^{+}\,\int d^{2}\mathbf{k}\,\frac{g^{3}N_{c}\rho^{a}(-\mathbf{k})\mathbf{k}^{i}}{32\pi^{7/2}\mathbf{k}^{2}\sqrt{k^{+}}}\\
&\times\left(3\ln\left(\frac{\Lambda}{k^{+}}\right)\left[-\frac{2}{\epsilon}+\ln\left(\frac{\mathbf{k}^{2}}{\mu_{\overline{MS}}^{2}}\right)\right]+\frac{\pi^{2}}{3}+2\ln^{2}\left(\frac{\Lambda}{k^{+}}\right)\right)\left|g_{i}^{a}(k)\right\rangle .
\end{split}\end{equation}\\
 \\
  $\bullet$ \textit{\textbf{Two $\rho$ part}}\\
  This part can be deduced directly from (\ref{stasixsim}):
     \begin{eqnarray}\label{psipsila}
  &&\left|\psi_{g\:\rho\rho}^{6}\right\rangle\\
  &&=-\int_{\Lambda}^{e^{\delta\mathsf{Y}}\Lambda}dk^{+}\,\int d^{2}\mathbf{k}\, d^{2}\mathbf{p}\,\int_{\frac{\Lambda}{k^{+}}}^{1-\frac{\Lambda}{k^{+}}}d\xi\,\frac{ig^{3}f^{bad}\left\{ \rho^{b}(-\mathbf{p}),\,\,\rho^{a}(-\mathbf{k}+\mathbf{p})\right\} }{32\pi^{9/2}\xi(\mathbf{k}-\mathbf{p})^{2}\left((1-\xi)\mathbf{p}^{2}+\xi(\mathbf{k}-\mathbf{p})^{2}\right)\mathbf{k}^{2}\sqrt{k^{+}}}\nonumber\\
 &&\times\left[\left(-2(1-\xi)\mathbf{k}\cdot \mathbf{p}+\frac{2(1-\xi^{2})}{\xi}\mathbf{p}^{2}\right)\mathbf{k}^{l}+\left(\mathbf{k}^{2}-\frac{1+\xi}{1-\xi}(\mathbf{k}-\mathbf{p})^{2}+\frac{\xi-2}{\xi}\mathbf{p}^{2}\right)\mathbf{p}^{l}\right]\left|g^{d}_{l}(k)\right\rangle.\nonumber \end{eqnarray}
 By adding (\ref{psipsila}) with the result which is obtained after changing $p\rightarrow k-p$ and dividing by 2, we can write $\left|\psi_{g\:\rho\rho}^{6}\right\rangle$ equivalently as:
       \begin{eqnarray}
&&\left|\psi_{g\:\rho\rho}^{6}\right\rangle=-\int_{\Lambda}^{e^{\delta\mathsf{Y}}\Lambda}dk^{+}\,\int d^{2}\mathbf{k}\, d^{2}\mathbf{p}\,\int_{\frac{\Lambda}{k^{+}}}^{1-\frac{\Lambda}{k^{+}}}d\xi\,\frac{ig^{3}f^{bad}\left\{ \rho^{b}(-\mathbf{p}),\,\rho^{a}(-\mathbf{k}+\mathbf{p})\right\} }{64\pi^{9/2}\xi(1-\xi)\mathbf{p}^{2}(\mathbf{k}-\mathbf{p})^{2}\mathbf{k}^{2}\sqrt{k^{+}}}\\
&&\times\left[\left(-2(1-\xi)\mathbf{k}\cdot \mathbf{p}+\frac{2(1-\xi^{2})}{\xi}\mathbf{p}^{2}\right)\mathbf{k}^{l}+\left(\mathbf{k}^{2}-\frac{1+\xi}{1-\xi}(\mathbf{k}-\mathbf{p})^{2}+\frac{\xi-2}{\xi}\mathbf{p}^{2}\right)\mathbf{p}^{l}\right]\left|g^{d}_{l}(k)\right\rangle. \nonumber \end{eqnarray}
After integration over $\xi$ with the aid of (\ref{int.10}) and (\ref{int.11ab}), the last result becomes:
     \begin{equation}\begin{split}\label{psipsilb}
&\left|\psi_{g\:\rho\rho}^{6}\right\rangle=-\int_{\Lambda}^{e^{\delta\mathsf{Y}}\Lambda}dk^{+}\,\int d^{2}\mathbf{k}\, d^{2}\mathbf{p}\,\frac{ig^{3}f^{bad}\left\{ \rho^{b}(-\mathbf{p}),\,\rho^{a}(-\mathbf{k}+\mathbf{p})\right\} }{32\pi^{9/2}\mathbf{p}^{2}(\mathbf{k}-\mathbf{p})^{2}\mathbf{k}^{2}\sqrt{k^{+}}}\\
&\times\left[\ln\left(\frac{\Lambda}{k^{+}}\right)\mathbf{k}\cdot \mathbf{p}\mathbf{k}^{l}+\left(\frac{k^{+}}{\Lambda}-1-\ln\left(\frac{\Lambda}{k^{+}}\right)\right)\mathbf{p}^{2}\left(\mathbf{k}^{l}-\mathbf{p}^{l}\right)\right.\\
&\left.-\left(\ln\left(\frac{\Lambda}{k^{+}}\right)\mathbf{k}^{2}+\left(\frac{k^{+}}{\Lambda}-1-\ln\left(\frac{\Lambda}{k^{+}}\right)\right)(\mathbf{k}-\mathbf{p})^{2}\right)\mathbf{p}^{l}\right]\left|g_{l}^{d}(k)\right\rangle. 
   \end{split}\end{equation}
  Notice that under the change $p\,\rightarrow\, k\,-\, p$ the second summand inside the rectangled brackets becomes exactly the same as the fourth summand. We write the result which is obtained after this change in (\ref{sixtw}).
  
    \subsection{Supplement for Computation of $\left|\psi_{g}^{7}\right\rangle $} \label{swf9}
$\left|\psi_{g}^{7}\right\rangle $ is defined in (\ref{gseven}). After inserting the matrix elements, (\ref{g}) and (\ref{g2}), we arrive at:
\begin{eqnarray}\label{stseve}
&&\left|\psi_{g}^{7}\right\rangle =-\frac{1}{2}\int_{\Lambda}^{e^{\delta\mathsf{Y}}\Lambda}dk^{+}\, dp^{+}\, dq^{+}\, dr^{+}\,\int d^{2}\mathbf{k}\, d^{2}\mathbf{p}\, d^{2}\mathbf{q}\, d^{2}\mathbf{r}\frac{1}{\left(\frac{\mathbf{q}^{2}}{2q^{+}}\right)\left(\frac{\mathbf{r}^{2}}{2r^{+}}+\frac{\mathbf{p}^{2}}{2p^{+}}\right)\left(\frac{\mathbf{k}^{2}}{2k^{+}}\right)}\nonumber\\
&&\times\left(\delta^{dc}\delta_{kl}\delta^{(3)}(q-k)\,\frac{g\mathbf{r}^{j}\rho^{b}(\mathbf{r})}{4\pi^{3/2}|r^{+}|^{3/2}}+\delta^{db}\delta_{jl}\delta^{(3)}(k-r)\,\frac{g\mathbf{q}^{k}\rho^{c}(\mathbf{q})}{4\pi^{3/2}|q^{+}|^{3/2}}\right)\nonumber\\
&&\times\left(\delta^{ab}\delta_{ij}\delta^{(3)}(r-p)\,\frac{g\mathbf{q}^{k}\rho^{c}(-\mathbf{q})}{4\pi^{3/2}|q^{+}|^{3/2}}+\delta^{ac}\delta_{ik}\delta^{(3)}(p-q)\,\frac{g\mathbf{r}^{j}\rho^{b}(-\mathbf{r})}{4\pi^{3/2}|r^{+}|^{3/2}}\right)\\
&&\times\left(\frac{g\rho^{a}(-\mathbf{p})\mathbf{p}^{i}}{4\pi^{3/2}|p^{+}|^{3/2}}\right)\left|g_{l}^{d}(k)\right\rangle .\nonumber\end{eqnarray}
After simplification of the last expression we arrive at:
\begin{equation}\begin{split}
&\left|\psi_{g}^{7}\right\rangle  =-\int_{\Lambda}^{e^{\delta\mathsf{Y}}\Lambda}dk^{+}\, dp^{+}\,\int d^{2}\mathbf{k}\, d^{2}\mathbf{p}\,\frac{g^{3}\mathbf{p}^{2}\mathbf{k}^{i}\rho^{b}(\mathbf{p})}{64\pi^{9/2}\left(\frac{\mathbf{k}^{2}}{2k^{+}}\right)\left(\frac{\mathbf{k}^{2}}{2k^{+}}+\frac{\mathbf{p}^{2}}{2p^{+}}\right)|p^{+}|^{3}|k^{+}|^{3/2}}\\
&\times\left(\frac{\rho^{b}(-\mathbf{p})\,\rho^{a}(-\mathbf{k})}{\frac{\mathbf{k}^{2}}{2k^{+}}}\,+\,\frac{\rho^{a}(-\mathbf{k})\,\rho^{b}(-\mathbf{p})}{\frac{\mathbf{p}^{2}}{2p^{+}}}\right)\left|g_{i}^{a}(k)\right\rangle. \\
\end{split}\end{equation}
By using the algebra of $\rho$ operators, as in (\ref{rhoalge}), we can write the last expression as:

\begin{equation}\begin{split}\label{fkd}
&\left|\psi_{g}^{7}\right\rangle  =-\int_{\Lambda}^{e^{\delta\mathsf{Y}}\Lambda}dk^{+}\, dp^{+}\,\int d^{2}\mathbf{k}\, d^{2}\mathbf{p}\,\frac{2g^{3}\mathbf{k}^{i}\rho^{b}(\mathbf{p})}{64\pi^{9/2}\left(\frac{\mathbf{k}^{2}}{2k^{+}}\right)|p^{+}|^{2}|k^{+}|^{3/2}}\\
&\times\left(\frac{\rho^{b}(-\mathbf{p})\,\rho^{a}(-\mathbf{k})}{\frac{\mathbf{k}^{2}}{2k^{+}}}+\frac{if^{abc}\rho^{c}(-\mathbf{k}-\mathbf{p})}{\frac{\mathbf{k}^{2}}{2k^{+}}+\frac{\mathbf{p}^{2}}{2p^{+}}}\right)\left|g_{i}^{a}(k)\right\rangle. \\
\end{split}\end{equation}
  which we can also write after changing variables as in (\ref{chanvar}) as:
  
\begin{equation}\begin{split}
\left|\psi_{g}^{7}\right\rangle &=-\int_{\Lambda}^{e^{\delta\mathsf{Y}}\Lambda}dk^{+}\,\int d^{2}\mathbf{k}\, d^{2}\mathbf{p}\,\int_{\frac{\Lambda}{k^{+}}}^{\frac{e^{\delta\mathsf{Y}}\Lambda}{k^{+}}}d\xi\,\frac{g^{3}\mathbf{k}^{i}\rho^{b}(\mathbf{p})}{8\pi^{9/2}\sqrt{k^{+}}\xi\mathbf{k}^{2}}\\
&\times\left(\frac{\rho^{b}(-\mathbf{p})\rho^{a}(-\mathbf{k})}{\xi\mathbf{k}^{2}}+\frac{if^{abc}\rho^{c}(-\mathbf{k}-\mathbf{p})}{\xi\mathbf{k}^{2}+\mathbf{p}^{2}}\right)\left|g_{i}^{a}(k)\right\rangle ,\\
\end{split}\end{equation}
 We can now isolate the contribution to one $\rho$ via the prescription in (\ref{decompose}).\\\\
  $\bullet$ \textit{\textbf{One $\rho$ part}}\\
Based on (\ref{fkd}), we see that the term with one $\rho$ operator reads:
  \begin{equation}\begin{split}\label{7oner1}
  &\left|\psi_{g\:\rho}^{7}\right\rangle \,=\,\mu^{2\epsilon}\int_{\Lambda}^{e^{\delta\mathsf{Y}}\Lambda}dk^{+}\,\int d^{d}\mathbf{k}\, d^{d}\mathbf{p}\,\int_{\frac{\Lambda}{k^{+}}}^{\frac{e^{\delta\mathsf{Y}}\Lambda}{k^{+}}}d\xi\,\frac{g^{3}\, N_{c}\,\rho^{a}(-\mathbf{k})\, \mathbf{k}^{i}}{16\pi^{9/2}\sqrt{k^{+}}\xi \mathbf{k}^{2}\left(\xi \mathbf{k}^{2}\,+\, \mathbf{p}^{2}\right)}\left|g_{i}^{a}(k)\right\rangle. \\
   \end{split}\end{equation}
   Integration over $\mathbf{p}$ by using (\ref{dim1}) yields:
    \begin{equation}\begin{split}\label{7oner2}
  &\left|\psi_{g\:\rho}^{7}\right\rangle \,=\,\mu^{2\epsilon}\int_{\Lambda}^{e^{\delta\mathsf{Y}}\Lambda}dk^{+}\,\int\frac{d^{d}\mathbf{k}}{(2\pi)^{d}}\,\int_{\frac{\Lambda}{k^{+}}}^{\frac{e^{\delta\mathsf{Y}}\Lambda}{k^{+}}}d\xi\,\\
&\times\frac{g^{3}N_{c}\rho^{a}(-\mathbf{k})\mathbf{k}^{i}}{\sqrt{\pi k^{+}}\xi\mathbf{k}^{2}}\frac{1}{(4\pi)^{d/2}}\Gamma\left(1-\frac{d}{2}\right)\left(\frac{1}{\xi\mathbf{k}^{2}}\right)^{1-\frac{d}{2}}\left|g_{i}^{a}(k)\right\rangle .\\
\end{split}\end{equation}
  Expanding with the aid of (\ref{gamexp}) taking the $\epsilon\,\rightarrow\,0$ limit and, the last result becomes:
    \begin{equation}\label{7oner3}
 \left|\psi_{g\:\rho}^{7}\right\rangle =-\int_{\Lambda}^{e^{\delta\mathsf{Y}}\Lambda}dk^{+}\,\int d^{2}\mathbf{k}\,\int_{\frac{\Lambda}{k^{+}}}^{\frac{e^{\delta\mathsf{Y}}\Lambda}{k^{+}}}d\xi\,\frac{g^{3}N_{c}\rho^{a}(-\mathbf{k})\mathbf{k}^{i}}{16\pi^{7/2}\sqrt{k^{+}}\xi \mathbf{k}^{2}}\left[-\frac{2}{\epsilon}+\ln\left(\frac{\xi \mathbf{k}^{2}}{\mu_{\overline{MS}}^{2}}\right)\right]\left|g_{i}^{a}(k)\right\rangle .
   \end{equation}
After integration over $\xi\,\equiv\,\frac{p^{+}}{k^{+}}$ according to (\ref{int.7}):
 \begin{equation}\begin{split}
\left|\psi_{g\:\rho}^{7}\right\rangle &\equiv\,-\int_{\Lambda}^{e^{\delta\mathsf{Y}}\Lambda}dk^{+}\,\int d^{2}\mathbf{k}\,\frac{g^{3}N_{c}\rho^{a}(-\mathbf{k})\mathbf{k}^{i}}{32\pi^{7/2}\sqrt{k^{+}}\mathbf{k}^{2}}\\
&\times\left(2\delta\mathsf{Y}\left[-\frac{2}{\epsilon}+\ln\left(\frac{\mathbf{k}^{2}}{\mu_{\overline{MS}}^{2}}\right)\right]+\ln^{2}\left(\frac{\Lambda e^{\delta\mathsf{Y}}}{k^{+}}\right)-\ln^{2}\left(\frac{\Lambda}{k^{+}}\right)\right)\left|g_{i}^{a}(k)\right\rangle .
  \end{split}\end{equation}\\
\section{Supplement for Section 4}\label{supsect4}

\subsection{Supplement for Computation of $\Sigma_{q\bar{q}}$} \label{sfk1}
Starting from equation (\ref{starqq}), and introducing new variables according to (\ref{chanvar}) and (\ref{chanvar2}), we obtain:
 \begin{eqnarray}\label{siqq1}
&&\Sigma_{q\bar{q}}^{NLO}=\int_{\mathbf{x},\mathbf{y},\mathbf{z},\mathbf{z}^{\prime}}\int_{\Lambda}^{e^{\delta\mathsf{Y}}\Lambda}\frac{dk^{+}}{k^{+}}\,\int_{0}^{1}d\xi\,\int d^{2}\mathbf{k}\, d^{2}\widetilde{\mathbf{p}}\, d^{2}\mathbf{u}\, d^{2}\widetilde{\mathbf{v}}\,\\
&&\times\frac{g^{4}N_{f}\, J_{L}^{a}(\mathbf{x})\, Tr\left[S^{\dagger}(\mathbf{z})t^{a}S(\mathbf{z}^{\prime})t^{b}\right]\, J_{R}^{b}(\mathbf{y})}{512\pi^{10}\left(\xi(1-\xi)\mathbf{k}^{2}+\widetilde{\mathbf{p}}^{2}\right)\left(\xi(1-\xi)\mathbf{u}^{2}+\widetilde{\mathbf{v}}^{2}\right)}\left(\frac{(1-2\xi)^{2}\widetilde{\mathbf{p}}^{i}\widetilde{\mathbf{v}}^{j}+\delta^{ij}\widetilde{\mathbf{p}}\cdot\widetilde{\mathbf{v}}-\widetilde{\mathbf{p}}^{j}\widetilde{\mathbf{v}}^{i}}{\mathbf{k}^{2}\mathbf{u}^{2}}\mathbf{k}^{i}\mathbf{u}^{j}\right.\nonumber\\
&&\left.+2\xi(1-\xi)(1-2\xi)\left(\frac{\mathbf{k}\cdot\widetilde{\mathbf{p}}}{\mathbf{k}^{2}}+\frac{\mathbf{u}\cdot\widetilde{\mathbf{v}}}{\mathbf{u}^{2}}\right)+4\xi^{2}(1-\xi)^{2}\right)e^{-i\widetilde{\mathbf{v}}\cdot Z+i\mathbf{u}\cdot\left(Y^{\prime}-\xi Z\right)+i\widetilde{\mathbf{p}}\cdot Z-i\mathbf{k}\cdot\left(X^{\prime}-\xi Z\right)}.\nonumber\end{eqnarray}
We can now perform Fourier transformation by using (\ref{fourier.3}) $-$ (\ref{fourier.4}). It is possible to simplify the denominators using:
\begin{equation}\begin{split}
\xi(1-\xi)Z^{2}+(X^{\prime}-\xi Z)^{2}=(1-\xi)(X^{\prime})^{2}+\xi X^{2}.
 \end{split}\end{equation}
Then, we arrive at:
 \begin{equation}\begin{split}\label{siqq2}
 &\Sigma_{q\bar{q}}^{NLO}\,=\,-\,\delta\mathsf{Y}\int_{\mathbf{x},\mathbf{y},\mathbf{z},\mathbf{z}^{\prime}}\int_{0}^{1}d\xi\,\frac{g^{4}N_{f}\, J_{L}^{a}(\mathbf{x})\, Tr\left[S^{\dagger}(\mathbf{z})t^{a}S(\mathbf{z}^{\prime})t^{b}\right]\, J_{R}^{b}(\mathbf{y})}{32\pi^{6}Z^{4}\left((1-\xi)(X^{\prime})^{2}+\xi X^{2}\right)\left((1-\xi)(Y^{\prime})^{2}+\xi Y^{2}\right)}\\
&\times\left(4\xi(1-\xi)\left(X^{\prime}\cdot Z-\xi Z^{2}\right)\,\left(Y^{\prime}\cdot Z-\xi Z^{2}\right)-\left(X^{\prime}-\xi Z\right)\cdot\left(Y^{\prime}-\xi Z\right)Z^{2}\right.\\
&\left.-2\xi(1-\xi)(1-2\xi)\left(X^{\prime}\cdot Z+Y^{\prime}\cdot Z-2\xi Z^{2}\right)Z^{2}-4\xi^{2}(1-\xi)^{2}Z^{4}\right).\\
\end{split}\end{equation}
Rewriting the scalar products:
  \begin{equation}\begin{split}\label{scalar}
&X^{\prime}\cdot Z=\frac{1}{2}\left((X^{\prime})^{2}-X^{2}+Z^{2}\right),\qquad Y^{\prime}\cdot Z=\frac{1}{2}\left((Y^{\prime})^{2}-Y^{2}+Z^{2}\right),\\
&X^{\prime}\cdot Y^{\prime}=\frac{1}{2}\left((X^{\prime})^{2}+(Y^{\prime})^{2}-(X-Y)^{2}\right),
 \end{split}\end{equation}
we arrive at:
  \begin{eqnarray}\label{siqq3}
&&\Sigma_{q\bar{q}}^{NLO}=-\delta\mathsf{Y}\int_{\mathbf{x},\mathbf{y},\mathbf{z},\mathbf{z}^{\prime}}\int_{0}^{1}d\xi\,\frac{g^{4}N_{f}\, J_{L}^{a}(\mathbf{x})\, Tr\left[S^{\dagger}(\mathbf{z})t^{a}S(\mathbf{z}^{\prime})t^{b}\right]\, J_{R}^{b}(\mathbf{y})}{64\pi^{6}Z^{4}\left((1-\xi)(X^{\prime})^{2}+\xi X^{2}\right)\left((1-\xi)(Y^{\prime})^{2}+\xi Y^{2}\right)}\nonumber\\
&&\times\left(-\left[(1-\xi)(X^{\prime})^{2}+\xi X^{2}+(1-\xi)(Y^{\prime})^{2}+\xi Y^{2}-(X-Y)^{2}\right]Z^{2}\right.\\
&&\left.+2\xi(1-\xi)\left((X^{\prime})^{2}-X^{2}\right)\left((Y^{\prime})^{2}-Y^{2}\right)\right)\nonumber\end{eqnarray}
It is possible to integrate over $\xi$ by using the integrals (\ref{int.2}), (\ref{int.3}), and (\ref{int.4}), the result is shown in (\ref{sigqq}).

\subsection{Supplement for Computation of $\Sigma_{JSSJ}$} \label{sfk3}
Starting from equation (\ref{starjssj}), integrating over $\xi$, and separating the result according to (\ref{ajssjsig}), we get:
 \begin{equation}\begin{split}\label{kofd}
&\Sigma_{JSSJ}^{NLO}=\frac{g^{4}}{32\pi^{6}}\int_{\mathbf{x},\, \mathbf{y},\, \mathbf{z},\, \mathbf{z}^{\prime}}\int_{\Lambda}^{e^{\delta\mathsf{Y}}\Lambda}\frac{dk^{+}}{k^{+}}\, f^{abc}f^{def}\, J_{L}^{a}(\mathbf{x})\, S_{A}^{be}(\mathbf{z})\, S_{A}^{cf}(\mathbf{z}^{\prime})\, J_{R}^{d}(\mathbf{y})\\
&\times\left[-\frac{1}{2Z^{4}}+\frac{\varDelta(\mathbf{x},\, \mathbf{y},\, \mathbf{z},\, \mathbf{z}^{\prime})}{(X^{\prime})^{2}X^{2}((X^{\prime})^{2}-X^{2})^{2}((X^{\prime})^{2}Y^{2}-X^{2}(Y^{\prime})^{2})}\ln\left(\frac{X^{2}}{(X^{\prime})^{2}}\right)\right.\\
&\left.-\frac{\varDelta(\mathbf{y},\, \mathbf{x},\, \mathbf{z},\, \mathbf{z}^{\prime})}{(Y^{\prime})^{2}Y^{2}((Y^{\prime})^{2}-Y^{2})^{2}((X^{\prime})^{2}Y^{2}-X^{2}(Y^{\prime})^{2})}\ln\left(\frac{Y^{2}}{(Y^{\prime})^{2}}\right)\right],\\
\end{split}\end{equation}
and
 \begin{equation}\begin{split}\label{kofdsq}
&\Sigma_{JSSJ}^{(\delta\mathsf{Y})^{2}}=-\frac{g^{4}}{32\pi^{6}}\int_{\mathbf{x},\, \mathbf{y},\, \mathbf{z},\, \mathbf{z}^{\prime}}\int_{\Lambda}^{e^{\delta\mathsf{Y}}\Lambda}\frac{dk^{+}}{k^{+}}\, f^{abc}f^{def}\, J_{L}^{a}(\mathbf{x})\, S_{A}^{be}(\mathbf{z})\, S_{A}^{cf}(\mathbf{z}^{\prime})\, J_{R}^{d}(\mathbf{y})\\
&\times\left(\frac{tr\left(\Lambda(\mathbf{x},\, \mathbf{z},\, \mathbf{z}^{\prime})\,\Lambda^{T}(\mathbf{y},\, \mathbf{z},\, \mathbf{z}^{\prime})\right)}{(X^{\prime})^{2}(Y^{\prime})^{2}}+\frac{tr\left(\Lambda(\mathbf{x},\, \mathbf{z}^{\prime},\, \mathbf{z})\,\Lambda^{T}(\mathbf{y},\, \mathbf{z}^{\prime},\, \mathbf{z})\right)}{X^{2}Y^{2}}\right)\ln\left(\frac{\Lambda}{k^{+}}\right).\\
 \end{split}\end{equation}
The following definitions were used:
    \begin{equation}\begin{split}
 \Lambda^{jl}(\mathbf{x},\, \mathbf{z},\, \mathbf{z}^{\prime})=\frac{(X^{\prime})^{j}Z^{l}}{Z^{2}}+\frac{X^{l}(X^{\prime})^{j}}{2X^{2}}\;,\qquad\qquad\Theta^{jl}(\mathbf{x},\, \mathbf{z},\, \mathbf{z}^{\prime})=\frac{X^{2}-(X^{\prime})^{2}}{2Z^{2}}\delta^{jl}\;,
  \end{split}\end{equation}
  \begin{eqnarray}
&&\varDelta(\mathbf{x},\, \mathbf{y},\, \mathbf{z},\, \mathbf{z}^{\prime})\\
&&=\left(\left(\Lambda^{jl}(\mathbf{y},\, \mathbf{z},\, \mathbf{z}^{\prime})\, X^{2}+\Lambda^{lj}(\mathbf{y},\, \mathbf{z}^{\prime},\, \mathbf{z})\,(X^{\prime})^{2}\right)(X^{2}-(X^{\prime})^{2})-\Theta^{jl}(\mathbf{y},\, \mathbf{z},\, \mathbf{z}^{\prime})\, X^{2}(X^{\prime})^{2}\right)\nonumber\\
&&\times\left(\left(\Lambda^{jl}(\mathbf{x},\, \mathbf{z},\, \mathbf{z}^{\prime})\, X^{2}+\Lambda^{lj}(\mathbf{x},\, \mathbf{z}^{\prime},\, \mathbf{z})\,(X^{\prime})^{2}\right)(X^{2}-(X^{\prime})^{2})-\Theta^{jl}(\mathbf{x},\, \mathbf{z},\, \mathbf{z}^{\prime})\, X^{2}(X^{\prime})^{2}\right).\nonumber\end{eqnarray}
from which we deduce:
 \begin{eqnarray}\label{fsdoar}
&&\frac{\varDelta(\mathbf{x},\,\mathbf{y},\,\mathbf{z},\,\mathbf{z}^{\prime})}{X^{2}(X^{\prime})^{2}(X^{2}-(X^{\prime})^{2})^{2}}\,=\,-\frac{Y^{2}-(Y^{\prime})^{2}}{2Z^{2}(X^{2}-(X^{\prime})^{2})}\left(\frac{X^{2}\, X^{\prime}\cdot Z-(X^{\prime})^{2}\, X\cdot Z}{Z^{2}}+X\cdot X^{\prime}\right.\nonumber\\
&&\left.-\frac{X^{2}(X^{\prime})^{2}}{Z^{2}}\right)+\frac{1}{(X^{\prime})^{2}Z^{2}}\left(\frac{X^{2}\, Z^{2}\, X^{\prime}\cdot Y^{\prime}-\,(X^{\prime})^{2}\, Y^{\prime}\cdot Z\, X\cdot Z}{Z^{2}}+X^{\prime}\cdot Y^{\prime}\, X\cdot Z\right.\nonumber\\
&&\left.-Y^{\prime}\cdot Z\,\frac{X^{2}(X^{\prime})^{2}}{2Z^{2}}\right)-\frac{1}{X^{2}Z^{2}}\left(\frac{X^{2}\,(X^{\prime})\cdot Z\, Y\cdot Z-(X^{\prime})^{2}Z^{2}X\cdot Y}{Z^{2}}+X\cdot Y\, X^{\prime}\cdot Z\right)\nonumber\\
&&\left.-Y\cdot Z\,\frac{X^{2}(X^{\prime})^{2}}{2Z^{2}}\right)+\frac{1}{2}\left(\frac{1}{(X^{\prime})^{2}Y^{2}}+\frac{1}{X^{2}(Y^{\prime})^{2}}\right)\left(\frac{X^{2}\, X^{\prime}\cdot Y^{\prime}\, Y\cdot Z-(X^{\prime})^{2}\, X\cdot Y\, Y^{\prime}\cdot Z}{Z^{2}}\right.\nonumber\\
&&\left.+X\cdot Y\, X^{\prime}\cdot Y^{\prime}-\frac{X^{2}(X^{\prime})^{2}}{2Z^{2}}Y\cdot Y^{\prime}\right). \end{eqnarray}
We open the brackets and replace the scalar products according to  (\ref{scalar}) and
     \begin{equation}\begin{split}
&X\cdot Z=\frac{1}{2}\left((X^{\prime})^{2}-X^{2}-Z^{2}\right)\;;\quad\quad Y\cdot Z=\frac{1}{2}\left((Y^{\prime})^{2}-Y^{2}-Z^{2}\right)\;;\\
&X\cdot X^{\prime}=\frac{1}{2}\left(X^{2}+(X^{\prime})^{2}-Z^{2}\right).\\
\end{split}\end{equation}
After these simplifications, the expression (\ref{fsdoar}) can be brought to the form:
  \begin{eqnarray}\label{asdfio}
&&\frac{\varDelta(\mathbf{x},\,\mathbf{y},\,\mathbf{z},\,\mathbf{z}^{\prime})}{X^{2}(X^{\prime})^{2}(X^{2}-(X^{\prime})^{2})^{2}}\,\nonumber\\
&&=\,\frac{1}{8}\left[\left(X^{2}(Y^{\prime})^{2}-(X^{\prime})^{2}Y^{2}\right)\left(2\frac{X^{2}(Y^{\prime})^{2}+(X^{\prime})^{2}Y^{2}-4(X-Y)^{2}Z^{2}}{Z^{4}(X^{2}(Y^{\prime})^{2}-(X^{\prime})^{2}Y^{2})}\right.\right.\nonumber\\
&&+\frac{(X-Y)^{4}}{X^{2}(Y^{\prime})^{2}-(X^{\prime})^{2}Y^{2}}\left(\frac{1}{X^{2}(Y^{\prime})^{2}}+\frac{1}{Y^{2}(X^{\prime})^{2}}\right)+\frac{(X-Y)^{2}}{Z^{2}}\left(\frac{1}{X^{2}(Y^{\prime})^{2}}-\frac{1}{Y^{2}(X^{\prime})^{2}}\right)\nonumber\\
&&\left.-\frac{2}{(X^{2}-(X^{\prime})^{2})Z^{2}}\left(\frac{X^{2}+(X^{\prime})^{2}}{Z^{2}}+\frac{Z^{2}-X^{2}}{2(X^{\prime})^{2}}+\frac{Z^{2}-(X^{\prime})^{2}}{2X^{2}}-3\right)\right)-\frac{(Y^{\prime})^{2}}{Z^{2}}\nonumber\\
&&+\frac{(Y^{\prime})^{4}X^{2}}{(X^{\prime})^{2}Z^{2}Y^{2}}+\frac{Y^{4}(X^{\prime})^{2}}{Z^{2}X^{2}(Y^{\prime})^{2}}-\frac{Y^{2}}{Z^{2}}-\frac{(X^{\prime})^{2}Y^{2}}{Z^{2}(Y^{\prime})^{2}}+\frac{X^{2}}{Z^{2}}+\frac{(X^{\prime})^{2}}{Z^{2}}-\frac{X^{2}(Y^{\prime})^{2}}{Z^{2}Y^{2}}\nonumber\\
&&-\frac{(X^{\prime})^{2}(X-Y)^{2}}{X^{2}Z^{2}}+\frac{(Y^{\prime})^{2}(X-Y)^{2}}{Z^{2}Y^{2}}+\frac{Y^{2}(X-Y)^{2}}{Z^{2}(Y^{\prime})^{2}}-\frac{X^{2}(X-Y)^{2}}{(X^{\prime})^{2}Z^{2}}-\frac{Y^{2}(X-Y)^{2}}{X^{2}(Y^{\prime})^{2}}\nonumber\\
&&\left.+\frac{(X-Y)^{2}}{(X^{\prime})^{2}}+\frac{(X-Y)^{2}}{X^{2}}-\frac{(Y^{\prime})^{2}(X-Y)^{2}}{(X^{\prime})^{2}Y^{2}}\right].\end{eqnarray}
Comparing (\ref{asdfio}) with (\ref{JSSJ}) and noticing that $\widetilde{K}(\mathbf{x},\, \mathbf{y},\, \mathbf{z},\, \mathbf{z}^{\prime})$ can be expressed in the following way:
\begin{eqnarray}
&&\widetilde{K}(\mathbf{x},\,\mathbf{y},\,\mathbf{z},\,\mathbf{z}^{\prime})=\frac{\alpha_{s}^{2}}{16\pi^{4}\left((X^{\prime})^{2}Y^{2}-X^{2}(Y^{\prime})^{2}\right)}\left(\frac{(Y^{\prime})^{2}}{Z^{2}}-\frac{(Y^{\prime})^{4}X^{2}}{(X^{\prime})^{2}Z^{2}Y^{2}}-\frac{Y^{4}(X^{\prime})^{2}}{Z^{2}X^{2}(Y^{\prime})^{2}}\right.\nonumber\\
&&+\frac{Y^{2}}{Z^{2}}+\frac{(X^{\prime})^{2}Y^{2}}{Z^{2}(Y^{\prime})^{2}}-\frac{X^{2}}{Z^{2}}-\frac{(X^{\prime})^{2}}{Z^{2}}+\frac{X^{2}(Y^{\prime})^{2}}{Z^{2}Y^{2}}+\frac{(X^{\prime})^{2}(X-Y)^{2}}{X^{2}Z^{2}}-\frac{(Y^{\prime})^{2}(X-Y)^{2}}{Z^{2}Y^{2}}\nonumber\\
&&-\frac{Y^{2}(X-Y)^{2}}{Z^{2}(Y^{\prime})^{2}}+\frac{X^{2}(X-Y)^{2}}{(X^{\prime})^{2}Z^{2}}+\frac{Y^{2}(X-Y)^{2}}{X^{2}(Y^{\prime})^{2}}-\frac{(X-Y)^{2}}{(X^{\prime})^{2}}-\frac{(X-Y)^{2}}{X^{2}}\nonumber\\
&&\left.+\frac{(Y^{\prime})^{2}(X-Y)^{2}}{(X^{\prime})^{2}Y^{2}}\right)\ln\left(\frac{X^{2}}{(X^{\prime})^{2}}\right)+\left(\mathbf{x}\leftrightarrow\mathbf{y}\right),\end{eqnarray}
we arrive at the conclusion that:
\begin{eqnarray}\label{fdasop}
&&\frac{g^{4}}{32\pi^{6}}\left[-\frac{1}{2Z^{4}}+\frac{\varDelta(\mathbf{x},\, \mathbf{y},\, \mathbf{z},\, \mathbf{z}^{\prime})}{(X^{\prime})^{2}X^{2}((X^{\prime})^{2}-X^{2})^{2}((X^{\prime})^{2}Y^{2}-X^{2}(Y^{\prime})^{2})}\ln\left(\frac{X^{2}}{(X^{\prime})^{2}}\right)\right.\\
&&\left.-\frac{\varDelta(\mathbf{y},\, \mathbf{x},\, \mathbf{z},\, \mathbf{z}^{\prime})}{(Y^{\prime})^{2}Y^{2}((Y^{\prime})^{2}-Y^{2})^{2}((X^{\prime})^{2}Y^{2}-X^{2}(Y^{\prime})^{2})}\ln\left(\frac{Y^{2}}{(Y^{\prime})^{2}}\right)\right]\,=\,-K_{JSSJ}(\mathbf{x},\, \mathbf{y},\, \mathbf{z},\, \mathbf{z}^{\prime}).\nonumber\end{eqnarray}
 By inserting (\ref{fdasop}) to (\ref{kofd}), it is possible to write the result as in (\ref{ajssjsig}).

   \subsection{Supplement for Computation of $\Sigma_{JJSJ}$} \label{sfk5}
$\Sigma_{JJSJ}$ appears in (\ref{sojjsj}). Here we will present the explict form for $\Sigma_{JJSJ}^{NLO}$ and $\Sigma_{JJSJ}^{(\delta\mathsf{Y})^{2}}$ as introduced in (\ref{splitjjsj}). Let us start with $\Sigma_{JJSJ}^{NLO}$:
  \begin{equation}\begin{split}
&\Sigma_{JJSJ}^{NLO}\,\equiv\,-\frac{g^{4}f^{bde}}{128\pi^{7}}\delta\mathsf{Y}\int_{\mathbf{w},\,\mathbf{x},\,\mathbf{y},\,\mathbf{z}}\,\frac{W^{i}}{W^{2}}\left[J_{L}^{d}(\mathbf{x})\, J_{L}^{e}(\mathbf{y})\, S_{A}^{ba}(\mathbf{z})\, J_{R}^{a}(\mathbf{w})\right.\\
&\left.-J_{L}^{a}(\mathbf{w})\, S_{A}^{ab}(\mathbf{z})\, J_{R}^{d}(\mathbf{x})\, J_{R}^{e}(\mathbf{y})\right]\bigg[\int d^{2}\mathbf{k}\, d^{2}\mathbf{p}\, e^{-i\mathbf{k}\cdot X+i\mathbf{p}\cdot\left(X-Y\right)}\\
&\times\left(\frac{\mathbf{p}^{i}}{\mathbf{p}^{2}(\mathbf{k}-\mathbf{p})^{2}}-\frac{\mathbf{k}^{i}}{\mathbf{k}^{2}(\mathbf{k}-\mathbf{p})^{2}}+\frac{\mathbf{k}^{i}}{\mathbf{k}^{2}\mathbf{p}^{2}}\right)\ln\left(\frac{(\mathbf{k}-\mathbf{p})^{2}}{\mathbf{k}^{2}}\right)-\left(X\leftrightarrow Y\right)\bigg].\\
 \end{split}\end{equation}
After the change $p\,\rightarrow\, k\,-\, p$, we find:
   \begin{eqnarray}\label{rimta}
&&\Sigma_{JJSJ}^{NLO}\,=\,-\frac{g^{4}f^{bde}}{128\pi^{7}}\delta\mathsf{Y}\int_{\mathbf{w},\,\mathbf{x},\,\mathbf{y},\,\mathbf{z}}\,\frac{W^{i}}{W^{2}}\left[J_{L}^{d}(\mathbf{x})\,J_{L}^{e}(\mathbf{y})\,S_{A}^{ba}(\mathbf{z})\,J_{R}^{a}(\mathbf{w})\right.\nonumber\\
&&\left.-J_{L}^{a}(\mathbf{w})\,S_{A}^{ab}(\mathbf{z})\,J_{R}^{d}(\mathbf{x})\,J_{R}^{e}(\mathbf{y})\right]\bigg[\int d^{2}\mathbf{k}\,d^{2}\mathbf{p}\:e^{-i\mathbf{k}\cdot Y-i\mathbf{p}\cdot\left(X-Y\right)}\\
&&\left.\left(\frac{\mathbf{k}^{i}-\mathbf{p}^{i}}{\mathbf{p}^{2}(\mathbf{k}-\mathbf{p})^{2}}-\frac{\mathbf{k}^{i}}{\mathbf{k}^{2}\mathbf{p}^{2}}+\frac{\mathbf{k}^{i}}{\mathbf{k}^{2}(\mathbf{k}-\mathbf{p})^{2}}\right)\ln\left(\frac{\mathbf{p}^{2}}{\mathbf{k}^{2}}\right)-\left(X\leftrightarrow Y\right)\right].\nonumber\end{eqnarray}
  We can now use integral (\ref{ian}), and write:
   \begin{equation}\begin{split}
 &\Sigma_{JJSJ}^{NLO}\,=\,\frac{ig^{4}f^{bde}}{64\pi^{5}}\delta\mathsf{Y}\,\int_{\mathbf{w},\, \mathbf{x},\, \mathbf{y},\, \mathbf{z}}\,\frac{W^{i}}{W^{2}}\left(\frac{X^{i}}{X^{2}}-\frac{Y^{i}}{Y^{2}}\right)\ln\left(\frac{Y^{2}}{(X-Y)^{2}}\right)\ln\left(\frac{X^{2}}{(X-Y)^{2}}\right)\\
&\times\left[J_{L}^{d}(\mathbf{x})\, J_{L}^{e}(\mathbf{y})\, S_{A}^{ba}(\mathbf{z})\, J_{R}^{a}(\mathbf{w})-J_{L}^{a}(\mathbf{w})\, S_{A}^{ab}(\mathbf{z})\, J_{R}^{d}(\mathbf{x})\, J_{R}^{e}(\mathbf{y})\right].\\
\end{split}\end{equation}
In addition we have:
 \begin{eqnarray}\label{rimta2}
&&\Sigma_{JJSJ}^{(\delta\mathsf{Y})^{2}}\,\\
&&=\,\frac{g^{4}f^{bde}}{256\pi^{7}}(\delta\mathsf{Y})^{2}\int_{\mathbf{w},\, \mathbf{x},\, \mathbf{y},\, \mathbf{z}}\,\frac{W^{i}}{W^{2}}\left[J_{L}^{d}(\mathbf{x})\, J_{L}^{e}(\mathbf{y})\, S_{A}^{ba}(\mathbf{z})\, J_{R}^{a}(\mathbf{w})-J_{L}^{a}(\mathbf{w})\, S_{A}^{ab}(\mathbf{z})\, J_{R}^{d}(\mathbf{x})\, J_{R}^{e}(\mathbf{y})\right]\nonumber\\
&&\times\left[\int d^{2}\mathbf{k}\, d^{2}\mathbf{p}\, e^{-i\mathbf{k}\cdot X+i\mathbf{p}\cdot\left(X-Y\right)}\left(\frac{2\mathbf{k}^{i}}{\mathbf{k}^{2}\mathbf{p}^{2}}+\frac{\mathbf{p}\cdot(\mathbf{p}-\mathbf{k})\, \mathbf{k}^{i}}{\mathbf{k}^{2}\mathbf{p}^{2}(\mathbf{k}-\mathbf{p})^{2}}+\frac{2\mathbf{p}^{i}}{\mathbf{p}^{2}(\mathbf{k}-\mathbf{p})^{2}}\right)-\left(X\leftrightarrow Y\right)\right].\nonumber\end{eqnarray}
 The Fourier transforms can be found in (\ref{frfr1}), (\ref{frfr2}) and (\ref{frfr3}), then we arrive at:
  \begin{eqnarray}\label{krqjjsjn}
&&\Sigma_{JJSJ}^{(\delta\mathsf{Y})^{2}}=\frac{ig^{4}f^{bde}}{128\pi^{6}}(\delta\mathsf{Y})^{2}\int_{\mathbf{w},\,\mathbf{x},\,\mathbf{y},\,\mathbf{z},\,\mathbf{z}^{\prime}}\,\left[\left(\frac{2X^{\prime}\cdot Y^{\prime}\,Y\cdot W}{(X^{\prime})^{2}(Y^{\prime})^{2}Y^{2}W^{2}}+\frac{X^{\prime}\cdot Y^{\prime}\,Z\cdot W}{(X^{\prime})^{2}(Y^{\prime})^{2}Z^{2}W^{2}}\right.\right.\nonumber\\
&&\left.\left.+\frac{2X^{\prime}\cdot Z\,Y\cdot W}{(X^{\prime})^{2}Z^{2}Y^{2}W^{2}}\right)-(X\leftrightarrow Y)\right]\left[J_{L}^{d}(\mathbf{x})\,J_{L}^{e}(\mathbf{y})\,S_{A}^{ba}(\mathbf{z})\,J_{R}^{a}(\mathbf{w})\right.\\
&&\left.-J_{L}^{a}(\mathbf{w})\,S_{A}^{ab}(\mathbf{z})\,J_{R}^{d}(\mathbf{x})\,J_{R}^{e}(\mathbf{y})\right].\nonumber\end{eqnarray}

  \subsection{Supplement for Computation of $\Sigma_{JJSSJJ}$}\label{sijjjj}
From (\ref{starjjssjj}) we deduce that:
\begin{equation}\begin{split}
&\Sigma_{JJSSJJ}\,=\,\int_{\mathbf{w},\, \mathbf{v},\, \mathbf{x},\, \mathbf{y},\, \mathbf{z},\, \mathbf{z}^{\prime}}\int_{\Lambda}^{e^{\delta\mathsf{Y}}\Lambda}dk^{+}\,\int_{\frac{\Lambda}{k^{+}}}^{1-\frac{\Lambda}{k^{+}}}d\xi\,\frac{g^{4}}{128\pi^{6}\xi(1-\xi)}\\
&\times\left(\frac{X\cdot Y\, W^{\prime}\cdot V^{\prime}}{X^{2}Y^{2}(W^{\prime})^{2}(V^{\prime})^{2}}\, J_{L}^{a}(\mathbf{w})\, J_{L}^{b}(\mathbf{x})\, S_{A}^{ad}(\mathbf{z})\, S_{A}^{bc}(\mathbf{z}^{\prime})\, J_{R}^{d}(\mathbf{v})\, J_{R}^{c}(\mathbf{y})\right.\\
&+\frac{X\cdot Y\, W^{\prime}\cdot V^{\prime}}{X^{2}Y^{2}(W^{\prime})^{2}(V^{\prime})^{2}}\, J_{L}^{b}(\mathbf{x})\, J_{L}^{a}(\mathbf{w})\, S_{A}^{ad}(\mathbf{z})\, S_{A}^{bc}(\mathbf{z}^{\prime})\, J_{R}^{d}(\mathbf{v})\, J_{R}^{c}(\mathbf{y})\\
&+\frac{X\cdot Y\, W^{\prime}\cdot V^{\prime}}{X^{2}Y^{2}(W^{\prime})^{2}(V^{\prime})^{2}}\, J_{L}^{a}(\mathbf{w})\, J_{L}^{b}(\mathbf{x})\, S_{A}^{ad}(\mathbf{z})\, S_{A}^{bc}(\mathbf{z}^{\prime})\, J_{R}^{c}(\mathbf{y})\, J_{R}^{d}(\mathbf{v})\\
&\left.+\frac{X\cdot Y\, W^{\prime}\cdot V^{\prime}}{X^{2}Y^{2}(W^{\prime})^{2}(V^{\prime})^{2}}\, J_{L}^{b}(\mathbf{x})\, J_{L}^{a}(\mathbf{w})\, S_{A}^{ad}(\mathbf{z})\, S_{A}^{bc}(\mathbf{z}^{\prime})\, J_{R}^{c}(\mathbf{y})\, J_{R}^{d}(\mathbf{v})\right).\\
\end{split}\end{equation}
Integrating over $\xi$ and reordering the $J$ operators:
\begin{equation}\begin{split}
&\Sigma_{JJSSJJ}\,=\,\int_{\Lambda}^{e^{\delta\mathsf{Y}}\Lambda}dk^{+}\,\frac{g^{4}}{64\pi^{6}}\ln\left(\frac{k^{+}}{\Lambda}\right)\\
&\times\left(\int_{\mathbf{w},\, \mathbf{v},\, \mathbf{x},\, \mathbf{y},\, \mathbf{z},\, \mathbf{z}^{\prime}}\frac{4X\cdot Y\, W^{\prime}\cdot V^{\prime}}{X^{2}Y^{2}(W^{\prime})^{2}(V^{\prime})^{2}}\, J_{L}^{a}(\mathbf{w})\, J_{L}^{b}(\mathbf{x})\, S_{A}^{ad}(\mathbf{z})\, S_{A}^{bc}(\mathbf{z}^{\prime})\, J_{R}^{d}(\mathbf{v})\, J_{R}^{c}(\mathbf{y})\right.\\
&\left.-\int_{\mathbf{x},\, \mathbf{y},\, \mathbf{z},\, \mathbf{z}^{\prime}}\frac{X\cdot Y\, X^{\prime}\cdot Y^{\prime}}{X^{2}Y^{2}(X^{\prime})^{2}(Y^{\prime})^{2}}\, f^{bar}\, f^{cde}J_{L}^{r}(\mathbf{x})\, S_{A}^{bc}(\mathbf{z}^{\prime})\, S_{A}^{ad}(\mathbf{z})J_{R}^{e}(\mathbf{y})\right).\\
 \end{split}\end{equation}
The result after integration over $k^{+}$ is shown in equation (\ref{losjjssjj}).

 \end{document}